\documentclass[letterpaper,11pt]{article}
\usepackage[utf8]{inputenc}
\usepackage[T1]{fontenc}
\usepackage[english]{babel}
\usepackage{lmodern}
\usepackage{amsmath,amsfonts,amssymb,mathtools}
\usepackage{thmtools,thm-restate,bm}
\usepackage{amsthm}
\usepackage{microtype,booktabs,longtable,tabularx,multirow,array,xspace}
\usepackage[margin=1in]{geometry}
\usepackage[dvipsnames]{xcolor}
\usepackage{enumitem,float}
\usepackage{xurl}
\newcolumntype{P}[1]{>{\raggedright\arraybackslash}p{#1}}
\floatstyle{ruled}
\newfloat{algorithm}{tbp}{loa}
\floatname{algorithm}{Algorithm}
\usepackage[colorlinks=true,pdfpagemode=UseNone,urlcolor=RoyalBlue,linkcolor=RoyalBlue,citecolor=OliveGreen,pdfstartview=FitH]{hyperref}
\usepackage[nameinlink,capitalize,noabbrev]{cleveref}
\setlist{itemsep=3pt,topsep=5pt}
\newtheorem{theorem}{Theorem}[section]
\theoremstyle{definition}
\newtheorem{definition}[theorem]{Definition}
\theoremstyle{plain}
\newtheorem{lemma}[theorem]{Lemma}

\newtheorem{corollary}[theorem]{Corollary}
\newtheorem{proposition}[theorem]{Proposition}

\theoremstyle{remark}
\newtheorem{remark}[theorem]{Remark}
\theoremstyle{plain}

\newcommand{\bin}{\{0,1\}}

\newcommand{\F}{\mathsf{F}}

\newcommand{\E}{\mathbf{E}}
\renewcommand{\Pr}{\mathbf{Pr}}
\newcommand{\eps}{\varepsilon}
\newcommand{\poly}{\mathsf{poly}}
\newcommand{\polylog}{\mathrm{polylog}}
\newcommand{\Supp}{\mathsf{Supp}}
\newcommand{\supp}{\mathsf{Supp}}
\newcommand{\rank}{\mathsf{rank}}

\newcommand{\Span}{\mathsf{Span}}
\newcommand{\avgH}{\widetilde{H}_\infty}
\newcommand{\Hinf}{H_\infty}
\newcommand{\pref}{\operatorname{prefix}}
\newcommand{\wt}{\operatorname{wt}}

\newcommand{\Adv}{\mathsf{Adv}}

\newcommand{\Enc}{\mathsf{Enc}}
\newcommand{\Samp}{\mathsf{Samp}}
\newcommand{\ACB}{\mathsf{ACB}}
\newcommand{\SCB}{\mathsf{SCB}}

\newcommand{\anmExt}{\mathsf{anmExt}}

\newcommand{\head}{\mathsf{H}}

\providecommand{\Tr}{\operatorname{Tr}}

\numberwithin{equation}{section}
\hypersetup{bookmarksnumbered=true}
\hypersetup{pdfauthor={Xin Li and Yan Zhong}}

\title{Non-Malleable Affine Extractors with Small Error\\and Complexity Lower Bounds}
\author{Xin Li\thanks{Department of Computer Science, Johns Hopkins University. Emails: \href{mailto:lixints@cs.jhu.edu}{\texttt{lixints@cs.jhu.edu}},
\href{mailto:yzhong36@jhu.edu}{\texttt{yzhong36@jhu.edu}}.}
\and
Yan Zhong\footnotemark[1]}
\date{September 21, 2026}
\hypersetup{pdftitle={Non-Malleable Affine Extractors with Small Error and Complexity Lower Bounds}}
\begin{document}
\pagenumbering{gobble}
\hypersetup{pageanchor=false}
\maketitle
\enlargethispage{\baselineskip}
\begin{abstract}
A non-malleable affine extractor produces nearly uniform bits from an
affine source even given its output after affine tampering with no
fixed points. We construct such extractors
with linear output length and exponentially small error at every constant
entropy rate, against any fixed number $t$ of tamperings. Previously, explicit non-malleable affine extractors with such small error are only known to work for entropy rate $1-\gamma$ where $\gamma>0$ is some small constant (Li, FOCS '23). Our extractors can be pushed to work for entropy $C_{\eta,t}n/\log n$, with output length $n^{1-\eta}$,
and error $2^{-n^{1-\eta}}$, for every fixed $0<\eta<1$ and $t$. We apply our non-malleable affine extractors and related constructions to weakly read-once linear branching programs, non-oblivious decision trees, and resolution over parities.

\begin{itemize}
\item Our linear-entropy non-malleable affine extractors yield Boolean
functions with correlation $2^{-\Omega(n)}$ against weakly read-once
linear branching programs of size $2^{\Omega(n)}$. A similar argument
gives exponential hardness from the directional affine extractors of
Li--Zhong (CCC '24). 
\item For non-oblivious degree-$r$ decision trees
(Golovnev--Gurumukhani, ECCC '26), we obtain depth at least
$E/2^{r-2}-O_r(1)$ %
by using an affine extractor with entropy threshold below $n/2$, error
$2^{-E}$, and $(2^{r-2}-1)$-non-malleability under translations,
for fixed $r\ge2$ and $1<E\le n$.
Thus exponentially small error gives linear non-oblivious depth.
\item For non-oblivious $\ell$-local decision trees, Li's sumset
extractor (FOCS '23) gives depth
$\Omega_\delta((n/\ell)\log\ell)$ for growing $\ell\le n^{1-\delta}$,
where $0<\delta<1$ is fixed.
In the same range, directional affine extractors at fixed entropy rate
below $1/2$ and error $2^{-\Omega(n)}$ give correlation
$2^{-\Omega(n/\sqrt\ell)}$ against trees of depth at most
$c(n/\ell)\log\ell/\log\log\ell$, for a sufficiently small constant
$c>0$. %
\end{itemize}

Furthermore, our $t$-non-malleable affine extractors derandomize the lossless
lifting of Efremenko--Itsykson (STOC '26): for every fixed
$0<\xi<1$, a constant $t=t(\xi)$ gives explicit polynomial-size
unsatisfiable CNFs on $N$ variables requiring
$\operatorname{Res}(\oplus)$ size $2^{(1-\xi)N}$ at resolution depth
at most $N$. They also give overlapping gadgets preserving the
size--depth tradeoff of Itsykson--Podolskii--Shekhovtsov (CCC '26),
under unrelated local affine coordinate changes.

Separately, by reducing the number of variables through parity
substitutions (condensation), we construct polynomial-size CNFs
on $N$ variables with polynomial-size ordinary-resolution proofs and
$\operatorname{Res}(\oplus)$ tradeoff $d\log(2S)=\Omega(N^2)$.
This removes the $\log^2N$ loss of Itsykson--Podolskii--Shekhovtsov
(CCC '26) for such formulas without using extractors.
\end{abstract}

\clearpage
\begingroup
\tableofcontents
\endgroup
\clearpage

\pagenumbering{arabic}
\hypersetup{pageanchor=true}

\section{Introduction}\label{sec:introduction}

An affine extractor is a deterministic function that transforms a
distribution that is uniform on an unknown affine subspace of
sufficiently large dimension into a nearly uniform distribution.
Affine extractors have been extensively studied as a basic problem in
deterministic extraction from structured weak
sources~\cite{GR05,Bou07,Li11b,Li16,CGL21,Li23,LZdir}. The central questions
are how little entropy suffices for explicit extraction, how much
randomness can be extracted, and how small the error can be.

Affine extractors also have a longstanding connection to explicit lower
bounds. Even the weaker object of affine dispersers is useful:
a function that remains nonconstant on every sufficiently large affine
subspace cannot be simplified too quickly by affine restrictions.
Starting with Demenkov and Kulikov~\cite{DK11}, and continuing through
Find, Golovnev, Hirsch, and Kulikov~\cite{FGHK16} and Li and
Yang~\cite{LY22}, this approach established circuit size lower bounds
for affine dispersers, culminating in a $3.1n-o(n)$ lower bound on the
number of gates in general Boolean circuits.

Variants of affine extractors have also found applications in several other models, for example, linear branching programs, introduced by Gryaznov, Pudl\'ak, and Talebanfard~\cite{GPT22}; and decision trees with local or low-degree queries, introduced by Golovnev and Gurumukhani~\cite{GG26}. We now discuss these models below.

\paragraph{Linear branching programs.} Linear branching programs are branching programs where each internal node computes some linear function of the input. They generalize several well studied model such as decision trees, parity decision trees, and standard branching programs. They are also closely related to proof systems such as resolution
over parities. To study strong average case hardness in this model, Gryaznov, Pudl\'ak, and
Talebanfard~\cite{GPT22} introduced \emph{directional affine extractors}, which are affine extractors whose output remains close to uniform even conditioned on the output of the extractor on any nonzero affine shift. \cite{GPT22} shows that such directional affine extractors give strong average-case lower bounds for a special class of linear branching programs known as strongly read-once linear
branching programs, which require that the linear subspaces spanned by past and future queries at
each node to intersect only trivially. Subsequent works of Chattopadhyay and Liao~\cite{CL23}, and Li and
Zhong~\cite{LZdir} strengthened these bounds. 

\cite{GPT22} also defines another kind of read-once linear branching programs, known as weakly read-once linear branching programs. Unlike strongly read-once linear branching programs, the weakly read-once condition only
requires that at each node, the current query lies outside the span of all queries on
all paths leading to that node. It therefore permits overlap between
past and future query spaces, removing the separation available in the
strongly read-once model. Obtaining worst-case and average-case lower bounds for explicit Boolean functions
in this model was posed as an open problem by Gryaznov, Pudl\'ak, and
Talebanfard~\cite{GPT22} and reiterated by Li and Zhong~\cite{LZdir}.

Linear branching programs also model the problem of clause search for unsatisfiable Boolean
formulas: given a Boolean formula and an unsatisfied assignment, find a falsified clause. Gryaznov,
Pudl\'ak, and Talebanfard~\cite{GPT22} connected weakly read-once linear
branching programs for this search problem to a regularity restriction
of resolution over parities. Efremenko, Garl\'ik, and
Itsykson~\cite{EGI24} obtained exponential clause-search lower bounds
for strongly read-once linear branching programs, and Alekseev and
Itsykson~\cite{AI25} subsequently established exponential bounds for
weakly read-once linear branching programs. However these search lower bounds
do not directly establish hardness of explicit Boolean functions for the corresponding linear branching programs.

\paragraph{Decision trees with local or low-degree queries.} Introduced by Golovnev and Gurumukhani~\cite{GG26}, decision trees with local or low-degree queries generalize standard decision trees by allowing nonlinear queries
of the input. A non-oblivious (or adaptive) $\ell$-local decision tree queries a
Boolean function of at most $\ell$ input coordinates at each node;
both the function and the coordinates may depend on previous answers.
A non-oblivious degree-$r$ decision tree instead queries polynomials of
degree at most $r$ over $\F_2$. The oblivious versions require
all nodes at the same depth to query the same function~\cite{GG26}.

These models are also motivated by circuit lower bounds. Sufficiently
strong lower bounds for non-oblivious local decision trees would imply
improved lower bounds for unrestricted circuits and superlinear lower
bounds for circuits of logarithmic depth~\cite{GG26}. For degree-$2$
decision trees, nearly maximal non-oblivious depth is a necessary property
of dispersers for large degree-$2$ varieties, whose explicit
construction would improve unrestricted circuit lower bounds.

Golovnev and Gurumukhani~\cite{GG26} obtained nearly maximal lower
bounds in the oblivious setting, using sumset dispersers and condensers (which are dispersers and condensers that work for the sum of two independent weak sources)
for local decision trees and directional affine dispersers for
degree-$2$ decision trees. However, the important question of getting non-trivial lower bounds in the non-oblivious model is not addressed in ~\cite{GG26} and specifically raised as open questions.

In this paper we study a stronger version of directional affine extractors, known as \emph{non-malleable affine extractors}. Introduced by
Chattopadhyay and Li~\cite{CL17} to construct non-malleable codes against affine tampering, these are affine extractors whose output remains close to uniform even conditioned on the output of the extractor on any non-trivial affine transformation. Thus the directional affine extractor is a special case of non-malleable affine extractor. Previously, the best known construction of such non-malleable affine extractors work for entropy $(1-\gamma)n$ for some constant $\gamma>0$ with error $2^{-\Omega(n)}$ where $n$ is the length of the input \cite{Li23}, or for entropy $\polylog(n)$ but with error $1/\poly(n)$ \cite{LZ24}. Here, we give improved constructions of such extractors with exponentially small error, and show that such extractors and related objects can be used to give improved hardness results in the above two models. Specifically, we give strong average case hardness results for weakly read once linear branching programs, and non-trivial lower bounds for  non-oblivious decision trees with local queries and low degree queries.

We further use our non-malleable affine extractors to obtain lower bounds in the proof system of bounded-depth resolution over parities.
A lifting construction replaces each variable of a base formula by a
function of new Boolean variables, called a gadget, and transfers
suitable lower bounds for the base formula to the resulting formula.
Efremenko and
Itsykson~\cite{EI26} obtained near-maximal size lower bounds for
bounded-depth resolution over parities by lossless lifting. However, their lifting uses a
randomly chosen family of gadgets. Derandomizing this step is an important open question, and we use our non-malleable affine extractors to provide such a derandomization. %

\subsection{Our Results}\label{sec:our-results}

We identify $\{0,1\}^n$ with $\F_2^n$. The min-entropy of a random variable
$X$ is $\Hinf(X)=-\log\max_x\Pr[X=x]$; an $(n,k)$-source is a distribution
on $n$ bits with min-entropy at least $k$. All logarithms are base two.
We write $U_m$ for the uniform distribution on $m$ bits and
$X\approx_\eps Y$ when the statistical distance between $X$ and $Y$ is at
most $\eps$. In each displayed ideal distribution, the new uniform
variable is independent of the other coordinates.

A construction is \emph{explicit} if its function family is computable
by a deterministic polynomial-time algorithm. For a fixed entropy rate,
the polynomial and the constants in the output and error bounds may
depend on that rate. We give a single polynomial-time algorithm for the
sublinear-entropy construction as well. The tampering functions are
arbitrary within the stated family and need not be efficiently computable.

\subsubsection{Non-Malleable Affine Extractors}

An affine source is the uniform distribution on an affine subspace of
$\F_2^n$. Its min-entropy equals the dimension of that subspace. A map
$T$ has no fixed points if $T(x)\ne x$ for every input $x$.

\begin{definition}[Non-malleable affine extractor~\cite{CL17}]
\label{def:anmext}
A function $F:\F_2^n\to\F_2^m$ is a $(k,\eps)$ non-malleable affine
extractor if, for every affine source $X$ of dimension at least $k$ and
every affine map $T:\F_2^n\to\F_2^n$ with no fixed points,
\[
 (F(X),F(T(X)))\approx_\eps(U_m,F(T(X))).
\]
\end{definition}

Equivalently, one may require $T$ to have no fixed point only on
$\Supp(X)$; see \Cref{pre:affine-nm-equivalence}. We allow all affine
maps satisfying this condition, including noninvertible maps.
For $t\ge1$, \emph{$t$-non-malleability} requires the output to remain
nearly uniform given all $t$ tampered outputs together, following the
multiple-tampering terminology of~\cite{CRS14}. We give the formal
affine definition in \Cref{pre:t-affine-nm}.

\begin{theorem}[Non-malleable affine extractors for linear entropy]
\label{intro:linear}\label{thm:affine-linear}
For every constant $0<\rho\le1$, there are constants
$c_\rho,e_\rho>0$ and an explicit family
$\anmExt_n:\F_2^n\to\F_2^{m_n}$ with output length $m_n\ge c_\rho n$
and the following property for all sufficiently large $n$.
For every affine source $X$ with $\Hinf(X)\ge\rho n$ and every affine
map $T$ with $T(x)\ne x$ for all $x\in\Supp(X)$,
\[
 (\anmExt_n(X),\anmExt_n(T(X)))
 \approx_{2^{-e_\rho n}}(U_{m_n},\anmExt_n(T(X))).
\]
\end{theorem}

By tracking the dependence on the entropy rate, we also obtain a
construction for sublinear entropy.

\begin{theorem}[Non-malleable affine extractors for sublinear entropy]
\label{intro:sublinear}\label{thm:affine-sublinear}
For every fixed $0<\eta<1$ and integer $h\ge1$, there is a constant
$C_{\eta,h}>0$ and a uniformly polynomial-time computable family
$\anmExt_n:\F_2^n\to\F_2^{m_n}$ with
$m_n\ge\lfloor n^{1-\eta}\rfloor$ and the following property for all
sufficiently large $n$. For every affine source $X$ of entropy at least
\[
 k=\left\lceil\frac{C_{\eta,h}n}{\log n}\right\rceil
\]
and affine maps $T_1,\ldots,T_h$ with no fixed point on $\Supp(X)$,
\[
 \bigl(\anmExt_n(X),(\anmExt_n(T_j(X)))_{j\in[h]}\bigr)
 \approx_{2^{-n^{1-\eta}}}
 \bigl(U_{m_n},(\anmExt_n(T_j(X)))_{j\in[h]}\bigr).
\]
\end{theorem}

The sublinear-entropy theorem also improves the entropy threshold
$Cn(\log\log\log n)^2/\log\log n$ of the directional affine extractor construction
in~\cite{LZdir}, while allowing general affine tampering.
\Cref{tab:comparison} summarizes the construction parameters.

\begin{table}[t]
\centering\small
\renewcommand{\arraystretch}{1.2}
\begin{tabularx}{\textwidth}{@{}P{0.25\textwidth}P{0.35\textwidth}>{\raggedright\arraybackslash}X@{}}
\toprule
Construction & Entropy requirement & Output / error\\
\midrule
Li~\cite{Li23} & $(1-\gamma)n$ & $\Omega(n)$ / $2^{-\Omega(n)}$\\
Li--Zhong~\cite{LZ24} & $k\ge\log^C n$ & $k^{\Omega(1)}$ / $n^{-\Omega(1)}$\\
\Cref{intro:linear} & $k\ge\rho n$ & $\Omega_\rho(n)$ / $2^{-\Omega_\rho(n)}$\\
\Cref{intro:sublinear} & $k\ge C_{\eta,h}n/\log n$ & $\lfloor n^{1-\eta}\rfloor$ / $2^{-n^{1-\eta}}$\\
\bottomrule
\end{tabularx}
\caption{Selected deterministic polynomial-time constructions of non-malleable
affine extractors. Here $\rho>0$ is an arbitrary fixed entropy
rate, $\eta\in(0,1)$ and $h\ge1$ are fixed, and $\gamma>0$ in the first row is a
sufficiently small absolute constant. The first two rows and \Cref{intro:linear} state
single-tampering parameters; \Cref{intro:sublinear} allows $h$ tamperings.}
\label{tab:comparison}
\end{table}

\paragraph{Non-malleability against several affine tamperings.}
For every fixed integer $h\ge1$ and entropy rate $0<\rho\le1$,
\Cref{thm:fixed-multi-affine} gives an explicit extractor
$F:\F_2^n\to\F_2^m$ with $m=\Omega_{\rho,h}(n)$ and error
$\varepsilon=2^{-\Omega_{\rho,h}(n)}$ that is $h$-non-malleable.
For every affine source $X$ of entropy at least $\rho n$ and affine
maps $T_1,\ldots,T_h$ satisfying $T_j(x)\ne x$ for every
$x\in\Supp(X)$, we have
\[
 (F(X),F(T_1(X)),\ldots,F(T_h(X)))
 \approx_\varepsilon(U_m,F(T_1(X)),\ldots,F(T_h(X))).
\]
Thus $F(X)$ is nearly uniform and independent of all the tampered
outputs together. The sources $T_j(X)$ need not have high entropy,
and two tampered sources may take the same value. Only equality with
the original source is excluded.

For the proof, we use the equivalent jointly affine formulation in
\Cref{pre:affine-nm-equivalence}. The inputs $X^0,\ldots,X^h$ may be
affine functions of a common uniform random vector, subject to
$H_\infty(X^0)\ge\rho n$ and $\Pr[X^0=X^j]=0$ for every $j>0$.
No entropy requirement is imposed on the other inputs. Fixing the
independent affine randomness not determined by $X^0$ reduces this
formulation to affine tamperings of $X^0$, without changing its entropy
or the error bound. \Cref{thm:joint-affine-tuple} proves this formulation
by modifying the correlation breaker to handle all $h$ tamperings together.

The number of tamperings may also grow with $n$. For fixed
$0<\rho\le1$ and $0<\eta<1$, \Cref{thm:joint-affine-quantitative}
allows
\[
 h+1\le c_{\rho,\eta}
       \sqrt{\frac{\log n}{\log\log n}}
\]
while retaining entropy threshold $\rho n$, output length at least
$\lfloor n^{1-\eta}\rfloor$, and error at most $2^{-n^{1-\eta}}$.
The construction is uniformly polynomial time in this range.

\subsubsection{Lower Bounds for Linear Branching Programs}

A linear branching program queries a linear form in the input bits at
each internal node. In the weakly read-once model of Gryaznov, Pudl\'ak,
and Talebanfard~\cite[Definition~1]{GPT22}, the query at a node lies outside the span
of all queries on all paths from the source to that node. Their strongly
read-once condition additionally requires the past and future query
spaces to have trivial intersection. We count all vertices in the size
of a branching program.

Non-malleability under nonzero translations already suffices for exponential
average-case hardness.

\begin{definition}[Directional affine extractor~{\cite[Definition~2]{LZdir}}]
\label{def:directional-intro}
A function $F:\F_2^n\to\F_2^m$ is a $(k,\varepsilon)$ directional
affine extractor if, for every affine source $X$ of dimension at least
$k$ and every nonzero $a\in\F_2^n$,
\[
 (F(X),F(X+a))\approx_\varepsilon(U_m,F(X+a)).
\]
\end{definition}

For a Boolean function $f:\F_2^n\to\F_2$, its \emph{directional
 derivative} in direction $a$ is $D_a f(x)=f(x)+f(x+a)$, with addition
in $\F_2$. We say that $f$ has \emph{directional derivative bias at
most $\beta$ at entropy $k$} if
\[
 \left|\E(-1)^{D_a f(X)}\right|\le\beta
\]
for every affine source $X$ of dimension at least $k$ and every
nonzero $a\in\F_2^n$. Every nonzero output parity
$f_u(x)=\langle u,F(x)\rangle$ of a $(k,\varepsilon)$ directional
affine extractor satisfies this bound with $\beta=2\varepsilon$:
the defining joint-distribution guarantee makes $D_a f_u(X)$
$\varepsilon$-close to $U_1$, whose bias is zero.

\begin{theorem}[Directional affine extractors imply average-case hardness]
\label{thm:directional-hardness-intro}
Let $f:\F_2^n\to\F_2$ have directional derivative bias at most
$\beta$ at entropy $k$. If $1\le q<n$ and $2q-n\ge k$, then every size-$s$ weakly
read-once linear branching program $P$ satisfies
\[
 \left|\E_{x\sim U_n}(-1)^{f(x)+P(x)}\right|
 \le\sqrt{2s\bigl(\beta+2^{-(n-q)}\bigr)}.
\]
Consequently, if $F$ is a $(k,\varepsilon)$ directional affine
extractor and $k\le n-2$, each nonzero output parity of $F$ has
correlation at most
$\sqrt{2s(2\varepsilon+2^{-\lfloor(n-k)/2\rfloor})}$ with $P$.
\end{theorem}

Affine non-malleability improves the term depending on $n-k$ from approximately
$2^{-(n-k)/2}$ to $2^{-(n-k)}$. This improvement persists, with a
one-bit loss, when the program also reads independent uniform bits.

\begin{theorem}[Affine non-malleability implies average-case hardness]
\label{intro:weak}\label{thm:affine-hardness-intro}
Let $1\le k<n$, and let $F:\F_2^n\to\F_2^m$ be a $(k,\eps)$ non-malleable
affine extractor. For every $u\in\F_2^m\setminus\{0\}$ and every
size-$s$ weakly read-once linear branching program $P$,
\[
 \left|\E_{x\sim U_n}(-1)^{\langle u,F(x)\rangle+P(x)}\right|
 \le\sqrt{2s\bigl(2\eps+2^{-(n-k)}\bigr)}.
\]
If $k\le n-2$, $\varepsilon\le1/2$, and $R\sim U_t$ is independent
of $X\sim U_n$, then, for every $t\ge0$ and every size-$s$ weakly read-once linear branching program $P$ on $(X,R)$,
\[
 \left|\E_{X,R}(-1)^{\langle u,F(X)\rangle+P(X,R)}\right|
 \le\sqrt{2s\left(2\varepsilon+
 (1-2\varepsilon)2^{-(n-k-1)}\right)}.
\]
In both statements, non-malleability against globally fixed-point-free
affine involutions suffices; an involution is a map $T$ satisfying
$T(T(x))=x$ for every $x$.
\end{theorem}

The two parts are proved in
\Cref{thm:lbp-affine-hardness} and \Cref{thm:lbp-affine-auxiliary}, respectively;
the auxiliary-input extension is in \Cref{sec:lbp-appendix}.

With $2r-1$ simultaneous affine tamperings, the $2r$-th moment improves
this term further. For every integer $r\ge1$,
\Cref{thm:lbp-higher-moment} gives
\[
 \operatorname{Corr}(\langle u,F\rangle,P)
 \le\left[2s\left(2\varepsilon+(2r-1)!!\,2^{-r(n-k)}\right)\right]^{1/(2r)}.
\]
Here $(2r-1)!!=\prod_{j=1}^{r}(2j-1)$ is the double factorial,
which counts the partitions of $2r$ positions into unordered pairs.
Thus error $2^{-E}$ and constant prediction advantage require size
$2^{\min\{E,r(n-k)\}-O_r(1)}$ for fixed $r$. The error exponent is
unchanged; the improvement is in the entropy-gap term. This exponent
also holds with independent uniform auxiliary inputs when
$k+\binom{2r}{2}<n$ (\Cref{thm:lbp-auxiliary-higher-moment}).
Since the error constant of our construction depends on the number of
tamperings, increasing $r$ need not improve every instantiation.

\begin{corollary}[Exponential average-case hardness]
\label{intro:hardness}\label{cor:exponential-average-hardness}
There are constants $c_1,c_2>0$ and an explicit Boolean function family
$f_n:\F_2^n\to\F_2$ such that, for all sufficiently large $n$, every
weakly read-once linear branching program $P$ of size at most $2^{c_1n}$
satisfies
\[
 \Pr[P(U_n)=f_n(U_n)]\le\frac12+2^{-c_2n}.
\]
\end{corollary}

The cubic trace construction of Gryaznov, Pudl\'ak, and
Talebanfard~\cite[Theorem~15]{GPT22}, combined with the directional
reduction, already proves this corollary. The directional affine extractors
of Li and Zhong~\cite{LZdir} and the non-malleable affine extractor
of Li~\cite[Theorem~1.12]{Li23} give further instantiations. Thus the
qualitative lower bound follows from published extractors independently
of our constructions.

For the cubic trace function, a more precise analysis of the derivative
ranks gives correlation at most $\sqrt{5s}\,2^{-n/15}$ when $15\mid n$,
and hence exact-computation size at least $2^{2n/15}/5$
(\Cref{cor:lbp-cubic-trace}). Padding gives size
$2^{2n/15-O(1)}$ at every sufficiently large input length.
This improves the exponent $n/9-O(1)$ obtained by substituting its
directional derivative bound into \Cref{thm:directional-hardness-intro}.

The vector-valued guarantee also bounds the probability of predicting a
short list containing the extractor output. Suppose $k\le n-2$,
$\varepsilon\le1/2$, and each sink outputs a list of at most $J$ values.
If $p=J/2^m\le1/2$, then, with arbitrary independent uniform auxiliary inputs,
\[
 \Pr[F(X)\in P(X,R)]
 \le p+\sqrt{2s(1-p)\left(\varepsilon+p2^{-(n-k-1)}\right)}.
\]
See \Cref{thm:lbp-list-prediction} and \Cref{thm:lbp-list-auxiliary} for the precise
statements and \Cref{thm:lbp-list-higher-moment} for a higher-moment
version without auxiliary inputs.

\subsubsection{Low-Degree Decision Trees}

A degree-$2$ decision tree~\cite[Definitions~3.1--3.2]{GG26}
queries a polynomial of degree at most two over $\F_2$ at each internal
node, with queries allowed to depend on previous answers. An oblivious tree uses the
same query at all nodes of a given depth. Write $D_2$ and
$D_2^{\mathrm{obl}}$ for the respective exact deterministic depths.
Golovnev and Gurumukhani~\cite{GG26} prove nearly maximal oblivious
lower bounds using directional affine dispersers and ask for analogous
non-oblivious lower bounds.

Small error gives a non-oblivious lower bound of a different form.
If $F$ is a $(k,2^{-E})$ directional affine extractor,
$k\le\lceil n/2\rceil-1$, and $f$ is any nonzero output parity, then
every depth-$d$ non-oblivious degree-$2$ decision tree $P$ satisfies
\[
 \Pr[P(U_n)=f(U_n)]\le\frac12+2^{d-E}.
\]
In particular, $D_2(f)\ge E-1$. This estimate already follows from
the ordinary affine extractor guarantee at the stated entropy threshold; affine
non-malleability is not needed. \Cref{aq:section} also gives the exact
lower bounds $D_2(f)\ge\lceil\log(n-k+2)\rceil$ and
$D_2(f)\ge\lceil(n-k+1)/(k+1)\rceil$.

The construction in \Cref{alg:affine-nm} admits a short non-oblivious evaluation: query the
initial-seed matrices, whose entries have degree two, then compute each
iteration using affine queries. These queries follow the successive
fixings used in the extractor analysis. Choosing the scale lengths to
match the requested output and error gives the following consequence.

\begin{theorem}[Non-oblivious and oblivious degree-$2$ decision-tree depth]
\label{intro:quadratic-adaptivity}
There is an absolute constant $C>0$ with the following
properties for all sufficiently large $n$.
\begin{enumerate}
\item For every fixed $0<\eta<1$, there is an explicit non-malleable
affine extractor $F_n:\F_2^n\to\F_2^{m_n}$ with output length
$m_n=\lceil n^{1-\eta}\rceil$, error $\varepsilon_n\le2^{-n^{1-\eta}}$,
and entropy threshold
\[
 k_n=\left\lceil\frac{Cn\log\log n}{\log n}\right\rceil.
\]
Every nonzero output parity $f_n$ satisfies
\[
 \begin{aligned}
 n^{1-\eta}-1&\le D_2(f_n)\le D_2(F_n)\le n^{1-\eta+o(1)},\\
 D_2^{\mathrm{obl}}(f_n)&\ge n-k_n+1.
 \end{aligned}
\]
\item For every fixed $a,e>0$, there is an explicit non-malleable
affine extractor $G_n:\F_2^n\to\F_2^{\lceil a\log n\rceil}$
with entropy threshold
$k_n=\lceil Cn\log\log\log n/\log\log n\rceil$
and error at most $n^{-e}$, such that every nonzero output parity $g_n$
satisfies
\[
 \begin{aligned}
 \Omega(\log n)&\le D_2(g_n)\le D_2(G_n)\le(\log n)^{1+o(1)},\\
 D_2^{\mathrm{obl}}(g_n)&\ge n-k_n+1=n-o(n).
 \end{aligned}
\]
\end{enumerate}
\end{theorem}

The proof and a comparison at matched entropy and error appear in
\Cref{aq:comparison}. Both constructions use the same algorithm as
\Cref{intro:linear} and \Cref{intro:sublinear}, with different scale lengths.
Li and Zhong~\cite[Theorem~1.2]{LZ24} already constructed non-malleable
affine extractors with inverse-polynomial error, even at
polylogarithmic entropy. The second part above establishes a
separation between non-oblivious and oblivious depth for the construction in
\Cref{alg:affine-nm}; it does not assert this non-oblivious upper bound
for the construction of Li and Zhong~\cite{LZ24}.
Thus affine non-malleability at sublinear entropy does not by itself
imply nearly maximal non-oblivious degree-$2$ decision-tree depth. The non-oblivious lower-bound
question of Golovnev and Gurumukhani~\cite[Section~8]{GG26} remains open.

\paragraph{Higher-degree queries.}
\label{intro:other-trees}

The same correlation argument applies to queries of every fixed degree
$r\ge2$. Write $D_r$ for exact non-oblivious depth with degree-$r$ queries.
If a Boolean function has correlation at most $2^{-\lambda}$
with every polynomial of degree at most $r$, then non-oblivious trees need
at least $2^\lambda$ leaves and depth $\lambda$, while oblivious trees
need depth $2\lambda$. For our extractors, $(2^{r-2}-1)$-non-malleability against translations,
with entropy threshold below $n/2$ and error $2^{-E}$, $1<E\le n$, gives non-oblivious
depth $E/2^{r-2}-O_r(1)$ and twice this oblivious lower bound.
For $r=2$, the ordinary affine extractor guarantee suffices; for $r=3$,
non-malleability under nonzero translations suffices. Our constructions give linear depth
for each fixed $r$ at a sufficiently small fixed entropy rate.
The number of tamperings and all construction constants may depend
on $r$. The quantitative multi-tampering theorem also permits growing
$r$: at fixed entropy rate below $1/2$, it gives error $2^{-n^{1-\eta}}$
and non-oblivious depth at least
$(n^{1-\eta}-\log_2 3)/2^{r-2}$ when
$2^{r-2}\le c_{\rho,\eta}\sqrt{\log n/\log\log n}$.
See \Cref{ldt:growing-degree-extractor} for the precise conditions.

At fixed entropy rate, we can also evaluate our construction at a
depth controlled by its output length and error exponent.

\begin{theorem}[Non-Oblivious Depth at Fixed Entropy Rate]
\label{intro:fixed-degree-depth}
Fix $0<\rho\le1$ and an integer $h\ge1$. For all sufficiently large
$n$, every integer $m\ge1$ and real $E\ge1$ with
$m+E\le c_{\rho,h}n$, for a sufficiently small constant
$c_{\rho,h}>0$, admit an explicit $m$-bit non-malleable affine
extractor $F_h$ at entropy $\lceil\rho n\rceil$, with error at most
$2^{-E}$ against $h$ affine tamperings, such that
\[
 D_2(F_h)\le O_{\rho,h}(m+E+\log n).
\]
Suppose, in addition, that the parameters satisfy the following conditions.
\begin{itemize}
\item Entropy rate: $\rho<1/2$.
\item Query degree and number of tamperings: $r\ge2$ is fixed and
$h\ge\max\{1,2^{r-2}-1\}$.
\item Output length and error exponent: $m\le E$ and $E\ge\log n$.
\end{itemize}
Then every nonzero output parity $f$ satisfies
\[
 \frac{E}{2^{r-2}}-O_r(1)
 \le D_r(f)\le D_r(F_h)\le O_{\rho,h}(E).
\]
Thus both $D_r(f)$ and $D_r(F_h)$ are $\Theta_{\rho,h,r}(E)$.
\end{theorem}

The upper bound computes the entire output exactly, using only
degree-$2$ queries. It therefore applies to every query degree
$r\ge2$. \Cref{aq:multi-upper} and \Cref{ldt:matching-depth} prove the theorem.
For example, at fixed entropy rate below $1/2$ and with
$m\le E$, error $n^{-\Theta(1)}$ gives depth $\Theta(\log n)$,
while error $2^{-\Theta(n)}$ gives depth $\Theta(n)$,
within the stated feasible range.

\subsubsection{Non-Oblivious Local Decision Trees}

A non-oblivious $\ell$-local decision tree~\cite[Definitions~3.1--3.2]{GG26}
may query any Boolean function of at most $\ell$ input coordinates.
Both the function and its support may depend on all previous answers.
We obtain superconstant overhead over $n/\ell$ from two types of
extractors. Sumset extractors give the larger depth bound in the
common locality range. Directional affine extractors give exponentially small
correlation bounds for every nonzero output parity of our construction.

\begin{theorem}[Non-Oblivious Local Depth]
\label{intro:local-depth}
The following statements hold for growing locality $\ell$.
\begin{itemize}
\item \emph{Sumset extractors.}
Fix $0<\delta<1$ and $0<\varepsilon<1/2$. There are an explicit
Boolean family $S_n$, independent of $\ell$, and a constant
$c_\delta>0$ such that, for $\ell\le n^{1-\delta}$, every non-oblivious
$\ell$-local tree $T$ of depth at most $c_\delta(n/\ell)\log\ell$ satisfies
\[
 \operatorname{Corr}(S_n,T)\le2\varepsilon(1+o(1)).
\]
In particular, $\operatorname{LDT}_\ell(S_n)
=\Omega_\delta((n/\ell)\log\ell)$.
\item \emph{Directional affine extractors.}
Fix $0<\rho<1/2$ and $a>0$. Suppose $F_n:\F_2^n\to\F_2^{m_n}$
is a directional affine extractor at dimension $\lceil\rho n\rceil$
with error at most $2^{-an}$. There is $C_\rho>0$ such that,
for $\ell\le n/(C_\rho\log n)$, every nonzero output parity $f_n$
and every non-oblivious $\ell$-local tree $T$ of depth at most
\[
 \left\lfloor\frac{n}{4\ell}\frac{\log\ell}{\log\log\ell}\right\rfloor
\]
satisfy $\operatorname{Corr}(f_n,T)\le2^{-\Omega(n/\sqrt\ell)}$.
Consequently,
$\operatorname{LDT}_\ell(f_n)
=\Omega((n/\ell)\log\ell/\log\log\ell)$.
\end{itemize}
\end{theorem}

Here $\operatorname{LDT}_\ell$ denotes exact deterministic depth, and
correlation is with respect to the uniform input. The first part uses
the sumset extractor of Li~\cite[Theorem~1.7]{Li23} at entropy
$C_\varepsilon\log n$. Its exact-computation consequence uses only
sumset dispersion. The second part applies both to the directional
affine extractors of Li and Zhong~\cite{LZdir} and to our non-malleable
affine extractors in \Cref{intro:linear}. Only single-tampering
non-malleability is needed, since nonzero translations are
fixed-point-free affine maps. The proofs appear in
\Cref{thm:local-sumset-logarithmic} and \Cref{thm:local-directional-logarithmic}.

The sumset family also retains useful finite-locality bounds:
\Cref{cor:local-explicit-family} gives, simultaneously for
$3\le\ell\le n$,
\[
 \operatorname{LDT}_\ell(S_n)\ge\frac{2n}{\ell+2}
 -O\!\left(\frac{\log n}{\ell}+\sqrt{\frac n\ell}+1\right),
\]
as well as $n-O(\log n)$ at locality two and
$n/2-O(\log n)$ at locality four. At locality sixteen the bound is
$n/9-O(\sqrt n)$. These improve the leading coefficient in the
classical PPSZ bound~\cite{PPSZ05} at locality four and at every
fixed locality at least eleven; \Cref{tab:local-comparison} records
selected constants. The logarithmic overhead in
\Cref{intro:local-depth} improves the growing-locality consequence
of this finite bound in the common range.

\begin{table}[tbp]
\centering\small
\begin{tabularx}{\textwidth}{@{}P{0.24\textwidth}P{0.27\textwidth}>{\raggedright\arraybackslash}X>{\raggedright\arraybackslash}X@{}}
\toprule
Model & Hard function or hypothesis & Non-oblivious depth & Oblivious depth\\
\midrule
Degree-$2$ & Non-malleable affine extractor in \Cref{intro:quadratic-adaptivity}(ii);
inverse-polynomial error &
$(\log n)^{1+o(1)}$ upper bound & $n-o(n)$ lower bound\\
Degree-$r$, fixed $r\ge2$ & $(2^{r-2}-1)$-non-malleability for translations; error $2^{-E}$;
entropy below $n/2$ & $E/2^{r-2}-O_r(1)$ & $E/2^{r-3}-O_r(1)$\\
Degree-$r$, fixed $r\ge2$ & Our $h$-non-malleable affine extractor with error $2^{-E}$;
fixed entropy rate & $\Theta_{\rho,h,r}(E)$ & $E/2^{r-3}-O_r(1)$\\
$4$-local & Sumset/affine disperser of Li~\cite{Li23}
& $n/2-O(\log n)$ & ---\\
$16$-local & The same disperser & $n/9-O(\sqrt n)$ & ---\\
$\ell$-local, $\ell\to\infty$, $\ell=o(n)$ & The same disperser
& $(2-o(1))n/\ell$ & ---\\
$\ell$-local, $\ell\to\infty$, $\ell\le n^{1-\delta}$
& Sumset extractor of Li~\cite{Li23}
& $\Omega_\delta((n/\ell)\log\ell)$ & ---\\
$\ell$-local, $\ell\to\infty$, $\ell\le n/(C_\rho\log n)$
& Directional affine or non-malleable affine extractors; fixed rate $\rho<1/2$, error $2^{-\Omega(n)}$
& $\Omega\!\left(\frac{n\log\ell}{\ell\log\log\ell}\right)$ & ---\\
\bottomrule
\end{tabularx}
\caption{Decision-tree consequences. Entries are lower bounds unless
marked otherwise. The second row uses ordinary affine extractors for $r=2$
and directional affine extractors for $r=3$, with $1<E\le n$. The third row
assumes \Cref{intro:fixed-degree-depth}, including $m\le E$ and
$E\ge\log n$. The first row uses \Cref{intro:quadratic-adaptivity}.
In the last two rows, $0<\delta<1$ and $\rho$ are
fixed; \Cref{intro:local-depth} also gives average-case bounds.
Dashes indicate bounds not listed here.}
\label{tab:decision-tree-results}
\end{table}

Both local-tree reductions give superconstant overhead for growing
locality in their stated ranges. They do not settle the near-maximal
non-oblivious local depth and size targets, or the near-maximal
non-oblivious degree-$2$ depth target, posed in~\cite[Section~8]{GG26}.

\subsubsection{Deterministic Lossless Lifting}

Efremenko and Itsykson~\cite{EI26} proved lossless lifting using
distinct random gadgets. We give a deterministic construction using
our non-malleable affine extractor against a fixed number of
tamperings.

For every fixed $0<\xi<1$, \Cref{thm:deterministic-lossless} gives a
deterministic polynomial-time construction of an infinite family of
unsatisfiable CNFs on $N$ Boolean variables, with polynomially many
clauses of width $O_\xi(\log N)$, such that:
\begin{itemize}
\item every $\operatorname{Res}(\oplus)$ refutation has width at least
$(1-\xi)N$;
\item every $\operatorname{Res}(\oplus)$ refutation of resolution depth
at most $N$ has size at least $2^{(1-\xi)N}$.
\end{itemize}
Here $N$ counts all Boolean variables, and the polynomial bounds may
depend on $\xi$. The logarithmic-size gadgets are given by complete
truth tables, so no extension variables are needed.
The size bound holds more generally for the path-free fragment in
\Cref{formula:lifting}. We give the deterministic gadget construction
in \Cref{sec:deterministic-lossless} and the full proof in
\Cref{app:lossless-lifting}.

\subsubsection{Overlapping Gadgets and Resolution over Parities}

We obtain a size--depth tradeoff for resolution over parities using
\emph{overlapping gadgets}, meaning gadgets that share input variables.
Starting from a Tseitin formula on a constant-degree expander, we
substitute a gadget for each edge variable. Each input block belongs to
exactly two gadgets. A second graph describes these intersections: its
vertices are gadgets and its edges are shared input blocks.

\begin{theorem}[Direct overlapping size--depth tradeoff]
\label{thm:lifting-intro}
There are constants $c,C>0$ and a deterministic polynomial-time
construction of unsatisfiable CNFs $\Psi_m$ on
$N=\Theta(m\log m)$ variables, of polynomial size and width
$O(\log m)$, such that every $\operatorname{Res}(\oplus)$ refutation
of $\Psi_m$ of depth $d$ and size $S$ satisfies
\[
 d\left(\log S+C\frac{N}{\log N}\right)\ge cN^2.
\]
In particular, for every fixed $a>0$, refutations of depth at most $aN$
have size $2^{\Omega_a(N)}$. The same conclusion applies to top-regular
and bottom-regular refutations.
\end{theorem}
We recall these regularity conditions in
\Cref{subsec:proof-complexity-prelim}.

The additive term measures the cost of requiring gadget outputs to lie
in the family of partial assignments maintained by the strategy for the
base formula. For each fixed set of assigned base variables, this family
forms an affine space. We bound the conditioning cost by the number of
independent equations defining that space; see
\Cref{thm:directional-overlapping-lifting,lem:overlap-tseitin-legal}.

In the non-malleable version, a matching in the graph of overlaps isolates
independent input blocks. Non-malleability bounds the resulting
correlations under unrelated invertible affine changes of coordinates
at the two endpoints. This argument does not require the graph of
overlaps to be an expander.

To remove the additive term, we choose gadgets from our
multiple-tampering extractor that can be restricted to majority functions
on disjoint triples. The lifting theorem of Itsykson, Podolskii, and
Shekhovtsov~\cite[Theorem~6.1]{IPS26} then gives
\[
 d=\Omega\!\left(\min\left\{H,
                  \frac{w^2}{\log(2S)}\right\}\right)
\]
whenever every ordinary-resolution refutation of the base formula has
width at least $w$ or depth at least $H$; see
\Cref{thm:overlap-ips-transfer}.

For fixed overlap degree $\Delta$, blocks of length at most $2^{C\Delta}$
suffice, for an absolute $C$. Thus the gadgets have constant size and
the final number of variables is linear in that of the base formula.
On the bases constructed below, this gives
$d\log(2S)\ge2^{-O(\Delta)}N^2$; see
\Cref{lem:overlap-block-length,cor:overlap-quantitative-product}.
The improvement over \cite[Theorem~6.4]{IPS26} for formulas with short
ordinary-resolution proofs comes from the parity substitutions in the
base formula, which we describe next.

\subsubsection{A Quadratic Tradeoff with Short Ordinary-Resolution Proofs}

\begin{theorem}[Quadratic size--depth tradeoff]
\label{intro:quadratic}
For every sufficiently large $N$, there is a uniformly polynomial-time
constructible, polynomial-size unsatisfiable CNF $\Psi_N$ on exactly
$N$ variables with a polynomial-size ordinary-resolution refutation,
such that every $\operatorname{Res}(\oplus)$ refutation of size $S$
and resolution depth $d$ satisfies
\[
                         d\log(2S)=\Omega(N^2).
\]
All variables, including padding variables, are counted.
\end{theorem}

Itsykson, Podolskii, and Shekhovtsov~\cite[Theorem~6.4]{IPS26} proved
$d\log S=\Omega(N^2/\log^2 N)$ for a family with polynomial-size
ordinary-resolution proofs. \Cref{intro:quadratic} removes that
$\log^2 N$ loss. In particular, it gives superpolynomial size at
every depth $o(N^2/\log N)$. The quadratic product bound
of~\cite[Corollary~1.3]{IPS26} applies to another family, without the short ordinary-proof guarantee;
our claim concerns the former comparison. The improvement comes from successively reducing the number of
variables in the base formula through parity substitutions
(condensation): the original variables are replaced by parities of
a smaller collection of shared variables. The lifting theorem
and its constant-size majority gadget are those of \cite[Theorem~6.1]{IPS26}; no
extractor property is used in this construction.

A recent preprint of Braun~\cite[Theorem~1.1]{Braun26} claims a size
lower bound $2^{\Omega(n/\log^2 n)}$ for unrestricted
$\operatorname{Res}(\oplus)$ refutations of the bit pigeonhole
principle with $n+1$ pigeons and $n$ holes, where $n$ is a power of two.
The formula has $N=(n+1)\log_2 n$ variables, so the claimed bound is
$2^{\Omega(N/\log^3 N)}$ in our notation. This concerns a different
question from \Cref{intro:quadratic}: our formulas have
polynomial-size ordinary-resolution proofs, but every polynomial-size
$\operatorname{Res}(\oplus)$ proof must have depth
$\Omega(N^2/\log N)$. At depth $O(N)$,
\Cref{intro:quadratic} gives size $2^{\Omega(N)}$.
\Cref{thm:deterministic-lossless} gives the near-maximal exponent
$(1-\xi)N$ for a different explicit family at depth at most $N$.
Braun's claimed bound applies at every depth. 

\Cref{tab:proof-complexity-overview} records these distinctions.
The overlapping transfer preserves the bound of
\cite[Theorem~6.1]{IPS26}; condensation improves the tradeoff for
formulas with short ordinary-resolution proofs; and the lossless
construction derandomizes the near-maximal bounded-depth bound
of~\cite{EI26}.
\begin{table}[!ht]
\centering
\small
\setlength{\tabcolsep}{4pt}
\renewcommand{\arraystretch}{1.18}
\begin{tabularx}{\linewidth}{@{}P{.22\linewidth}P{.30\linewidth}>{\raggedright\arraybackslash}X@{}}
\toprule
Result & Lower bound & Construction and scope \\
\midrule
\cite[Corollary~1.3]{IPS26}
& $dL=\Omega(N^2)$
& Explicit constant-size gadgets. No polynomial-size ordinary-resolution
proof guarantee for this family. \\
\addlinespace
\cite[Theorem~6.4]{IPS26}
& $dL=\Omega(N^2/\log^2N)$
& Explicit; polynomial-size ordinary-resolution proofs; no extractor. \\
\addlinespace
Direct overlapping lift (\Cref{thm:lifting-intro})
& $d\bigl(L+\frac{CN}{\log N}\bigr)\ge cN^2$
& Directional affine or non-malleable affine extractors; overlapping
logarithmic-size gadgets. \\
\addlinespace
Overlapping size--depth transfer (\Cref{cor:overlap-quantitative-product})
& $dL\ge2^{-CD}N^2$
& Non-malleable affine extractors; edge-block length $t\le2^{CD}$ for fixed
sharing degree $D$. Applies \cite[Theorem~6.1]{IPS26} unchanged. \\
\addlinespace
Condensed base (\Cref{intro:quadratic})
& $dL=\Omega(N^2)$
& Explicit; polynomial-size ordinary-resolution proofs; no extractor.
Improves the bound for the short-proof family in \cite[Theorem~6.4]{IPS26}. \\
\addlinespace
\cite{EI26}
& $\log S\ge(1-\xi)N$\newline for $d\le N$
& Random gadget family. \\
\addlinespace
Deterministic lossless lifting
(\Cref{thm:deterministic-lossless})
& $\log S\ge(1-\xi)N$\newline for $d\le N$
& Deterministic polynomial time; non-malleable affine extractors;
logarithmic clause width. \\
\addlinespace
Braun~\cite{Braun26} (claimed)
& $\log S=\Omega(N/\log^3N)$
& Explicit bit pigeonhole formulas; unrestricted depth and regularity. \\
\bottomrule
\end{tabularx}
\caption{Proof-complexity comparisons. Here $S$ and $d$ are the size and
resolution depth of a $\operatorname{Res}(\oplus)$ refutation,
$L=\log(2S)$, and $N$ counts all variables. All formula families have
polynomial size. In the transfer row, $C$ is absolute and $D$ is fixed;
the bound remains $\Omega(N^2)$ for every such $D$. That row uses the
condensed bases in \Cref{quad:main}; the general width--depth statement
is \Cref{thm:overlap-ips-transfer}.
The two lossless-lifting rows hold for every fixed $0<\xi<1$ and
record the depth-$N$ consequences of path-free lower bounds.
Their polynomial bounds may depend on $\xi$. The last row records the claim in
\cite[Theorem~1.1]{Braun26}, normalized to the total variable count.}
\label{tab:proof-complexity-overview}
\end{table}

\subsection{Related Work}\label{sec:related-work}

\paragraph{Seedless and Non-Malleable Extractors.}
Affine and sumset extractors are models of seedless extraction from
structured sources. Non-malleable extraction additionally requires the
output to remain nearly uniform given outputs on related inputs.
Chattopadhyay and Li~\cite{CL17} introduced non-malleable affine extractors to
construct non-malleable codes against affine tampering.
Li~\cite{Li23} obtained linear output and exponentially small error at a particular
constant entropy rate, while Li and Zhong~\cite{LZ24} constructed non-malleable
affine extractors for polylogarithmic entropy with polynomially
small error. Our constructions address the small-error
regime, obtaining linear output and exponentially small error at every
positive constant entropy rate, together with guarantees against
several affine tamperings. They build on the directional affine
extractor of Li and Zhong~\cite{LZdir}, extending robustness from
translations to general affine maps; \Cref{tab:comparison} compares the
parameters. These constructions complement our lower-bound reductions,
which also apply to existing seedless extractors: the directional
construction of Li and Zhong~\cite{LZdir} gives exponential average-case hardness
for weakly read-once linear branching programs, and the
logarithmic-entropy sumset extractor of Li~\cite{Li23} gives superconstant
depth overhead for non-oblivious local decision trees.

\paragraph{Linear branching programs.}
Gryaznov, Pudl\'ak, and Talebanfard~\cite{GPT22} used directional affine
extractors to obtain average-case lower bounds for strongly read-once
linear branching programs. Chattopadhyay and Liao~\cite{CL23},
Li~\cite{Li23}, and Li and Zhong~\cite{LZdir,LZ24} obtained improved
size--advantage tradeoffs in this model. The strongly read-once condition
requires the past and future query spaces to have trivial intersection;
the weakly read-once condition only requires a node's query to lie
outside its past query space. Our reductions apply under the latter
condition.

\begin{table}[t]
\centering\small
\renewcommand{\arraystretch}{1.2}
\begin{tabularx}{\textwidth}{@{}P{0.25\textwidth}P{0.30\textwidth}P{0.24\textwidth}>{\raggedright\arraybackslash}X@{}}
\toprule
Work & Program size & Correlation & Model\\
\midrule
Gryaznov--Pudl\'ak--Talebanfard~\cite{GPT22} & $2^{n/3-o(n)}$ & $2^{-o(n)}$ & Strongly read-once\\
Chattopadhyay--Liao~\cite{CL23} & $2^{n-\log^{O(1)}n}$ & $n^{-\Omega(1)}$ & Strongly read-once\\
Li~\cite{Li23} & $2^{n-O(\log n)}$ & Any fixed positive constant & Strongly read-once\\
Li--Zhong~\cite{LZdir} & $2^{n-\widetilde O(n/\log\log n)}$ & $2^{-n^{\Omega(1)}}$ & Strongly read-once\\
\Cref{thm:directional-hardness-intro} and \Cref{intro:weak} & $2^{\Omega(n)}$ & $2^{-\Omega(n)}$ & Weakly read-once\\
\Cref{cor:lbp-cubic-trace}, cubic trace & $2^{(2/15-\eta)n}$ & $O(2^{-\eta n/2})$ & Weakly read-once\\
\bottomrule
\end{tabularx}
\caption{Selected average-case lower bounds for Boolean linear branching
programs. The $\widetilde O$ in the Li--Zhong row hides a
$(\log\log\log n)^2$ factor. Published constructions suffice for the
two weakly read-once rows. In the cubic trace row, $0<\eta<2/15$ is
fixed; rounding changes only the absolute constant. The reduction using non-malleable affine extractors improves the entropy-gap
parameter and extends to independent uniform auxiliary inputs
(\Cref{thm:lbp-affine-auxiliary}). The
near-maximal size bounds in this table concern strongly read-once linear branching programs.}
\label{tab:branching-comparison}
\end{table}

The reductions in \Cref{thm:directional-hardness-intro} and \Cref{intro:weak}
give exponential average-case hardness from the published extractors
of~\cite{GPT22,Li23,LZdir}. They do not give size $2^{n-o(n)}$ for
weakly read-once linear branching programs from our current parameters: the exponent is
limited by the error exponent and, in the second-moment argument, the
entropy gap. Higher moments multiply the latter by a fixed factor but
leave the error exponent unchanged. The
near-maximal bounds in \Cref{tab:branching-comparison}, as well as the
affine-mixedness argument of~\cite[Theorem~11]{GPT22}, concern strongly read-once linear branching programs.

Search lower bounds for weakly read-once linear branching programs, arising from
regular resolution over parities and bounded-depth proofs, concern a
different task: finding a falsified clause of an unsatisfiable formula.
They do not by themselves give a total Boolean function that is hard to
predict on a uniform input. A reduction between the two tasks must also
preserve the read-once restriction.

\paragraph{Resolution over parities.}
Itsykson and Sokolov~\cite{IS20} proved exponential lower bounds for
tree-like resolution over parities. Efremenko, Garl\'ik, and
Itsykson~\cite{EGI24} proved exponential lower bounds for bottom-regular
refutations of the bit pigeonhole principle. Bhattacharya,
Chattopadhyay, and Dvo\v{r}\'ak~\cite{BCD24} gave formulas that are easy for
ordinary resolution but hard for bottom-regular resolution over
parities.

Alekseev and Itsykson~\cite{AI25} obtained top-regular lower bounds and
used random walks with restarts for bounded-depth refutations, including
depth $O(N\log\log N)$. Efremenko and Itsykson~\cite{EI25} introduced
amortized closure and extended the depth range to $O(N\log N)$.
Itsykson and Knop~\cite{IK26} proved supercritical size--depth tradeoffs
for formulas with short unrestricted proofs: short proofs require
superlinear depth, although every $N$-variable contradiction has a
depth-$N$ refutation when size is unrestricted. Bhattacharya, Byramji,
Chattopadhyay, and Impagliazzo~\cite{BBCI26} obtained, for every fixed
$\eta>0$, size $2^{\Omega(N^\eta)}$ at depth $O(N^{2-\eta})$ for
explicit formulas with $O(N)$ clauses of width six.

Itsykson, Podolskii, and Shekhovtsov~\cite{IPS26} proved a quadratic
size--depth tradeoff using constant-size $1$-stifling gadgets:
their output can be forced to either value by fixing all but any one
of their input bits. The tradeoff in~\cite[Theorem~6.4]{IPS26} is
within a $\log^2 N$ factor of quadratic
for formulas with short ordinary-resolution proofs. Our two comparisons
with this work are different: the overlapping construction extends the
scope of their lifting theorem, while condensation removes the
$\log^2 N$ loss for the latter family.

Efremenko and Itsykson~\cite{EI26} obtained near-maximal lower bounds
$2^{(1-\xi)N}$ at depth $N$ using distinct random gadgets. These bounds are unconditional. Our construction replaces the random gadgets by evaluations of a
non-malleable affine extractor with a distinct fixed prefix, or tag,
identifying each gadget. We prove that one fixed family realizes every
output tuple on every affine source required by the lifting argument.
This is the simultaneous surjectivity requirement.

Alekseev and Gaevoy~\cite{AG26} obtained lower bounds
for constrained bit pigeonhole formulas at every fixed polynomial
depth: their results are unconditional for reversible resolution over
parities and conditional for general resolution over parities. They
also obtain an unconditional nearly quadratic depth lower bound for
the full system.

Overlapping substitutions have appeared in hardness condensation,
including the expander-based construction of Razborov~\cite{Raz16},
regular-resolution lower bounds of Vinyals, Elffers, Johannsen, and
Nordstr\"om~\cite{VEJN20}, and depth tradeoffs of Fleming, Pitassi,
and Robere~\cite{FPR22}. We use affine extractors to analyze nonlinear overlapping gadgets
under parity constraints on their inputs.

\subsection{Overview of the Techniques}\label{sec:overview}

\subsubsection{Non-Malleable Affine Extractors}

\paragraph{The Li--Zhong construction.}
The construction of Li and Zhong~\cite[Algorithm~8 in the full version]{LZdir}
proceeds in iterations, each using one input block together with the whole
source. Linear somewhere condensers produce lists of linear images, and a
bilinear extractor applied to pairs of these images forms a matrix. An
affine somewhere-random extractor, which extracts from a jointly affine
tuple containing a uniform row, is applied to this matrix to obtain an
initial seed $H_i(x)$. Linear seeded extraction then produces a row
$Z_i(x)$, whose bits select coordinates of a linear encoding of $x$.
The row and sampled coordinates supply a seed for non-malleable
extraction, followed by affine correlation breaking and linear seeded
extraction. Finally, products on disjoint groups of bits and linear-code
maps are applied to the iteration outputs, and the resulting strings are
combined by $\mathsf{XOR}$.

\paragraph{From translations to affine maps.}
The advice is a short string supplied to the correlation breaker to
distinguish the two computations. Under the correlation breaker's source
and seed hypotheses, distinct advice strings give an output that is
nearly uniform even conditioned on the other output. The principal new
difficulty is to generate such advice for general affine tampering.
Under a translation, the difference between the original and tampered
inputs is fixed. For an affine tampering $x'=Mx+b$, this difference
depends on $x$ and can be correlated with the seed used to sample its
encoding. We adapt the code-sampling approach of Li~\cite[Section~3]{Li23}.
Writing the seed as $z=z_1\circ z_2$, we define
\[
 \Adv(x,z)=z_1\circ z_2\circ
 \Enc(x)|_{\Samp(E_0(z_1))}\circ L_0(x,z_2).
\]
Here $E_0$ is an affine extractor and $L_0$ is a linear seeded
extractor. After fixing $z_2$, the sampled coordinates handle the case
in which the difference of the two inputs has small remaining entropy;
the output of $L_0$ handles the case of large remaining entropy.
The guarantee permits jointly affine inputs, including a tampered input
of low entropy; see \Cref{branch:advice}.

Once the advice strings differ, we use the first phase of the affine
correlation-breaker construction of Chattopadhyay and
Liao~\cite[Section~6]{CL22}, instantiated with the strong correlation
breaker of Li~\cite[Theorem~5.2]{Li23}. This replaces the intermediate seeded
non-malleable extractor and row-indexed correlation breaker.
We retain the initial-seed construction. Only these initial seeds
need a degree bound: $\deg H_i\le D$, where
$D=(2/\kappa)^{O(1)}$ at working entropy rate $\kappa$.
After an initial seed is fixed, the remaining output is determined by
successively fixing a short sequence of affine functions. Each function
depends on the values fixed previously. Thus the length of this
sequence, rather than the degree of the whole computation, controls
the entropy lost when fixing other iterations.

\paragraph{Shorter initial seeds.}
We reduce the length of the initial seed by replacing the single
linear extraction that produces $Z_i$ with a sequence of linear
extractions. Starting with $O(u_i+\log n)$ seed bits, the output
length doubles until it reaches $O(s_i+\log n)$. Here $s_i$ is the
iteration's output length and $2^{-u_i}$ is its required error.
The seed-length bound of Chattopadhyay and
Liao~\cite[Lemma~3.13]{CL22} permits each doubling.
After the current seed tuple is fixed, the remaining affine sources
have a common direction independent of the values fixed. A rank
argument bounds the fraction of seeds for which the next output is
not exactly uniform. We apply this argument separately at every
step, before fixing that step's seed. Since the lengths double, the
total number of bits fixed is $O(s_i+\log n)$.

\paragraph{Combining the iterations.}
Let $W_i(x)$ be the $s_i$-bit output of iteration $i$.
Choose an odd block length $b=16D^2+1$. Let $V_i(w)$ consist of
majorities on disjoint groups of $b$ coordinates, and let $M_i$ be a generator
matrix of a constant-rate, constant-distance binary linear code.
Writing $\Phi_i(w)=M_iV_i(w)$, the final output is
\begin{equation}\label{intro:construction}
 \anmExt(x)=\bigoplus_i\Phi_i(W_i(x)).
\end{equation}
The code ensures that each nonzero output parity involves many disjoint
majorities. The correlation theorem of Chattopadhyay
et al.~\cite[Theorem~1.1]{CHLLTV26} bounds its correlation with every degree-$D$ polynomial by $2^{-\zeta_Ds_i}$,
where $\zeta_D=\Theta(D^{-2})$; see \Cref{amp:section}.
Only the initial seeds need bounded degree; the degree of this
majority amplifier does not enter the argument.

The analysis selects an iteration whose block has sufficient conditional
rank. The earlier blocks can be fixed while leaving enough entropy for
this iteration; for later iterations, we guess the initial seeds. We then
fix the affine functions determining the outputs of the other iterations.
On each remaining affine source, the analysis of the selected iteration
provides further affine fixings. Except for fixings of small total
probability, the selected output is affine and exactly uniform, while the
output of the same iteration on the other input is fixed. Here we use the fact that a
nonuniform affine distribution has statistical distance at least $1/2$
from uniform. This guarantee is established before conditioning on
whether the guessed seeds are correct.

We parametrize the remaining affine source by the selected output and a
complementary set of coordinates. Fixing the latter leaves the selected
output uniform and expresses both inputs as affine functions of it.
Consequently, the actual seed bits become degree-$D$ polynomials in the
selected output. We expand the indicator that all seed guesses are
correct in Fourier characters of the seed bits. The coefficients have
total absolute value one, and each character becomes a degree-$D$
polynomial phase in the selected output, so the majority correlation
bound applies.
If $\varepsilon_i$ bounds the exceptional probability and $Q_i$ is the
number of guessed seed bits, the resulting Fourier bound is
\begin{equation}\label{intro:error}
 \beta=\max_i2^{Q_i}
       \bigl(\varepsilon_i+2^{-\zeta_D s_i}\bigr).
\end{equation}
The statistical error is at most $2^{m-1}\beta$. Set
$u_i=\Theta(\zeta_Ds_i)$. Because the initial seed now has length
$O(u_i+\log n)$, a constant geometric decrease in the $u_i$ makes
$Q_i\le u_i/4$. The ratio is independent of $\kappa$.
Together with $O(1/\kappa)$ iterations, this gives a common
output and error exponent $n2^{-O(1/\kappa)}$. Taking
$\kappa=\Theta_\eta(1/\log n)$ yields the single-tampering case of
\Cref{intro:sublinear};
the parameter choices appear in \Cref{sec:parameters}.

The selected block may belong to either input. When both sources have
sufficient entropy and both output parities are nonzero, the jointly
affine guarantee lets us interchange the two inputs. We prove the ordinary
affine extractor guarantee separately when the parity on the tampered output
has coefficient vector zero, or in
the case of a low-entropy tampered image.

For several tamperings, we retain the additional rounds of the affine
correlation breaker of Chattopadhyay and Liao~\cite[Section~6]{CL22}.
The same successive fixings determine
their outputs. In the analysis, an affine partition separates unequal
tampered inputs and groups inputs that coincide, outside a set of small
probability. No entropy condition
is imposed on the tampered inputs. For $q=h+1$ inputs, the sequence
of linear extractions fixes at most $q$ times as many bits, and the
geometric ratio can be chosen $\Theta(1/q)$. Keeping separate
accounts of the entropy lost from the entire input and from each
individual block gives $O(q^2/\rho)$
iterations and the quantitative bound of
\Cref{thm:joint-affine-quantitative}. The details appear in
\Cref{sec:multiple-tampering}.

\subsubsection{Weakly Read-Once Linear Branching Programs}

A difficulty is that the inputs reaching a node need not form an
affine source. We first analyze correlations on larger affine spaces
and only then restrict to the inputs reaching the node.
We normalize the program so that all paths reaching a node have the
same past query space, and take a cut at a fixed rank of this space.
For the analysis, complete each future path with independent linear queries
until its answers and the past answers determine the input uniquely.
For fixed future answers $b$, the resulting linear equations have a unique
solution $x=\sigma_b(a)$, affine in the answers $a$ to a basis of the past
query space. We call this parametrization an \emph{affine section}. Distinct sections are related by
a fixed-point-free affine involution. Affine non-malleability bounds
correlations between the function on these sections. A second-moment
estimate and Cauchy--Schwarz permit restriction to the actual past
answers, and summing over the cut proves \Cref{intro:weak}.

Non-malleability under nonzero translations also suffices if we first fix the displacement
between two sections, at an additional entropy cost. This accounts for
the different dependence on the entropy threshold in the two reductions. With independent
auxiliary inputs, fixing those inputs need not preserve the read-once
condition. Instead, we count independent linear combinations of the past queries in
which all auxiliary coefficients cancel, and use this count to choose the
cut. We then project the affine sections onto the original input
coordinates and separate intersecting projected sections by one affine
constraint.
This costs one entropy bit. The vector-valued version uses the same
second-moment argument with a centered list-membership indicator.
These two extensions are proved in
\Cref{subsec:lbp-auxiliary} and \Cref{subsec:lbp-list-prediction}.

For several affine tamperings, we expand a $2r$-th moment. Products
with a section of odd multiplicity have small expectation by
$(2r-1)$-non-malleability. The other products correspond to pairings of the
section indices, which replace the collision probability
$2^{-(n-k)}$ by $O_r(2^{-r(n-k)})$. H\"older's inequality then replaces
Cauchy--Schwarz. With auxiliary inputs, projected sections may meet;
we separate values that occur only once by a bounded number of affine
constraints and use inclusion--exclusion.

The cubic trace function of Gryaznov, Pudl\'ak, and
Talebanfard~\cite[Theorem~15]{GPT22} permits a separate improvement. Its derivative
in a nonzero direction $c\in\F_{2^{n/3}}^3$ has a polar form whose
radical\footnote{For a degree-$2$ polynomial $q$, its polar form is
$B(u,v)=q(u+v)+q(u)+q(v)+q(0)$. Its radical is the subspace of
vectors $u$ for which $B(u,v)=0$ for every $v$.} is the field line $\F_{2^{n/3}}c$. We count displacements for which this line
has a large intersection with the direction of the affine space
obtained by fixing the displacement.
Averaging over the displacements improves the derivative estimate.
Balancing it with the collision probability gives the exponent $2n/15$.
The details appear in \Cref{subsec:lbp-instantiations}.

\subsubsection{Decision Trees}

Let $\delta_r(f)$ be the largest correlation of $f$ with a degree-$r$
polynomial. The indicator of a leaf has a Fourier expansion in its path
queries with coefficient $\ell_1$-norm one. Every term still has degree
at most $r$, even when the queries depend on previous answers. A tree with $L$ leaves
therefore has correlation at most $L\delta_r(f)$ with $f$.
For degree two, a partition into affine spaces on which the polynomial
is constant gives $\delta_2(f)\le2\varepsilon$ from the ordinary affine
extractor guarantee below entropy $n/2$. For higher degree, $r-2$ additive
derivatives leave a sign $(-1)^{p(x)}$ with $\deg p\le2$ and extractor values on an affine
cube. Non-malleability against all its nonzero translations together gives
\[
 \delta_r(f)\le
 \bigl(2\varepsilon+O_r(2^{-n})\bigr)^{1/2^{r-2}}.
\]
The dependence on the error exponent follows from this inequality.

Conversely, degree-$2$ queries compute the initial-seed matrices of our
construction. The successive affine fixings then give a sequence of affine queries,
each determined by previous answers, that computes all iteration outputs. The majorities, code maps, and $\mathsf{XOR}$
require no further queries. At fixed entropy rate and with a fixed
number of tamperings, choosing the smallest scale proportional to
$m+E+\log n$ gives depth $O_{\rho,h}(m+E+\log n)$.

For non-oblivious local trees, we consider the supports of all queries
on one root-to-leaf path. Color these supports red or blue. Let $I$
contain the coordinates incident only to red supports, let $J$ contain
those incident only to blue supports, and put the remaining coordinates
in $S$. No query meets both $I$ and $J$. After the $S$ coordinates are
fixed, the leaf is a rectangle $A\times B$. The partition may depend
on the path, so the argument allows each query to depend on all previous answers.

For sumset extractors, an unbiased coloring gives two large sides.
The read-$\ell$ concentration bound of Gavinsky, Lovett, Saks, and
Srinivasan~\cite{GLSS15} makes both sides large simultaneously.
We express each normalized rectangle indicator as a signed sum of
products of distributions with sufficient entropy. Summing the
absolute coefficients over leaves gives a bound close to one,
rather than the number of leaves. This yields
$\Omega_\delta((n/\ell)\log\ell)$ depth from logarithmic-entropy
sumset extractors with constant error.

For directional affine extractors, a biased coloring keeps $J$ larger than
the extractor's entropy threshold while leaving a smaller set $I$.
Two distinct assignments to $I$ give translates of the same affine
source on $J$, so their extractor outputs have correlation at most
$2\varepsilon$. A second-moment argument followed by Cauchy--Schwarz
bounds a leaf's signed contribution by
$\sqrt{p_\tau(2^{-|I|}+2\varepsilon)}$, where $p_\tau$ is its probability.
If every leaf's partition has $|I|\ge t$, summing over leaves yields
$\operatorname{Corr}(f,T)^2\le2^d(2^{-t}+2\varepsilon)$.
The separator size gives the depth and correlation estimates in
\Cref{intro:local-depth}. Every nonzero output parity of a non-malleable
affine extractor inherits the required directional affine extractor guarantee.
The earlier path-by-path entropy balancing argument gives the sharper
constants at fixed locality in \Cref{tab:local-comparison}.

\subsubsection{Proof Complexity}

\paragraph{Deterministic lossless lifting.}
The required gadget property is simultaneous surjectivity. For an
affine source $X=(X_1,\ldots,X_r)$ with $\ell$-bit blocks, assume
\[
 H_\infty(X_I)\ge\eta\ell|I|
 \qquad\text{for every }I\subseteq[r].
\]
We must hit every output pattern using one fixed family of gadgets.
For distinct fixed tags $c_i$, let
$g_i(x)=\pref_a\bigl(\anmExt(c_i\circ x)\bigr)$, where
$\anmExt$ is the non-malleable affine extractor constructed in this
paper and $a$ is constant.

The main structural lemma groups a positive fraction of the blocks
into disjoint clusters of bounded size. Each cluster has a designated
block with many affine directions independent of all other clusters.
The remaining blocks can be partitioned into a bounded number of
classes, against each of which every collection of designated blocks
retains proportional joint rank. Cauchy--Schwarz removes the product
factors indexed by blocks in these classes.
Non-malleability bounds the contribution of each cluster, while the
joint-rank condition bounds the probability of simultaneous collisions
among the affine inputs created by Cauchy--Schwarz.

For each tuple of output parity coefficients
$v=(v_1,\ldots,v_r)\in(\F_2^a)^r$, let
$S=\{i:v_i\ne0\}$ be its support.
Applying the argument separately to each such support gives
\[
 \left|\E(-1)^{\sum_i v_i\cdot g_i(X_i)}\right|
 \le 2^{-\gamma\ell|S|}
\]
for a constant $\gamma>0$. Choosing $\ell=C\log r+O(1)$ makes
the sum of the absolute values of all nontrivial Fourier coefficients
less than one, so Fourier inversion gives positive probability to every
output pattern.
The clusters are used only in the proof; the construction computes
tagged extractor values without knowing the source.

We combine these gadgets with an explicit sparse inconsistent linear
system. The lifting argument fixes selected input blocks to preserve
sufficient joint entropy in the remaining blocks; surjectivity then
allows the fixed gadget outputs to follow a strategy that assigns base
variables without falsifying a base equation. \Cref{app:lossless-lifting}
gives this argument, the sparse base, and the construction of the clusters.

\paragraph{Overlapping gadgets.}
For the non-malleable gadgets, we control the correlations caused by
overlap through a matching argument. Put an input block on each edge of a
constant-degree graph, and let each vertex gadget depend on its incident
blocks. Choose a large matching among the remaining free edges and fix
the other blocks. Each gadget then depends on at most one remaining block,
so products of gadget signs factor over the matching edges. At the two
endpoints of an edge, prefixing the extractor inputs by different fixed
strings makes them unequal at every point. Non-malleability therefore
bounds their correlation even under unrelated invertible affine changes
of coordinates. The same estimates hold after multiplication by the sign
of any parity of the input bits, allowing Fourier analysis under affine
restrictions.

As parity queries accumulate, we fix additional input blocks until no
set of remaining blocks supports too many independent constraints.
The product estimates then ensure that every output pattern on the
gadgets with free input blocks still occurs. We bound the number of
gadgets whose inputs become fully fixed, so their outputs can be chosen
within a family of partial assignments to the base formula that avoids
falsifying a clause and allows further variables to be assigned. This
lets us answer parity queries while preserving the strategy for the base
formula. We use these answers to find long backward paths in a refutation.
A counting argument finds nodes at which this process can restart while
retaining few additional query equations, yielding the size--depth
tradeoff in \Cref{thm:lifting-intro}. Its additive term comes from
conditioning the gadget outputs
to belong to the chosen family of partial assignments.

For the transfer without this term, we use a constant-size
multiple-tampering extractor. After fixing blocks outside a matching,
non-malleability ensures that a prescribed table occurs for the two
endpoint gadgets. Choosing two majority functions on disjoint triples
restricts the formula to the majority substitution required by
\cite[Theorem~6.1]{IPS26}. This preserves that theorem's tradeoff for
our overlapping gadgets; it does not improve the lifting theorem.

\paragraph{Condensation.}

The improvement for formulas with short ordinary-resolution proofs uses
no extractor. Repeated parity substitutions reduce the number of
variables in the base formula of Buss and Thapen~\cite{BT24} while preserving the
lower bound that every
refutation has either large width or large depth. To control clause
support, we retain successive CNF encodings and track support through
each substitution instead of composing the substitutions at once.
We merge the compressed parts one at a time, using a parity substitution
of constant row weight after each merge. The order limits how many
further substitutions act on parts with larger support bounds, so each
original clause still involves only logarithmically many variables.
The resulting base has a linear number of variables and logarithmic
clause support. Applying the
lifting theorem of Itsykson, Podolskii, and
Shekhovtsov~\cite[Theorem~6.1]{IPS26} then gives the
quadratic tradeoff in \Cref{intro:quadratic}.

\subsection{Organization of the Paper}\label{sec:organization}
\Cref{sec:prelim} collects the probabilistic tools and extractor primitives.
\Cref{sec:affine-nm} gives the non-malleable affine extractor
constructions. \Cref{sec:lbp} gives the lower bounds for linear
branching programs. \Cref{sec:decision-trees} gives the degree-$2$
separation between non-oblivious and oblivious depth and the bounds for higher-degree and local
decision trees, including matching non-oblivious depth bounds for the
multiple-tampering construction.
\Cref{sec:proof-complexity} gives deterministic lossless lifting,
the overlapping-gadget transfer, and the condensation tradeoff for
resolution over parities. \Cref{app:lossless-lifting} contains the
complete proof of deterministic lossless lifting.
\Cref{sec:lbp-appendix} contains the auxiliary-input and list-prediction
extensions, the higher-moment bounds for linear branching programs,
and the detailed calculations
for their explicit instantiations.

\section{Preliminaries}\label{sec:prelim}

All vector spaces are over $\F_2$, and all logarithms are to base two.
We identify $\bin^n$ with $\F_2^n$ and write $[n]=\{1,\ldots,n\}$.
Capital letters denote random variables and lower-case letters denote their
values. We write $\Supp(X)$ for the support of $X$, $x\circ y$ for
concatenation, and $\pref_r(x)$ for the first $r$ bits of $x$.
The Hamming weight $\wt(x)$ is the number of nonzero coordinates of $x$.
For $v,x\in\F_2^r$, write
$\chi_v(x)=(-1)^{\langle v,x\rangle}$ for the linear character indexed
by $v$; the character is trivial when $v=0$.
A polynomial phase is the sign function $x\mapsto(-1)^{p(x)}$
associated with a polynomial $p$ over $\F_2$; its degree refers to
the degree of $p$.
The variable $U_r$ denotes the uniform distribution on $\bin^r$; when it
appears in an ideal joint distribution, it is independent of the other
coordinates. The algebraic degree of a Boolean function is the degree of
its unique multilinear representation over $\F_2$. Affine substitution
does not increase algebraic degree.

\subsection{Statistical Distance, Entropy, and Conditioning}

For distributions $P,Q$ on a finite set $\Omega$, their statistical distance is
\[
 \Delta(P,Q)=\frac12\sum_{\omega\in\Omega}|P(\omega)-Q(\omega)|.
\]
We write $P\approx_\epsilon Q$ when this distance is at most $\epsilon$.
Statistical distance does not increase under deterministic maps. If two joint
distributions have the same marginal on $Z$, their distance is the average,
over $Z$, of their conditional distances. In particular,
\begin{equation}\label{pre:conditional-distance}
 \Delta((W,Z),(U_r,Z))
 =\E_z\Delta((W\mid Z=z),U_r).
\end{equation}
When we fix $Z=z$, we mean that we pass to the conditional distribution
given $Z=z$. Eqn.~\eqref{pre:conditional-distance} expresses the joint error as
the average error over these fixings, weighted by $\Pr[Z=z]$; it does not assert the same error
bound for every fixing. In particular, conditional error bounds average
over the cells of an affine partition without a factor depending on the
number of cells.

The min-entropy of $X$ is
$H_\infty(X)=-\log\max_x\Pr[X=x]$. An $(n,k)$-source is a random
variable on $\bin^n$ with min-entropy at least $k$. Its entropy rate is
$H_\infty(X)/n$. The average conditional min-entropy of $X$ given $Z$ is
\[
 \avgH(X\mid Z)
 =-\log\E_z\max_x\Pr[X=x\mid Z=z].
\]
We use the following standard entropy bounds in the correlation-breaker
analysis; see~\cite{DORS08}.

\begin{lemma}\label{pre:entropy}
If $V$ has a range of size at most $2^b$, then
\[
 \avgH(X\mid Z,V)\ge\avgH(X\mid Z)-b.
\]
Moreover, if $\avgH(X\mid Z)\ge k+q$, then
\[
 \Pr_z[H_\infty(X\mid Z=z)<k]\le2^{-q}.
\]
\end{lemma}
\begin{proof}
For the first assertion, write
\[
 2^{-\avgH(X\mid Z,V)}
 =\sum_{z,v}\max_x\Pr[X=x,Z=z,V=v]
 \le 2^b\sum_z\max_x\Pr[X=x,Z=z]
 =2^b 2^{-\avgH(X\mid Z)}.
\]
The second assertion is Markov's inequality applied to
$\max_x\Pr[X=x\mid Z]$, whose expectation is at most $2^{-k-q}$.
\end{proof}

We write $A\perp B\mid Z$ for conditional independence. If this relation
holds, additionally fixing a function of $(A,Z)$ and a function of $(B,Z)$
preserves independence of the two sides conditioned on all these values.
This follows by factoring their conditional probability mass functions.
The relation is exact, even when statistical approximations are used elsewhere
in the proof.

\begin{definition}[Strong seeded extractor]
A function $L:\bin^N\times\bin^d\to\bin^r$ is a strong $(k,\epsilon)$
seeded extractor if, for every $X$ with $H_\infty(X)\ge k$ and independent
uniform $Y\in\bin^d$,
\[
 (L(X,Y),Y)\approx_\epsilon(U_r,Y).
\]
We call $L$ a \emph{linear seeded extractor} if $L(\cdot,y)$ is linear
for every fixed seed $y$, as in~\cite{Li11b,LZdir}.
No linearity in the seed is required.
\end{definition}

By \Cref{pre:entropy}, a strong
$(k,\epsilon)$ extractor has average conditional error at most
$\epsilon+2^{-q}$ when the source's average conditional min-entropy is
at least $k+q$ and its seed is uniform independently of the source and
the conditioning variable.

\begin{lemma}[Independence merging]\label{pre:merge}
Suppose $(A,A')\perp(Y,Y')\mid Z$ and
\[
 (Y,Y',Z)\approx_\delta(U_d,Y',Z).
\]
Let $L:\bin^N\times\bin^d\to\bin^r$ be a strong
$(k,\epsilon)$ extractor. If
\[
 \avgH(A\mid Z)\ge k+r+\log(1/\epsilon),
\]
then
\[
 (L(A,Y),L(A',Y'),Y,Y',Z)
 \approx_{\delta+2\epsilon}
 (U_r,L(A',Y'),Y,Y',Z).
\]
\end{lemma}
\begin{proof}
First suppose $Y$ is uniform independently of $(Y',Z)$, retaining the given
conditional law of $(A,A')$. Fix $Z,Y'$ and the output $L(A',Y')$.
Independence and \Cref{pre:entropy} give average remaining entropy of
$A$ at least $k+\log(1/\epsilon)$. The seed $Y$ remains independent
uniform, so the average-case strong-extractor guarantee gives error at most $2\epsilon$.

For the general seed law, fix $(y,y',z)$ and use the law of $(A,A')$ given
$Z=z$. The distance of $(L(A,y),L(A',y'))$ from $(U_r,L(A',y'))$ is
a function of $(y,y',z)$ taking values in $[0,1]$.
Changing the marginal of $(Y,Y',Z)$ to $(U_d,Y',Z)$ changes its expectation
by at most $\delta$. Conditional independence ensures that this conditional law remains
unchanged. This proves the assertion, including the
retention of both seeds. This is the $t=1$ case of
the independence-merging lemma in~\cite[Lemma 3.26]{CL22}.
\end{proof}

\subsection{Affine Sources, Fibers, and Sections}

An affine subspace has the form $A=x_0+V$, where $V$ is a linear
subspace. We call $V$ its \emph{direction}; its dimension is
$\dim A=\dim V$. An affine map has the form $J(x)=Mx+b$, with
linear part $J_{\rm lin}=M$. Its rank on $A$ means $\dim M(V)$.
An \emph{affine embedding} is an injective affine map. Restricting
a function to an affine subspace, or expressing it in affine
coordinates on that subspace, is an \emph{affine restriction}.

\begin{definition}[Affine fibers and sections]\label{pre:fibers-sections}
Let $A=x_0+V$ be an affine subspace and let $J$ be an affine map.
\begin{itemize}
\item For $a\in J(A)$, the \emph{fiber} over $a$ is
$A_a=\{x\in A:J(x)=a\}$. If $x_a\in A_a$, then
\[
 A_a=x_a+(V\cap\ker J_{\rm lin}).
\]
Thus all nonempty fibers have the same direction and dimension
$\dim V-\dim J_{\rm lin}(V)$.
\item An \emph{affine section} of $J|_A$ is an affine map
$\sigma:J(A)\to A$ such that $J(\sigma(a))=a$ for every $a\in J(A)$.
It is a right inverse of $J|_A$: its image contains exactly one point
of each fiber. A section is injective, and its image has dimension
$\dim J(A)$.
\end{itemize}
\end{definition}

For example, for $J:\F_2^2\to\F_2$ given by $J(x_1,x_2)=x_1$,
the fiber over $a$ is $\{(a,0),(a,1)\}$. The maps
$\sigma(a)=(a,0)$ and $\tau(a)=(a,1)$ are disjoint affine sections:
each chooses one point in every fiber. In general, two sections of
the same map are called disjoint when their images are disjoint.
Equivalently, they choose different points in every fiber, since an
intersection point would have the same image under $J$.

An affine section always exists: choose a complementary subspace $W$
to $V\cap\ker J_{\rm lin}$ in $V$, and invert $J$ on $x_0+W$.
Here \emph{complementary} means
$V=(V\cap\ker J_{\rm lin})\oplus W$, so every vector in $V$ has a
unique expression as a sum from the two subspaces.

An affine source is uniform on a coset $x_0+V$ of a linear subspace.
Its entropy is $\dim V$, denoted by $H(X)$; for affine sources this
agrees with min-entropy. An affine extractor for entropy $k$ and error
$\epsilon$ is a function $E:\F_2^N\to\bin^r$ satisfying
$E(X)\approx_\epsilon U_r$ for every affine source $X$ with $H(X)\ge k$.

A tuple $(X^0,\ldots,X^h)$ is \emph{jointly affine} if its concatenation
is an affine source. Equivalently, there are a common uniform random
vector $Z$ and affine maps such that $X^j=A_jZ+b_j$ for every $j$.
In particular, $(X,T_1(X),\ldots,T_h(X))$ is jointly affine whenever
$X$ is an affine source and each $T_j$ is affine. The general definition
does not require the other variables to be functions of $X^0$.
Requiring each variable separately to be an affine source would not
impose this condition on their joint distribution.
Every affine image of an affine source is uniform on an affine subspace.
If $X$ is uniform on $A$, then fixing $J(X)=a$ makes $X$ uniform on
the fiber $A_a$. Thus an \emph{affine fiber} in a conditioning
argument is the set of inputs consistent with the fixed values.
For jointly affine variables, conditional entropy is the dimension of the
corresponding conditional support; this dimension is the same on every
nonempty fiber.

\begin{lemma}[Conditioning affine sources]\label{pre:affine}
Let $U$ be uniform on $u_0+V$, and let $J$ and $L$ be affine maps.
Every nonempty fiber $U\mid J(U)=j$ is uniform on a translate of
$V\cap\ker J_{\rm lin}$, where $J_{\rm lin}$ denotes the linear part.
Consequently, its conditional image entropy is independent of $j$ and
\[
 H(L(U)\mid J(U))\ge H(L(U))-H(J(U)).
\]
In particular, imposing at most $b$ affine constraints reduces the rank of
every affine image by at most $b$.
\end{lemma}
\begin{proof}
Every consistent system $J(u)=j$ gives a coset of the stated kernel.
The codimension of that kernel in $V$ is $H(J(U))$. The dimension of a
linear image decreases by at most the codimension of a restriction of its
domain, proving the inequality.
\end{proof}

Applying the affine-conditioning argument of Li~\cite{Li11b} to the
concatenated source gives the following form; see also
\cite[Lemma~8 in the full version]{LZdir}.

\begin{lemma}[Affine conditioning]\label{pre:reservoir}
Let $(Q,Q')$ be jointly affine, and let $J$ be an affine map on the joint
space with output length $r$. There are independent jointly affine tuples
$(A,A')$ and $(B,B')$ such that
\[
 (Q,Q')=(A,A')+(B,B'),
\]
$J(Q,Q')$ is determined by $(B,B')$, and
\[
 H(A)\ge H(Q)-r.
\]
If $V$ is the direction of the joint support, one may take $(A,A')$
uniform on $V\cap\ker J_{\rm lin}$ and $(B,B')$ uniform on a translate
of a complementary subspace of $V$.
\end{lemma}
\begin{proof}
Decompose the joint direction into that kernel and a complementary subspace,
and sample their coordinates independently and uniformly. Assign the affine
offset to the complement. The kernel has codimension at most $r$, so its
selected projection loses at most $r$ dimensions by \Cref{pre:affine}.
The selected projection need not be injective, and no functional relation
between $Q$ and $Q'$ is used.
\end{proof}

\begin{remark}\label{pre:query-def}
We shall repeatedly fix affine functions of a source $U$ in succession.
Each bit has the form $\lambda(U)+b$, where $\lambda$ is linear. The
choice of $\lambda,b$, and whether to stop, depend only on the values
fixed previously and on fixed external parameters. At each step we
condition separately on every value of positive probability.
The resulting supports form an \emph{affine partition}: a partition
into affine subspaces. Its parts may have different directions or
dimensions, because the functions fixed later can depend on earlier
values. The function recording the complete sequence need not be
affine; the next lemma shows that each of its nonempty fibers is
nevertheless an affine subspace.
\end{remark}

\begin{lemma}\label{pre:queries}
Suppose we successively fix at most $L$ affine functions of an affine
source $U$ as above. Conditioned on any possible sequence of values,
$U$ is an affine source whose support has codimension at most $L$ in
the original support. The entropy of any fixed affine image of $U$
decreases by at most $L$.
\end{lemma}
\begin{proof}
Once the sequence of values is specified, each fixing imposes one
affine equation. Their conjunction defines the corresponding conditional
support. Apply \Cref{pre:affine}.
\end{proof}

The entropy bound in \Cref{pre:queries} holds for every sequence of
values of positive probability. When subsequently applying an extractor
theorem to the conditional sources, we average their errors over the
fixed values using Eqn.~\eqref{pre:conditional-distance}. There is no factor
depending on the number of possible values.

\begin{lemma}[Common bad-seed set]\label{pre:rank}
Let $L:\F_2^N\times\bin^d\to\F_2^r$ be a strong linear seeded
$(k,\epsilon)$ extractor. For a fixed linear subspace $V\subseteq\F_2^N$
of dimension at least $k$, put
\[
 \mathcal B_V=\{y:\rank(L(\cdot,y)|_V)<r\}.
\]
Then $\Pr_{U_d}[\mathcal B_V]\le2\epsilon$. Every seed outside this set
gives an exactly uniform output on every translate of $V$.
If $Y\approx_\eta U_d$, then
\[
 \Pr[Y\in\mathcal B_V]\le\eta+2\epsilon.
\]
\end{lemma}
\begin{proof}
Let $U_V$ denote the uniform distribution on $V$. For a fixed seed,
$L(U_V,y)$ is uniform on a linear image. A proper
image has distance at least $1/2$ from $U_r$. Average over an independent
uniform seed in the strong-extractor guarantee. Translation of the source
only translates its image, and statistical distance bounds the change in
the probability of the fixed bad-seed set.
\end{proof}

In applications, we use \Cref{pre:rank} separately on each current
affine part. For any fixed affine map whose values are to be fixed
next, \Cref{pre:affine} gives a common kernel for all its nonempty
fibers. The bad-seed set is determined by this kernel before those
values are fixed. It may depend on values fixed at earlier steps.
Full rank then gives exact uniformity on every surviving conditional
source, even though the source and the seed need not be independent.
The probability of a bad seed must be bounded before fixing the
seed itself.

\subsection{\texorpdfstring{$t$}{t}-Non-Malleable Affine Extractors}

We use the standard $t$-non-malleability formulation~\cite{CRS14},
specialized to affine sources and affine tampering.

\begin{definition}[$t$-Non-Malleable Affine Extractor]
\label{pre:t-affine-nm}
For $t\ge1$, a function $F:\F_2^n\to\F_2^m$ is a
$(k,\epsilon)$ $t$-non-malleable affine extractor if, for every affine
source $X$ with $H(X)\ge k$ and affine maps $T_1,\ldots,T_t$ satisfying
$T_j(x)\ne x$ for every $x\in\Supp(X)$,
\begin{equation}\label{pre:t-affine-nm-condition}
 \bigl(F(X),(F(T_j(X)))_{j\in[t]}\bigr)
 \approx_\epsilon
 \bigl(U_m,(F(T_j(X)))_{j\in[t]}\bigr).
\end{equation}
The uniform variable is independent of all comparison outputs together.
The maps may coincide and need not be invertible. Repeating maps gives
the same guarantee for fewer than $t$ tamperings.
For $t=0$, we use the convention that this means an ordinary
$(k,\epsilon)$ affine extractor.
\end{definition}

\begin{lemma}[Equivalent Jointly Affine Formulation]
\label{pre:affine-nm-equivalence}
\Cref{pre:t-affine-nm} is equivalent, with the same $k,m,\epsilon$,
to requiring
\[
 \bigl(F(X^0),(F(X^j))_{j\in[t]}\bigr)
 \approx_\epsilon
 \bigl(U_m,(F(X^j))_{j\in[t]}\bigr)
\]
for every jointly affine tuple $(X^0,\ldots,X^t)$ with
$H(X^0)\ge k$ and $\Pr[X^0=X^j]=0$ for $j>0$.
\end{lemma}
\begin{proof}
Let $A$ be the joint support, $\pi_0$ its first-coordinate projection,
and $S=\pi_0(A)$. Choose an affine section $\sigma:S\to A$ as in
\Cref{pre:fibers-sections}. If $V$ is the direction of $A$, put
$K=V\cap\ker\pi_0$. The tuple has the representation
\[
 (X^0,\ldots,X^t)=\sigma(X^0)+W,
\]
where $W$ is uniform on $K$ and independent of $X^0$: the map
$(x,w)\mapsto\sigma(x)+w$ is a bijection from $S\times K$ to $A$.

Fixing $W=w$ preserves the law and entropy of $X^0$. Each remaining
coordinate is an affine function $T_{j,w}(X^0)$; extend this function
affinely from $S$ to $\F_2^n$. The pointwise inequalities persist on
every fiber. Apply Eqn.~\eqref{pre:t-affine-nm-condition} and average
over $w$, with no change in the error. Conversely, every tuple
$(X,T_1(X),\ldots,T_t(X))$ in the definition is jointly affine.
\end{proof}

One may equivalently require the tampering maps to have no fixed point
on the entire ambient space. To see this, let $\pi_S:\F_2^n\to S$
be an affine retraction, so $\pi_S(x)=x$ on $S$. For an affine map
$T$ with no fixed point on $S$, define
\[
 \widetilde T(x)=x+T(\pi_S(x))-\pi_S(x).
\]
This affine map agrees with $T$ on $S$, and its difference from $x$
is nonzero everywhere. Such a retraction is obtained by choosing a
complement to the direction of $S$. Thus the two fixed-point conventions
give the same extractor requirement for unrestricted affine maps.

\subsection{Ingredients from Previous Work on Extractors}

\begin{theorem}[Linear seeded extractors]\label{pre:linear}
There is an absolute constant $C_L$ such that, for every pair of integers
$r,q\ge1$ and every $N\ge C_L(r+q)$, there is an explicit strong linear
seeded extractor on $N$-bit sources with the following parameters:
\begin{itemize}
\item output length $r$ and error $2^{-q}$;
\item entropy requirement at most $C_L(r+q)$;
\item seed length at most $C_L(r+q+\log N)$.
\end{itemize}
\end{theorem}
This is the extractor of~\cite{CGL21}, in the parameter form stated in
\cite[Lemma 2.12]{LZ24}. Padding the seed with ignored bits or truncating
the output preserves both the strong property and linearity.

We also use its tunable seed-length bound to increase the output
length in successive applications of a linear seeded extractor.
\begin{theorem}[A linear seeded extractor with a tunable seed length]
\label{pre:linear-tunable}
There is an absolute constant $C$ such that, for positive integers
$t,m,p$ and $N\ge C(m+p)$, there is an explicit strong linear
seeded extractor on $N$ bits with
\begin{itemize}
\item output length $m$;
\item entropy requirement at most $C(m+p)$ and error $2^{-p}$;
\item seed length at most
$C(m/t+\log N+p+\log^2(t)(\log m+p))$.
\end{itemize}
\end{theorem}
This is \cite[Lemma~3.13]{CL22}, attributed there to~\cite{CGL21}.
In particular, fix a sufficiently large absolute $t$. Whenever
$r\ge C'(p+\log N)$ and the entropy requirement is met, the
extractor can output $2r$ bits using at most $r$ seed bits.

\begin{definition}[Strong correlation breaker with advice~{\cite[Definition~3.3]{Li23}}]
A function $\SCB:\bin^N\times\bin^r\times\bin^a\to\bin^t$ is a
strong correlation breaker with advice for source entropy $r$ and error
$\epsilon$ if the following holds. Let $(V,V')$ and $(R,R')$ be arbitrary tuples satisfying
\begin{itemize}
\item $(V,V')\perp(R,R')$;
\item $H_\infty(V)\ge r$;
\item $R=U_r$.
\end{itemize}
For distinct fixed advice strings $\alpha,\alpha'$, one has
\[
 (\SCB(V,R,\alpha),\SCB(V',R',\alpha'),R,R')
 \approx_\epsilon(U_t,\SCB(V',R',\alpha'),R,R').
\]
\end{definition}

\begin{theorem}[Correlation breaker with advice~\cite{Li23}]
\label{pre:scb}
There are absolute constants $C_S,c_S>0$ such that, whenever
$r\ge C_S(a+\log(N/\epsilon))$ and $r\le N$, there is an explicit
strong correlation breaker with advice as above with output length
$\lfloor c_Sr\rfloor$.
\end{theorem}
This parameter form is given in~\cite[Theorem 5.2]{Li23}. The inputs
$V'$ and $R'$ may be arbitrarily correlated with $V$ and $R$,
respectively; they need not be deterministic functions of those inputs.

\begin{theorem}[Bounded-degree affine extractors]\label{pre:affine-ext}
For every fixed $\alpha>0$ there are constants $\gamma_\alpha,
\xi_\alpha>0$ and an integer $D_\alpha$ and an explicit affine extractor
$E:\F_2^N\to\F_2^{\lfloor\gamma_\alpha N\rfloor}$ for entropy
$\alpha N$, with error at most $2^{-\xi_\alpha N}$, such that every
output bit has algebraic degree at most $D_\alpha$.
\end{theorem}
This follows by ignoring the tampered output in the guarantee for the
directional affine extractor in~\cite[Theorem 43 and Lemma 53]{LZdir}. The final
linear encoding preserves the degree bound on each output bit. We use this
theorem at the fixed rate $\alpha=1/2$ in the advice generator. The initial
seeds for the sublinear-entropy construction instead use the ingredients in
\Cref{branch:seed-ingredients}.

We use the following families of explicit binary linear codes. Fix
rational constants $R_0,\delta_0>0$ such that, for every sufficiently
large $t$ and $1\le m\le R_0t$, there is a generator matrix
$M\in\F_2^{m\times t}$ whose nonzero codewords have weight at least
$\delta_0t$. Here the rate is $m/t$, and the relative distance is the
minimum nonzero codeword weight divided by $t$. We also use an encoding
$\Enc:\F_2^N\to\F_2^{O(N)}$ of constant relative distance.

To sample codeword positions, we use the averaging sampler of
Vadhan~\cite{Vad04}, in the form stated in~\cite[Theorem~2.7]{LZ24}.
It uses $O(q+\log N)$ uniform bits to sample $O(q)$ positions of an
$O(N)$-bit string. For any fixed string containing a constant fraction
of ones, the probability that all sampled positions contain zeros is at
most $2^{-q}$. Both the codes and the sampler, including their parameter
choices, are uniformly computable in polynomial time.

\subsection{Proof Complexity Preliminaries}
\label{subsec:proof-complexity-prelim}

\paragraph{Resolution over Parities.}
A linear clause is a disjunction of affine linear equations over
$\F_2$. Resolution over parities~\cite[Section~2.2]{IS20}, denoted $\operatorname{Res}(\oplus)$,
has the resolution rule
\[
 \frac{C\vee(\ell=0)\qquad D\vee(\ell=1)}{C\vee D},
\]
where $\ell$ is linear, and permits semantic weakening from $C$ to $D$
whenever $C$ implies $D$. A refutation is a
directed acyclic graph of inferences ending in the empty clause. Its size
counts line occurrences, and its resolution depth counts resolution
inferences on a longest path from the empty clause backwards to an axiom;
weakening steps do not contribute to this depth. The width of a clause is
the number of equations it contains, and the width of a refutation is
the maximum clause width. The falsifying assignments
of a non-tautological linear clause form an affine subspace. The rank of
the clause is the rank of a linear system defining this subspace; it
may be smaller than the clause width.

\paragraph{Tseitin Formulas.}
For a graph $G=(V,E)$ and vertex charges $b_v\in\F_2$,
the Tseitin formula has a Boolean variable $x_e$ for each edge and
encodes the equations
\[
 \sum_{e\ni v}x_e=b_v\qquad(v\in V)
\]
as a CNF by forbidding the assignments that violate each equation.
If the total charge is odd, that is, $\sum_{v\in V}b_v=1$, adding all
equations gives $0=1$, since
each edge is counted twice. Thus the formula is unsatisfiable.
On a graph of constant degree, this encoding has constant clause width.

\paragraph{Query Games and Winning Strategies.}
In the query game for ordinary resolution, Prover maintains a partial
assignment, asks Adversary for the value of an unassigned variable, and
may forget assigned values. Prover wins by falsifying a clause.
A winning strategy for Adversary prevents this within the prescribed
memory and round bounds; a winning strategy for Prover forces it. The parity version stores a consistent system of affine
equations, queries linear forms, and forgets information by replacing
the system with one it implies; see~\cite[Section~6]{EI26} and
\cite[Section~3]{IPS26}. We use the following form of an Adversary
strategy, in which a query is allowed only when fewer than $w$ variable
values are stored.

\begin{definition}[Legal partial assignments]
\label{def:overlap-legal-family}
Let $\varphi$ be an unsatisfiable CNF on $m$ variables and let
$1\le w\le m$. A family $\mathcal P$ of partial assignments has capacity
$w$ if it contains the empty assignment, is closed under restriction, and
satisfies the following conditions:
\begin{enumerate}
\item No member of $\mathcal P$ falsifies a clause of $\varphi$.
\item If $\rho\in\mathcal P$ has domain of size less than $w$ and $i$ is
unassigned, then at least one value of $i$ extends $\rho$ to a member of
$\mathcal P$.
\end{enumerate}
We restrict the family to domains of size at most $w$. For
$A\subseteq[m]$, $|A|\le w$, write $\mathcal P_A$ for its assignments
with domain exactly $A$. The family is affine if every $\mathcal P_A$ is
an affine subspace of $\F_2^A$.
\end{definition}

The extension condition supplies an answer to each query, and closure
under restriction permits forgetting. The requirement that the family
be affine is an additional property of the families used below.

\paragraph{Safety, Closure, and Surjectivity.}
Partition the input coordinates into $r$ blocks of length $\ell$.
Let $F$ be the row space of the linear parts of a consistent system of
affine equations, and let $E_I$ be the space of linear forms supported
on the blocks in $I\subseteq[r]$. Following~\cite[Section~4]{EI26},
for a threshold $k\ge0$, call $F$ \emph{$k$-safe} if
\[
 \dim(F\cap E_I)\le k|I|\qquad(I\subseteq[r]).
\]
Thus no set of blocks supports too many independent constraints.
If $V=F^\perp$ is the direction of the solution space, then
\begin{equation}
 \dim\pi_I V=\ell|I|-\dim(F\cap E_I).
 \label{eq:deterministic-safety}
\end{equation}
Indeed, the annihilator of $\pi_I V$ inside $E_I$ is $F\cap E_I$.
Safety is therefore equivalent to
$\dim\pi_I V\ge(\ell-k)|I|$ for every $I$. Thus the uniform
solution has at least $(\ell-k)|I|$ bits of entropy on each block set
$I$.

In the lifting arguments, a \emph{closure} selects blocks to handle
explicitly so that the projected equation space on the remaining
blocks is safe. Its size and its behavior under adding or forgetting
equations control the cost of the simulation. We give the particular
closures where they are used; see also~\cite[Section~4]{EI26} and
\cite[Section~2.8]{IPS26}.

Let $G:\F_2^{r\ell}\to\mathcal Y_1\times\cdots\times\mathcal Y_m$
be the joint map defined by the gadgets. The required
\emph{simultaneous surjectivity} condition is that, for every nonempty
affine solution space $A$ whose equation row space is safe,
\[
 G(A)=\mathcal Y_1\times\cdots\times\mathcal Y_m.
\]
This realizes every prescribed output tuple on $A$; it neither requires
uniform output nor follows from surjectivity of the gadgets
individually. This is the gadget property used to extend legal
assignments in lossless lifting~\cite[Sections~5--6]{EI26}.

\paragraph{Stifling Gadgets.}
A Boolean function $g:\F_2^b\to\F_2$ is \emph{$1$-stifling} if,
for every coordinate $i$ and every $a\in\F_2$, one can fix the other
$b-1$ coordinates so that $g$ is constantly $a$, regardless of the
value at coordinate $i$~\cite[Definition~2.3]{IPS26}.
Three-bit majority has this property: fix the other two bits to $a$.

\paragraph{Regularity and Path-Free Refutations.}
Given a refutation, reverse its proof graph and label each node by the negation of its
clause. At a node resolving a linear form $f$, query $f$ and select
the falsified premise;
at a weakening node follow its sole predecessor. The next affine label
is a consequence of the current label and answer. The source label is
empty, sinks imply the negation of an initial clause, and label rank
is at most clause width. This conversion preserves size and resolution depth.

The regularity conditions are stated in this reversed query DAG.
For a query node $v$, let $f_v$ be its query. A refutation is
\emph{top-regular} if $f_v$ lies outside the span of all queries
occurring before $v$ on source-to-$v$ paths. It is
\emph{bottom-regular} if, for each outgoing edge $(v,w)$, $f_v$ lies
outside the span of all queries on paths from $w$ to a sink.
Each span is taken over the union of the indicated paths, not
separately on each path; see~\cite[Section~2.3]{AI25}.

Following~\cite{EI26}, call the resulting DAG \emph{$(a,b)$-path-free}
if no source-to-sink
path splits into a prefix of query rank at least $a$ and a suffix of
query rank at least $b$, measured separately.
Resolution depth less than $a+b$ implies $(a,b)$-path-freeness.

\section{Non-Malleable Affine Extractors}\label{sec:affine-nm}

We give the construction and prove \Cref{intro:linear} and \Cref{intro:sublinear}.
We first specify the correlation breaker, advice generator, and initial
seeds, then combine the iteration outputs and set the parameters.
\Cref{sec:multiple-tampering} extends the construction to several
affine tamperings and tracks the dependence on their number.

\subsection{An Affine Correlation Breaker}
\label{sec:compact-cb}

We begin with the affine correlation breaker used in the construction.
\Cref{merge:general} uses the following two properties of this subroutine.
\begin{itemize}
\item \textbf{Fixing the output with few affine constraints.}
For fixed seed and advice, we can determine the output by successively
fixing a small number of bits. Each bit is an affine function of the input,
and this function may depend on the values fixed previously.
We use this property to fix
the outputs of the other iterations; see \Cref{cb:two-transcripts}.
\item \textbf{Exact uniformity after fixing intermediate variables.}
For the selected iteration, under the hypotheses of
\Cref{cb:affine-fibers}, we fix intermediate variables and the tampered
output. Except for fixings of small total probability, the output on the
original source is then exactly uniform and is an affine function of the
remaining source.
Every nonempty conditional source remains affine. Here the number of
bits fixed need not be small.
\end{itemize}

Both properties follow from the first phase of the affine correlation
breaker of Chattopadhyay and Liao~\cite[Section~6]{CL22}, using the
correlation breaker with advice constructed by Li~\cite[Theorem~5.2]{Li23}. We specify the
intermediate variables fixed in its analysis. For jointly affine inputs,
these fixings leave an affine source on which the output is affine. The
correlation-breaker guarantee bounds the total probability of fixings for
which this output map is not surjective, yielding the exact-uniformity
statement in \Cref{cb:affine-fibers}.

Recall the strong correlation-breaker guarantee in
\Cref{pre:scb}. Let $V,V'\in\bin^N$, and let $(V,V')$ and $(R,R')$
be independent tuples. Assume $H_\infty(V)\ge r$ and $R=U_r$.
No entropy assumption is imposed on $V'$ or $R'$.
By \Cref{pre:scb}, the output length $t$ can be any sufficiently
small constant multiple of $r$, provided $r\ge C(a+\log(N/\epsilon))$.
For distinct fixed advice strings $\alpha,\alpha'\in\bin^a$,
the conclusion is
\begin{equation}\label{cb:eq:standard}
 \Delta\bigl((\SCB(V,R,\alpha),\SCB(V',R',\alpha'),R,R'),
 (U_t,\SCB(V',R',\alpha'),R,R')\bigr)\le\epsilon.
\end{equation}

\subsubsection{Construction}

Fix output length $w$, advice length $a$, and $\epsilon=2^{-q}$, where
$q\ge1$. Assume $a+w+q+\log n\le c_0n$ for a sufficiently small
absolute constant $c_0>0$. We choose it small enough that the seed length
$d$ and entropy threshold $K$ below are at most $n$.
For absolute constants $K_0,K_1$ to be chosen below, set
$r=K_0(a+w+q+\log n)$ and $d=K_1r$.
Use the following strong linear seeded extractors, both supplied by
\Cref{pre:linear} with error $\epsilon$.
\begin{itemize}
\item For the first extraction, take
$L_1:\bin^n\times\bin^e\longrightarrow\bin^r$ with seed length
$e=O(r+q+\log n)$ and entropy threshold $k_1=O(r+q)$.
\item For the final extraction, take
$L_2:\bin^n\times\bin^t\longrightarrow\bin^w$ with seed length
$t=O(w+q+\log n)$ and entropy threshold $k_2=O(w+q)$.
\end{itemize}
Choose the constants and entropy threshold $K$ in the following order.
\begin{itemize}
\item Choose the absolute constant $K_0$ so that
Eqn.~\eqref{cb:eq:standard} supplies $t$ output bits with an $r$-bit seed.
\item Choose $K_1$ so that
\begin{equation}\label{cb:eq:budgets}
 d-2e-2r\ge r+q.
\end{equation}
\item Take the entropy threshold $K$ large enough
that
\begin{equation}\label{cb:eq:reservoir-budget}
 \begin{aligned}
 K&\ge k_1,\\
 K-2r-w&\ge k_2+q.
 \end{aligned}
\end{equation}
\end{itemize}
All five lengths $d,K,e,r,t$ are $O(a+w+q+\log n)$, with absolute
implied constants.

\begin{algorithm}[t]
\caption{$\ACB(x,y,\alpha)$}\label{alg:acb}
\textbf{Input:} Strings $x\in\bin^n$, $y\in\bin^d$, and advice
$\alpha\in\bin^a$.\par
\textbf{Output:} A string in $\bin^w$.\par
\textbf{Subroutines and parameters:} The strong linear seeded extractors
$L_1,L_2$ and the strong correlation breaker $\SCB$ specified above, with
error $\epsilon=2^{-q}$ and lengths satisfying Eqn.~\eqref{cb:eq:budgets}
and Eqn.~\eqref{cb:eq:reservoir-budget}.
\begin{enumerate}
\item Take $S_1=\pref_e(y)$ and compute $R_1=L_1(x,S_1)$.
\item Compute $S_2=\SCB(y,R_1,\alpha)$.
\item Output $L_2(x,S_2)$.
\end{enumerate}
\end{algorithm}

For reference, the intermediate variables satisfy
\begin{equation}\label{cb:eq:construction}
 \begin{aligned}
 S_1&=\pref_e(y),\\
 R_1&=L_1(x,S_1),\\
 S_2&=\SCB(y,R_1,\alpha),\\
 \ACB(x,y,\alpha)&=L_2(x,S_2).
 \end{aligned}
\end{equation}
This is the first phase of the construction in~\cite[Section~6]{CL22},
with $L_2$ chosen to output $w$ bits. The additional phases there handle
several tampered inputs and output extension; neither is needed here.

\begin{lemma}[Correlation breaking with intermediate variables]\label{cb:retained}
There is an absolute constant $C_{\rm cb}$ such that the following holds for
the construction in Eqn.~\eqref{cb:eq:construction}. Let $A,A',B,B'\in\bin^n$
and $Y,Y'\in\bin^d$ be random variables such that $(A,A')$ is
independent of $\mathcal C=(B,B',Y,Y')$. Put $X=A+B$ and
$X'=A'+B'$, and assume $H_\infty(A)\ge K$.  Fix distinct advice strings $\alpha,\alpha'$.  The
second computation uses $(X',Y',\alpha')$, and its intermediate variables
are marked by primes. There is a tuple $\mathcal T$ determining
\[
 B,B',Y,Y',R_1,R_1',S_2,S_2',\ACB(X',Y',\alpha')
\]
such that, if $Y=U_d$, then
\begin{equation}\label{cb:eq:retained}
 \Delta\bigl((\ACB(X,Y,\alpha),\mathcal T),(U_w,\mathcal T)\bigr)
 \le C_{\rm cb}\epsilon.
\end{equation}
The constant $C_{\rm cb}=20$ suffices.  More generally, if
$H_\infty(Y)\ge d-\lambda$, the right-hand side can be replaced by
$C_{\rm cb}2^\lambda\epsilon$.
\end{lemma}

The distance in Eqn.~\eqref{cb:eq:retained} is the average, over the
values fixed for $\mathcal T$, of the distance between the conditional
output distribution and uniform.

\begin{proof}
We follow the conditioning argument for the affine correlation breaker,
allowing the two input sources to be arbitrarily correlated,
as in~\cite[Theorem~3.5]{Li23}.

Write $R_2=\ACB(X,Y,\alpha)$ and
$R_2'=\ACB(X',Y',\alpha')$. For $j=1,2$, define
\[
 \begin{aligned}
 R_{j,A}&=L_j(A,S_j), & R_{j,B}&=L_j(B,S_j),\\
 R_{j,A}'&=L_j(A',S_j'), & R_{j,B}'&=L_j(B',S_j').
 \end{aligned}
\]
Linearity gives $R_j=R_{j,A}+R_{j,B}$ and
$R_j'=R_{j,A}'+R_{j,B}'$, as in the affine correlation-breaker analysis
in~\cite[proof of Theorem~37]{LZdir}. Initially assume $Y=U_d$, and set
\[
 \begin{aligned}
 T_0&=(S_1,S_1',R_{1,B},R_{1,B}'),\\
 T_1&=(T_0,R_1,R_1'),\\
 T_2&=(T_1,S_2,S_2').
 \end{aligned}
\]

\paragraph{First extraction.}
The tuple $T_0$ is a function of $\mathcal C$.  Conditional on $T_0$, the
tuple $(A,A')$ remains independent of $\mathcal C$.  Since $S_1$ is
a uniform seed independent of $A$, the strong-extractor guarantee gives
\begin{equation}\label{cb:eq:first-uniform}
 \Delta((R_1,T_0),(U_r,T_0))\le\epsilon.
\end{equation}
Indeed, conditional on $\mathcal C$, the law of $R_{1,A}$ depends only on
$S_1$; retaining all of $\mathcal C$ does not increase its average
error.  Adding the fixed value $R_{1,B}$ preserves this error.

For each fixed $T_0$, the tuple $(R_1,R_1')$ is a function of $(A,A')$ plus
known shifts.  Consequently
\begin{equation}\label{cb:eq:markov-before-scb}
 (R_1,R_1')\quad\hbox{and}\quad(Y,Y')
 \quad\hbox{are independent conditional on }T_0.
\end{equation}
Moreover, conditioning on $T_0$ decreases the average conditional min-entropy of
$Y$ by at most $2e+2r$ bits, so
\[
 \avgH(Y\mid T_0)\ge d-2e-2r\ge r+q.
\]
Thus $H_\infty(Y\mid T_0=u)\ge r$ for all values $u$ except a set of
total probability at most $\epsilon$.

\paragraph{The standard correlation breaker.}
For a value $u$ of $T_0$, let
$\delta_u=\Delta((R_1\mid T_0=u),U_r)$.  In the conditional law, replace the
marginal of $R_1$ by uniform while keeping the conditional law of $R_1'$ given
$R_1$; choose arbitrary conditional laws at values outside the original
support.  This changes the joint seed tuple by distance $\delta_u$ and, by
Eqn.~\eqref{cb:eq:markov-before-scb}, preserves its independence of $(Y,Y')$.
For values $u$ with $H_\infty(Y\mid T_0=u)\ge r$,
Eqn.~\eqref{cb:eq:standard} applies.  Returning to the original seed law costs at
most $2\delta_u$, once for the actual distribution and once for its ideal
distribution.  Average Eqn.~\eqref{cb:eq:first-uniform} and charge at most
$\epsilon$ for the remaining values of $u$.  We obtain
\begin{equation}\label{cb:eq:second-seed}
 \Delta((S_2,S_2',T_1),(U_t,S_2',T_1))\le4\epsilon.
\end{equation}

Fixing $(R_1,R_1')$ after $T_0$ imposes conditions only on $(A,A')$.
Thus $(A,A')$ and $\mathcal C$ are still independent conditional on $T_1$.  With $T_1$ fixed, the tuple $(S_2,S_2')$ is a function of
$\mathcal C$.  It follows in particular that
\begin{equation}\label{cb:eq:exact-markov}
 (A,A')\quad\hbox{and}\quad\mathcal C
 \quad\hbox{are independent conditional on }T_2.
\end{equation}
All these independence statements hold exactly in the original distribution;
the preceding replacements were used only to bound statistical distance.

\paragraph{Second extraction.}
The tuple $(R_1,R_1')$ has length $2r$, and $T_0$ is independent of
$(A,A')$. Conditioning on these variables therefore gives
\[
 \avgH(A\mid T_1)\ge K-2r.
\]
Apply the independence-merging argument of \Cref{pre:merge} to
$(A,A')$ and $(S_2,S_2')$ conditional on $T_1$.  For completeness, its
specialization here is as follows.  Replace the conditional marginal of
$S_2$ given $(S_2',T_1)$ by uniform, retaining the law of $(S_2',T_1)$ and
the independent conditional law of $(A,A')$.  Eqn.~\eqref{cb:eq:second-seed}
bounds the distance of the complete input laws by $4\epsilon$.  In this
auxiliary law $S_2$ is independent of
$(A,A',S_2',T_1)$. Conditioning further on $S_2'$ does not decrease the
average conditional min-entropy of $A$ given $T_1$, because of conditional
independence. Conditioning on the $w$-bit string
$R_{2,A}'=L_2(A',S_2')$ decreases it by at most $w$ bits.
Therefore
\[
 \avgH(A\mid T_1,S_2',R_{2,A}')\ge K-2r-w\ge k_2+q.
\]
The strong-extractor guarantee for $L_2$, averaged over this conditioning,
has error at most $2\epsilon$.  Returning to the original law costs at
most twice $4\epsilon$.  In the original distribution we thus have
\begin{equation}\label{cb:eq:reservoir-output}
 \Delta((R_{2,A},R_{2,A}',T_2),(U_w,R_{2,A}',T_2))\le10\epsilon.
\end{equation}
This is the single-tampering counterpart of Eqn.~(9)
in~\cite[Section~6]{CL22}, with a conservative error bound.

\paragraph{Including $(B,B',Y,Y')$ in the conditioning tuple.}
Conditional on $T_2$, both $R_{2,A}$ and $R_{2,A}'$ are functions of
$(A,A')$.  The exact independence in~Eqn.~\eqref{cb:eq:exact-markov}
therefore permits adjoining all of $\mathcal C$ to
Eqn.~\eqref{cb:eq:reservoir-output} with no additional error.  Take
\[
 \mathcal T=(T_2,\mathcal C,R_{2,A}').
\]
The shift $R_{2,B}$ and the second output
$R_2'=R_{2,A}'+R_{2,B}'$ are determined by $\mathcal T$.  Adding the known
shift $R_{2,B}$ to $R_{2,A}$ proves
Eqn.~\eqref{cb:eq:retained}, since $C_{\rm cb}=20$.  Equivalently, the
tuple can include $R_2'$ instead of $R_{2,A}'$, since $\mathcal C$ and
$S_2'$ determine $R_{2,B}'$.

\paragraph{A seed with bounded deficiency.}
The entropy deficiency of a $d$-bit seed $Y$ is $d-H_\infty(Y)$;
the hypothesis in the lemma bounds this quantity by $\lambda$.
For a general seed distribution, define an auxiliary distribution by making $Y$
uniform, keeping the conditional law of $(B,B',Y')$ given $Y$, and drawing
$(A,A')$ independently with its original law. Extend the conditional law
of $(B,B',Y')$ arbitrarily at unsupported values of $Y$. The original law
has density
\[
 2^d\Pr[Y=y]\le2^\lambda
\]
relative to this auxiliary distribution. Since $Y$ is included in $\mathcal T$,
the two distributions have identical conditional distributions of $R_2$ on
every value of $\mathcal T$ in the original support. The average conditional
distance from uniform in the original distribution is at most $2^\lambda$
times that in the auxiliary distribution. Applying the
uniform-seed conclusion in the auxiliary distribution proves the claimed
bound for a seed of min-entropy at least $d-\lambda$.
The auxiliary distribution need not be affine.
\end{proof}

\subsubsection{Exact Uniformity on Affine Fibers}

\begin{corollary}[Exact uniformity on affine fibers]\label{cb:affine-fibers}
Under the hypotheses of \Cref{cb:retained}, assume in addition that
$(A,A',B,B',Y,Y')$ is jointly affine and
$H_\infty(Y)\ge d-\lambda$ for some $\lambda\ge0$. Then every nonempty
fiber obtained by fixing the tuple $\mathcal T$ from its proof is affine, and
$W=\ACB(X,Y,\alpha)$ is affine on it. The total probability of fibers on which $W$ is
not exactly uniform on $\bin^w$ is at most
\begin{equation}\label{cb:eq:bad-fibers}
 2C_{\rm cb}2^\lambda\epsilon.
\end{equation}
On every other fiber, $\ACB(X',Y',\alpha')$ is fixed and $W$ is an exactly
uniform affine $w$-bit variable.
\end{corollary}

\begin{proof}
Use the tuple with the second output $R_2'$ in place of
$R_{2,A}'$.  First fix the affine variables $B,B',Y,Y'$.  This fixes
$S_1,S_1',R_{1,B},R_{1,B}'$. With the two first seeds fixed, both
extractor outputs $R_1,R_1'$ are affine in the underlying uniform parameter.  Once their
values have also been fixed, the possibly nonlinear computations defining
$S_2,S_2'$ are determined and impose no further constraints.  Fixing
$R_2'=L_2(X',S_2')$ is one final system of affine constraints.  This proves
that each nonempty fiber of $\mathcal T$ is affine; the output
$L_2(X,S_2)$ is affine there as well.  Its conditional law is uniform on
an affine subspace of $\bin^w$.  If that subspace is proper, its distance
from $U_w$ is at least $1/2$.  The average conditional distance is bounded
by \Cref{cb:retained}, so the total probability of fibers on which the
image is a proper affine subspace is at most twice that bound.
\end{proof}

\begin{remark}\label{cb:two-transcripts}
The tuple $\mathcal T$ includes all of $\mathcal C$. It can be long and
depend on the affine-source decomposition. Fixing $\mathcal T$ gives the
affine partition on which $\ACB(X,Y,\alpha)$ is exactly uniform, apart
from the exceptional fibers in \Cref{cb:affine-fibers}.

To fix the outputs of the other iterations in \Cref{merge:general}, we
use fewer variables. For fixed $y,\alpha$, first fix the $r$ bits of
$R_1$. These are linear functions of the input because $S_1$ is fixed.
The value of $S_2$ is then determined, so the $w$ bits of the final output
are also linear functions of the input and can be fixed next. Thus at
most $r+w$ bits need to be fixed, and each linear function is determined
by $y,\alpha$ and the values fixed previously. This procedure does not
use the decomposition $(X,X')=(A,A')+(B,B')$. The tuple $\mathcal T$
and this sequence of fixings serve different purposes in the proof.
\end{remark}

\subsection{Advice Generation and One Iteration}
\label{sec:branch}

Each iteration begins with the initial-seed construction of Li and
Zhong~\cite[Algorithm~8, Steps~1--5, in the full version]{LZdir}, which
combines one input block with the whole input. We retain this construction
and modify the advice generator and correlation breaker to handle jointly
affine inputs.

The argument for combining the iterations uses two properties: the
initial-seed function has bounded degree, and, once this seed is fixed,
the output can be determined by successively fixing a small number of
affine functions of the input, each chosen according to the values fixed
previously. When analyzing a jointly affine pair, we may impose the
entropy assumptions on either member.

\subsubsection{An Advice Generator for Jointly Affine Sources}

Our advice generator combines the code-sampling construction
of~\cite[Section~3]{Li23} with a linear seeded extractor. The advice
contains the seed, the sampled codeword bits, and the extractor output.
To bound the probability of a collision, we condition on the part of the
seed used by the linear extractor and consider the entropy of the
difference between the two inputs. Codeword sampling handles the case
of small remaining entropy; the linear extractor handles the case of
large remaining entropy.

\begin{lemma}[Advice generator]\label{branch:advice}
There is an absolute constant $c_0>0$ with the following property. For every
integer $p\ge1$ satisfying $p+\log n\le c_0n$, there is an explicit function
\[
 \Adv:\bin^n\times\bin^{r_0}\longrightarrow\bin^a
\]
with seed and advice lengths $r_0,a=O(p+\log n)$ such that the following
holds. Let $(Q,Q',Z,Z')$ be jointly affine, with
$Q,Q'\in\F_2^n$, $Z,Z'\in\bin^{r_0}$, $Z=U_{r_0}$, and
$Q\ne Q'$ at every point of the joint support. Then
\begin{equation}\label{branch:advice-error}
 \Pr[\Adv(Q,Z)=\Adv(Q',Z')]\le2^{-p}.
\end{equation}
For an affine input tuple $(Q,Z)$, we can fix the advice by first fixing
$Z$ and then fixing linear functions of $Q$, with at most $a$ bits fixed
in total. The two advice strings can therefore be fixed jointly with at
most $2a$ bits fixed. The same function may be applied to longer rows by
ignoring the suffix after position $r_0$.
\end{lemma}

\begin{proof}
Put $q_0=p+\lceil\log n\rceil+4$ and $e=2^{-q_0}$. Let
\[
 L_0:\bin^n\times\bin^{d_0}\longrightarrow\bin^{q_0}
\]
be a strong linear seeded extractor with error $e$, entropy threshold
$k_0=O(q_0)$, and seed length $d_0=O(q_0+\log n)$, using
\Cref{pre:linear}. Take $z=z_1\circ z_2$, where $|z_2|=d_0$ and
$|z_1|=C_0q_0$, and set $r_0=C_0q_0+d_0$. The absolute integer $C_0$ is large enough that
$|z_1|\ge2k_0$ and a rate-$1/2$ affine extractor $E_0$ on $|z_1|$ bits,
supplied by \Cref{pre:affine-ext}, has error at most $e$ and enough
output for the sampler below. These requirements are compatible because
the rate-$1/2$ extractor has linear output and exponentially small error.

Let $\Enc$ be a binary linear encoding of constant relative distance.
The sampler $\Samp$ uses $O(q_0+\log n)$ uniform seed bits to sample
$O(q_0)$ codeword positions. For every fixed codeword of constant relative
weight, it misses the support with probability at most $e$. Define
\begin{equation}\label{branch:advice-formula}
 \Adv(x,z)=z_1\circ z_2\circ
   \Enc(x)|_{\Samp(E_0(z_1))}\circ L_0(x,z_2).
\end{equation}
Here the restriction of $\Enc(x)$ lists its bits at the sampled
positions, and $a$ is the total length of the concatenation.
All lengths are $O(p+\log n)$. Choosing $c_0$ small enough ensures the
feasibility of $L_0$ on $n$ bits.

Write $\Delta=Q+Q'$. Joint affineness implies that the conditional entropy
\[
 v=H(\Delta\mid Z_2)
\]
is the dimension of a conditional affine support, and hence is the same
on every nonempty $Z_2$ fiber. The uniformity
of $Z$ makes $Z_1,Z_2$ independent and individually uniform.

If $v\le k_0$, fix $Z_2$ and $\Delta$. On each nonempty fiber, $Z_1$ is
an affine source of entropy at least
\[
 |Z_1|-v\ge |Z_1|/2.
\]
The vector $\Delta$ is nonzero by hypothesis. Thus $\Enc(\Delta)$ is
a fixed codeword of constant relative weight. The extractor followed by
the sampler misses its support with probability at most $2e$ on each
fiber. A collision in~Eqn.~\eqref{branch:advice-formula} implies $Z_1=Z_1'$
and $Z_2=Z_2'$, hence identical sample positions on the two sides. It
therefore implies this miss event, whose unconditional probability is at
most $2e\le2^{-p}$.

If $v>k_0$, the conditional sources $\Delta\mid Z_2=z_2$ have a common
direction of dimension $v$. \Cref{pre:rank} supplies a fixed set of
bad seeds of uniform measure at most $2e$. Outside this set,
$L_0(\Delta,z_2)$ is exactly uniform on $\bin^{q_0}$, despite the
dependence of the affine offset on $z_2$. Consequently
\[
 \Pr[L_0(\Delta,Z_2)=0]\le2e+2^{-q_0}
   =3e\le2^{-p}.
\]
Advice equality implies equality of $Z_2,Z_2'$ and of the last advice
components, and hence implies this zero event because
$L_0(\cdot,z_2)$ is linear for every fixed seed $z_2$.
This proves~Eqn.~\eqref{branch:advice-error}. In both cases the probability
is bounded before conditioning on equality of the two initial seeds.

To bound the number of bits fixed, first fix $z_1,z_2$ in each call. This
determines the sampler positions and the seed of $L_0$. Every remaining
advice bit is then a linear function of $x$. Fixing these bits fixes the
entire advice, with at most its length in bits fixed. Each resulting
nonempty fiber is affine by \Cref{pre:queries}.
\end{proof}

\subsubsection{Growing a Uniform Affine Row}

The initial seed need not be as long as the final row used by the
advice generator and correlation breaker. We grow a short uniform row
by successive linear seeded extractions. The intermediate fixings
remain affine.

Fix an error exponent $p\ge1$ and put $v=p+\lceil\log n\rceil+1$.
By \Cref{pre:linear-tunable}, there are absolute constants $C_0,C_1$
such that, whenever $r\ge C_0v$, a strong linear seeded extractor can
output any $m\le2r$ bits with error $2^{-p}$, seed length at most $r$,
and entropy threshold at most $C_1(m+p)$. Indeed, choose the auxiliary
parameter in that theorem to be a sufficiently large absolute constant;
its seed bound is at most $r/4+O(v)$. Increase $C_0$ to absorb the
second term. Unused seed coordinates are ignored.

Choose a starting length $r_*=\lceil C_0v\rceil$ and a target
$R\ge r_*$. Let $r_0=r_*$ and $r_j=\min\{2r_{j-1},R\}$ until
$r_J=R$. Take strong linear seeded extractors $L_j$ with these output
lengths and error $2^{-p}$. The seed length $d$ of $L_0$ is $O(v)$,
by \Cref{pre:linear}; the seed length of $L_j$, $j>0$, is at most
$r_{j-1}$. Define
\begin{equation}\label{branch:grow}
 \begin{aligned}
 Y_0(x,h)&=L_0(x,h),\\
 Y_j(x,h)&=L_j(x,Y_{j-1}(x,h)),\\
 \mathsf{Grow}(x,h)&=Y_J(x,h).
 \end{aligned}
\end{equation}
The first seed has length $O(p+\log n)$, independently of $R$.
For fixed $h$, fixing each row in order determines the next seed.
The total number of linear bits fixed is at most
$\sum_{j=0}^Jr_j<3R$.

\begin{lemma}[Growing an Affine Row]\label{branch:row-growth}
Let $(Q^0,\ldots,Q^{q_*-1},Y_0^0,\ldots,Y_0^{q_*-1})$ be jointly
affine, where each $Q^j$ has length $n$, every row has length $r_*$,
and $Y_0^0=U_{r_*}$.
Suppose
\[
 H(Q^0)\ge2q_*R+C_1(R+p).
\]
Apply the maps $L_1,\ldots,L_J$ in Eqn.~\eqref{branch:grow} to each
input, starting with its given row. There is an affine partition,
obtained by fixing fewer than $2q_*R$ bits, such that outside total
probability $2J2^{-p}$ the final row on $Q^0$ is exactly uniform on
$\bin^R$ and all final rows are affine on the corresponding part.
The final rows themselves are not fixed in this partition.
\end{lemma}
\begin{proof}
Work on a part where the current rows are affine and the row at
index $0$ is uniform. The nonempty fibers obtained by fixing all
current rows have a common direction. Its projection onto $Q^0$
has dimension at least the current entropy of $Q^0$ minus $q_*r$,
where $r$ is the current row length. This direction is independent
of the values being fixed.

For this direction, \Cref{pre:rank} bounds the uniform measure of
seeds for which the next linear extractor does not have full rank
by $2\cdot2^{-p}$. We call quantities computed on the original input
$Q^0$ \emph{honest}, to distinguish them from those on the comparison
inputs. The current honest row is uniform before it is fixed, so the
same bound holds for its marginal seed, even when the other rows are
correlated with it. On every other fiber the next
honest row is uniform, and all next rows are affine because their
seeds are fixed. Iterate this argument. The number of linear constraints imposed is at most
\[
 q_*\sum_{j=0}^{J-1}r_j<2q_*R,
\]
so the entropy hypothesis supplies every extractor threshold.
Average the failure probabilities at each stage and sum them.
\end{proof}

\subsubsection{Low-Degree Initial Seeds}

In this setting, a \emph{linear somewhere condenser} is a list of linear
maps such that every source under consideration has at least one image
with the specified entropy rate. An \emph{affine somewhere-random source}
is a jointly affine tuple of equal-length rows with at least one row
uniform on the full bit cube; the rows need not be independent.
We apply linear somewhere condensers to a local block and the whole input,
take bilinear products of all pairs of rows, and apply an extractor for
affine somewhere-random sources.

\begin{lemma}[Ingredients for the initial seed]\label{branch:seed-ingredients}
There are absolute constants $\beta>0$ and $A_0\ge1$ with the following
properties. Let $0<\kappa\le1/2$ and $\ell=\Theta(\kappa n)$, in the
parameter ranges considered in \Cref{sec:parameters}.
\begin{enumerate}[label=(\roman*)]
\item There are uniformly explicit linear maps
\[
 \begin{aligned}
 C_b&:\F_2^\ell\longrightarrow\F_2^q &&(b\in[u_0]),\\
 T_a&:\F_2^n\longrightarrow\F_2^q &&(a\in[v_0]),
 \end{aligned}
\]
with $u_0,v_0,n/q\le(2/\kappa)^{A_0}$, such that an affine source of entropy
at least $\kappa\ell/8$ has a $C_b$ image of entropy at least
$(1/2+\beta)q$. An affine source of entropy at least $\kappa n/8$ has a
$T_a$ image with the same guarantee.
\item For every $r\le\beta q/4$, there is a uniformly explicit bilinear
map $\mathsf{IP}_r:\F_2^q\times\F_2^q\to\F_2^r$ such that, for independent
affine sources $U,V$ of entropy at least $(1/2+\beta)q$,
\[
 \Delta((\mathsf{IP}_r(U,V),V),(U_r,V))\le\epsilon_{\rm IP},
\]
where $\epsilon_{\rm IP}\le2^{-c_{\rm IP}q}$ for an absolute
$c_{\rm IP}>0$.
\item An affine somewhere-random source with $T$ rows of length $r$ and
one exactly uniform row, in the range
$r/(2T)^{A_0\log(2T)}\ge(2T)^4$, admits a uniformly explicit extractor
$\mathsf{SRExt}$ with output length at least $r/(2T)^{A_0\log(2T)}$,
error at most $2^{-r/(2T)^{A_0\log(2T)}}$, and coordinate degree at most
$(2T)^{A_0}$, with harmless integer rounding in the output length.
\end{enumerate}
\end{lemma}
\begin{proof}
Part (i) uses the linear somewhere condenser of
\cite[Theorem~18]{LZdir}. Its number of rows and reciprocal output length
are polynomial in the reciprocal entropy rate. The two input lengths
can be matched using the additional $O(\log(1/\kappa))$ applications of
$\mathsf{BasicCond}$ in Steps 1--2 of Algorithm 8 there. Pad the local
and global input lengths separately to powers of two, and use the common
condenser entropy rate $\kappa/16$. The longer input requires
$O(\log(1/\kappa))$ additional halving steps. The output family at each
step includes the two coordinate projections, so some row preserves the
good entropy rate. Both bounds remain polynomial in $1/\kappa$.
Part (ii) is the bilinear inner-product extractor of
\cite[Theorem~38]{LZdir}, restricted to the indicated output length.
Part (iii) is the affine somewhere-random extractor of Li~\cite{Li11b},
in the form recorded in \cite[Theorem~39]{LZdir}. The original
construction~\cite[Algorithms 6.3.9--6.3.10 and Theorem 6.3.12]{Li11a}
uses $O(\log T)$ stages of constant-degree operations, giving coordinate
degree $T^{O(1)}$. Its output and error losses are $T^{O(\log T)}$.
The row-length assumption ensures the size requirements at every level.
The exponents are absolute and do not require $T$ to be constant.
Increasing $A_0$ covers all three parts and their uniform implementations.
\end{proof}

Choose the initial-seed parameters as follows.
\begin{itemize}
\item Local block: fix a projection $P:\F_2^n\to\F_2^\ell$.
\item Iteration output length and error parameter: choose
$s\ge u\ge\lceil\log n\rceil+2$.
\item Initial-seed length: take the seed length $d=O(u+\log n)$ of
$L_0$ in Eqn.~\eqref{branch:grow}, with $p=8u$.
\item Number of seed-matrix rows: $T:=u_0v_0$.
\item Seed-matrix row length: choose $r_{\rm SR}$ large enough that part (iii) of
\Cref{branch:seed-ingredients} supplies $d$ bits with error at most
$2^{-8u}$. A sufficiently large absolute multiple of
$(u+\log n+(2T)^4)(2T)^{A_0\log(2T)}$ suffices. We require
$r_{\rm SR}\le\beta q/4$, as verified in \Cref{sec:parameters}.
\end{itemize}
Define
\begin{align}
 \mathsf{SR}(x)&=
 (\mathsf{IP}_{r_{\rm SR}}(T_a(x),C_b(Px)))_{a\in[v_0],\,b\in[u_0]},
 \label{branch:seed-matrix}\\
 \head(x)&=\pref_d\mathsf{SRExt}(\mathsf{SR}(x)).
 \label{branch:seed-function}
\end{align}
The matrix has $Tr_{\rm SR}$ bits. After increasing an absolute
constant $A$, use the following bounds.
\begin{subequations}\label{branch:seed-cost}
\begin{itemize}
\item Length overhead:
\begin{equation}
 H_\kappa:=(2/\kappa)^{A\log(2/\kappa)}.
\end{equation}
\item Initial-seed degree bound:
\begin{equation}
 D_\kappa:=(2/\kappa)^A.
\end{equation}
\end{itemize}
\end{subequations}
Write $A_{s,u}=Tr_{\rm SR}$. In the range $u\ge(2T)^4$, we may
choose the parameters so that $A_{s,u}\le H_\kappa(u+\log n)$ and
every coordinate of $\head$ has degree at most $D_\kappa$.
Indeed, the entries of $\mathsf{SR}$ have degree two. Once $Px$ is fixed,
they are linear in $x$, so fixing its $A_{s,u}$ entries imposes at most $A_{s,u}$
linear constraints. We do not require $\head$ itself to extract from arbitrary
affine sources on $n$ bits.

\begin{algorithm}[H]
\caption{The initial-seed function $\head(x)$}\label{alg:initial-seed}
\textbf{Input:} A string $x\in\F_2^n$.\par
\textbf{Output:} An initial seed $h\in\bin^d$.\par
\textbf{Subroutines and parameters:} The block projection $P$, maps
$C_b,T_a,\mathsf{IP}_{r_{\rm SR}}$, and $\mathsf{SRExt}$ chosen above.
\begin{enumerate}[leftmargin=2em]
\item Compute $(C_b(Px))_{b\in[u_0]}$ and $(T_a(x))_{a\in[v_0]}$.
\item For each $(a,b)\in[v_0]\times[u_0]$, compute
$\mathsf{IP}_{r_{\rm SR}}(T_a(x),C_b(Px))$ and form $\mathsf{SR}(x)$.
\item Output the first $d$ bits of $\mathsf{SRExt}(\mathsf{SR}(x))$.
\end{enumerate}
\end{algorithm}

\subsubsection{One Iteration}

Choose the remaining ingredients and their parameters in the following order.
\begin{itemize}
\item Advice generator: instantiate \Cref{branch:advice} with
error parameter $p:=4s$.
Write $a$ for its advice length and $r_0$ for its required row prefix.
\item Correlation breaker: set
\begin{equation}\label{branch:internal-error}
 q_C:=2a+4s+\lceil\log(4C_{\rm cb})\rceil+10.
\end{equation}
Instantiate \Cref{cb:retained} with advice length $a$, output length
$s$, and internal error $2^{-q_C}$. Write $d_C$ for its seed length
and $K_C$ for its required min-entropy of $A$.
\item Outer extraction: instantiate Eqn.~\eqref{branch:grow} with
$p=8u$ and target $R\ge\max\{r_0,d_C,r_*\}$, choosing
$R=O(s+\log n)$. The initial seed has the length $d=O(u+\log n)$
specified above.
\end{itemize}
The row-growing step starts with $O(u+\log n)$ seed bits and supplies
the $O(s+\log n)$-bit row required by the advice generator and correlation
breaker. The error $2^{-8u}$ is used by $\mathsf{SRExt}$ and the linear
extractors in $\mathsf{Grow}$; the advice generator and correlation
breaker use errors $2^{-4s}$ and $2^{-q_C}$, respectively.

\begin{algorithm}[H]
\caption{One iteration: the function $W$}\label{alg:branch}
\textbf{Input:} A string $x\in\F_2^n$.\par
\textbf{Output:} A string $W(x)\in\bin^s$.\par
\textbf{Subroutines and parameters:} The initial-seed function $\head$,
$\mathsf{Grow}$, $\Adv$, and $\ACB$, with the lengths specified above.
\begin{enumerate}[leftmargin=2em]
\item Compute $h=\head(x)$ and $z=\mathsf{Grow}(x,h)$.
\item Compute $\alpha=\Adv(x,\pref_{r_0}z)$ and $y=\pref_{d_C}z$.
\item Output $W(x)=\ACB(x,y,\alpha)\in\bin^s$.
\end{enumerate}
\end{algorithm}
For $z_h(x)=\mathsf{Grow}(x,h)$, define
\[
 G(x,h)=\ACB(x,\pref_{d_C}z_h(x),\Adv(x,\pref_{r_0}z_h(x))).
\]
Thus
\begin{equation}\label{branch:pipeline}
 W(x)=G(x,\head(x)).
\end{equation}
The definition applies to every prescribed seed $h$, whether or not
$h=\head(x)$. \Cref{merge:general} uses the degree bound for $\head$
and the number of bits needed to fix $G(\cdot,h)$ separately; no degree
bound is required of $G$.

\subsubsection{Properties of One Iteration}

\begin{proposition}[Properties of one iteration]\label{branch:interface}
There is an absolute constant $C_*\ge1$ for which the following holds
for \Cref{alg:branch}, with $s\ge u\ge\max\{\lceil\log n\rceil+2,(2T)^4\}$.
Assume $\epsilon_{\rm IP}\le2^{-8s}$,
$2\ell\le\kappa n/8$, and
\begin{equation}\label{branch:budget}
 2\ell+2A_{s,u}+C_*(s+\log n)\le\kappa n/4.
\end{equation}
\begin{enumerate}[label=(\roman*)]
\item For every fixed initial seed $h$, we can fix $G(x,h)$ by successively
fixing at most $L\le C_*(s+\log n)$ affine functions of $x$. Each function
is determined by $h$ and the values fixed previously. If $Px$ is fixed,
we can fix $W(x)$ in the same way with at most $A_{s,u}+L$ bits fixed.
\item Let $(Q,Q')$ be jointly affine, with $Q\ne Q'$ at every point of
the joint support. Suppose
\begin{equation}\label{branch:entropy}
 \begin{aligned}
 H(PQ)&\ge\kappa\ell/8,\\
 H(Q)&\ge\kappa n/3.
 \end{aligned}
\end{equation}
Its joint support has a partition into affine subspaces such that, except
on subspaces of total probability at most
\begin{equation}\label{branch:exceptional}
 \epsilon_{s,u}\le2\epsilon_{\rm IP}
 +(2\lceil\log n\rceil+3)2^{-8u}+2\cdot2^{-4s}\le2^{-u},
\end{equation}
$W(Q)$ is an exactly uniform affine $s$-bit variable and $W(Q')$ is
fixed on each subspace.
\end{enumerate}
The conclusion also holds after interchanging $Q$ and $Q'$ in both the
entropy assumptions and the conclusion. The same
constant $C_*$ bounds the initial-seed length by $C_*(u+\log n)$
and all additional entropy requirements by $C_*(s+\log n)$.
\end{proposition}
\begin{proof}
Increase $C_*$ to dominate every length and entropy requirement below.
Eqn.~\eqref{branch:budget} then ensures $4s+\log n\le c_0n$.
Every fixing is made separately on each nonempty fiber; exceptional
probabilities are averaged over the preceding fixings.

\paragraph{Fixing the output for a fixed initial seed.}
First fix the intermediate rows in $\mathsf{Grow}(x,h)$ in order,
using fewer than $3R$ linear bits, then fix the advice as in
\Cref{branch:advice}. This determines the correlation-breaker seed and
advice. In Eqn.~\eqref{cb:eq:construction}, fix $R_1$, which determines
$S_2$, and then fix $L_2(x,S_2)$. At each step, the seed has already been
fixed, so these are affine functions of $x$. The number of bits fixed is
$O(s+\log n)$. If $Px$ is fixed but the initial seed is not, first fix
the $A_{s,u}$ linear entries of $\mathsf{SR}(x)$. These determine $\head(x)$.

\paragraph{A uniform row after fixing the two blocks.}
Put $J=(PQ,PQ')$. An affine decomposition gives
\[
 (Q,Q')=A(J)+(B,B'),
\]
where $A_Q(j)$ is the $Q$-coordinate of $A(j)$, and $(B,B')$ is uniform
on the kernel direction and independent of $J$. In particular,
$H(B)\ge H(Q)-2\ell\ge\kappa n/8$.
Choose a full-input condenser row $T_a(B)$ of entropy at least
$(1/2+\beta)q$, and a local row $V=C_b(PQ)$ with the same guarantee.
They are independent. The strong guarantee of $\mathsf{IP}_{r_{\rm SR}}$
bounds its average conditional distance from uniform by
$\epsilon_{\rm IP}$. For fixed $V$, the output is a linear image of an
affine source; unless its rank is full, its distance from uniform is at
least $1/2$. Bad values of $V$ therefore have probability at most
$2\epsilon_{\rm IP}$.

On the fiber obtained by fixing $J=j$, the selected row of $\mathsf{SR}(Q)$ is
\[
 \mathsf{IP}_{r_{\rm SR}}(T_a(B),V(j))+
 \mathsf{IP}_{r_{\rm SR}}(T_a(A_Q(j)),V(j)).
\]
The second term is fixed. Thus this row is exactly uniform outside
fibers of $J$ of total probability at most $2\epsilon_{\rm IP}$.
This bound uses the marginal law of $V$ and requires no residual entropy
in $PQ$ after $PQ'$ is fixed.

\paragraph{Fixing the seed matrices and growing the row.}
For every nonexceptional fixing of the block pair $J$, both seed
matrices are affine and $\mathsf{SR}(Q)$ has a uniform row. Hence
$\head(Q)$ is $2^{-8u}$-close to uniform. The fibers obtained by
fixing both complete matrices have a common direction, whose
projection onto $Q$ has dimension at least
$H(Q)-2\ell-2A_{s,u}$.
For this direction, the bad-seed set for $L_0$ has uniform measure
at most $2\cdot2^{-8u}$. Its definition is independent of the matrix
values. The marginal near-uniformity of the initial seed therefore
bounds the mass of bad matrix fibers by
\begin{equation}\label{branch:outer-bad}
 3\cdot2^{-8u}.
\end{equation}

Fix the two matrices. On every other fiber, the first row pair is
affine and its honest row is exactly uniform. Apply
\Cref{branch:row-growth} with $q_*=2$. It fixes fewer than $4R$
additional bits and fails with probability at most
$2\lceil\log n\rceil2^{-8u}$. For sufficiently large $C_*$, Eqn.~\eqref{branch:budget} leaves enough
entropy after these fixings to apply each extractor.
On each surviving part the final tuple $(Q,Q',Z,Z')$ is jointly
affine and $Z=U_R$. In particular, this exact uniformity is obtained
before fixing the advice.

\paragraph{Advice and the correlation breaker.}
Apply \Cref{branch:advice} and fix both advice strings by successively
fixing at most $2a$ affine functions. Discard the individual fibers on which they coincide,
with total conditional probability at most $2^{-4s}$. On each remaining fiber the
advice strings are fixed and distinct, while
$Y=\pref_{d_C}Z$ has entropy at least $d_C-2a$.
Apply \Cref{pre:reservoir} with the jointly affine final rows $(Z,Z')$.
Writing the current source as
$(Q,Q')=(A_0,A_0')+(B_0,B_0')$, the pair $(A_0,A_0')$ is independent of
$(B_0,B_0',Y,Y')$, with
\begin{equation}\label{branch:reservoir-entropy}
 H(A_0)\ge H(Q)-2\ell-2A_{s,u}-4R-2a-2R\ge K_C.
\end{equation}
The last inequality follows from Eqn.~\eqref{branch:budget} and the
choice of $C_*$. \Cref{cb:affine-fibers}, with seed deficiency at most
$2a$, supplies the required finer affine partition except on a set of conditional probability at most
$2C_{\rm cb}2^{2a-q_C}\le2^{-4s}$.
Averaging over fibers and adding the exceptional probability from fixing
the block pair, the bound in Eqn.~\eqref{branch:outer-bad}, the failure probabilities
from growing the row, and the
last two exceptional probabilities proves Eqn.~\eqref{branch:exceptional}. Interchanging $Q$ and $Q'$
throughout the argument gives the conclusion with $W(Q')$ uniform and
$W(Q)$ fixed, under the entropy assumptions on $Q'$.
\end{proof}

Part (i) bounds the number of bits fixed for the nonselected iterations.
For the selected iteration, we use the partition in part (ii). The number
of bits fixed to obtain this partition need not be small: the entropy
required for its conditioning argument has already been ensured.

\subsection{Amplification by Majority}\label{amp:section}

We replace the products on disjoint groups of bits in the directional
affine extractor of Li and Zhong~\cite{LZdir} by majority functions. The correlation
bound of Chattopadhyay, Hatami, Lee, Lovett, Tal, and Viola~\cite{CHLLTV26}
allows us to use majority blocks whose length is quadratic in the degree
bound for the initial-seed functions; see \Cref{amp:robust}.

Fix absolute constants $R_0,\delta_0>0$ for an explicit family of binary
linear codes. For $m\le R_0t$, let $M\in\F_2^{m\times t}$ satisfy
\begin{equation}\label{amp:code}
 |aM|\ge\delta_0t\qquad(a\in\F_2^m\setminus\{0\}).
\end{equation}
Decrease the constants if necessary to cover all admissible lengths.
For an integer degree bound $D\ge1$, choose the parameters as follows.
\begin{itemize}
\item Majority block length: $c:=16D^2+1$.
\item Number of blocks: $t:=\lfloor s/c\rfloor$, where $s\ge2c$.
\item Output length: any positive integer $m\le R_0t$.
\end{itemize}
Writing $\operatorname{Maj}_c$ for majority on $c$ bits, define
\begin{equation}\label{amp:definition}
 \begin{aligned}
 V_j(w)&=\operatorname{Maj}_c(w_{(j-1)c+1},\ldots,w_{jc}),\\
 \Phi(w)&=M(V_1(w),\ldots,V_t(w))^{\mathsf T}.
 \end{aligned}
\end{equation}
We call $\Phi$ the majority amplifier. Unused coordinates are ignored. Set
\begin{equation}\label{amp:zeta}
 \zeta_D:=\frac{\delta_0}{2c}.
\end{equation}

\begin{lemma}[Correlation with bounded-degree polynomial phases]\label{amp:robust}
For every nonzero $a\in\F_2^m$ and every polynomial
$P:\F_2^s\to\F_2$ of degree at most $D$,
\[
 \left|\E_{W\sim U_s}(-1)^{a\cdot\Phi(W)+P(W)}\right|
 \le2^{-\zeta_Ds}.
\]
The statement allows $D$ to grow with $s$.
\end{lemma}
\begin{proof}
Let $I$ be the support of $aM$. Fix the coordinates outside its
majority blocks, including the unused coordinates. The remaining
coordinates are uniform, and the restriction of $P$ still has degree
at most $D$. By~\cite[Theorem~1.1]{CHLLTV26}, its correlation with
 the $\mathsf{XOR}$ of these $|I|$ majorities is at most
\[
 (2D/\sqrt c)^{|I|}\le2^{-|I|}.
\]
Average over the fixed coordinates and use
$|I|\ge\delta_0\lfloor s/c\rfloor\ge\delta_0s/(2c)$.
\end{proof}

Only the degree of $P$ is bounded in this lemma. No upper bound on
 the degree of $\Phi$ is used in the combining argument. The map is
 evaluated by counting the ones in each block and applying the code map.

\subsection{Combining the Outputs}\label{merge:section}\label{sec:merger}

We now combine the outputs from \Cref{branch:interface}. For each
$i\in[B]$, let $x_i\in\F_2^\ell$ be the $i$th consecutive block of
$x\in\F_2^n$, and let $\head_i:\F_2^n\to\F_2^{d_i}$ be the initial-seed
function of iteration $i$ from Eqn.~\eqref{branch:seed-function}. As in
Eqn.~\eqref{branch:pipeline}, write
\[
 G_i:\F_2^n\times\F_2^{d_i}\longrightarrow\F_2^{s_i},
\]
where $G_i(x,h)$ evaluates the iteration using $h$ as its initial seed,
and put $W_i(x)=G_i(x,\head_i(x))$.
For every fixed $h$, the output of $G_i(x,h)$ is determined by successively
fixing at most $L_i$ affine functions of $x$. Each function depends only on
$h$ and the values fixed previously, independently of the source distribution.
Once $x_i$ is fixed, the coordinates of its seed matrix are affine functions
of $x$; fixing these $A_i$ bits determines $\head_i(x)$. The seed depends on
the whole input; fixing the local block alone does not fix it.

For a majority block length $c$, put $t_i=\lfloor s_i/c\rfloor$.
Let $M_i$ be the generator matrix of a constant-rate,
constant-distance binary code. Define
\[
 \Phi_i(w)=M_i\left(\operatorname{Maj}_c(w_{(j-1)c+1},\ldots,w_{jc})\right)_{j=1}^{t_i},
\]
where the vector of majority values is a column. The combining map is
\[
 \anmExt(x)=\bigoplus_{i=1}^B\Phi_i(W_i(x)).
\]
Thus we apply majority on disjoint groups of bits, apply the code
maps, and $\mathsf{XOR}$ the resulting strings. This replaces the disjoint
products used in~\cite{LZdir}. The guarantee relies on the joint affine
structure and conditioning properties of \Cref{branch:interface}.

The proof selects an iteration with enough conditional entropy and
fixes the outputs of all other iterations. For the later iterations, it
first guesses the initial seeds, so their outputs can be fixed by affine
constraints. The selected iteration then supplies a uniform affine
output, apart from the exceptional fixings in \Cref{branch:interface}.
We keep the event that all seed guesses are correct inside the
expectation, expand its indicator in polynomial phases, and apply
\Cref{amp:robust}. The theorem allows both the number of iterations and
the degree of the initial-seed functions to grow with $n$.

\subsubsection{The Construction}

\begin{algorithm}[H]
\caption{The non-malleable affine extractor $\anmExt(x)$}\label{alg:affine-nm}
\textbf{Input:} A string $x\in\F_2^n$.\par
\textbf{Output:} A string in $\bin^m$.\par
\textbf{Subroutines and parameters:} The local blocks, output lengths,
initial-seed functions $\head_i$, maps $G_i$, and amplifiers
$\Phi_i$, with parameters satisfying \Cref{merge:general}.
\begin{enumerate}[leftmargin=2em]
\item For each $i\in[B]$, compute $h_i=\head_i(x)$ and $w_i=G_i(x,h_i)$.
\item Compute $z_i=\Phi_i(w_i)$ for each $i\in[B]$.
\item Output $z_1\oplus\cdots\oplus z_B$.
\end{enumerate}
\end{algorithm}

\begin{theorem}[Combining the outputs]\label{merge:general}
Fix $0<\kappa\le1/2$ and positive integers $n,B,\ell$ with $B\ell\le n$.
Partition the first $B\ell$ coordinates of an $n$-bit string into
$B$ consecutive blocks of length $\ell$.  Suppose
\begin{equation}\label{merge:rounding}
 \begin{aligned}
 B&\ge64/\kappa,\\
 2(n-B\ell)&\le\kappa n/4.
 \end{aligned}
\end{equation}
For each $i\in[B]$, let
$\head_i:\F_2^n\to\F_2^{d_i}$ be an initial-seed function from Eqn.~\eqref{branch:seed-function}, with
algebraic degree at most $D$ and seed-matrix length $A_i$.  Let $G_i$ be the map evaluated in iteration $i$
as in \Cref{branch:interface}, with output length $s_i$ and exceptional
probability at most $\varepsilon_i$. For every fixed initial seed, its output
is determined by successively fixing at most $L_i$ affine functions as above.
Let $C_*$ be
an absolute constant that bounds all parameters of an iteration as in that
proposition. The selected source must have entropy at least $\kappa n/3$, and its
$i$th block must have entropy at least $\kappa\ell/8$. Assume the
bilinear and somewhere-random primitive parameters are feasible at every
scale, as in \Cref{branch:interface}, and
\begin{equation}\label{merge:budgets}
 \begin{aligned}
 2\sum_{i=1}^B (A_i+L_i)&\le\frac{\kappa\ell}{16},\\
 \max_i\bigl(2\ell+2A_i+C_*(s_i+\log n)\bigr)&\le\frac{\kappa n}{4}.
 \end{aligned}
\end{equation}
Set $c=16\lceil D\rceil^2+1$, assume $s_i\ge2c$ for all $i$,
and choose a positive integer $m$ satisfying
\begin{equation}\label{merge:code-budget}
 m\le R_0\min_i\lfloor s_i/c\rfloor.
\end{equation}
For each iteration let $\Phi_i:\F_2^{s_i}\to\F_2^m$ be the amplifier
in~Eqn.~\eqref{amp:definition}.  Define
\begin{equation}\label{merge:formula}
 \anmExt(x)=\bigoplus_{i=1}^B
    \Phi_i\bigl(G_i(x,\head_i(x))\bigr).
\end{equation}
Write $Q_i=2\sum_{j>i}d_j$ for the total length of the later initial
seeds on the two inputs, and set
\begin{equation}\label{merge:beta}
 \beta=\max_{i\in[B]}2^{Q_i}
       \bigl(\varepsilon_i+2^{-\zeta_Ds_i}\bigr).
\end{equation}
Then $\anmExt$ is an ordinary affine extractor for entropy $\kappa n$
with error at most $\tfrac12\sqrt{2^m-1}\,\beta$.  Moreover, for every
affine source $X$ with $H(X)\ge2\kappa n$ and every affine map $T$ having
no fixed point in the support of $X$,
\begin{equation}\label{merge:nm-error}
 \Delta\bigl((\anmExt(X),\anmExt(TX)),
          (U_m,\anmExt(TX))\bigr)
 \le2^{m-1}\beta.
\end{equation}
The ideal uniform variable is independent of the second coordinate.
If the initial-seed functions and maps $G_i$ are uniformly computable in
polynomial time with these parameters, so is $\anmExt$.
\end{theorem}

\subsubsection{A Good Block in a Jointly Affine Pair}

\begin{lemma}[Selecting a good block]\label{merge:good-block}
Under the assumptions of \Cref{merge:general}, let $(X,X')$ be jointly
affine, supported on distinct pairs, with
\begin{equation}\label{merge:pair-entropies}
 H(X),H(X')\ge\kappa n.
\end{equation}
There are $g\in[B]$ and $\sigma\in\{0,1\}$ such that, after fixing any
joint prefix $(X_{<g},X'_{<g})$ and then successively fixing any sequence of
at most $2\sum_i(A_i+L_i)$ affine functions, each chosen according to the
previously fixed values, the conditional sources $Z^0,Z^1$ on every
nonempty part satisfy
\[
 \begin{aligned}
 H(Z_g^\sigma)&\ge3\kappa\ell/16,\\
 H(Z^\sigma)&\ge\kappa n/3.
 \end{aligned}
\]
The choices of $g,\sigma$
depend only on the original conditional ranks.
\end{lemma}
\begin{proof}
Neither projection is required to determine the other. For $i\in[B]$
let
\[
 h_i=H(X_i,X_i'\mid X_{<i},X_{<i}').
\]
Here $X_{<i}=(X_1,\ldots,X_{i-1})$, and similarly for $X'_{<i}$.
Since the pair is jointly affine, these conditional entropies are the same
on every nonempty fiber of the conditioning variables.  The omitted coordinates have total length
$2(n-B\ell)$ in the pair.  Hence
\[
 \sum_{i=1}^Bh_i
 \ge H(X,X')-2(n-B\ell)
 \ge3\kappa n/4
 >\kappa B\ell/2.
\]
Let $g$ be the first index with $h_g\ge\kappa\ell/2$.
Then
\begin{equation}\label{merge:prefix-rank}
 H(X_{<g},X_{<g}')=\sum_{i<g}h_i<\kappa n/2.
\end{equation}
Fix a value $v$ in the support of $(X_{<g},X'_{<g})$. Each of the two
sources still has entropy at least $\kappa n/2$.  At least one of the two conditional
block entropies at index $g$ is at least $\kappa\ell/4$, since their
sum is at least $h_g$.  Call that side $\sigma\in\{0,1\}$, with $0$
denoting $X$ and $1$ denoting $X'$.  The choice of $g$ and $\sigma$
can be made from the original conditional ranks, independently of $v$.

Successively fixing at most $2\sum_i(A_i+L_i)$ further affine functions
preserves the affine structure on every nonempty cell. On each cell the
selected block and whole input, denoted by $Z_g^\sigma$ and $Z^\sigma$, satisfy
\begin{align}\label{merge:leaf-ranks}
 H(Z_g^\sigma)&\ge\kappa\ell/4-\kappa\ell/16
                =3\kappa\ell/16,\notag\\
 H(Z^\sigma)&\ge\kappa n/2-\kappa\ell/16
                \ge\kappa n/3.
\end{align}
These inequalities hold in the conditional distribution at each
nonempty cell. In particular, the selected block has the entropy required
by its local condenser. Since $B\ge64/\kappa$ and $B\ell\le n$,
$2\ell\le\kappa n/32$. The second inequality in Eqn.~\eqref{merge:budgets}
supplies the remaining entropy budget. Thus \Cref{branch:interface}
applies on every such cell.
\end{proof}

\subsubsection{Bounding Correlations between the Two Outputs}

\begin{lemma}[Fourier correlations between the two outputs]\label{merge:mixed-correlations}
Under the assumptions of \Cref{merge:general}, every jointly affine
pair $(X,X')$ supported on distinct points and satisfying
Eqn.~\eqref{merge:pair-entropies} obeys
\[
 \left|\E(-1)^{a_0\cdot\anmExt(X)+a_1\cdot\anmExt(X')}\right|\le\beta
 \qquad(a_0,a_1\in\F_2^m\setminus\{0\}).
\]
\end{lemma}
\begin{proof}
Choose $g,\sigma$ by \Cref{merge:good-block}. Fix character indices $a_0,a_1\in\F_2^m\setminus\{0\}$ and a prefix value
$v$ as above.  We bound the corresponding Fourier coefficient.  Write $Z^0=X$, $Z^1=X'$, and
let $\mathcal H_{>g}(Z^0,Z^1)\in\F_2^{Q_g}$ concatenate the two
initial seeds at every index greater than $g$. For a possible value
$h\in\F_2^{Q_g}$, write $h_j^\tau\in\F_2^{d_j}$ for the component
corresponding to iteration $j$ on side $\tau\in\{0,1\}$, and define
\begin{equation}\label{merge:forced}
 F_h^\tau(z)=
  \bigoplus_{j\le g}\Phi_j(G_j(z,\head_j(z)))
  \;\oplus\!\bigoplus_{j>g}\Phi_j(G_j(z,h_j^\tau)),
 \qquad \tau\in\{0,1\},
\end{equation}
Thus $F_h^\tau$ uses the prescribed seeds $h_j^\tau$ in iterations
$j>g$ and the actual seeds in iterations $j\le g$. On the event
$\mathcal H_{>g}=h$, it agrees with $\anmExt$.  Therefore
\begin{align}\label{merge:decompose}
 &\E\bigl[(-1)^{a_0\cdot\anmExt(Z^0)+a_1\cdot\anmExt(Z^1)}
            \mid v\bigr]\notag\\
 &\quad=\sum_{h\in\F_2^{Q_g}}
   \E\bigl[\mathbf1[\mathcal H_{>g}=h]
       (-1)^{a_0\cdot F_h^0(Z^0)+a_1\cdot F_h^1(Z^1)}\mid v\bigr].
\end{align}
The expectation in each summand is conditioned only on $v$; the
possibly nonlinear event $\mathcal H_{>g}=h$ remains inside the expectation.

Fix $h$. For $j<g$, both local blocks are determined by $v$.
Fix their two seed matrices, whose $2A_j$ coordinates are affine after
the local blocks have been fixed. Then determine their outputs by
successively fixing at most $2L_j$ further affine functions. For $j>g$,
use the prescribed seeds in $h$ and determine the two outputs by fixing
at most $2L_j$ affine functions in the same way. This gives an affine
partition $\mathcal L_{v,h}$. Define its weights by
\[
 p_L=\Pr[(Z^0,Z^1)\in L\mid v].
\]
They satisfy $\sum_{L\in\mathcal L_{v,h}}p_L=1$.
The number of bits fixed is at most
$2\sum_{j<g}(A_j+L_j)+2\sum_{j>g}L_j\le2\sum_{j\ne g}(A_j+L_j)$. On each cell $L$,
all outputs in~Eqn.~\eqref{merge:forced}, except the two outputs
at index $g$, are fixed.

By~Eqn.~\eqref{merge:leaf-ranks} and \Cref{branch:interface}, each
cell $L$ admits a further partition into affine subspaces. Call a subcell
exceptional if it fails the conclusion of that proposition; the total
conditional probability of these subcells is at most $\varepsilon_g$.
On every other subcell $C$, the map $W_g(Z^\sigma)$ is affine and uniform
on $\F_2^{s_g}$, while $W_g(Z^{1-\sigma})$ is constant.

Fix a nonexceptional subcell $C=z_C+V_C$. Let $K\subseteq V_C$ be the
kernel of the linear part of the selected output map, and choose a
complement $J$ so that $V_C=K\oplus J$. A uniform point of $C$ has the
unique representation $z_C+k+j$, where $k\in K$ and $j\in J$ are
independent and uniform. Refine $C$ by fixing $k$.  The variable
$W=W_g(Z^\sigma)$ remains uniform on $\F_2^{s_g}$, and the entire
pair $(Z^0,Z^1)$ is now an affine function of $W$.  Consequently every coordinate of $\mathcal H_{>g}$ is a polynomial
of degree at most $D$ in $W$: it is a coordinate of some $\head_j$
composed with an affine function of $W$.

On this refined cell the phase in~Eqn.~\eqref{merge:decompose} has the form
\[
 (-1)^{b_C+a_\sigma\cdot\Phi_g(W)},
\]
where $b_C\in\F_2$ is fixed.  Expand the remaining consistency
indicator as
\begin{equation}\label{merge:indicator}
 \mathbf1[\mathcal H_{>g}(W)=h]
 =2^{-Q_g}\sum_{\lambda\in\F_2^{Q_g}}
       (-1)^{\lambda\cdot(\mathcal H_{>g}(W)+h)}.
\end{equation}
The exponent of each phase on the right is a polynomial of degree at most $D$.
Since $a_\sigma\ne0$, \Cref{amp:robust} and the fact that the
absolute coefficients in~Eqn.~\eqref{merge:indicator} sum to one imply
\begin{equation}\label{merge:good-cell}
 \left|\E\left[
  \mathbf1[\mathcal H_{>g}=h]
  (-1)^{b_C+a_\sigma\cdot\Phi_g(W)}\right]\right|
 \le2^{-\zeta_Ds_g}.
\end{equation}

We now average over the partitions, still without conditioning on
$\mathcal H_{>g}=h$. Let $q_{C\mid L}$ be the conditional probability
of subcell $C$ within cell $L$. The absolute contribution of exceptional
subcells is at most
\[
 \sum_Lp_L\sum_{C\text{ exceptional}}q_{C\mid L}
 \le\varepsilon_g.
\]
For the other subcells, averaging~Eqn.~\eqref{merge:good-cell} first over the
fixings of $k$, then over subcells $C$ and cells $L$, contributes at most
$2^{-\zeta_Ds_g}$.  Thus each summand of~Eqn.~\eqref{merge:decompose} has absolute value at
most
$\varepsilon_g+2^{-\zeta_Ds_g}$.  Summing over the $2^{Q_g}$ guesses
and averaging over $v$ gives
\begin{equation}\label{merge:mixed-bias}
 \left|\E(-1)^{a_0\cdot\anmExt(X)+a_1\cdot\anmExt(X')}\right|
 \le2^{Q_g}(\varepsilon_g+2^{-\zeta_Ds_g})\le\beta.
\end{equation}
The cell probabilities sum to one, so the number of cells causes
no additional loss. The factor $2^{Q_g}$ comes from summing over the
possible initial seeds of the later iterations.

\end{proof}

\subsubsection{From Correlation Bounds to Non-Malleability}

\begin{proof}[Proof of \Cref{merge:general}]
The argument just given requires only that the character on the
selected side be nonzero.  To prove the ordinary affine extractor guarantee, take an
affine source $X$ with $H(X)\ge\kappa n$, choose any nonzero fixed
$u\in\F_2^n$, and consider the artificial pair $(X,X+u)$.  It has no
diagonal points, and its conditional joint block ranks equal the
conditional block ranks of $X$. In particular, the $X$ side may
always be selected at the first good block.  Apply the same proof with
characters $(a,0)$ for $a\ne0$.  This gives
\begin{equation}\label{merge:ordinary-bias}
 |\E(-1)^{a\cdot\anmExt(X)}|\le\beta.
\end{equation}
Parseval's identity and Cauchy--Schwarz give statistical error at most
$\tfrac12\sqrt{2^m-1}\,\beta$.

Now let $H(X)\ge2\kappa n$ and let $T$ be an affine map with no fixed
point on the support of $X$.  If $H(TX)\ge\kappa n$, the pair
$(X,TX)$ satisfies~Eqn.~\eqref{merge:pair-entropies}.  Eqn.~\eqref{merge:mixed-bias} treats characters with both coordinates
nonzero, while~Eqn.~\eqref{merge:ordinary-bias} treats characters $(a,0)$
with $a\ne0$.  In the difference of the two distributions in
Eqn.~\eqref{merge:nm-error}, every Fourier coefficient with first
character zero vanishes.  Each of the remaining
$(2^m-1)2^m$ character expectations has absolute value at most
$\beta$.  Hence
\[
 \Delta\bigl((\anmExt(X),\anmExt(TX)),(U_m,\anmExt(TX))\bigr)
 \le\tfrac12\sqrt{(2^m-1)2^m}\,\beta
 \le2^{m-1}\beta.
\]

If $H(TX)<\kappa n$, fix the complete affine variable $TX$.  Every
nonempty fiber is an affine source, with
\[
 H(X\mid TX)=H(X)-H(TX)\ge\kappa n.
\]
The ordinary affine-extractor guarantee applies on each fiber, while
$\anmExt(TX)$
is fixed.  Averaging the resulting bounds proves
Eqn.~\eqref{merge:nm-error} also in this case.  This proves both statistical guarantees in \Cref{merge:general}.

To evaluate the construction, we compute the $B$ outputs,
apply their amplifiers, and take the $\mathsf{XOR}$. Selecting $g$, prescribing the
later seeds, and partitioning the source are operations used only in the
analysis. Thus the construction is explicit whenever its components are
explicit, as claimed.
\end{proof}

\subsubsection{Choosing the Output Lengths}

The following consequence isolates the numerical condition used in
both entropy regimes.

\begin{corollary}\label{merge:decay}
Under the hypotheses of \Cref{merge:general}, choose positive numbers
$u_i\le\zeta_Ds_i$ and let $u_{\min}=\min_i u_i$. Assume
\[
 \begin{aligned}
 \varepsilon_i&\le2^{-u_i},\\
 2\sum_{j>i}d_j&\le u_i/4\quad(i\in[B]),
 \end{aligned}
\]
and choose the output length so that $m\le u_{\min}/8$.
Then the non-malleability error is at most $2^{-5u_{\min}/8}$.
\end{corollary}
\begin{proof}
The two terms in Eqn.~\eqref{merge:beta} give
$\beta\le2\max_i2^{-3u_i/4}=2^{1-3u_{\min}/4}$.
Multiply by $2^{m-1}$.
\end{proof}

By \Cref{branch:interface}, the initial-seed length is $O(u_i+\log n)$,
while the iteration output length and the number of bits fixed are
$O(s_i+\log n)$. When $u_i\ge\log n$, choosing the $u_i$ to decrease
geometrically with a sufficiently small absolute ratio satisfies the
summation condition in \Cref{merge:decay}. The choice
$s_i=\lceil u_i/\zeta_D\rceil$ ensures $u_i\le\zeta_Ds_i$.
The geometric ratio is independent of the initial-seed degree.

\subsection{Setting the Parameters}\label{sec:parameters}

We now choose the parameters in \Cref{merge:general}, keeping their
dependence on the entropy rate explicit.

\subsubsection{The Common Schedule}\label{param:schedule}

For target entropy $\rho n$, choose the block parameters as follows.
\begin{itemize}
\item Working rate: $\kappa:=\rho/2$.
\item Reserved coordinates: $r_{\rm res}:=\lfloor\kappa n/16\rfloor$.
\item Number of local blocks: $B:=\lceil64/\kappa\rceil$.
\item Block length: $\ell:=\lfloor(n-r_{\rm res})/B\rfloor$.
\end{itemize}
All iterations receive the whole input. Let $T$ be the number of
seed-matrix rows and $q$ their condenser input length. Increasing an
absolute integer $A$ in \Cref{branch:seed-ingredients}, we have
\begin{subequations}\label{param:seed-costs}
\begin{itemize}
\item Row count and input length:
\begin{equation}
 T,n/q\le(2/\kappa)^A.
\end{equation}
\item Length overhead:
\begin{equation}
 H_\kappa:=(2/\kappa)^{A\log(2/\kappa)}.
\end{equation}
\item Initial-seed degree bound:
\begin{equation}
 D_\kappa:=\lceil(2/\kappa)^A\rceil.
\end{equation}
\end{itemize}
\end{subequations}
Take the majority block length $c=16D_\kappa^2+1$ and write
\begin{equation}\label{param:b}
 b:=\zeta_{D_\kappa}=\frac{\delta_0}{2c}.
\end{equation}
In particular, $\log(1/b)=O(\log(2/\kappa))$.

Choose an absolute dyadic ratio $\lambda$ such that
\begin{equation}\label{param:eta}
 \frac1{256C_*}\le\lambda\le\frac1{128C_*},
\end{equation}
where $C_*$ is large enough for \Cref{branch:interface}. Set
\begin{itemize}
\item First error parameter: $u_1:=S$, for an integer $S$ chosen below.
\item Subsequent error parameters:
\begin{equation}\label{param:scales}
 u_{i+1}:=\lfloor\lambda u_i\rfloor\qquad(1\le i<B).
\end{equation}
\item Iteration output lengths: $s_i:=\lceil u_i/b\rceil$.
\end{itemize}
Rounding gives
\begin{equation}\label{param:rounding}
 \begin{aligned}
 \lambda^{i-1}S-2&\le u_i\le\lambda^{i-1}S,\\
 u_j&\le\lambda^{j-i}u_i\quad(j>i).
 \end{aligned}
\end{equation}
Assume for the moment that
$u_B\ge\max\{\lceil\log n\rceil+2,(2T)^4\}$.
Writing $A_i=A_{s_i,u_i}$, the iteration bounds are
\begin{equation}\label{param:branch-costs}
 \begin{aligned}
 d_i&\le C_*(u_i+\log n),\\
 L_i&\le C_*(s_i+\log n),\\
 A_i+L_i&\le H_\kappa(s_i+\log n).
 \end{aligned}
\end{equation}
Here and below increasing $A$ absorbs absolute constants. We use
$u_i\le s_i$, the matrix bound $A_i\le H_\kappa(u_i+\log n)$,
and $s_i\le2u_i/b$. Consequently
\begin{align}
 2\sum_i(A_i+L_i)&\le16H_\kappa S/b,
                  \label{param:queries}\\
 Q_i:=2\sum_{j>i}d_j&\le
       \frac{4C_*\lambda}{1-\lambda}u_i\le u_i/4.
                  \label{param:tail}
\end{align}
The matrix lengths enter the first bound; only the short initial
seeds enter the second.

Let $\overline S=\lceil S/b\rceil$. Choose the seed rows as in
\Cref{sec:branch}, and ensure
\begin{equation}\label{param:branch-error}
 \epsilon_{\rm IP}\le2^{-8\overline S}.
\end{equation}
Then \Cref{branch:interface} gives $\epsilon_i\le2^{-u_i}$.
Choose an absolute rational constant
$\nu\le\min\{1/8,R_0/(2\delta_0)\}$ and take
\begin{equation}\label{param:output}
 1\le m\le\lfloor\nu u_B\rfloor.
\end{equation}
Since $u_i\ge1$ and $bc=\delta_0/2$, the code-dimension bounds hold
(after decreasing $\nu$ if necessary). \Cref{merge:decay} yields
\begin{equation}\label{param:final-error}
 \epsilon_{\rm NM}\le2^{-5u_B/8}.
\end{equation}
It remains to satisfy the entropy constraints
\begin{equation}\label{param:branch-budget}
 \begin{aligned}
 2\sum_i(A_i+L_i)&\le\kappa\ell/16,\\
 2\ell+2A_i+C_*(s_i+\log n)&\le\kappa n/4,
 \end{aligned}
\end{equation}
 together with $r_{{\rm SR},i}\le\beta q/4$.

\subsubsection{Dependence on the Entropy Rate}\label{param:linear}

\begin{theorem}\label{thm:affine-rate-quantitative}
There is an absolute constant $C$ with the following property.
For $0<\rho\le1$, put $E_\rho=n2^{-C/\rho}$. If
$E_\rho\ge C\log n$, there is a uniformly explicit non-malleable
affine extractor on $n$ bits with entropy threshold
$\lceil\rho n\rceil$, at least $\lfloor E_\rho\rfloor$ output bits,
and error at most $2^{-E_\rho}$.
\end{theorem}
\begin{proof}
Choose a sufficiently small absolute rational $c_0>0$ and set
\begin{equation}\label{param:linear-head}
 S=\left\lfloor\frac{c_0b\kappa^{A+2}n}{H_\kappa}\right\rfloor.
\end{equation}
Then $\overline S\le c_0\kappa^{A+2}n/H_\kappa+1$.
The choice of $c_0$ and Eqn.~\eqref{param:queries} give
\begin{equation}\label{param:linear-budgets}
 \begin{aligned}
 2\sum_i(A_i+L_i)&\le\kappa\ell/16,\\
 \max_i r_{{\rm SR},i}&\le\beta q/4,\\
 8\overline S&\le c_{\rm IP}q.
 \end{aligned}
\end{equation}
Indeed, the first right-hand side is $\Theta(\kappa^2n)$,
whereas the corresponding cost is $O(c_0\kappa^{A+2}n)$.
The row bound follows from $q\ge n(\kappa/2)^A$ and
$r_{{\rm SR},i}\le H_\kappa(u_i+\log n)$; the last bound follows
from the same lower bound on $q$. There is constant slack in each
inequality. Also $2\ell\le\kappa n/32$, and all other terms in
Eqn.~\eqref{param:branch-budget} are
$O(H_\kappa\overline S)=O(c_0\kappa^{A+2}n)$.
Thus the entropy constraints in Eqn.~\eqref{param:branch-budget} and
the error bound in Eqn.~\eqref{param:branch-error} hold.

The accumulated losses satisfy
\begin{align}
 \log H_\kappa&=O(\log^2(2/\kappa)),\notag\\
 \log(1/b)&=O(\log(2/\kappa)),\notag\\
 B\log(1/\lambda)&=O(1/\kappa).
 \label{param:geometric-loss}
\end{align}
Thus, for absolute $C_0,C_1$, the unrounded final scale is at least
\begin{equation}\label{param:last-scale}
 \lambda^{B-1}\frac{c_0b\kappa^{A+2}n}{H_\kappa}
 \ge n2^{-C_0/\kappa-C_1\log^2(2/\kappa)}
 \ge n2^{-C_2/\rho}.
\end{equation}
Choose the theorem constant $C>C_2$ sufficiently large.
The hypothesis $E_\rho\ge C\log n$ leaves a factor
$2^{(C-C_2)/\rho}$ between the last expression and $E_\rho$.
It absorbs rounding, the fixed output and error constants, and every
polynomial lower bound in $1/\kappa$, including $(2T)^4$ and $2c$.
In particular, $u_B\ge\max\{\lceil\log n\rceil+2,(2T)^4\}$,
$\nu u_B\ge E_\rho$, and $5u_B/8\ge E_\rho$.
This also justifies the minimum-length assumptions used above.
The rounding condition is checked below. Apply
\Cref{merge:general} and \Cref{merge:decay} and truncate the output if needed.
\end{proof}

\begin{proof}[Proof of \Cref{thm:affine-linear}]
For fixed $\rho>0$, \Cref{thm:affine-rate-quantitative} has
$E_\rho=\Omega_\rho(n)$. Its size hypothesis holds for sufficiently
large $n$, giving linear output and exponentially small error.
\end{proof}

\subsubsection{Entropy of Order \texorpdfstring{$n/\log n$}{n/log n}}
\label{param:sublinear}

\begin{proof}[Proof of \Cref{thm:affine-sublinear} for One Tampering]
Fix $0<\eta<1$ and take $\rho=C_\eta/\log n$.
For a sufficiently large constant $C_\eta$, the quantity
$E_\rho$ in \Cref{thm:affine-rate-quantitative} is at least
$n^{1-\eta/2}$ for all sufficiently large $n$. Its hypotheses
hold, and truncation gives at least $\lfloor n^{1-\eta}\rfloor$
output bits with error at most $2^{-n^{1-\eta}}$.
The extension to every fixed number of tamperings follows from
\Cref{thm:joint-affine-quantitative} below.
\end{proof}

The same estimate gives output length and error exponent
$n^{1-o(1)}$ whenever $\rho^{-1}=o(\log n)$. In particular, every
fixed inverse-polylogarithmic rate $(\log n)^{-\tau}$ with $0<\tau<1$
is allowed. 

\subsubsection{Rounding and Uniform Evaluation}\label{param:explicitness}

The omitted coordinates have length
$r'=n-B\ell<r_{\rm res}+B$. In the admitted range,
$B=O(1/\kappa)$ and $B=o(\kappa n)$, so
\[
 \begin{aligned}
 2r'&\le\kappa n/8+2B\le\kappa n/4,\\
 H(X,X')-2r'&\ge3\kappa n/4>\kappa B\ell/2.
 \end{aligned}
\]
This proves the rounding condition in \Cref{merge:general}.
Eqn.~\eqref{param:rounding} accounts for the integer error scales.

All choices can be made uniformly. Replace a supplied positive
rational rate by a dyadic lower approximation within a factor of two,
and increase the absolute constants accordingly. The hypothesis of
\Cref{thm:affine-rate-quantitative} implies $1/\rho=O(\log n)$.
Thus the row counts, degrees, majority block lengths, and all primitive
lengths are polynomially bounded, while
$H_\kappa=2^{O((\log\log n)^2)}$. The row-growing sequence has at most
$\lceil\log n\rceil$ stages. All component algorithms have absolute
polynomial running-time bounds in these parameters. Majority is
computed by counting input bits; no polynomial expansion is used.
The source-dependent affine partitions and seed guesses occur only
in the proof.

\subsection{Several Affine Tamperings}
\label{sec:multiple-tampering}

To handle several affine tamperings, we restore the independence-merging
rounds of the correlation breaker of Chattopadhyay and
Liao~\cite[Section~6]{CL22}. These rounds allow the output on the original
source to remain nearly uniform given all the tampered outputs jointly.
We track their entropy requirements through the iteration and the final
combination, obtaining the quantitative dependence on the number of
tamperings below.

We use the equivalent formulation in \Cref{pre:affine-nm-equivalence},
where the inputs are affine functions of a common uniform random vector.
The tampered inputs may coincide with one another. To account for this,
we partition the joint support into affine subspaces on which each pair
of tampered inputs is either always equal or always distinct, apart from
a small exceptional set.

\begin{theorem}[Non-malleable affine extractors for jointly affine inputs]
\label{thm:joint-affine-tuple}
For fixed $h\ge1$ and $0<\rho\le1$, there is an explicit family
$F_n:\F_2^n\to\F_2^{m_n}$, with $m_n=\Omega_{\rho,h}(n)$, such that
for all sufficiently large $n$ and every jointly affine tuple
$(X^0,\ldots,X^u)$, $u\le h$, with $H_\infty(X^0)\ge\rho n$ and
$\Pr[X^0=X^j]=0$ for every $j>0$,
\[
 (F_n(X^0),F_n(X^1),\ldots,F_n(X^u))
 \approx_{2^{-\Omega_{\rho,h}(n)}}
 (U_{m_n},F_n(X^1),\ldots,F_n(X^u)).
\]
Here $U_{m_n}$ is independent of all the other outputs.
No entropy condition is imposed on $X^1,\ldots,X^u$, and these
random variables may equal one another. The condition
$\Pr[X^0=X^j]=0$ requires only that each $X^j$ differ from $X^0$
on every outcome of their joint distribution; it does not require
disjoint supports or different marginal distributions.
\end{theorem}

\begin{corollary}[A fixed number of affine tamperings]
\label{thm:fixed-multi-affine}
For every fixed $h\ge1$ and fixed positive entropy rate, there is
an explicit $h$-non-malleable affine extractor with linear output length
and exponentially small error. Each tampering is affine and has no
fixed point on the source support. The maps need not be invertible,
and distinct maps may have coincident outputs.
\end{corollary}

The corollary follows by taking $X^j=T_j(X^0)$. In this case,
the condition in the theorem says that $T_j(x)\ne x$ for every
$x\in\operatorname{Supp}(X^0)$. The theorem bounds the average
statistical distance from uniform after fixing the outputs on
$X^1,\ldots,X^u$, with each fixing weighted by its probability;
it does not assert the bound separately for every fixing.
We first prove \Cref{lem:multi-retained} on conditioning on several
outputs, then quantify the dependence on $\rho$ and $h$. The fixed-parameter
theorem will follow from \Cref{thm:joint-affine-quantitative}.

\subsubsection{The Additional Correlation-Breaker Rounds}

Pad the list of tampered inputs by repetitions to
$H=2^{\lceil\log h\rceil}$, and let $\nu=\log H$.
Fix error $\epsilon=2^{-q}$ with $q\ge w$, where $w$ is the desired
output length. All seed lengths, intermediate output lengths, and
entropy thresholds below are $O_h(a+w+q+\log n)$, with sufficiently
large constants. The required sum is assumed to be a sufficiently
small constant multiple of $n$.

Use a strong seeded extractor $E$ and extractors $L$ that are linear
in their sources. Each subroutine has error $\epsilon$. Choose the intermediate
lengths as follows.
\begin{itemize}
\item The long rows have length $|W_i|=r$, with $r\ge w$.
\item The short rows have length $|P_i|=|V_i|=d_y$.
\item The $Q$ variables have length $d_x$.
\end{itemize}
Start with the first phase from \Cref{sec:compact-cb}:
\[
 \begin{aligned}
 S_1&=\pref_eY,\\
 R_1&=L_1(X,S_1),\\
 S_2&=\SCB(Y,R_1,\alpha),\\
 W_0&=L_r(X,S_2).
 \end{aligned}
\]
For $i=1,\ldots,\nu$, compute
\begin{align*}
 P_{i-1}&=\pref_{d_y}W_{i-1},\\
 Q_{m,i-1}&=E(Y,P_{i-1}),\\
 V_i&=L_m(W_{i-1},Q_{m,i-1}),\\
 Q_{r,i}&=E(Y,V_i),\\
 W_i&=L_r(X,Q_{r,i}).
\end{align*}
Output the first $w$ bits of $W_\nu$ and denote this algorithm by $\ACB_h$.

\begin{lemma}[Conditioning on several outputs]
\label{lem:multi-retained}
Write $X^j=A^j+B^j$ for $0\le j\le H$, and assume that
$\mathcal A=(A^0,\ldots,A^H)$ is independent of
$\mathcal C=(B^0,\ldots,B^H,Y^0,\ldots,Y^H)$.
Suppose $H_\infty(A^0)\ge K$, $Y^0=U_d$, and the fixed advice
satisfies $\alpha^0\ne\alpha^j$ for all $j>0$.
There are choices $K,d=O_h(a+w+q+\log n)$ and a tuple
$\mathcal T$ containing $\mathcal C$ and determining the outputs
at indices $1,\ldots,H$, such that
\[
 \Delta((\ACB_h(X^0,Y^0,\alpha^0),\mathcal T),
                  (U_w,\mathcal T))\le C_h\epsilon.
\]
For $H_\infty(Y^0)\ge d-\lambda$, the bound is
$C_h2^\lambda\epsilon$.
If the full tuple $(A^j,B^j,Y^j)_{j=0}^H$ is jointly affine, every nonempty
set obtained by fixing $\mathcal T$ is affine, and the output
at index $0$ is affine on that set. Except for fixings of total probability at most
$2C_h2^\lambda\epsilon$, this output is exactly uniform after
conditioning on the fixing.
For fixed seed and advice, the output is determined by successively fixing
at most $O_h(a+w+q+\log n)$ affine functions of the input. Each function
depends only on the seed, advice, and previously fixed values, independently
of the source distribution.
\end{lemma}

\begin{proof}
For fixed seeds, linearity decomposes each output of a linear seeded
extractor into two summands. The subscripts $A$ and $B$ denote evaluation
on the corresponding source summands with the seed fixed. For example,
$W_i^j=W_{i,A}^j+W_{i,B}^j$. For the prefix $P_i^j$, these subscripts
denote the corresponding prefixes of $W_{i,A}^j$ and $W_{i,B}^j$;
for $V_i^j$, they denote the evaluations of $L_m$ on the two summands
of $W_{i-1}^j$ with seed $Q_{m,i-1}^j$ fixed.
Let the tuples of variables fixed through round $i$ be
\begin{align*}
 J_0&=(S_1^{[0,H]},R_{1,B}^{[0,H]},R_1^{[0,H]},S_2^{[0,H]}),\\
 J_i&=(J_{i-1},P_{i-1,B}^{[0,H]},P_{i-1}^{[0,H]},
       Q_{m,i-1}^{[0,H]},V_{i,B}^{[0,H]},V_i^{[0,H]},Q_{r,i}^{[0,H]}).
\end{align*}
The superscript $[0,H]$ denotes the tuple of variables at indices
$0,\ldots,H$. For $T\subseteq\{1,\ldots,H\}$, write
$W_{i,A}^{T}=(W_{i,A}^j)_{j\in T}$, and let $E_i$ denote the error bound
after round $i$. The two induction invariants are
\begin{align}
 (W_{i,A}^0,W_{i,A}^{T},J_i)
   &\approx_{E_i}(U_r,W_{i,A}^{T},J_i)
       &&(|T|\le2^i),\label{eq:multi-uniform-invariant}\\
 \mathcal A&\perp\mathcal C\mid J_i.
       \label{eq:multi-independence-invariant}
\end{align}
The second statement is exact. Each variable fixed at this step is,
conditional on the previously fixed variables, a function of
either $\mathcal A$ or $\mathcal C$.
For each such output, first fix the summand determined by
$\mathcal C$; fixing the output then fixes the summand determined by
$\mathcal A$. This proves Eqn.~\eqref{eq:multi-independence-invariant}
without assuming independence among $A^0,\ldots,A^H$.

For the first phase put
$T_0=(S_1^{[0,H]},R_{1,B}^{[0,H]})$ and
$T_1=(T_0,R_1^{[0,H]})$. Conditional on $T_0$, the tuple
$R_1^{[0,H]}$ is independent of $\mathcal C$. Apply the strong
single-tampering correlation breaker to the pair with indices
$0,j$, keeping
$R_1^0,R_1^j,T_0$. The remaining $R_1$ outputs may be adjoined:
given the variables already included they are independent of $\mathcal C$,
which determines $S_2^0,S_2^j$ together with those values.
Thus $S_2^0$ is nearly uniform and independent of $(S_2^j,T_1)$.
Conditional on $T_1$, $\mathcal A$ is independent of the entire $S_2$ tuple.
Apply the independence-merging lemma of~\cite[Lemma~3.26]{CL22}, with
index set $\{j\}$ on the seed side and an empty index set on the source
side. Its conclusion includes all seed values and gives
Eqn.~\eqref{eq:multi-uniform-invariant} for $T=\{j\}$. Since $j$ is
arbitrary, this proves the $i=0$ case of the induction.

For a later round split a set of at most $2^i$ indices from
$\{1,\ldots,H\}$ into two sets of size at most $2^{i-1}$. The four extraction steps
transfer uniformity to $Q_{m,i-1}$, combine the guarantees for the two index sets
in $V_i$, and restore the output length through $Q_{r,i},W_i$.
These are the four applications of the same independence-merging
lemma in~\cite[Section~6, Eqns.~(10)--(11)]{CL22};
Eqn.~\eqref{eq:multi-independence-invariant} supplies their exact
conditional independence premise. The variant allowing different,
arbitrarily correlated sources $A^j$ is also stated
in~\cite[Definition~3.4 and Theorem~3.5]{Li23}.
The conservative recurrence
$E_i\le8E_{i-1}+20\epsilon$, with $E_0=O(\epsilon)$,
gives $E_\nu\le C_h\epsilon$.

To check the entropy requirements, the variables in $J_\nu$ contain at most
$(H+1)O(|R_1|+e+\nu(d_x+d_y))$ bits of information about $\mathcal A$.
Choose $K$ larger than this amount plus the extractor thresholds,
$Hr$, and $q$ additional bits of entropy slack. Choose $d$ analogously,
accounting for the bits fixed from $\mathcal C$ before extraction
from $Y^0$ and leaving $q$ bits above the required entropy thresholds.
Since $h$ is fixed, all lengths have the stated order. For each entropy
check on $A^0$ or $Y^0$, let $U$ denote that source and $Z$ the tuple
fixed at that step. The chosen lengths ensure
$\avgH(U\mid Z)\ge k_{\rm ext}+q$, where $k_{\rm ext}$ is the
required extractor entropy. By \Cref{pre:entropy},
\[
 \Pr_{z\sim Z}\bigl[H_\infty(U\mid Z=z)<k_{\rm ext}\bigr]
 \le 2^{-q}=\epsilon.
\]
A union bound over the $O_h(1)$ entropy checks bounds the total
probability of exceptional fixings by $O_h(\epsilon)$, which is
absorbed into $C_h\epsilon$.

At the last round take
$\mathcal T=(J_\nu,\mathcal C,W_\nu^1,\ldots,W_\nu^H)$.
By Eqn.~\eqref{eq:multi-independence-invariant}, adjoining $\mathcal C$
to Eqn.~\eqref{eq:multi-uniform-invariant} adds no error. The
contributions from $B^0,\ldots,B^H$ are now fixed shifts. This proves the uniform-seed
claim. For a seed with min-entropy at least $d-\lambda$, make $Y^0$
uniform while preserving
the conditional law of the remaining variables in $\mathcal C$ given $Y^0$.
The original law has density at most $2^\lambda$ relative to this
law. Since $Y^0$ is retained, the conditional output distributions
for each value of $\mathcal T$ in the original support agree. Their average distance is
therefore multiplied by at most $2^\lambda$.

For affine inputs, first fix $\mathcal C$. Each subsequent output
of a linear seeded extractor is affine in the underlying source parameter
once its seed is fixed; the nonlinear seed computations are determined
by previously fixed values. Thus each set obtained by fixing $\mathcal T$ is affine,
and the final output at index $0$ is affine on that set. A proper affine image has distance at least $1/2$ from
uniform, proving the exceptional-mass bound. With seed and advice fixed,
successively fix the required linear seeded-extractor outputs in their
order of computation. At each step the seed is already determined, so
each output coordinate is an affine function of the input. Their total
length has the claimed bound. The tuple $\mathcal T$ used in the
conditioning argument contains additional variables; no such length
bound is asserted for $\mathcal T$.
\end{proof}

\subsubsection{Quantitative Correlation-Breaker Costs}

We record the dependence on the number of comparisons in
\Cref{lem:multi-retained}. This is also needed when that number grows.
Put $q_*=h+1$, $v=\lceil\log(2q_*)\rceil$, and let $2^{-p}$ be the
error of each extractor subroutine, with $p\ge w$. Write $z=p+\lceil\log n\rceil$.
The parameter choices in~\cite[Section~6]{CL22}, using
\Cref{pre:scb}, may be taken to satisfy
\begin{itemize}
\item Short intermediate lengths:
$d_y=O(z)$ and $d_x=O(v^2z)$.
\item Long intermediate row: $r=O(q_*z)$, chosen large enough that
$r\ge w$ and $r\ge(q_*+c)d_y$ for the constant required by
independence merging.
\item First-phase lengths: $|R_1|,e=O(a+v^2z)$.
\end{itemize}
The linear extractor bound used here is \Cref{pre:linear-tunable};
for a row of length $O(q_*z)$, it supplies a seed of length
$O(v^2z)$. All constants in these statements are absolute.

Fixing the intermediate variables in the proof of
\Cref{lem:multi-retained} loses at most
$O(q_*[a+v^2z+v(d_x+d_y)])$ bits of entropy from either independent tuple.
Adding the entropy required by the final extractors and $p$ bits of slack
gives
\begin{equation}\label{eq:multi-quantitative-cb}
 \begin{aligned}
 K&\le C\bigl(q_*a+(q_*^2+q_*v^3)z\bigr),\\
 d&\le C\bigl(q_*a+q_*v^3z\bigr).
 \end{aligned}
\end{equation}
For a fixed seed and advice, the total number of bits fixed by the
successive linear functions is $O(a+(v^2+q_*v)z)$. Finally, the recurrence in the proof gives
$E_\nu\le256q_*^3 2^{-p}$ after padding to fewer than $2q_*$
comparisons. These estimates follow by summing the displayed lengths
in the $O(\log q_*)$ rounds; they use no independence among the
comparison inputs.

In an iteration with output length $s$, retain advice error
$2^{-4s}$, so $a=O(s+\log n)$, and choose
\begin{equation}\label{eq:multi-internal-error}
 p=q_*a+4s+\lceil\log(1024q_*^3)\rceil+20.
\end{equation}
The deficiency from fixing all advice strings is at most $q_*a$.
Eqn.~\eqref{eq:multi-quantitative-cb} and
Eqn.~\eqref{eq:multi-internal-error} show that the required outer row,
the fixed-seed evaluation cost, and all additional entropy costs are
bounded by $Cq_*^6(s+\log n)$, with an absolute $C$.

Now choose a separate error parameter $u\le s$, with $u\ge C\log n$.
Use the short initial seed and the row-growing sequence of \Cref{sec:branch},
with subroutine error $2^{-8u}$, to supply the required outer row.
Its target length is polynomial in $q_*$ times $s+\log n$, but its
initial seed still has length
\begin{equation}\label{eq:multi-short-head}
 d_{s,u}\le C(u+\log n)
\end{equation}
with an absolute constant independent of $q_*$. The extra fixing cost
in \Cref{branch:row-growth} is less than $2q_*R$, which is absorbed
by increasing the absolute polynomial bound above. Thus, for some
absolute integer $C_0$, both the evaluation cost and all entropy
requirements for the selected source are at most $Cq_*^{C_0}(s+\log n)$.

The initial seed is nearly uniform before its matrices are fixed.
Fixing the matrices gives the first uniform affine row, as in the
single-comparison proof. Apply \Cref{branch:row-growth} to the full
tuple. Its final row is exactly uniform on every surviving affine
part, so the scale-$s$ advice and correlation-breaker guarantees apply
there without change. Advice collisions have total probability at most
$h2^{-4s}$. The sequence has at most $\lceil\log n\rceil$ stages.
Together with the matrix and breaker failures, this gives iteration
error at most $2^{-u}$ whenever $u\ge C\log n$ and the bilinear
error is at most $2^{-8s}$. The constants remain absolute.

The same algorithm supports any list of at most $h$ comparisons.
A shorter nonempty list is padded in the analysis by repeated
comparisons; their advice strings need differ only from the selected
honest advice. With no comparison, adjoin constant auxiliary inputs
with a different fixed advice in the proof of \Cref{lem:multi-retained}.
This proves the ordinary affine extractor guarantee for the same function.
No construction parameter is changed after a source-dependent fixing.

\subsubsection{The Quantitative Theorem}

\begin{theorem}\label{thm:joint-affine-quantitative}
There is an absolute constant $C$ with the following property. Given
$n$, $h\ge1$, and $0<\rho\le1$, put $q_*:=h+1$. Define
\[
 \Psi:=\dfrac{q_*^2}{\rho}\log(2q_*)+
                  \log^2\!\left(\dfrac{2q_*^2}{\rho}\right)
\]
and set $E_*:=n2^{-C\Psi}$.
If $E_*\ge C\log n$, there is a uniformly polynomial-time function
$F:\F_2^n\to\F_2^m$, with $m\ge\lfloor E_*\rfloor$, such that
for every jointly affine tuple $(X^0,\ldots,X^u)$, $u\le h$, with
$H(X^0)\ge\rho n$ and $\Pr[X^0=X^j]=0$ for $j>0$,
\[
 (F(X^0),F(X^1),\ldots,F(X^u))
 \approx_{2^{-E_*}}(U_m,F(X^1),\ldots,F(X^u)).
\]
The uniform variable is independent of the comparison outputs.
No entropy assumption is imposed on the comparisons, which may
coincide with one another. The running-time exponent is absolute.
\end{theorem}

\begin{proof}[Proof for Pairwise Distinct Inputs]
First suppose all inputs differ pairwise at every point. We construct
the function for honest rate $\rho_0$; the theorem will use
$\rho_0=\rho/2$. Set
\begin{itemize}
\item Whole-input working rate: $\kappa:=\rho_0/(4q_*)$.
\item Local working rate: $\theta:=\kappa/q_*$.
\item Block count and length: $B:=\lceil256/\theta\rceil$ and
$\ell:=\lfloor n/B\rfloor$.
\item Reciprocal rate parameter: $\mathcal R:=2/\theta=8q_*^2/\rho_0$.
\end{itemize}
We distinguish the whole-input rate from the local rate: fixing all
$q_*$ local blocks costs $q_*\ell\le\kappa n/256$, although the
local block may have only $\theta\ell/8$ entropy. The seed ingredients
are instantiated at rate $\theta$; after the blocks are fixed, the
whole source still has more than their required $\theta n/8$ entropy.

Consider an output Fourier coefficient whose coefficient at index
$0$ is nonzero. Omit zero-coefficient inputs, and repeatedly fix and
remove a comparison whose current rank is less than $\kappa n$.
These operations fix fewer than $h\kappa n$ bits in total, leaving
more than $3\rho_0n/4$ honest entropy. Every remaining whole input
has rank at least $\kappa n$.

Let $h_i$ be the joint conditional rank of block $i$ given the
preceding joint blocks. The omitted-coordinate loss is at most
$q_*B$. In the parameter range verified below this is at most
$\kappa n/4$, so $\sum_i h_i\ge3\kappa n/4$. This subtraction bounds
the entropy of the local blocks; the omitted coordinates are not fixed.
The first $g$ with $h_g\ge\kappa\ell/2$ has preceding joint rank
less than $\kappa n/2$. After fixing that prefix, all active whole
inputs retain rank at least $\kappa n/2$, and one active local block
has rank at least $\theta\ell/2$.

The nonselected outputs are fixed as in \Cref{merge:mixed-correlations},
using prescribed initial seeds after index $g$. It suffices to ensure
\begin{equation}\label{eq:multi-fixing-budget}
 q_*\sum_i(A_i+L_i)\le\theta\ell/4.
\end{equation}
Then the selected block retains rank at least $\theta\ell/4$ and
its whole input retains rank at least $\kappa n/3$.
The proof for the selected iteration just given applies under
\begin{equation}\label{eq:multi-branch-budget}
 q_*\ell+q_*A_i+Cq_*^{C_0}(s_i+\log n)\le\kappa n/4.
\end{equation}
This bounds the total entropy lost by fixing the local blocks, matrices,
intermediate rows, and advice, while leaving the entropy required by the correlation
breaker. The available entropy here is measured by the whole-input rate
$\kappa$.

Let $q_{\rm row}$ be the condenser input length. Increase an absolute
integer $A$ and define $\mathcal H=2^{A\lceil\log \mathcal R\rceil^2}$
so that the seed ingredients satisfy
\begin{equation}\label{eq:multi-seed-costs}
 \begin{aligned}
 T,n/q_{\rm row},D&\le \mathcal R^A,\\
 A_i+L_i&\le\mathcal H(s_i+\log n).
 \end{aligned}
\end{equation}
Polynomial factors
in $q_*$ are absorbed into $\mathcal H$, since $q_*^2\le \mathcal R$.
Take $D=\lceil \mathcal R^A\rceil$, majority block length $c=16D^2+1$,
and $b=\delta_0/(2c)$. Choose
\begin{itemize}
\item A dyadic ratio $\lambda$ between $1/(C_2q_*)$ and
$2/(C_2q_*)$, for a sufficiently large absolute $C_2$.
\item For the initial error scale, set
\begin{equation}\label{eq:multi-start-scale}
 S:=\left\lfloor\frac{bn}{C_3\mathcal R^{A+C_0+4}\mathcal H}\right\rfloor.
\end{equation}
\item Set $u_1=S$ and define the subsequent error scales by
$u_{i+1}=\lfloor\lambda u_i\rfloor$.
\item Use iteration lengths $s_i=\lceil u_i/b\rceil$.
\end{itemize}
For $u_B\ge C\log n$, Eqn.~\eqref{eq:multi-short-head} gives
\begin{equation}\label{eq:multi-tail-budget}
 q_*\sum_{j>i}d_j\le u_i/4.
\end{equation}
The sums of the $s_i$ and of their rounding errors are
$O(S/b)$. Thus Eqn.~\eqref{eq:multi-seed-costs} and
Eqn.~\eqref{eq:multi-start-scale} verify
Eqn.~\eqref{eq:multi-fixing-budget} and
Eqn.~\eqref{eq:multi-branch-budget}, for a sufficiently large $C_3$.
For example, the first cost is at most $Cq_*\mathcal HS/b$,
whereas $\theta\ell/4=\Theta(n/\mathcal R^2)$.
The same choice gives
$r_{{\rm SR},i}\le\beta q_{\rm row}/4$ and
$8\max_i s_i\le c_{\rm IP}q_{\rm row}$, since
$q_{\rm row}\ge n/\mathcal R^A$. Hence every iteration has error at most
$2^{-u_i}$.

The proof of \Cref{merge:mixed-correlations} now applies to the
remaining tuple. After fixing the kernel coordinates of the selected
output map as in that proof, all later initial-seed coordinates still have degree at most
$D$. Every Fourier coefficient under consideration has absolute value
at most
\begin{equation}\label{eq:multi-phase-bound}
 \beta=\max_i2^{q_*\sum_{j>i}d_j}
                    (2^{-u_i}+2^{-bs_i})
       \le2^{1-3u_B/4}.
\end{equation}
We average the conditional bounds over the fixings that removed inputs
of small rank and over the subsequent affine partitions, so their numbers
of parts do not multiply the bound. Take
$m\le\nu u_B/q_*$ for a sufficiently small absolute $\nu>0$.
The code-dimension constraints hold, and Parseval gives
\begin{equation}\label{eq:multi-output-error}
 \Delta((F(X^0),Z),(U_m,Z))
 \le\tfrac12\sqrt{(2^m-1)2^{(q_*-1)m}}\,\beta
 \le2^{-5u_B/8},
\end{equation}
where $Z$ denotes the comparison outputs.

Finally,
\begin{equation}\label{eq:multi-schedule-loss}
 \begin{aligned}
 B\log(1/\lambda)&=O\!\left(\frac{q_*^2}{\rho_0}\log(2q_*)\right),\\
 \log\mathcal H&=O(\log^2 \mathcal R).
 \end{aligned}
\end{equation}
All remaining factors in $u_B/q_*$ are polynomial in $\mathcal R$.
It follows that the available output length and error exponent are
at least
\[
 n2^{-C_4[(q_*^2/\rho_0)\log(2q_*)+\log^2(2q_*^2/\rho_0)]}.
\]
Choose the theorem constant larger than $C_4$ by a sufficiently
large absolute factor. The resulting slack dominates every polynomial
in $\mathcal R$, verifying $u_B\ge\max\{C\log n,(2T)^4\}$, all
intermediate-row requirements, and the omitted-coordinate condition
$q_*B\le\kappa n/4$. It also absorbs floors and the constants in
Eqn.~\eqref{eq:multi-output-error}.

The size hypothesis implies $q_*^2/\rho=O(\log n)$.
All row counts and lengths are polynomially bounded, with
$\mathcal H=2^{O((\log\log n)^2)}$.
The row-growing sequence has $O(\log n)$ stages and the breaker has
$O(\log h)$ rounds. The primitive algorithms have absolute polynomial
running-time bounds. This proves the pairwise-distinct case uniformly
in $h$ and $\rho$.
\end{proof}

\subsubsection{Equalities among the Tampered Inputs}

\begin{lemma}[Partitioning according to equalities among tampered inputs]
\label{lem:collision-partition}
Let $(X^0,\ldots,X^h)$ be jointly affine and let
$P=\binom h2$. For every integer $L\ge1$, there is a partition
obtained by successively fixing at most $PL$ affine functions, each
determined by the values fixed previously, such that, outside total
mass $P2^{-L}$, for every pair $1\le i<j\le h$, either $X^i=X^j$ throughout
the corresponding part or $X^i\ne X^j$ throughout that part.
Each nonempty part is affine, and conditioning on any such part
reduces the entropy of $X^0$ by at most $PL$.
\end{lemma}

\begin{proof}
Process each pair $X^i,X^j$ with $1\le i<j\le h$ on each
current part of the partition. The difference
of a pair is uniform on an affine image. If zero is outside this
image, the pair is already distinct. Otherwise the image is a
linear space of dimension $d$. Successively fix independent coordinates
of this image until a coordinate has value one, certifying distinctness
on the resulting part. If $d\le L$, fixing all $d$ coordinates to zero
certifies equality. If $d>L$, declare the part with $L$ initial zeros
exceptional; its conditional mass is $2^{-L}$. Earlier certificates persist
under restriction. Summing the costs and exceptional masses proves
the lemma.
\end{proof}

\begin{proof}[Completion of \Cref{thm:joint-affine-quantitative}]
Use the pairwise-distinct construction at rate $\rho_0=\rho/2$.
If $P=\binom h2>0$, take
$L=\lfloor\rho n/(4P)\rfloor$ in \Cref{lem:collision-partition}.
On every nonexceptional part retain one representative of each
comparison equality class. The honest source still has entropy at
least $3\rho n/4$, the representatives differ pointwise, and none
can equal the honest input. Apply the proved guarantee for the same
function built for the original upper bound $h$. Duplicate outputs
to restore the list and average over the parts.

The additional error is $P2^{-L}$, with exponent
$\Omega(\rho n/q_*^2)$. Increasing the theorem constant makes this
smaller than half the target error and leaves the other half for the
tuple construction. Indeed $E_*=n2^{-C\Psi}$ is smaller by an
arbitrarily large constant factor than $\rho n/q_*^2$, uniformly in
the admitted range. When $P=0$, no partition is needed.
All partitioning and equality-class choices occur only in the proof.
\end{proof}

\begin{proof}[Proof of \Cref{thm:joint-affine-tuple}]
For fixed $\rho,h$, the quantity $\Psi$ is constant and
$E_*=\Omega_{\rho,h}(n)$. Apply
\Cref{thm:joint-affine-quantitative} for sufficiently large $n$.
\end{proof}

\begin{corollary}\label{cor:affine-logarithmic-rate}
For every fixed $h\ge1$ and $0<\eta<1$, there is a constant
$C_{\eta,h}$ such that entropy $C_{\eta,h}n/\log n$ suffices for
an explicit non-malleable affine extractor against $h$ tamperings,
with at least $\lfloor n^{1-\eta}\rfloor$ output bits and error
at most $2^{-n^{1-\eta}}$. The jointly affine tuple guarantee holds
with the same parameters.
\end{corollary}
\begin{proof}
For fixed $h$, $\Psi=O_h(1/\rho)$.
Set $\rho=C_{\eta,h}/\log n$ and choose the constant large enough
that $E_*\ge n^{1-\eta/2}$ for all sufficiently large $n$.
This proves the remaining case of \Cref{thm:affine-sublinear}.
\end{proof}

\begin{corollary}[Prescribed Output and Error]\label{cor:affine-error-tuning}
For every fixed $h\ge1$ and $0<\rho\le1$, there is
$G_{\rho,h}\le2^{C_h/\rho}$ with the following property.
For positive integers $m,E$ satisfying $E\ge\log n$ and
$G_{\rho,h}(m+E)\le n$, the construction can be chosen to output
$m$ bits with error at most $2^{-E}$ on all jointly affine tuples
in \Cref{thm:joint-affine-tuple}. Its degree bounds and the number
of affine bits fixed have the same dependence on the parameters as
in that proof.
\end{corollary}
\begin{proof}
Retain the block count, degree of the initial-seed function, and
majority length above, using working honest rate $\rho/2$. Let $P_{\rho,h}$ dominate all minimum
row lengths, including $(2T)^4$, and all polynomial factors in
$h$ and $1/\rho$. It can be chosen polynomial in $1/\rho$ for fixed
$h$. Choose the final error scale to be at least
$C_hP_{\rho,h}(m+E+\log n)$ and set the first scale by multiplying
this quantity, with rounding slack, by $\lambda^{-(B-1)}$.
This ensures all minimum lengths, the Fourier output bound, and error
at most $2^{-E-1}$ before the collision partition.

The length-dependent costs in the upper-bound constraints are at most
linear in the first iteration length. To satisfy these constraints,
it suffices that $n/(m+E+\log n)$ be at least a sufficiently large
polynomial in $1/\rho$ times $\mathcal H\lambda^{-(B-1)}/b$.
For fixed $h$, the logarithm of this bound is
$O_h(\rho^{-1}+\log^2(2/\rho))=O_h(1/\rho)$.
Increase the bound if necessary to make the collision error at most
$2^{-E-1}$. Since $E\ge\log n$, a suitable
$G_{\rho,h}\le2^{C_h/\rho}$ makes $G_{\rho,h}(m+E)\le n$
sufficient for all these requirements.
The row-growing and correlation-breaker algorithms are unchanged;
only their integer length parameters are adjusted.
\end{proof}

For fixed $\rho>0$ and $\eta>0$, the quantitative theorem permits
\[
 h+1\le c_{\rho,\eta}\sqrt{\frac{\log n}{\log\log n}}
\]
with a sufficiently small positive constant, while retaining output
length and error exponent at least $n^{1-\eta}$. This follows by
substituting the displayed bound into $\Psi$.

\section{Linear Branching Programs}
\label{sec:lbp}

We show that both directional affine extractors and non-malleable affine extractors imply average-case hardness for weakly read-once linear branching programs. The reductions use the same decomposition of a computation into affine sections. Directional affine extractors apply after conditioning on the displacement between two sections, whereas non-malleable affine extractors apply directly. The qualitative exponential lower bound already follows from the published directional affine extractors of~\cite{GPT22,LZdir}.

\subsection{The Model and the Decorrelation Condition}
\label{subsec:lbp-model}

A linear branching program on $\F_2^n$ is a finite directed acyclic graph with a unique source and sinks labeled by bits. Every internal node $v$ queries a linear form $q_v:\F_2^n\to\F_2$ and has two outgoing edges, labeled $0$ and $1$. Evaluation follows the edge labeled $q_v(x)$. We count all vertices in its size and allow the two outgoing edges to have the same destination. For Boolean functions $f,P$, write
\[
\operatorname{Corr}(f,P)
=\left|\E_{x\sim U_n}(-1)^{f(x)+P(x)}\right|.
\]
In particular, $\Pr[P(U_n)=f(U_n)]\le(1+\operatorname{Corr}(f,P))/2$.

\begin{definition}[Weakly read-once linear branching program~{\cite[Definition~1]{GPT22}}]
\label{def:lbp-weakly-read-once}
For a node $v$, let $\operatorname{Pre}(v)$ be the span of all linear forms queried on all directed source-to-$v$ paths, excluding $q_v$ itself. A linear branching program is \emph{weakly read-once} if
\[
q_v\notin\operatorname{Pre}(v)
\]
for every internal node $v$.
\end{definition}

Here $\operatorname{Pre}(v)$ contains queries from all paths leading to $v$, not just the path followed by a particular input. The weakly read-once condition implies that the queries on each path are linearly independent; the converse need not hold. Following~\cite{GPT22}, a program is \emph{full} if every path to an internal node $v$ has query span $\operatorname{Pre}(v)$.

We recall the directional derivative bias bound stated after \Cref{def:directional-intro}. This is the consequence of the directional affine extractor guarantee used in the proof. It also lets us apply the reduction to the Boolean construction of~\cite{GPT22}, whose guarantee is stated in terms of bias.

\begin{definition}[Directional derivative bias~{\cite[Definition~12]{GPT22}}]
\label{def:lbp-derivative-bias}
A Boolean function $f:\F_2^n\to\F_2$ has directional derivative bias at most $\beta$ at entropy $K$ if, for every affine source $X$ of dimension at least $K$ and every $c\in\F_2^n\setminus\{0\}$,
\begin{equation}
\left|\E(-1)^{D_c f(X)}\right|
=\left|\E(-1)^{f(X)+f(X+c)}\right|\le\beta.
\label{eq:lbp-derivative-bias}
\end{equation}
\end{definition}

If $F:\F_2^n\to\F_2^m$ is a directional affine extractor with statistical error $\varepsilon$, then every nonzero output parity $f_u(x)=\langle u,F(x)\rangle$ satisfies Eqn.~\eqref{eq:lbp-derivative-bias} with $\beta=2\varepsilon$. Indeed,
$(f_u(X),f_u(X+c))\approx_\varepsilon(U_1,f_u(X+c))$, and the sign of the sum has expectation zero under the ideal distribution. The reduction below uses only this bias bound.

The common intermediate condition compares two affine embeddings of $\F_2^k$ into $\F_2^n$. We require them to be disjoint sections of one linear map, in the sense of \Cref{pre:fibers-sections}: this map recovers the parameter $a$ from either embedded point.

\begin{definition}[Decorrelation across disjoint affine sections]
\label{def:lbp-section-decorrelation}
Let $1\le k<n$. A Boolean function $f:\F_2^n\to\F_2$ has $(k,\delta)$ section decorrelation if, for every rank-$k$ linear map $Q:\F_2^n\to\F_2^k$ and affine maps $\sigma,\tau:\F_2^k\to\F_2^n$ satisfying
\[
Q\circ\sigma=Q\circ\tau=\operatorname{id}_{\F_2^k},
\qquad \operatorname{Im}(\sigma)\cap\operatorname{Im}(\tau)=\varnothing,
\]
one has
\begin{equation}
\left|\E_{a\sim U_k}(-1)^{f(\sigma(a))+f(\tau(a))}\right|\le\delta.
\label{eq:lbp-section-decorrelation}
\end{equation}
\end{definition}

Both section images have dimension $k$, and the comparison matches $\sigma(a)$ with $\tau(a)$ using their common coordinates $a=Qx$.

\begin{lemma}[Affine non-malleability implies section decorrelation]
\label{lem:lbp-nonmalleable-sections}
Let $F:\F_2^n\to\F_2^m$ be a $(k,\varepsilon)$ non-malleable affine extractor. For every nonzero $u\in\F_2^m$, the function $f_u(x)=\langle u,F(x)\rangle$ has $(k,2\varepsilon)$ section decorrelation. It suffices to assume non-malleability against globally fixed-point-free affine involutions.
\end{lemma}

\begin{proof}
Fix $Q,\sigma,\tau$ as in \Cref{def:lbp-section-decorrelation}. Put
$\Delta(a)=\sigma(a)+\tau(a)$ and define
\[
T(x)=x+\Delta(Qx).
\]
The map $\Delta$ is affine, $Q\Delta(a)=0$, and $\Delta(a)\ne0$ for every $a$. Consequently $T$ is affine, has no fixed points, and satisfies $QT=Q$. Thus
\[
T^2(x)=x+\Delta(Qx)+\Delta(QT(x))=x,
\qquad T(\sigma(a))=\tau(a).
\]
Since $Q\sigma=\operatorname{id}$, the source $X=\sigma(U_k)$ has dimension $k$. Applying non-malleability and taking the $u$-parity of each output gives
\[
(f_u(\sigma(U_k)),f_u(\tau(U_k)))
\approx_\varepsilon(U_1,f_u(\tau(U_k))).
\]
The sign $(-1)^{z+z'}$ has expectation zero under the ideal distribution. Its expectations under two distributions differ by at most twice their statistical distance, proving the claim.
\end{proof}

\begin{lemma}[Conditioning on the displacement]
\label{lem:lbp-directional-sections}
Suppose $f:\F_2^n\to\F_2$ has directional derivative bias at most $\beta$ at entropy $K$. For every integer $q<n$ with $2q-n\ge K$, the function $f$ has $(q,\beta)$ section decorrelation.
\end{lemma}

\begin{proof}
Fix a rank-$q$ map $Q$ and disjoint sections $\sigma,\tau$ of $Q$. The affine displacement $D(a)=\sigma(a)+\tau(a)$ takes its values in $\ker Q$, so its linear part has rank at most $n-q$. Every nonempty fiber $D^{-1}(c)$ has dimension at least
\begin{equation}
q-(n-q)=2q-n\ge K.
\label{eq:lbp-displacement-entropy}
\end{equation}
Its value $c$ is nonzero, since the section images are disjoint. The map $\sigma$ is injective, so $\sigma(U_q)$ conditioned on $D(U_q)=c$ is uniform on an affine space of the same dimension. On that fiber, $\tau(a)=\sigma(a)+c$. Apply Eqn.~\eqref{eq:lbp-derivative-bias} and average over $c$.
\end{proof}

The sections need not be parallel. Their common right-inverse condition bounds the number of constraints needed to fix their displacement, giving the entropy cost in Eqn.~\eqref{eq:lbp-displacement-entropy}.

\subsection{Normalizing the Program and Choosing a Cut}
\label{subsec:lbp-cut}

We use the normalization of~\cite[Lemma 16]{GPT22}, which makes the program full by inserting queries whose answers are ignored. We include the argument to keep track of the cut size and to handle computations that terminate before $k$ queries have been asked.

\begin{lemma}[A cut of bounded size in a full program]
\label{lem:lbp-full-cut}
Let $P$ be a size-$s$ weakly read-once linear branching program on $n$ bits, and let $1\le k<n$. There is a full weakly read-once linear branching program computing $P$, with a set $V$ of at most $2s$ nodes such that:
\begin{enumerate}
\item Every computation reaches exactly one node in $V$.
\item For each $v\in V$, every path reaching $v$ has the same $k$-dimensional query span $W_v$.
\item If the rows of $Q_v:\F_2^n\to\F_2^k$ form a basis for $W_v$, there is a set $A_v\subseteq\F_2^k$ such that an input reaches $v$ exactly when $Q_vx\in A_v$. Its reaching probability is $p_v=|A_v|/2^k$, and $\sum_{v\in V}p_v=1$.
\end{enumerate}
The normalized program has $O(ns)$ vertices. The cut may contain original nodes, inserted nodes, and copies of sinks.
\end{lemma}

\begin{proof}
Discard vertices not reachable from the source. For every original internal node $v$, put $W_v=\operatorname{Pre}(v)$ in the original program. Process the edges in topological order. On an original edge $u\to v$ with internal destination,
\[
W_u+\operatorname{span}\{q_u\}\subseteq W_v.
\]
Insert a chain of queries completing a basis of the space on the left to a basis of $W_v$. Both outgoing edges of each inserted query lead to the next node. Inductively, every path reaching an original internal node $v$ now has query span exactly $W_v$. Each inserted query is independent of its preceding space, and $q_v$ remains outside $W_v$. The program is therefore weakly read-once and full.

Give each edge entering an original sink its own copy of that sink. On a terminal edge out of $u$, if $\dim W_u+1<k$, insert ignored independent queries until the prefix rank is $k$. Otherwise no padding is needed. In particular, an edge already at rank $k$ has its own output copy with a well-defined prefix space. The padding queries are chosen outside that edge's prefix space; they are not appended after a shared sink. If the original program is a constant sink, use one chain of $k$ ignored queries. Each edge receives at most $n$ queries, giving size $O(ns)$.

Take the first node of prefix rank $k$ on each computation. These nodes form a cut, since ranks grow by one at every query and every computation now makes at least $k$ queries. Associate an original cut node with itself, and a cut node inserted along an original edge with that edge's source. Each original node has at most two outgoing chains and hence accounts for at most two cut nodes. If the node itself is in the cut, neither outgoing chain contributes another. The constant-program case contributes one node. Thus $|V|\le2s$.

On every path reaching $v\in V$, the $k$ independent queries span $W_v$. Its inputs form one full fiber $Q_vx=a$: all inputs in this fiber follow that path because its queries and answers are fixed. Taking the union over paths gives $Q_v^{-1}(A_v)$ for some $A_v$. Since $Q_vU_n=U_k$, the reaching probability is $|A_v|/2^k$. The cut partitions the inputs, so the probabilities sum to one.
\end{proof}

At a cut node, the parameter $a=Q_vx$ records the answers to a basis of the past query space. A second parameter records a complete future transcript. Although the future queries depend on previous answers, fixing their answers makes the input an affine function of $a$.

\begin{lemma}[Affine sections from future transcripts]
\label{lem:lbp-future-sections}
Fix a node $v$ in the cut of \Cref{lem:lbp-full-cut}, let $Q=Q_v$, and put $d=n-k$. There are affine maps
\[
\sigma_b:\F_2^k\to\F_2^n
\qquad(b\in\F_2^d)
\]
such that:
\begin{enumerate}
\item $Q\sigma_b=\operatorname{id}$ for every $b$, and the section images are disjoint and partition $\F_2^n$.
\item The map $(a,b)\mapsto\sigma_b(a)$ is a bijection, so uniform input corresponds to independent uniform $a,b$.
\item There is a Boolean function $P_v:\F_2^d\to\F_2$ such that the subprogram started at $v$ outputs $P_v(b)$ on $\sigma_b(a)$, for every $a,b$. On inputs reaching $v$, this equals $P(x)$.
\end{enumerate}
\end{lemma}

\begin{proof}
Unfold the subprogram rooted at $v$ into a binary decision tree. Extend every terminating path by ignored independent queries until it has exactly $d$ queries from $v$ onward. Each added query is chosen outside the span of $W_v$ and the preceding suffix queries. This tree is used only for the analysis.

At an original query node $w$ on a suffix path, $W_v$ and all preceding suffix queries lie in $\operatorname{Pre}(w)$: concatenate the suffix with a source-to-$v$ path. The weak read-once condition therefore makes the next query independent of those forms. The added queries have the same property by construction. For every fixed $a$, every suffix transcript of length $d$ is thus feasible on $Qx=a$.

Fix a complete transcript $b$, and let $R_b$ be the matrix of the linear forms queried on that transcript. The rows of $Q$ and $R_b$ together form a basis, so
\begin{equation}
Qx=a,\qquad R_bx=b
\label{eq:lbp-section-system}
\end{equation}
have a unique solution $x=\sigma_b(a)$. Inverting this fixed matrix shows that $\sigma_b$ is affine. Conversely, every solution of Eqn.~\eqref{eq:lbp-section-system} follows transcript $b$. Each input has one complete transcript, proving disjointness, coverage, and the bijection. The transcript determines the original terminal output; define $P_v(b)$ to be this output. The argument holds for every $a$, including values outside $A_v$, because independence uses the full prefix space $W_v$.
\end{proof}

\subsection{The Correlation Bound}
\label{subsec:lbp-correlation}

\begin{theorem}[Correlation bound]
\label{thm:lbp-section-correlation}
Let $1\le k<n$, and suppose $f:\F_2^n\to\F_2$ has $(k,\delta)$ section decorrelation. Every size-$s$ weakly read-once linear branching program $P$ satisfies
\begin{equation}
\operatorname{Corr}(f,P)
\le\sqrt{2s\bigl(\delta+2^{-(n-k)}\bigr)}.
\label{eq:lbp-section-correlation}
\end{equation}
\end{theorem}

\begin{proof}
Use the cut of \Cref{lem:lbp-full-cut} and the sections of \Cref{lem:lbp-future-sections}. Put $d=n-k$ and $h=(-1)^f$. For each $v\in V$, write $g_v(b)=(-1)^{P_v(b)}$ for the suffix output sign and define
\[
H_v(a)=\E_{b\sim U_d}g_v(b)h(\sigma_b(a)).
\]
For $b\ne b'$, section decorrelation bounds
$|\E_a h(\sigma_b(a))h(\sigma_{b'}(a))|$ by $\delta$. For $b=b'$, the product is one. Hence
\begin{equation}
\begin{split}
\E_aH_v(a)^2
&=\E_{b,b'}g_v(b)g_v(b')\E_a h(\sigma_b(a))h(\sigma_{b'}(a))\\
&\le2^{-d}+(1-2^{-d})\delta\le2^{-d}+\delta.
\end{split}
\label{eq:lbp-section-second-moment}
\end{equation}
All averages over $a$ are uniform on the whole of $\F_2^k$. The contribution from inputs reaching $v$ is
\[
\E_x\mathbf1_{\{x\text{ reaches }v\}}(-1)^{f(x)+P(x)}
=\E_a\mathbf1_{A_v}(a)H_v(a).
\]
Cauchy--Schwarz bounds its absolute value by $\sqrt{p_v(2^{-d}+\delta)}$. Summing over the cut and applying Cauchy--Schwarz again gives
\[
\operatorname{Corr}(f,P)
\le\sqrt{2^{-d}+\delta}\sum_{v\in V}\sqrt{p_v}
\le\sqrt{(2^{-d}+\delta)|V|\sum_vp_v}
\le\sqrt{2s(2^{-d}+\delta)}.
\]
\end{proof}

\begin{remark}[The order of conditioning]
\label{rem:lbp-conditioning}
The reachability set $A_v$ may be arbitrary. We do not regard $U_k$ conditioned on $A_v$ as an affine source. Instead, fix two future transcripts and average over all uniform past answers, obtaining affine sections to which \Cref{lem:lbp-nonmalleable-sections} applies. Only after the second-moment estimate do we impose reachability, using Cauchy--Schwarz.
\end{remark}

\begin{theorem}[Directional affine extractors imply weakly read-once hardness]
\label{thm:lbp-directional-hardness}
Let $f:\F_2^n\to\F_2$ have directional derivative bias at most $\beta$ at entropy $K$. For every integer $q<n$ with $2q-n\ge K$ and every size-$s$ weakly read-once linear branching program $P$,
\begin{equation}
\operatorname{Corr}(f,P)\le\sqrt{2s\bigl(\beta+2^{-(n-q)}\bigr)}.
\label{eq:lbp-directional-hardness}
\end{equation}
In particular, if $F$ is a $(K,\varepsilon)$ directional affine extractor and $K\le n-2$, every nonzero output parity $f_u$ satisfies
\begin{equation}
\operatorname{Corr}(f_u,P)
\le\sqrt{2s\bigl(2\varepsilon+2^{-\lfloor(n-K)/2\rfloor}\bigr)}.
\label{eq:lbp-directional-parity}
\end{equation}
\end{theorem}

\begin{proof}
\Cref{lem:lbp-directional-sections} gives $(q,\beta)$ section decorrelation, so \Cref{thm:lbp-section-correlation} applies. For the second assertion, take $q=\lceil(n+K)/2\rceil$ and $\beta=2\varepsilon$. Then $q<n$ and $n-q=\lfloor(n-K)/2\rfloor$.
\end{proof}

\begin{theorem}[Affine non-malleability implies weakly read-once hardness]
\label{thm:lbp-affine-hardness}
Let $F:\F_2^n\to\F_2^m$ be a $(k,\varepsilon)$ non-malleable affine extractor, with $1\le k<n$. For every nonzero $u\in\F_2^m$ and every size-$s$ weakly read-once linear branching program $P$,
\begin{equation}
\operatorname{Corr}(f_u,P)
\le\sqrt{2s\bigl(2\varepsilon+2^{-(n-k)}\bigr)}.
\label{eq:lbp-affine-hardness}
\end{equation}
It suffices to assume non-malleability against globally fixed-point-free affine involutions.
\end{theorem}

\begin{proof}
\Cref{lem:lbp-nonmalleable-sections} gives $(k,2\varepsilon)$ section decorrelation. Apply \Cref{thm:lbp-section-correlation}. In particular,
\[
\Pr[P(U_n)=f_u(U_n)]
\le\frac12+\frac12\sqrt{2s\bigl(2\varepsilon+2^{-(n-k)}\bigr)}.
\]
\end{proof}

\paragraph{Comparison of the Hypotheses.}
For a $(K,\varepsilon)$ extractor, the correlation bound for every
nonzero output parity is at most
$\sqrt{2s(2\varepsilon+\text{collision term})}$. The two reductions give:
\begin{center}
\begin{tabular}{lll}
\hline
Hypothesis & Cut rank & Collision term\\
\hline
Directional affine extractor & $\lceil(n+K)/2\rceil$ & $2^{-\lfloor(n-K)/2\rfloor}$\\
Affine non-malleability & $K$ & $2^{-(n-K)}$\\
\hline
\end{tabular}
\end{center}
Thus agreement probability at least $1/2+\eta$ requires, respectively,
\begin{equation}
\begin{aligned}
s&\ge\frac{2\eta^2}{2\varepsilon+2^{-\lfloor(n-K)/2\rfloor}},\\
s&\ge\frac{2\eta^2}{2\varepsilon+2^{-(n-K)}}.
\end{aligned}
\label{eq:lbp-agreement-size}
\end{equation}
For constant advantage, both give exponential size lower bounds when the
error is exponentially small and $n-K=\Omega(n)$. Affine non-malleability improves the collision
term; this improves the final bound only when extractor error does not
dominate that term.

\paragraph{Several Affine Tamperings.}
If $F$ is a $(2r-1)$-non-malleable affine extractor,
where $r\ge1$ is an integer, a $2r$-th moment gives
\begin{equation}
\operatorname{Corr}(f_u,P)
\le\left[2s\left(2\varepsilon+(2r-1)!!\,2^{-r(n-k)}\right)\right]^{1/(2r)}.
\label{eq:lbp-higher-moment-summary}
\end{equation}
Indeed, in the moment expansion, $(2r-1)$-non-malleability controls every
tuple with a section of odd multiplicity. The remaining tuples have
all multiplicities even; their probability is at most
$(2r-1)!!\,2^{-r(n-k)}$. \Cref{thm:lbp-higher-moment} gives the proof.
For fixed $r$, error $\varepsilon=2^{-E}$, and constant prediction
advantage, the size lower bound is therefore
\[
 s\ge2^{\min\{E,r(n-k)\}-O_r(1)}.
\]
Only the entropy-gap term is multiplied by $r$; the error exponent
remains $E$. The error constant of \Cref{thm:fixed-multi-affine} depends
on the number of tamperings, so the best choice of $r$ depends on the
extractor parameters. For $k+\binom{2r}{2}<n$, the same exponent holds
with arbitrary independent uniform auxiliary inputs
(\Cref{thm:lbp-auxiliary-higher-moment}).

\paragraph{Explicit Functions.}
For the cubic trace function of~\cite{GPT22}, averaging the ranks of
the restricted derivatives gives a stronger bound than applying the
directional derivative estimate separately to every displacement.

\begin{corollary}[The cubic trace function]
\label{cor:lbp-cubic-trace}
For $n$ divisible by $15$, identify each of three $n/3$-bit blocks with $\F_{2^{n/3}}$, and define
\[
f_n(x,y,z)=\Tr_{\F_{2^{n/3}}/\F_2}(xyz).
\]
Here $\Tr_{\F_{2^{n/3}}/\F_2}(u)=\sum_{j=0}^{n/3-1}u^{2^j}$ is the field trace.
Every size-$s$ weakly read-once linear branching program satisfies
\begin{equation}
\operatorname{Corr}(f_n,P)\le\sqrt{5s}\,2^{-n/15}.
\label{eq:lbp-cubic-trace}
\end{equation}
Consequently, exact computation requires size at least $2^{2n/15}/5$.
For every fixed $0<\eta<2/15$, programs of size at most
$2^{(2/15-\eta)n}$ have correlation at most
$\sqrt5\,2^{-\eta n/2}$. These bounds allow any number of independent
uniform auxiliary bits. At every sufficiently large input length $N$,
the same cubic formula applied to the first $3\lfloor N/3\rfloor$ bits requires size
$2^{\lfloor2\lfloor N/3\rfloor/5\rfloor}/5=2^{2N/15-O(1)}$.
\end{corollary}

The function and derivative estimate in~\cite[Theorem~15]{GPT22} were used there to prove hardness for strongly read-once linear branching programs. Applying that estimate to our reduction gives a weakly read-once size lower bound of $2^{n/9-O(1)}$. To obtain $2^{2n/15-O(1)}$, we use a more precise average of the derivative biases over the displacements between affine sections.

The exponential lower bound also follows from the directional affine extractor of Li and Zhong~\cite[Theorem~8]{LZdir} or the non-malleable affine extractor of Li~\cite[Theorem~1.12]{Li23}. Our construction in \Cref{thm:affine-linear} provides further instances at every fixed positive entropy rate. The parameter calculations, including the extension of the cubic trace function to all input lengths, are given in \Cref{subsec:lbp-instantiations}.

Further extensions are proved in \Cref{sec:lbp-appendix}. Independent uniform
auxiliary inputs preserve the directional bound; for affine
non-malleability, the collision term becomes
$2^{-(n-K-1)}$, at a cost of one entropy bit
(\Cref{thm:lbp-directional-auxiliary} and \Cref{thm:lbp-affine-auxiliary}).
For full-output and list prediction, suppose $k\le n-2$,
$\varepsilon\le1/2$, and each sink outputs a list of at most $J$ values.
Write $p=J/2^m\le1/2$.
\Cref{thm:lbp-list-prediction} and \Cref{thm:lbp-list-auxiliary} give
\[
 \Pr[F(X)\in P(X,R)]
 \le p+\sqrt{2s(1-p)\left(\varepsilon+p2^{-(n-k-1)}\right)}.
\]
Without auxiliary inputs, $n-k-1$ improves to $n-k$.
The higher-moment version appears in \Cref{thm:lbp-list-higher-moment}.

The proof-complexity applications in \Cref{sec:proof-complexity} use separate lifting arguments for the clause-search problem of an unsatisfiable formula.

\section{Decision Trees}
\label{sec:decision-trees}

We prove lower bounds for non-oblivious decision trees with low-degree
or local queries, and compare them with the depth
needed to compute our extractors.

\subsection{Non-Oblivious Degree-2 Decision Trees}
\label{aq:section}

Allowing a degree-$2$ query to depend on previous answers can greatly reduce the depth needed to compute an extractor. We first prove lower bounds in terms of the extractor error, then give an efficient evaluation of our construction. Comparing this evaluation with the oblivious lower bound gives the separation below.

\subsubsection{Degree-2 Queries and Correlation}
\label{aq:model}

A degree-$2$ decision tree~\cite[Definitions~3.1--3.2]{GG26} queries
a multilinear polynomial of degree at most two over $\F_2$ at each
internal node. Queries may depend on
previous answers. We write $D_2(f)$ for the minimum depth of a tree
computing $f$, and $D_2^{\mathrm{obl}}(f)$ when the query is fixed at
each level. Leaves may carry vectors when $f$ has several output bits.

We use the translation non-malleability guarantee of directional affine
extractors~\cite{LZdir}: for every affine source $X$ of dimension at least
$k$ and every $a\ne0$,
\begin{equation}
 (F(X),F(X+a))\approx_\varepsilon(U_m,F(X+a)).
 \label{aq:joint}
\end{equation}
Every nonzero output parity $f$ is then an ordinary affine extractor
with error $\varepsilon$ and satisfies the directional derivative bias
bound stated after \Cref{def:directional-intro}:
\begin{equation}
 \left|\E(-1)^{f(X)+f(X+a)}\right|\le2\varepsilon.
 \label{aq:derivative}
\end{equation}
Affine non-malleability implies~Eqn.~\eqref{aq:joint}. The derivative bound
alone suffices for the first part of the next lemma; the second part
uses the ordinary affine extractor guarantee with the stated error.

\begin{lemma}[Correlation with quadratic polynomials]
\label{aq:phases}
Let $0\le\varepsilon\le1/2$ and let $q$ be quadratic.
Put $\Gamma_n^2=2\varepsilon+(1-2\varepsilon)2^{-n}$.
If $f$ satisfies~Eqn.~\eqref{aq:derivative} at dimension $k\le n-1$, then
\begin{equation}
 \left|\E_{x\sim U_n}(-1)^{f(x)+q(x)}\right|
 \le\Gamma_n.
 \label{aq:phase-directional}
\end{equation}
If $f$ is an ordinary affine extractor with error $\varepsilon$ at
dimension $k\le\lceil n/2\rceil-1$, the same correlation is at most
$2\varepsilon$.
\end{lemma}

\begin{proof}
For $a\ne0$, the derivative $D_aq(x)=q(x)+q(x+a)$ is affine.
Its nonempty level sets have dimension at least $n-1$. Applying
Eqn.~\eqref{aq:derivative} on these sets gives
$|\E(-1)^{D_af(x)+D_aq(x)}|\le2\varepsilon$. Therefore
\[
 \left(\E(-1)^{f(x)+q(x)}\right)^2
 =2^{-n}\sum_{a\in\F_2^n}
      \E_x(-1)^{D_af(x)+D_aq(x)}
 \le2^{-n}+(1-2^{-n})2\varepsilon.
\]
For the second assertion, the polar form
\[
 B_q(u,v)=q(u+v)+q(u)+q(v)+q(0)
\]
is an alternating bilinear form: $B_q(u,u)=0$ for every $u$. It has a
totally isotropic subspace $U$ of dimension at least $\lceil n/2\rceil$,
meaning that $B_q(u,v)=0$ for all $u,v\in U$.
On each coset of $U$, $q$ is affine. Split that coset into the
nonempty level sets of $q$. Each resulting affine space has dimension
at least $\lceil n/2\rceil-1$, and $q$ is constant there. The ordinary
affine extractor guarantee bounds the bias of $f+q$ by $2\varepsilon$
on each piece.
\end{proof}

\begin{proposition}[Average-case bounds]
\label{aq:correlation}
Let $T$ be a degree-$2$ decision tree with $L$ leaves. Under the first
hypothesis of \Cref{aq:phases},
\begin{equation}
 \operatorname{Corr}(f,T)\le\min\{1,L\Gamma_n\}.
 \label{aq:tree-directional}
\end{equation}
Under its second hypothesis,
\begin{equation}
 \operatorname{Corr}(f,T)\le\min\{1,2L\varepsilon\}.
 \label{aq:tree-ordinary}
\end{equation}
Consequently, a tree of uniform-input error $\delta<1/2$ has depth at
least $\log_2((1-2\delta)/\Gamma_n)$ in the first case and
$\log_2((1-2\delta)/(2\varepsilon))$ in the second.
\end{proposition}

\begin{proof}
Fix a leaf with path queries $q_1,\ldots,q_t$ and answers
$b_1,\ldots,b_t$. Although the tree is non-oblivious, these polynomials
are fixed on this path. Its indicator is
\[
 1_{\mathrm{leaf}}(x)
 =2^{-t}\sum_{a\in\F_2^t}
       (-1)^{a\cdot b+\sum_i a_iq_i(x)}.
\]
Every phase is quadratic, and the absolute coefficient sum is one.
Apply \Cref{aq:phases} and sum over the leaves. A depth-$d$ tree has
$L\le2^d$, whereas error at most $\delta$ gives correlation at least
$1-2\delta$.
\end{proof}

The same directional argument on an affine input space of dimension
$t\ge k+1$ gives phase correlation at most
$\sqrt{2\varepsilon+(1-2\varepsilon)2^{-t}}$. For uniform input on
the full space, Eqn.~\eqref{aq:tree-ordinary} is the stronger bound whenever
its entropy hypothesis holds.

\subsubsection{Exact Depth Bounds}
\label{aq:exact}

We record two structural bounds in addition to the error-dependent
bound. The first finds affine spaces in common zero sets of quadratic
polynomials; the second simulates a non-oblivious tree obliviously.

\begin{lemma}[Common constant spaces]
\label{aq:common-flat}
The nonempty common zero set of $r$ quadratic polynomials on
$\F_2^n$ contains an affine space of dimension at least
\[
 \max\left\{0,\left\lceil\frac{n-2r}{r+1}\right\rceil\right\}.
\]
Such a space can be chosen through any specified common zero.
\end{lemma}

\begin{proof}
Translate the specified zero to the origin. Let $U$ be an
inclusion-maximal linear space on which every polynomial vanishes,
and put $s=\dim U$. Write $B_i$ for the polar form of the $i$th
polynomial and set
\[
 H=\{v:B_i(v,u)=0\text{ for all }u\in U,\ i\in[r]\}.
\]
Then $U\subseteq H$ and $\dim H\ge n-rs$. On $H$, each polynomial
is constant on cosets of $U$, and so defines a quadratic polynomial
on $H/U$ vanishing at zero. If $\dim(H/U)>2r$, the number of their
common zeros is even: sum their common-zero indicator
$\prod_i(1+q_i)$ over this quotient modulo two. Its degree is at most
$2r$, and every monomial misses a coordinate. There would then be a
nonzero common zero, extending $U$. Hence
$n-(r+1)s\le\dim(H/U)\le2r$, as required.
\end{proof}

\begin{theorem}[Generic exact bounds]
\label{aq:generic-bounds}
Suppose $F:\F_2^n\to\F_2^m$ satisfies~Eqn.~\eqref{aq:joint} with
$0<\varepsilon<1/2$ and $1\le k\le\lceil n/2\rceil-1$.
Put $E=\log_2(1/\varepsilon)$. For every nonzero output parity $f$,
\begin{equation}
 D_2(f)\ge
 \max\left\{
 \lceil E-1\rceil,
 \left\lceil\frac{n-k+1}{k+1}\right\rceil,
 \lceil\log_2(n-k+2)\rceil
 \right\}.
 \label{aq:generic-max}
\end{equation}
For computation of the entire vector $F$, include $m$ in this maximum.
The last two terms in~Eqn.~\eqref{aq:generic-max} remain valid for
$1\le k\le n-1$; in this range the first can be replaced by
$\lceil\log_2(1/\Gamma_n)\rceil$.
\end{theorem}

\begin{proof}
The first term follows from \Cref{aq:correlation} with correlation
one. For the second, a nonempty depth-$t$ leaf is the common zero
set of $t$ quadratics after subtracting the answers. Since $f$ is
an affine disperser at dimension $k$, \Cref{aq:common-flat} implies
\[
 \left\lceil\frac{n-2t}{t+1}\right\rceil\le k-1,
 \qquad
 t\ge\frac{n-k+1}{k+1}.
\]

For the third term, we recall the oblivious argument
of~\cite[Lemma~7.1]{GG26}. If a quadratic map
$Q:\F_2^n\to\F_2^s$ has $s\le n-k$, choose distinct $u,v$ with
$Q(u)=Q(v)$ and put $a=u+v$. The affine system
\[
 Q(x)+Q(x+a)=0
\]
is consistent and has a solution space of dimension at least
$n-s\ge k$. Every function of $Q$ agrees with its translate by
$a$ throughout that space, contrary to
Eqn.~\eqref{aq:derivative} and $\varepsilon<1/2$. Thus
$D_2^{\mathrm{obl}}(f)\ge n-k+1$. Ask all the polynomials at the
internal nodes of a non-oblivious depth-$d$ tree in an oblivious order.
Their answers determine its output, and there are at most $2^d-1$
such nodes. Hence $2^d-1\ge n-k+1$.

Finally, the ordinary affine extractor guarantee on $U_n$ implies that
$F$ has at least
$(1-\varepsilon)2^m$ distinct outputs. A depth-$d$ tree has at most
$2^d$ outputs, so
$d\ge m+\log_2(1-\varepsilon)>m-1$. Integer depth gives $d\ge m$.
\end{proof}

The output-length term applies to the vector, not to a single output
parity. More generally, a tree computing $F$ with uniform-input error
$\delta$ satisfies
$d\ge m+\log_2(1-\varepsilon-\delta)$ when
$\varepsilon+\delta<1$. The structural arguments in
\Cref{aq:generic-bounds} concern exact computation; the average-case
bounds are those of \Cref{aq:correlation}.

\subsubsection{Feasible Error Parameters}
\label{aq:feasibility}

The affine extractor error cannot be chosen independently of the entropy.
The following elementary constraint is useful when comparing
Eqn.~\eqref{aq:generic-max} for different constructions.

\begin{lemma}[Lower bounds on extractor error]
\label{aq:error-floor}
A Boolean affine extractor with error $\varepsilon$ at dimension
$k<n$ satisfies
\begin{equation}
 \varepsilon\ge\frac12
 \sqrt{\frac{2^n-2^k}{2^k(2^n-1)}}.
 \label{aq:ordinary-floor}
\end{equation}
If $f$ satisfies~Eqn.~\eqref{aq:derivative} and $1\le k\le n-1$, then
\begin{equation}
 \varepsilon\ge\frac12
 \sqrt{\frac{2^{n-1}-2^{k-1}}
 {2^{k-1}(2^{n-1}-1)}}.
 \label{aq:directional-floor}
\end{equation}
In particular, $\log_2(1/\varepsilon)\le k/2+1$ in the second case.
\end{lemma}

\begin{proof}
Let $h:\F_2^n\to\{-1,1\}$ have mean $\mu$, and let $W$ be a
uniformly chosen affine $k$-space. Put $N=2^n$ and $K=2^k$.
Counting diagonal and distinct ordered pairs gives
\[
 \E_W\left(\E_{x\in W}h(x)\right)^2
 =\frac1K+\left(1-\frac1K\right)
       \frac{N\mu^2-1}{N-1}
 =\frac{N-K+N(K-1)\mu^2}{K(N-1)}
 \ge\frac{N-K}{K(N-1)}.
\]
For an affine extractor each bias has magnitude at most
$2\varepsilon$, proving~Eqn.~\eqref{aq:ordinary-floor}.

For the directional statement, fix $a\ne0$. The function
$(-1)^{f(x)+f(x+a)}$ is invariant under translation by $a$, and so
is a Boolean sign function on the $(n-1)$-dimensional quotient by
$\langle a\rangle$. Affine $k$-spaces whose direction contains
$a$ correspond to affine $(k-1)$-spaces in this quotient. Apply the
same squared-bias bound there. Since $k\le n-1$, its ratio is at
least $2^{-k}$, yielding the final assertion.
\end{proof}

For $k=o(n)$, Eqn.~\eqref{aq:directional-floor} gives the sharper
asymptotic inequality
$\log_2(1/\varepsilon)\le(k+1)/2+o(1)$. Thus the
error-dependent term in~Eqn.~\eqref{aq:generic-max} cannot by itself
give depth $n-o(n)$ at sublinear entropy. Our sublinear construction guarantees error $2^{-n^{1-\eta}}$,
where $n^{1-\eta}=o(k)$,
consistent with this constraint.

\begin{remark}[The full-entropy endpoint]
\label{aq:full-entropy}
The condition $k\le n-1$ in \Cref{aq:phases} is necessary. For
$n=2s$, the quadratic $f(x,y)=x\cdot y$ has global bias
$2^{-s}$ and balanced nonzero directional derivatives on $U_n$.
Its non-malleability error against any single nonzero translation on $U_n$ is exactly
$2^{-s-1}$: both marginal sign biases are $2^{-s}$, and their
product bias is zero. Nevertheless $D_2(f)=1$. This is a statement
about translation non-malleability; no guarantee against arbitrary
affine maps is asserted.
\end{remark}

\subsubsection{Computing the Construction by Non-Oblivious Queries}

The initial seeds may have degree greater than two, but their inputs
are the bilinear seed matrices. A degree-$2$ decision tree can
first query each bit of these matrices, then compute the seeds from
the answers. The remaining steps use affine queries, giving a small
depth bound for computing the whole construction.

\begin{proposition}[Non-oblivious degree-$2$ depth upper bound]\label{aq:upper}
For every feasible choice of the parameters in
\Cref{param:schedule}, the function $F=\anmExt$ of
\Cref{alg:affine-nm} can be computed exactly by a non-oblivious
degree-$2$ decision tree of depth
\begin{equation}\label{aq:query-sum}
 D_2(F)\le\sum_{i=1}^B(A_i+L_i)\le\frac{8H_\kappa S}{b},
\end{equation}
provided $u_B\ge\log n$. The same bound holds for every fixed
postprocessing of $F$.
\end{proposition}
\begin{proof}
In iteration $i$, query all $A_i$ bits of the matrix in
Eqn.~\eqref{branch:seed-matrix}. Each bit is bilinear in two linear
functions of the input and therefore has degree at most two.
The answers determine $\head_i(x)$. With this seed fixed,
\Cref{branch:interface} computes $G_i(x,\head_i(x))$ using at most
$L_i$ affine queries whose coefficients may depend on earlier answers. Once all iteration outputs are known,
the amplifiers and their $\mathsf{XOR}$ require no further queries. Summing the
query counts and applying Eqn.~\eqref{param:queries} proves the claim.
Here each successive fixing in \Cref{branch:interface} is implemented
by an affine query whose coefficients are determined by earlier answers.
In particular, after each growing-row output is known, the next
seeded extractor output is an affine function of the input. The
somewhere-random extractor and the strong correlation breaker act
on values already known, so neither needs to be linear.
\end{proof}

\subsubsection{Choosing the Output and Error Parameters}

We can choose the initial scale directly from the requested output
length and error. Retain $B,\ell,T,q,c,b,\lambda,\nu,H_\kappa$ from
\Cref{param:schedule}. Write $\beta_{\rm IP}$ for the constant
$\beta$ in \Cref{branch:seed-ingredients}, and let $C_{\min}$ be a
sufficiently large absolute constant.

\begin{proposition}[Prescribed output length and error]\label{aq:retuning}
Let $m\ge1$ be an integer and $E\ge1$. Choose the scales as follows.
\begin{subequations}\label{aq:scale-choice}
\begin{itemize}
\item Set the minimum scale to
\begin{equation}
 L_0=\left\lceil\max\{\lceil\log n\rceil+2,2c,(2T)^4,C_{\min}\}\right\rceil.
\end{equation}
\item Set the target final scale to
\begin{equation}
 t_0=\left\lceil\max\left\{\frac m\nu,
                    \frac{8E}{5},L_0\right\}\right\rceil.
\end{equation}
\item Put
\begin{equation}
 K=\lambda^{-(B-1)}.
\end{equation}
\item Set the initial scale to
\begin{equation}
 S=\lceil K(t_0+2)\rceil.
\end{equation}
\end{itemize}
\end{subequations}
Write $\overline S=\lceil S/b\rceil$ for the largest iteration
output length. Suppose that the block lengths and subroutine parameters satisfy
\begin{gather}
 2(n-B\ell)\le\kappa n/4,\notag\\
 16H_\kappa\overline S\le\kappa\ell/16,\notag\\
 H_\kappa(\overline S+\log n)\le\beta_{\rm IP}q/4,\notag\\
 8\overline S\le c_{\rm IP}q,\notag\\
 2\ell+(2H_\kappa+C_*)(\overline S+\log n)\le\kappa n/4.
 \label{aq:upper-feasibility}
\end{gather}
Then \Cref{alg:affine-nm}, with these scales, is an explicit
$m$-bit non-malleable affine extractor for entropy
$\lceil2\kappa n\rceil$ and error at most $2^{-E}$, with
\begin{equation}\label{aq:retuned-depth}
 D_2(F)\le\frac{8H_\kappa}{b}\lceil K(t_0+2)\rceil
 =O\left(\frac{H_\kappa K}{b}(m+E+L_0)\right).
\end{equation}
In particular, for an absolute constant $C$ one may take
\begin{equation}\label{aq:overhead}
 \begin{aligned}
 G_\kappa&=\frac{CH_\kappa K}{b},\\
 \log G_\kappa
 &=O\left(\kappa^{-1}+\log^2(2/\kappa)\right)=O(\kappa^{-1}).
 \end{aligned}
\end{equation}
Writing $\mathcal P_\kappa=2c+(2T)^4+C_{\min}+1=(2/\kappa)^{O(1)}$, we obtain
\[
 D_2(F)\le G_\kappa
       \bigl(m+E+\log n+\mathcal P_\kappa\bigr).
\]
\end{proposition}
\begin{proof}
Eqn.~\eqref{param:rounding} gives
\[
 t_0\le u_B\le t_0+3.
\]
Thus $m\le\lfloor\nu u_B\rfloor$, $s_B\ge2c$, and
$5u_B/8\ge E$. The minimum scale supplies the initial-seed and
somewhere-random extractor requirements in
\Cref{branch:seed-ingredients}. Since $s_i\le\overline S$, the
row-length and bilinear-error conditions follow from
Eqn.~\eqref{aq:upper-feasibility}; in particular,
$\epsilon_{\rm IP}\le2^{-8\overline S}$. The growing-row
construction and the unchanged scale-$s_i$ advice and
correlation-breaker parameters give $\epsilon_i\le2^{-u_i}$.

The remaining inequalities imply the parameter constraints of
\Cref{merge:general}. The later-seed estimate in
Eqn.~\eqref{param:tail} is unchanged by the choice of $S$.
Since $bs_i\ge u_i$, \Cref{merge:decay} gives error at most
$2^{-5u_B/8}\le2^{-E}$. Summing $s_i=\lceil u_i/b\rceil$
and using $u_i\ge\log n$ proves the query bound in
\Cref{aq:upper}. Here $\nu$ is an absolute positive constant,
so the definition of $t_0$ gives Eqn.~\eqref{aq:retuned-depth}.
Finally,
\[
 \begin{aligned}
 \log K&=(B-1)\log(1/\lambda)=O(\kappa^{-1}),\\
 \log(1/b)&=O(\log(2/\kappa)),\\
 \log H_\kappa&=O(\log^2(2/\kappa)).
 \end{aligned}
\]
The geometric ratio is independent of $\kappa$ because the guessed
initial seeds have length $O(u_i+\log n)$, rather than $O(s_i)$.
These estimates prove Eqn.~\eqref{aq:overhead}.
\end{proof}

For fixed $\kappa$, the parameter constraints hold for all sufficiently large $n$
whenever $m+E\le c_\kappa n$, for a sufficiently small constant
$c_\kappa>0$. In this range the upper bound is
$O_\kappa(m+E+\log n)$. When $\kappa$ varies, a sufficient
asymptotic condition, together with the block rounding, is
\[
 G_\kappa(m+E+\log n+\mathcal P_\kappa)
       =o\bigl(\min\{\kappa^2n,q\}\bigr).
\]
For example, if
$\kappa=\Theta(\log\log n/\log n)$, then $G_\kappa=n^{o(1)}$.
Taking $m=E=\lceil n^{1-\eta}\rceil$, for fixed $0<\eta<1$,
gives $D_2(F)\le n^{1-\eta+o(1)}$. All parameter constraints
hold, since $q,\kappa\ell=n/(\log n)^{O(1)}$.
These parameter choices preserve uniform polynomial-time computability. Since $\lambda$ is dyadic, we can compute $K$ and $S$ by rational and integer arithmetic and then use the implementation from \Cref{param:explicitness}. All subroutine parameters remain polynomial in $n$ for the entropy rates above and for the more slowly decreasing rate considered below.

\paragraph{The achieved error.}
The construction may achieve a smaller error than the requested
$2^{-E}$. The proved bounds give non-malleability error at most $2^{-E_*}$
and bias at most $\beta_*$ for every nonzero output parity, where
\begin{equation}\label{aq:achieved-error}
 \begin{aligned}
 E_*&:=\frac{5u_B}{8},\\
 \beta_*&:=2^{1-3u_B/4}.
 \end{aligned}
\end{equation}
By Eqn.~\eqref{merge:ordinary-bias} and \Cref{merge:decay}, every
nonzero output parity has bias at most $\beta_*$ on affine sources
of entropy at least $\lceil\kappa n\rceil$. If
$\lceil\kappa n\rceil\le\lceil n/2\rceil-1$, applying
\Cref{aq:correlation} with ordinary affine extractor error
$\beta_*/2$ gives, for every $a\ne0$,
\begin{equation}\label{aq:achieved-lower}
 D_2(a\cdot F)\ge\log(1/\beta_*)
             =\frac{3u_B}{4}-1=\frac65E_*-1.
\end{equation}
For each fixed $0<\kappa<1/2$ and sufficiently large $n$, the upper and lower bounds give $D_2(a\cdot F)=\Theta_\kappa(E_*)$. The additional entropy condition in Eqn.~\eqref{aq:achieved-lower} is needed only for the lower bound.

\subsubsection{Computing the Construction for Several Tamperings}

\begin{proposition}[Fixed Multiple Tamperings]\label{aq:multi-upper}
Fix $0<\rho\le1$ and an integer $h\ge1$. There are constants
$c_{\rho,h},C_{\rho,h}>0$ such that, for all sufficiently large $n$,
every integer $m\ge1$ and real $E\ge1$ with
$m+E\le c_{\rho,h}n$ admit a choice of the seed lengths and
intermediate output lengths in \Cref{sec:multiple-tampering}
for which the construction satisfies \Cref{thm:joint-affine-tuple}
with output length $m$ and error at most $2^{-E}$.
Thus its output on $X^0$ is nearly uniform and independent of
all the outputs on $X^1,\ldots,X^u$ together.
Its entire output can be computed exactly by a non-oblivious
degree-$2$ decision tree of depth
\begin{equation}\label{aq:multi-query-bound}
 D_2(F_h)\le C_{\rho,h}(m+E+\log n).
\end{equation}
For fixed $h$, one may take $C_{\rho,h}=2^{O_h(1/\rho)}$.
The same upper bound holds for every query degree $r\ge2$.
\end{proposition}

\begin{proof}
Fix the block lengths and a dyadic geometric ratio $\lambda\le1/2$
from the pairwise-distinct construction at entropy rate $\rho/2$.
All constants below may depend on $\rho,h$. Let $b_0,c$ be the amplifier parameters in
\Cref{sec:multiple-tampering} and $R_0$ be the code rate.
Choose the remaining parameters in order:
\begin{itemize}
\item Put $\alpha=5/8$.
\item Put $\nu_h=\min\{1/(4(h+1)),R_0/(4b_0c)\}$.
\item Choose sufficiently large constants $C_0,C_{\min}$, with
$C_{\min}$ at least the minimum primitive lengths, $2c$, and $(2T)^4$.
\item Set
\[
 t_0=\left\lceil\max\left\{C_0\log n,\frac m{\nu_h},
              \frac{E+1}{\alpha},C_{\min}\right\}\right\rceil.
\]
\item Set
\[
 S=\left\lceil\lambda^{-(B-1)}(t_0+2)\right\rceil.
\]
\end{itemize}
Starting with $u_1=S$, set
\[
 \begin{aligned}
 u_{i+1}&=\lfloor\lambda u_i\rfloor,\\
 s_i&=\lceil u_i/b_0\rceil.
 \end{aligned}
\]
The rounding estimate gives
$t_0\le u_B\le t_0+3$. Thus the output length is feasible and
the pairwise-distinct error is at most
$2^{-\alpha u_B}\le2^{-E-1}$. The seed lengths, matrix lengths,
and affine-query costs are $O_{\rho,h}(s_i+\log n)$.
The same fixed choice of $\lambda$ bounds the later-seed sums.
All remaining parameter constraints hold when $S+\log n$ is a sufficiently
small constant multiple of $n$, which follows by decreasing
$c_{\rho,h}$.

For $P=\binom h2>0$, apply \Cref{lem:collision-partition} with
$L=\lceil E+1+\log_2P\rceil$. Its exceptional mass is at most
$2^{-E-1}$. Decreasing $c_{\rho,h}$ also ensures $PL\le\rho n/2$,
so the conditional entropy of $X^0$ on each nonexceptional
part of the partition is at least $\rho n/2$. Apply the guarantee
for pairwise distinct inputs on that part, then duplicate the
outputs corresponding to identical inputs. Averaging gives
total error at most $2^{-E}$. For $h=1$, no collision partition
is needed.

To compute the function, query every bit of the bilinear seed
matrix in each iteration using degree-$2$ queries. The initial seed is then known, and the
advice and remaining computation use
$O_{\rho,h}(s_i+\log n)$ affine queries whose coefficients may depend on earlier answers, by
\Cref{branch:advice} and \Cref{lem:multi-retained}. In particular, every
nonlinear seed computation in the additional correlation-breaker
rounds is determined by previously computed extractor outputs. The amplifiers and
their $\mathsf{XOR}$ require no further queries. Hence
\[
 D_2(F_h)\le\sum_{i=1}^B(A_i+L_i)
       =O_{\rho,h}(S+\log n)
       =O_{\rho,h}(m+E+\log n).
\]
The collision partition is used only in the analysis. There are
$O_h(1/\rho)$ iterations, the ratio $\lambda$ depends only on
$h$, and the remaining multiplicative overhead is
$2^{O_h(\log^2(2/\rho))}$. This gives the stated dependence
of $C_{\rho,h}$ on the entropy rate.
\end{proof}

\subsubsection{Comparison at Matched Parameters}
\label{aq:comparison}

We compare the general lower bounds for directional affine extractors with the exact evaluation of \Cref{alg:affine-nm} using the scales in \Cref{aq:retuning}, and of its extension in \Cref{aq:multi-upper}. At fixed entropy rate $0<\rho<1$, the construction with linear output and error $2^{-\Omega_\rho(n)}$ has non-oblivious depth $\Theta_\rho(n)$ for every nonzero output parity, by Eqn.~\eqref{aq:achieved-lower}. The bounds therefore match up to constant factors. In the first part of \Cref{intro:quadratic-adaptivity}, their ratio is $n^{o(1)}$.

\begin{proof}[Proof of \Cref{intro:quadratic-adaptivity}]
Choose a sufficiently large absolute constant $C$. For the first
claim take $\kappa=C\log\log n/(2\log n)$ and
\[
 m=E=\lceil n^{1-\eta}\rceil.
\]
By Eqn.~\eqref{aq:overhead},
\[
 \log G_\kappa
 =O\left(\frac{\log n}{\log\log n}
              +(\log\log n)^2\right)=o(\log n).
\]
Thus \Cref{aq:retuning} gives $D_2(F_n)\le n^{1-\eta+o(1)}$.
Every scale and query budget is $n^{1-\eta+o(1)}$, while
$q$ and $\kappa\ell$ are $n/(\log n)^{O(1)}$. All primitive
minimum lengths are $(\log n)^{O(1)}$. The remaining term
$2\ell$ is at most $\kappa n/32$, and the reserved coordinates
satisfy Eqn.~\eqref{aq:upper-feasibility}. Hence all feasibility
conditions hold for sufficiently large $n$. The generic error bound
gives $D_2(f_n)\ge E-1$, and directional dispersion gives
$D_2^{\mathrm{obl}}(f_n)\ge n-k_n+1$.

For the second claim take $\kappa=C\log\log\log n/(2\log\log n)$, with output length and error exponent
\[
 \begin{aligned}
 m&=\lceil a\log n\rceil,\\
 E&=\lceil e\log n\rceil.
 \end{aligned}
\]
Now
\[
 \log G_\kappa
 =O\left(\frac{\log\log n}{\log\log\log n}
              +(\log\log\log n)^2\right)=o(\log\log n).
\]
Consequently $D_2(G_n)\le(\log n)^{1+o(1)}$.
The primitive minimum lengths are $(\log\log n)^{O(1)}=o(\log n)$,
and every scale and query budget is $(\log n)^{1+o(1)}$.
Since $q,\kappa\ell=n/(\log\log n)^{O(1)}$, feasibility again
holds. The error-dependent lower bound is $e\log n-O(1)$,
and the oblivious lower bound is $n-k_n+1$.
All primitive parameters and running times are polynomial in $n$.
Dyadic choices and integer rounding, as in \Cref{aq:retuning},
preserve uniform explicitness.
\end{proof}

\begin{table}[tbp]
\centering
\small
\setlength{\tabcolsep}{5pt}
\renewcommand{\arraystretch}{1.25}
\begin{tabularx}{\textwidth}{@{}P{.25\textwidth}P{.17\textwidth}P{.22\textwidth}X@{}}
\toprule
Entropy threshold $k$ & Error exponent $E$ & Lower bound on $D_2(f)$ & Upper bound on $D_2(F)$ \\
\midrule
$\rho n$, fixed $0<\rho<1/2$
 & $cn$ & $cn-1$ & $O_\rho(n)$ \\
$\rho n$, fixed $0<\rho<1/2$
 & $(\log n)^C$ & $\Omega((\log n)^C)$ & $O_\rho((\log n)^C)$ \\
$\Theta(n\log\log n/\log n)$
 & $n^\alpha$ & $n^\alpha-1$ & $n^{\alpha+o(1)}$ \\
$\Theta(n/(\log n)^\tau)$
 & $(\log n)^C$ & $\Omega((\log n)^C)$
 & $2^{O((\log n)^\tau)}$ \\
$\Theta(n\log\log\log n/\log\log n)$
 & $\Theta(\log n)$ & $\Omega(\log n)$ & $(\log n)^{1+o(1)}$ \\
\bottomrule
\end{tabularx}
\caption{Non-oblivious degree-$2$ decision-tree depth at matched parameters. The lower bound
holds for every nonzero parity $f$ of a directional or non-malleable
affine extractor with error at most $2^{-E}$. The upper bound
computes the entire $m$-bit output $F$ of \Cref{alg:affine-nm}
with the scale choice in \Cref{aq:retuning}; here $m\le E$.
All parameters are in the feasible asymptotic regime:
$C\ge1$, $0<\alpha<1$, and $0<\tau<1$ are fixed,
and $c>0$ is sufficiently small depending on $\rho$.
For the last three rows, every such parity also has oblivious depth
at least $n-k+1=n-o(n)$.}
\label{aq:table}
\end{table}

Applying the same lower bounds to the directional affine extractors of Li and Zhong~\cite[Theorems~1--2]{LZdir} gives non-oblivious depth $\Omega(n)$ at fixed entropy rate $\rho<1/2$, and $n^{\Omega(1)}$ at entropy $Cn(\log\log\log n)^2/\log\log n$. These follow from their error bounds $2^{-\Omega_\rho(n)}$ and $2^{-n^{\Omega(1)}}$, respectively. The upper bounds in \Cref{aq:table}, however, use the evaluation of our construction with the parameters of \Cref{aq:retuning}.

At a fixed error guarantee, the lower bound is the same for directional affine extractors and non-malleable affine extractors. The role of our construction here is to exhibit a separation between non-oblivious and oblivious queries to the same function. These examples do not resolve the question of nearly maximal non-oblivious depth posed in~\cite[Section~8]{GG26}.

\subsection{Decision Trees of Higher Degree}
\label{ldt:section}

A degree-$r$ decision tree queries a polynomial of degree at most $r$
over $\F_2$ at each internal node. We write $D_r(f)$ and
$D_r^{\mathrm{obl}}(f)$ for the minimum non-oblivious and oblivious depths,
respectively, and $L_r(f)$ for the minimum number of leaves. Degrees
refer to the multilinear representation of a Boolean function. The
degree-$2$ model is defined in \Cref{aq:model}.

We give two extensions to higher degree. Non-malleability against a fixed number of translations together bounds correlation with low-degree
polynomials, and hence gives average-case lower bounds for non-oblivious
trees. An argument finding an affine space of common zeros gives exact
bounds from affine dispersion alone.

\subsubsection{Correlation and Decision Trees}

For $f:\F_2^n\to\F_2$, define
\begin{equation}
 \begin{aligned}
 \delta_r(f)&=\max_{\deg p\le r}
       \left|\E_x(-1)^{f(x)+p(x)}\right|,\\
 \lambda_r(f)&=\log_2\frac1{\delta_r(f)}.
 \end{aligned}
 \label{ldt:correlation-parameter}
\end{equation}

\begin{proposition}[Correlation Transfer]
\label{ldt:transfer}
For a non-oblivious degree-$r$ decision tree $T$ with $L$ leaves,
\begin{equation}
 \operatorname{Corr}(f,T)\le\min\{1,L\delta_r(f)\}.
 \label{ldt:adaptive-transfer}
\end{equation}
For an oblivious degree-$r$ decision tree of depth $t$,
\begin{equation}
 \operatorname{Corr}(f,T)\le\min\{1,2^{t/2}\delta_r(f)\}.
 \label{ldt:oblivious-transfer}
\end{equation}
Consequently,
\begin{equation}
 L_r(f)\ge2^{\lambda_r(f)},\qquad
 D_r(f)\ge\lambda_r(f),\qquad
 D_r^{\mathrm{obl}}(f)\ge2\lambda_r(f).
 \label{ldt:exact-transfer}
\end{equation}
For uniform-input error at most $\eta<1/2$, replace
$\lambda_r(f)$ in the two depth bounds by
$\lambda_r(f)+\log_2(1-2\eta)$, and multiply the leaf bound by
$1-2\eta$.
\end{proposition}

\begin{proof}
The proof of \Cref{aq:correlation} applies in every degree: on a fixed
path, the leaf indicator is a linear combination of phases
$(-1)^{\sum_i a_iq_i}$ with absolute coefficient sum one. Each exponent
has degree at most $r$. Its signed mass is therefore at most
$\delta_r(f)$ in absolute value. Summing over leaves proves
Eqn.~\eqref{ldt:adaptive-transfer}.

In the oblivious case let $q_1,\ldots,q_t$ be the fixed queries.
There is a function $H:\F_2^t\to\{-1,1\}$ with
$(-1)^{T(x)}=H(q_1(x),\ldots,q_t(x))$. Expand $H$ in its Fourier
basis. Parseval and Cauchy--Schwarz give
$\sum_a|\widehat H(a)|\le2^{t/2}$. Every resulting phase is again a
degree-$r$ polynomial, proving Eqn.~\eqref{ldt:oblivious-transfer}.
This argument does not assume that the query answers are independent
or uniformly distributed. Finally, a depth-$t$ binary tree has at most
$2^t$ leaves, and uniform-input error at most $\eta$ gives correlation
at least $1-2\eta$.
\end{proof}

The correlation bound also controls expected depth, including for
randomized trees.

\begin{proposition}[Expected Queries and Query Tails]
\label{ldt:expected-queries}
Suppose $\delta_r(f)\le\delta<1$, where $\delta>0$, and put
$\Lambda=\log_2(1/\delta)$. Let a randomized non-oblivious degree-$r$
decision tree have uniform-input error at most $\eta<1/2$, and let
$Q$ be its number of queries. Then
\begin{equation}
 \E Q\ge\Lambda\frac{1-2\eta-\delta}{1-\delta},\qquad
 \Pr[Q\le t]\le2\eta+2^t\delta\quad(t\in\mathbb Z_{\ge0}).
 \label{ldt:expected-tail}
\end{equation}
Both probabilities include the internal randomness of the tree.
In particular, zero-error computation requires $\E Q\ge\Lambda$.
\end{proposition}

\begin{proof}
First fix the internal randomness. For each leaf $v$, let $p_v$
be its probability and $c_v=\E[(-1)^{f(X)}\mathbf1_v(X)]$.
The leaf expansion in \Cref{ldt:transfer} gives
$|c_v|\le\min\{p_v,\delta\}$. For $0\le p\le1$,
\[
 p\log_2(1/p)\ge
 \frac{\Lambda}{1-\delta}\bigl(\min\{p,\delta\}-\delta p\bigr).
\]
For $p\le\delta$, this follows from $\log_2(1/p)\ge\Lambda$;
for $p\ge\delta$, it follows from concavity of $p\log_2(1/p)$
and its chord between $\delta$ and $1$. The strings of answers leading to leaves form a prefix-free binary
code: no such string is a proper prefix of another. Hence
$\E Q\ge H(v)$, where $H(v)=\sum_v p_v\log_2(1/p_v)$ is the
Shannon entropy of the leaf distribution. Summing the preceding
inequality over leaves gives
\[
 |\operatorname{Corr}(f,T)|
 \le\delta+\frac{1-\delta}{\Lambda}\E Q.
\]
Average over the internal randomness and use success correlation
at least $1-2\eta$.

For the second assertion, at most $2^t$ leaves have depth at most
$t$. Their signed success mass is at most $2^t\delta$, and at
least $\Pr[Q\le t]$ minus twice the error mass on those leaves.
Averaging again proves the claim.
\end{proof}

Thus $\lambda_r(f)$ is a common parameter for the two models. For a
fixed function it is nonincreasing in $r$. We next estimate it using
non-malleability under several translations.

\subsubsection{Non-Malleability under Several Translations}

For $a\in\F_2^n$, put $D_ag(x)=g(x)+g(x+a)$, and write
$D_{\boldsymbol a}=D_{a_1}\cdots D_{a_s}$. We use the elementary
differencing inequality
\begin{equation}
 \left|\E_x(-1)^{g(x)}\right|^{2^s}
 \le \E_{a_1,\ldots,a_s}
           \left|\E_x(-1)^{D_{\boldsymbol a}g(x)}\right|.
 \label{ldt:differencing}
\end{equation}
For $s=1$, expand the square of the bias and take absolute values.
The general case follows by induction, applying Jensen's inequality
before the next squaring. This is the usual repeated-differencing
argument for low-degree correlation; see also~\cite{VW08}.

\begin{theorem}[Non-malleability under translations and low-degree correlation]
\label{ldt:joint-extraction}
Let $r\ge2$ and put $s=r-2$. Assume $s\le n$ and $0\le\varepsilon<1/2$. Suppose
$F:\F_2^n\to\F_2^m$ is an affine extractor with error
$\varepsilon$ at dimension $k$. When $s\ge1$, assume in addition that,
for every linearly independent $a_1,\ldots,a_s$ and every affine source
$X$ of dimension at least $k$,
\begin{equation}
 \bigl(F(X),(F(X+\omega\cdot\boldsymbol a))_{\omega\ne0}\bigr)
 \approx_\varepsilon
 \bigl(U_m,(F(X+\omega\cdot\boldsymbol a))_{\omega\ne0}\bigr),
 \quad \omega\in\F_2^s.
 \label{ldt:cube-extraction}
\end{equation}
Define
\begin{equation}
 \begin{aligned}
 K_{s,n}&=\left\lceil\frac{n+s}{2}\right\rceil-1,\\
 p_{s,n}&=1-\prod_{i=0}^{s-1}(1-2^{i-n}),\\
 B_{s,n}&=2\varepsilon+(1-2\varepsilon)p_{s,n},
 \end{aligned}
 \label{ldt:cube-parameters}
\end{equation}
where $p_{0,n}=0$. If $k\le K_{s,n}$, every nonzero output parity $f$
of $F$ satisfies
\begin{equation}
 \begin{aligned}
 \delta_r(f)&\le B_{s,n}^{1/2^s},\\
 D_r(f)&\ge\frac{\log_2(1/B_{s,n})}{2^s},\\
 D_r^{\mathrm{obl}}(f)&\ge
                \frac{\log_2(1/B_{s,n})}{2^{s-1}}.
 \end{aligned}
 \label{ldt:cube-bounds}
\end{equation}
For $s=0$, only the ordinary affine extractor guarantee is required.
\end{theorem}

\begin{proof}
The case $s=0$ is the ordinary affine extractor case of \Cref{aq:phases}.
Suppose $s\ge1$, fix $p$ of degree at most $r$, and apply
Eqn.~\eqref{ldt:differencing} to $g=f+p$. Independent directions occur
with probability $1-p_{s,n}$. For such directions, put
$A=\operatorname{span}\{a_1,\ldots,a_s\}$. The polynomial
$D_{\boldsymbol a}p$ has degree at most two and is invariant under
translation by $A$. It therefore descends to a degree-$2$ polynomial
on $\F_2^n/A$.

As in \Cref{aq:phases}, the quotient can be partitioned into affine
spaces on which this polynomial is constant, each of dimension at
least $\lceil(n-s)/2\rceil-1$. Their lifts have dimension at least
$K_{s,n}$. On any such lifted space, apply
Eqn.~\eqref{ldt:cube-extraction} and take the product of the output
parity signs at all $2^s$ vertices. The ideal expectation is zero,
since the output parity at the original source is replaced by a
uniform bit independent of all the other output parities. The absolute expectation is at most $2\varepsilon$.
The phase $D_{\boldsymbol a}p$ is constant on the piece, so the same
bound holds after multiplying by this phase. Averaging over pieces
bounds the inner absolute value in Eqn.~\eqref{ldt:differencing} by
$2\varepsilon$. For dependent directions use the bound one. This
proves $\delta_r(f)^{2^s}\le B_{s,n}$; the depth bounds follow from
\Cref{ldt:transfer}.
\end{proof}

Full affine non-malleability against $h=2^{r-2}-1$ maps implies
Eqn.~\eqref{ldt:cube-extraction}: all the nonzero cube translations
are fixed-point-free. The number of tamperings is essential here. The
ordinary affine extractor guarantee and non-malleability against one,
three, and seven tamperings give the respective degree bounds $r=2,3,4,5$. Writing
$E=\log_2(1/\varepsilon)$ and assuming $E\le n$, the resulting non-oblivious
depth bounds are, respectively,
\[
 E-O(1),\qquad E/2-O(1),\qquad E/4-O(1),\qquad E/8-O(1),
\]
and the oblivious bounds are twice these quantities. The exact
expression in Eqn.~\eqref{ldt:cube-bounds} includes the contribution
of dependent directions.

For degree three, the hypothesis in Eqn.~\eqref{ldt:cube-extraction}
is precisely directional affine non-malleability: only one nonzero
translation is involved. Thus a directional affine extractor of error
$\varepsilon$ at dimension $k\le\lfloor n/2\rfloor$ gives
\[
 \delta_3(f)\le
 \sqrt{2\varepsilon+(1-2\varepsilon)2^{-n}},\qquad
 D_3(f)\ge
 \frac12\log_2\frac1{2\varepsilon+(1-2\varepsilon)2^{-n}}
\]
for every nonzero output parity $f$. In particular, its non-oblivious
and oblivious depth bounds are $\min\{E,n\}/2-O(1)$ and
$\min\{E,n\}-O(1)$, respectively. Full affine non-malleability
is unnecessary for this specialization of \Cref{ldt:joint-extraction}.

For every fixed $r$ and fixed $\rho<1/2$,
\Cref{thm:fixed-multi-affine} supplies the required property at
$k=\lceil\rho n\rceil$ with error $2^{-c_{\rho,r}n}$ for some
$c_{\rho,r}>0$. Hence it gives linear depth and exponentially many
leaves for degree-$r$ decision trees, including constant-error
computation. The quantified construction below also permits the degree to grow. Alternatively, differentiating
$r-1$ times leaves an affine phase. With $2^{r-1}-1$ tampered sources,
the same proof works at every $k\le n-1$, using affine level sets in
place of the degree-$2$ partition.

\begin{corollary}[Matching Bounds at Fixed Entropy Rate]
\label{ldt:matching-depth}
Fix $r\ge2$, $0<\rho<1/2$, and
$h\ge\max\{1,2^{r-2}-1\}$. For
$1\le m\le E$, $E\ge\log n$, and
$m+E\le c_{\rho,h}n$, the construction of
\Cref{aq:multi-upper} satisfies, for every nonzero output parity $f$,
\[
 \frac{E}{2^{r-2}}-O_r(1)
 \le D_r(f)\le D_r(F_h)\le D_2(F_h)
 \le O_{\rho,h}(E).
\]
In particular, both $D_r(f)$ and $D_r(F_h)$ are
$\Theta_{\rho,h,r}(E)$.
\end{corollary}

\begin{proof}
Take $c_{\rho,h}\le1$. For sufficiently large $n$ the entropy
threshold $\lceil\rho n\rceil$ is below $n/2$.
\Cref{ldt:joint-extraction} gives the lower bound, and
\Cref{aq:multi-upper} gives the upper bound since
$m+\log n\le2E$. A tree computing $F_h$ also computes each
output parity, and degree-$2$ queries are allowed for every $r\ge2$.
\end{proof}

\begin{corollary}[Sublinear Entropy at Fixed Degree]
\label{ldt:sublinear-cubic}\label{ldt:sublinear-degree}
For every fixed $r\ge2$ and $0<\eta<1$, there is an explicit
non-malleable affine extractor with entropy threshold
$\lceil C_{\eta,r}n/\log n\rceil$, output length $\lfloor n^{1-\eta}\rfloor$,
and error at most $2^{-n^{1-\eta}}$, such that every nonzero
output parity $f$ satisfies
\[
 D_r(f)\ge\frac{n^{1-\eta}}{2^{r-2}}-O_r(1),\qquad
 D_r^{\mathrm{obl}}(f)\ge\frac{n^{1-\eta}}{2^{r-3}}-O_r(1).
\]
The same bounds hold for every fixed uniform-input error below $1/2$.
\end{corollary}

\begin{proof}
Use \Cref{thm:joint-affine-quantitative} with
$h=\max\{1,2^{r-2}-1\}$. Its common output and error exponent
is at least $n^{1-\eta}$ when $C_{\eta,r}$ is sufficiently large.
The entropy threshold is $o(n)$. Apply
\Cref{ldt:joint-extraction} and \Cref{ldt:transfer}; for $r=2$, the
ordinary affine extractor bound gives the sharper constant $E-1$.
\end{proof}

\begin{corollary}[Growing Query Degree]
\label{ldt:growing-degree-extractor}
Fix $0<\rho<1/2$ and $0<\eta<1$. There is a constant
$c_{\rho,\eta}>0$ such that, for all sufficiently large $n$
and integers $r\ge3$ satisfying
\[
 2^{r-2}\le c_{\rho,\eta}
            \sqrt{\frac{\log n}{\log\log n}},
\]
our non-malleable affine extractor has entropy threshold
$\lceil\rho n\rceil$, output length $\lfloor n^{1-\eta}\rfloor$,
and error at most $2^{-n^{1-\eta}}$, and every nonzero output
parity $f$ satisfies
\begin{equation}
 \begin{aligned}
 D_r(f)&\ge\frac{n^{1-\eta}-\log_2 3}{2^{r-2}},\\
 L_r(f)&\ge2^{(n^{1-\eta}-\log_2 3)/2^{r-2}},\\
 D_r^{\mathrm{obl}}(f)&\ge
                  \frac{n^{1-\eta}-\log_2 3}{2^{r-3}}.
 \end{aligned}
 \label{ldt:growing-degree-bounds}
\end{equation}
For uniform-input error at most $\gamma<1/2$, add
$\log_2(1-2\gamma)$ to the non-oblivious depth bound and twice this
quantity to the oblivious bound, and multiply the leaf bound by
$1-2\gamma$.
\end{corollary}

\begin{proof}
Set $h=2^{r-2}-1$ in \Cref{thm:joint-affine-quantitative}.
Its quantified tampering range gives the stated extractor parameters.
Put $a=r-2$ and $E=n^{1-\eta}$. For sufficiently large $n$,
$E\le n-a$ and $\lceil\rho n\rceil\le K_{a,n}$.
The probability of dependent differencing directions satisfies
\[
 p_{a,n}\le\sum_{j=0}^{a-1}2^{j-n}<2^{a-n}\le2^{-E}.
\]
Consequently $B_{a,n}\le3\cdot2^{-E}$ in
Eqn.~\eqref{ldt:cube-parameters}. Apply
\Cref{ldt:joint-extraction} and \Cref{ldt:transfer}.
\end{proof}

This permits query degrees up to
$\tfrac12\log_2\log n-\tfrac12\log_2\log\log n+O_{\rho,\eta}(1)$
while retaining a nearly linear error exponent. The conclusion
concerns every nonzero output parity of the extractor.

\subsubsection{Common Constant Spaces}

The following argument for finding an affine space of common zeros extends
\Cref{aq:common-flat}. Related bounds on constant affine spaces for
systems of low-degree polynomials are developed in~\cite{CT15}.
For integers $a\ge1$ and $u\ge0$, let
\begin{equation}
 W_a(u)=\sum_{j=0}^{a-1}(a-j)\binom uj.
 \label{ldt:flat-weight}
\end{equation}
Here $\binom uj=0$ when $j>u$.
In particular, $W_1(u)=1$, $W_2(u)=u+2$, and
$W_3(u)=3+2u+\binom u2$.

\begin{lemma}[An Affine Space Through a Prescribed Common Zero]
\label{ldt:common-flat}
Let $q_1,\ldots,q_t$ have degree at most $a$ on $\F_2^N$,
and let $x_0$ be a common zero. Their common zero set contains an
affine space through $x_0$ of dimension $u$ satisfying
\begin{equation}
 N\le u+tW_a(u).
 \label{ldt:common-flat-bound}
\end{equation}
\end{lemma}

\begin{proof}
Translate $x_0$ to zero and choose an inclusion-maximal linear
common-zero space $U$, of dimension $u$. In complementary coordinates
$v\in\F_2^{N-u}$ and coordinates $z\in\F_2^u$ on $U$,
write
\[
 q_i(v,z)=\sum_{J\subseteq[u]}c_{i,J}(v)\prod_{j\in J}z_j.
\]
The coefficient $c_{i,J}$ has degree at most $a-|J|$ and vanishes
at zero. The coefficients for $|J|=a$ vanish identically, and those
for $|J|>a$ are absent. The sum of the degree bounds of the remaining
coefficients is at most $tW_a(u)$.

If $N-u>tW_a(u)$, the common-zero indicator
$\prod_{i,J}(1+c_{i,J}(v))$, viewed over $\F_2$, has degree
less than $N-u$. Its sum over $v$ is zero: every monomial misses a
coordinate. Thus the number of common zeros of the coefficient
system is even. Zero is one such solution, so there is a nonzero
solution $v$. Every $q_i$ then vanishes on both $U$ and $v+U$, and
hence on the larger linear space $U+\operatorname{span}\{v\}$.
This contradicts maximality and proves
Eqn.~\eqref{ldt:common-flat-bound}.
\end{proof}

Recall that a Boolean affine disperser at dimension $k$ is nonconstant
on every affine space of dimension at least $k$. A directional affine
disperser has this property for every nonzero derivative $D_af$.

\begin{theorem}[Depth From Affine Dispersion]
\label{ldt:disperser-depth}
If $f$ is an affine disperser at dimension $1\le k\le n$, then
\begin{equation}
 D_r(f)\ge
 \left\lceil\frac{n-k+1}{W_r(k-1)}\right\rceil.
 \label{ldt:disperser-adaptive}
\end{equation}
If $f$ is a directional affine disperser at dimension $2\le k\le n$,
then
\begin{equation}
 D_r^{\mathrm{obl}}(f)\ge
 \left\lceil\frac{n-k+1}{W_{r-1}(k-2)}\right\rceil.
 \label{ldt:disperser-oblivious}
\end{equation}
\end{theorem}

\begin{proof}
A nonempty leaf of length $t$ is defined by a consistent system of
$t$ degree-$r$ equations. Every affine space contained in this leaf
has dimension at most $k-1$. By \Cref{ldt:common-flat} and the
monotonicity of $W_r$, we have $n\le k-1+tW_r(k-1)$.

For the second bound, let $q_1,\ldots,q_t$ be the oblivious queries.
The assertion is immediate if $t\ge n$, so assume $t<n$. Two
distinct inputs have the same query vector; let $a\ne0$ be their
difference. The equations $D_aq_i(x)=0$ are consistent, have degree
at most $r-1$, and descend to the quotient by
$\operatorname{span}\{a\}$. A common-zero affine space of dimension
$u$ in this quotient lifts to one of dimension $u+1$ on which the
query vectors at $x$ and $x+a$ coincide. Consequently $D_af=0$
there, and directional dispersion implies $u\le k-2$.
\Cref{ldt:common-flat} on the $(n-1)$-dimensional quotient now gives
$n-1\le k-2+tW_{r-1}(k-2)$.
\end{proof}

For $r=2$, these are the bound from a common affine space in
\Cref{aq:generic-bounds} and the directional bound of~\cite{GG26}.
For $r=3$, they read
\[
 D_3(f)\ge
 \left\lceil\frac{2(n-k+1)}{k^2+k+4}\right\rceil,
 \qquad
 D_3^{\mathrm{obl}}(f)\ge
 \left\lceil\frac{n-k+1}{k}\right\rceil.
\]
They have no error parameter, but are weaker than linear correlation
bounds when the entropy threshold is a fixed fraction of $n$.

\subsection{Non-Oblivious Local Decision Trees}
\label{sec:adaptive-local}

A non-oblivious $\ell$-local decision tree queries a Boolean function of at most
$\ell$ input coordinates at each internal node~\cite{GG26}. Both the
function and its support may depend on previous answers. Let
$\operatorname{LDT}_{\ell}(f)$ denote its minimum exact worst-case depth.
The oblivious model fixes the query at each level.
All logarithms in this section are to base two, except $\ln$.

We first derive exact lower bounds by maintaining a product $A\times B$ of sets on two disjoint coordinate blocks along a computation path. To obtain stronger bounds when $\ell$ grows, we choose the coordinate partition using all query supports on the path at once.

In the respective locality ranges below, a sumset extractor gives a logarithmic factor over $n/\ell$, while a directional affine extractor of exponentially small error gives a factor $\log\ell/\log\log\ell$ and a bound on correlation. The latter argument also applies to every nonzero output parity of our non-malleable affine extractors. Throughout, the queries may depend on previous answers and their supports may overlap arbitrarily.

\subsubsection{Preserving a Rectangle Along a Path}
\label{subsec:local-rectangle}

A Boolean sumset disperser at entropy $(k,k)$ is a function
$f:\F_2^n\to\F_2$ that is nonconstant on $A+B$ whenever
$|A|,|B|\ge2^k$~\cite{Li23,GG26}. A sumset extractor with error
$\varepsilon$ instead requires $f(X+Y)\approx_\varepsilon U_1$
for all independent sources of min-entropy at least $k$ each.
Error below $1/2$ implies dispersal.

\begin{theorem}[Rectangle depth bound]
\label{thm:local-rectangle}
Fix a partition $I\sqcup J=[n]$ with
$|I|,|J|\ge h=\lfloor n/2\rfloor$. Suppose $f$ is nonconstant on
every product $A\times B$, where
$A\subseteq\F_2^I$, $B\subseteq\F_2^J$, and $|A|,|B|\ge2^k$.
For $\ell\ge3$, put
\begin{equation}
 \begin{aligned}
 c_\ell&=\frac{\lfloor\ell^2/4\rfloor-1}{\ell-2},\\
 b_\ell&=(\ell-2)\sqrt{\frac{\ln2}{2}}.
 \end{aligned}
 \label{eq:local-constants}
\end{equation}
Every exact non-oblivious $\ell$-local tree of depth $d$ computing $f$
satisfies
\begin{equation}
 c_\ell d+b_\ell\sqrt d>h-k.
 \label{eq:local-rectangle}
\end{equation}
At locality three the stronger bound $d>h-k$ holds.
In particular, the hypotheses hold for every sumset disperser at
entropy $(k,k)$.
\end{theorem}

\begin{proof}
Maintain nonempty sets $A\subseteq\F_2^I$ and
$B\subseteq\F_2^J$ whose product reaches one node of the tree.
Initially both sets are full cubes. Suppose the current query uses
$a$ coordinates of $I$ and $b$ of $J$, with $a+b\le\ell$.
If $a,b>0$, fix the queried coordinates in $A$ to a most frequent
pattern, retaining at least a $2^{-a}$ fraction. The query now
depends only on $B$; retain its more frequent answer, keeping at
least half of $B$. Interchanging the sources gives another choice.
Thus the two available upper bounds on the losses in
$(\log|A|,\log|B|)$ are
\begin{equation}
 (a,1)\quad\hbox{and}\quad(1,b).
 \label{eq:local-loss-vectors}
\end{equation}
Each choice retains a product contained in an actual child.
A one-sided query costs at most $(1,0)$ or $(0,1)$; a constant
query costs nothing.

For $a+b>2$, choose the first vector in
Eqn.~\eqref{eq:local-loss-vectors} with probability
$(b-1)/(a+b-2)$, and the second otherwise. Each coordinate of
the resulting random vector $V$ has expectation
\[
 \frac{ab-1}{a+b-2}\le c_\ell.
\]
Indeed, at a fixed sum $s=a+b$ the product is largest at the
balanced integer pair, and
$(\lfloor s^2/4\rfloor-1)/(s-2)$ is nondecreasing for $s\ge3$.
If $a=b=1$, use $(1,1)$ deterministically. Each coordinate of
$V$ ranges over an interval of length at most $\ell-2$.

Let $D_I,D_J$ be cumulative upper bounds on the entropy losses,
and fix $\eta>0$. Use the potential
$\Phi=e^{\eta D_I}+e^{\eta D_J}$.
A random variable $Z$ in an interval of length $R$ satisfies
\[
 \E e^{\eta Z}\le
 \exp\bigl(\eta\E Z+\eta^2R^2/8\bigr).
\]
To see this, the second derivative of
$\log\E e^{\eta Z}$ is a variance under an exponentially
tilted distribution, hence at most $R^2/4$; integrate twice.
Therefore one of the two restrictions multiplies $\Phi$ by at most
\[
 \exp\bigl(\eta c_\ell+\eta^2(\ell-2)^2/8\bigr).
\]
Choose that restriction. The deterministic one-sided cases satisfy
the same estimate. Starting from $\Phi=2$, after at most $d$ queries
we obtain
\[
 \max\{D_I,D_J\}
 \le c_\ell d+\frac{\ln2}{\eta}
                 +\frac{d\eta(\ell-2)^2}{8}.
\]
For $d>0$, optimizing $\eta$ gives the right side of
Eqn.~\eqref{eq:local-rectangle}. If this quantity is at most
$h-k$, both sets have size at least $2^k$ at the resulting leaf,
contrary to the hypothesis on $f$. The case $d=0$ is immediate.

At locality three, every two-sided query has $\min\{a,b\}=1$.
One choice in Eqn.~\eqref{eq:local-loss-vectors} then costs at
most $(1,1)$, so each source loses at most $d$ bits.
Finally, embed $A$ and $B$ in their disjoint coordinate blocks,
with zeroes elsewhere. Their sumset is exactly $A\times B$.
\end{proof}

For $h\ge k$, Eqn.~\eqref{eq:local-rectangle} is equivalent to
\begin{equation}
 d>
 \left(
 \frac{\sqrt{b_\ell^2+4c_\ell(h-k)}-b_\ell}{2c_\ell}
 \right)^2.
 \label{eq:local-explicit-depth}
\end{equation}
The proof follows one path and allows arbitrary overlap between
query supports. It gives an exact depth bound; it does not give an
exponential leaf-count bound or a distributional lower bound.

\begin{proposition}[Small localities]
\label{prop:local-small}
If $f$ is an affine disperser at dimension $k$, then
\begin{equation}
 \operatorname{LDT}_{2}(f)\ge n-k+1,
 \qquad
 \operatorname{LDT}_{3}(f),\operatorname{LDT}_{4}(f)
 \ge\left\lfloor\frac{n-k}{2}\right\rfloor+1.
 \label{eq:local-small}
\end{equation}
\end{proposition}

\begin{proof}
Maintain an affine space $W$ of inputs reaching the current node.
Its projection onto the query support is an affine space of
dimension $r\le\ell$. A monochromatic affine subspace of this
image pulls back to a subspace of the same codimension in $W$.
For $r\le1$, fixing a point costs at most one dimension. For
$r=2,3$, some color contains two points, which form an affine line.
For $r=4$, some color contains at least eight points. Among their
$28$ unordered pairs two have the same nonzero sum, since only
$15$ sums are possible. These pairs are disjoint and their union
is an affine plane. Thus the codimension cost is at most one for
$\ell=2$, and at most two for $\ell=3,4$.
At a leaf the surviving affine space has dimension at least
$n-d$ or $n-2d$, respectively. Its dimension must be below $k$.
\end{proof}

\subsubsection{One Explicit Family for All Localities}
\label{subsec:local-instantiation}

\begin{corollary}
\label{cor:local-explicit-family}
There is an explicit Boolean family $f_n$, independent of $\ell$,
such that, uniformly for $3\le\ell\le n$,
\begin{equation}
 \operatorname{LDT}_{\ell}(f_n)
 \ge\frac{2n}{\ell+2}
 -O\left(\frac{\log n}{\ell}+\sqrt{\frac n\ell}+1\right).
 \label{eq:local-uniform}
\end{equation}
The same family satisfies
$\operatorname{LDT}_{2}(f_n)\ge n-O(\log n)$ and
$\operatorname{LDT}_{3}(f_n),\operatorname{LDT}_{4}(f_n)
\ge n/2-O(\log n)$.
For $\ell\to\infty$ and $\ell=o(n)$,
\begin{equation}
 \operatorname{LDT}_{\ell}(f_n)\ge(2-o(1))n/\ell.
 \label{eq:local-growing}
\end{equation}
\end{corollary}

\begin{proof}
The sumset extractor of Li~\cite{Li23} gives one Boolean family at
entropy $k=C\log n$ and any prescribed constant error. Fix an
error below $1/2$ and apply \Cref{thm:local-rectangle}.
If $d<(h-k)/c_\ell$, Eqn.~\eqref{eq:local-rectangle} implies
\[
 d>\frac{h-k}{c_\ell}
       -\frac{b_\ell}{c_\ell}\sqrt{\frac{h-k}{c_\ell}}.
\]
The same weak lower bound holds when $d\ge(h-k)/c_\ell$.
Now $c_\ell=\Theta(\ell)$, $b_\ell/c_\ell=O(1)$, and
\[
 c_\ell=\begin{cases}
 (\ell+2)/4,&\ell\text{ even},\\
 (\ell+2)/4-1/(4(\ell-2)),&\ell\text{ odd}.
 \end{cases}
\]
This proves Eqn.~\eqref{eq:local-uniform}, including the vacuous
small-$n$ cases after adjusting the absolute constant.
A sumset disperser is an affine disperser: take independent
uniform sources on an affine space and its direction space.
Thus \Cref{prop:local-small} applies to the same family.
Finally divide the error term in Eqn.~\eqref{eq:local-uniform}
by $n/\ell$ to obtain Eqn.~\eqref{eq:local-growing}.
\end{proof}

For comparison, the classical code argument of
Paturi, Pudl\'ak, Saks, and Zane~\cite[Theorems~4--5]{PPSZ05}
gives, for fixed $\ell\ge3$, exact non-oblivious depth
$\beta_\ell n-o(n)$ for explicit code indicators, where
\begin{equation}
 \beta_\ell=\sum_{j\ge1}\frac{1}{j((\ell-1)j+1)}.
 \label{eq:local-ppsz}
\end{equation}
Each accepting leaf is an $\ell$-CNF contained in the code:
express each local query constraint by clauses on its support.
An assignment satisfying a CNF is isolated if flipping any one bit
makes it falsify the CNF. The isolated-assignment bound limits the number of codewords at
one leaf, and the leaves must cover the code. This also proves
an exponential leaf-count bound, a conclusion not supplied by
\Cref{thm:local-rectangle} and \Cref{prop:local-small}.

\begin{table}[t]
\centering
\begin{tabular}{@{}lll@{}}
\toprule
Locality & Classical PPSZ depth coefficient & Depth bound here\\
\midrule
$3$ & $0.61370564$ & $n/2-O(\log n)$\\
$4$ & $0.44518188$ & $n/2-O(\log n)$\\
$11$ & $0.15346072$ & $9n/58-O(\sqrt n)$\\
$16$ & $0.10462126$ & $n/9-O(\sqrt n)$\\
\bottomrule
\end{tabular}
\caption{Exact non-oblivious depth bounds. The middle column means
$\beta_\ell n-o(n)$ for a code indicator; the last column uses
one logarithmic-entropy sumset disperser. The hard functions and
the available leaf-count conclusions differ.}
\label{tab:local-comparison}
\end{table}

In particular, the PPSZ bound is stronger at locality three.
The affine argument gives a larger depth coefficient at locality
four, and the rectangle argument does so at localities eleven
and sixteen. For growing locality,
$\ell\beta_\ell\to\sum_{j\ge1}j^{-2}=\pi^2/6$ by dominated
convergence. More explicitly, a Bose--Chaudhuri--Hocquenghem (BCH) code of distance
$\Theta(\log n)$ and redundancy $O(\log^2n)$ in the PPSZ bound
gives
\[
 \beta_\ell n
 -O\left(\frac{n}{(\ell-2)\log n}+\log^2n\right).
\]
Thus, in the common range $\ell\to\infty$ and
$\ell=o(n/\log^2n)$, the coefficients are $\pi^2/6$ and $2$.

\begin{remark}
\label{rem:local-rectangle-cost}
A single $\ell$-local query can force a loss of $\Omega(\ell)$ bits when the surviving inputs must form a rectangle. For
$q(x,y)=\langle x,y\rangle$ on two $r$-bit blocks,
every monochromatic rectangle $A\times B$ satisfies
$|A||B|\le2^r$. Indeed, constancy implies
$\langle a+a',b+b'\rangle=0$ for all $a,a'\in A$, $b,b'\in B$.
Hence $\Span(A-A)$ and $\Span(B-B)$ are orthogonal, and their
dimensions sum to at most $r$. Since $A$ and $B$ lie in cosets
of these spans, the bound follows. With $\ell=2r$, one query
can therefore force at least $\ell/2$ bits of total loss.
In the next argument, we choose the coordinate partition using all query supports on a path at once.
\end{remark}

\subsubsection{Logarithmic Overhead from Sumset Extractors}
\label{subsec:local-correlation}

Call a partition $I\sqcup J\sqcup S=[n]$ \emph{separated} for a
root-to-leaf path if no query support on the path meets both $I$ and
$J$. The partition may depend on the leaf. After fixing $x_S=s$,
the leaf is a rectangle $A_s\times B_s\times\{s\}$, since each
remaining query constraint depends only on $I$ or only on $J$.
The next lemma lets us apply the sumset extractor guarantee even when
one of these sets is smaller than the entropy threshold.

\begin{lemma}[Signed completion to the entropy threshold]
\label{lem:local-signed-completion}
Let $A\subseteq\F_2^m$, where $m\ge k+1$, and put
$\alpha=|A|/2^m$ and $\theta=2^{k-m}$.
The measure $2^{-m}\mathbf1_A$ is a signed combination of
probability distributions of min-entropy at least $k$, with absolute
coefficient sum at most $\min\{1,\alpha+2\theta\}$.
\end{lemma}

\begin{proof}
If $\alpha\ge\theta$, use $2^{-m}\mathbf1_A=\alpha U_A$.
The claim is immediate if $A$ is empty. Otherwise, when
$0<\alpha<\theta\le1/2$, let $R$ be uniform on
$\F_2^m\setminus A$ and put
\[
 P=\theta^{-1}2^{-m}\mathbf1_A+(1-\alpha/\theta)R.
\]
Both $P$ and $R$ have point masses at most $2^{-k}$: on $A$,
$P$ has mass exactly $2^{-k}$, and on its complement its mass is
$(\theta-\alpha)/(\theta(1-\alpha)2^m)\le2^{-k}$.
Now $2^{-m}\mathbf1_A=\theta P-(\theta-\alpha)R$, whose
absolute coefficient sum is $2\theta-\alpha\le\min\{1,\alpha+2\theta\}$.
\end{proof}

\begin{proposition}[Separated leaves]
\label{prop:local-separated-leaves}
Suppose every path of a depth-$d$ tree $T$ with $L$ leaves has a
separated partition with $|I|,|J|\ge m\ge k+1$.
If $f$ is a Boolean sumset extractor at entropy $(k,k)$ and error
$\varepsilon$, then
\begin{equation}
 \operatorname{Corr}(f,T)
 \le2\varepsilon(1+6L2^{k-m})
 \le2\varepsilon(1+6\cdot2^{d+k-m}).
 \label{eq:local-separated-correlation}
\end{equation}
If $f$ is only a sumset disperser at entropy $(k,k)$, exact computation
is impossible whenever $m\ge d+k$.
\end{proposition}

\begin{proof}
Fix a leaf $\tau$ and its separated partition. On a slice $x_S=s$,
write the leaf as $A_s\times B_s\times\{s\}$, with densities
$\alpha_s,\beta_s$ in their respective coordinate cubes.
Let $\theta=2^{k-m}\le1/2$.
Apply \Cref{lem:local-signed-completion} to both factors.
Their product has a signed decomposition with absolute coefficient sum
at most
\[
 (\alpha_s+2\theta)(\beta_s+2\theta)
 \le\alpha_s\beta_s+6\theta.
\]
Each product distribution is the sum of two independent entropy-$k$
sources: embed the first in $I\cup S$, with fixed $S$ coordinates,
and the second in $J$, with zeroes elsewhere.
The absolute bias of $f$ on each such sum is at most $2\varepsilon$.
Averaging over $s$ bounds the leaf's signed contribution by
$2\varepsilon(p_\tau+6\theta)$, where
$p_\tau=\Pr[U_n\text{ reaches }\tau]=\E_s\alpha_s\beta_s$.
Sum over the leaves and use $\sum_\tau p_\tau=1$.

For dispersion, some leaf has probability at least $2^{-d}$, and
some slice of that leaf satisfies $\alpha_s\beta_s\ge2^{-d}$.
Both factors therefore have density at least $2^{-d}$, and each set
has size at least $2^{m-d}\ge2^k$. Their product is an admissible
sumset contained in a monochromatic leaf, a contradiction.
\end{proof}

\begin{lemma}[Separating the path supports]
\label{lem:local-symmetric-separator}
For any collection of at most $d$ supports of size at most $\ell$,
put $\mu=(n/2)2^{-d\ell/n}$.
If $\mu\ge8\ell\ln4$, there is a separated partition with
$|I|,|J|\ge\lfloor\mu/2\rfloor$.
\end{lemma}

\begin{proof}
Add a singleton support for each coordinate absent from the collection.
If $t_v$ is the resulting incidence count of coordinate $v$, then
$t_v\ge1$ and $\sum_vt_v\le d\ell+n$.
Color the supports independently red or blue with equal probability.
Let $I$ consist of coordinates whose incident supports are all red,
and $J$ of those whose incident supports are all blue.
These sets are disjoint, and no original support meets both.
Convexity gives
\[
 \E|I|=\E|J|=\sum_v2^{-t_v}\ge n2^{-(d\ell+n)/n}=\mu.
\]
Each random support color affects at most $\ell$ coordinate
indicators. The read-$\ell$ tail bound of
Gavinsky, Lovett, Saks, and Srinivasan~\cite[Theorem~1.1]{GLSS15}
states that a sum of such indicators with expectation $\nu$ is
below $\nu/2$ with probability at most $e^{-\nu/(8\ell)}$.
Consequently each event $|I|<\mu/2$ and $|J|<\mu/2$ has
probability at most $1/4$. Some coloring avoids both events.
\end{proof}

\begin{theorem}[Logarithmic overhead from sumset extractors]
\label{thm:local-sumset-logarithmic}
Fix $0<\delta<1$ and $0<\varepsilon<1/2$.
Let $f_n$ be a Boolean sumset extractor at entropy
$(C\log n,C\log n)$ and error $\varepsilon$, for a fixed $C$.
There is $c_\delta>0$ such that every non-oblivious $\ell$-local
tree $T$ satisfies
\begin{equation}
 \operatorname{Corr}(f_n,T)\le2\varepsilon(1+o(1))
 \quad\text{if}\quad
 \operatorname{depth}(T)\le c_\delta\frac n\ell\log\ell,
 \label{eq:local-sumset-logarithmic}
\end{equation}
whenever $\ell\to\infty$ and $\ell\le n^{1-\delta}$.
In particular,
\[
 \operatorname{LDT}_{\ell}(f_n)
 =\Omega_\delta\left(\frac n\ell\log\ell\right).
\]
The explicit sumset extractor of~\cite{Li23} supplies one such family
for every prescribed constant $\varepsilon$.
\end{theorem}

\begin{proof}
Take
$c_\delta=\min\{1/4,\delta/(4(1-\delta))\}$ and
$d\le c_\delta(n/\ell)\log\ell$.
Then $\mu=(n/2)2^{-d\ell/n}\ge n/(2\ell^{c_\delta})$.
Since $\ell\le n^{1-\delta}$,
\[
 \frac\mu\ell\ge\frac12n^{1-(1-\delta)(1+c_\delta)}
 \longrightarrow\infty.
\]
Thus \Cref{lem:local-symmetric-separator} applies to every leaf path
with $m=\lfloor\mu/2\rfloor$.
Moreover,
\[
 \frac d\mu=O\left(\frac{\log\ell}{\ell^{1-c_\delta}}\right)=o(1),
 \qquad C\log n=o(\mu).
\]
Hence $m-d-C\log n\to\infty$, and
\Cref{prop:local-separated-leaves} proves the correlation bound and
excludes exact computation.
\end{proof}

The improvement over Eqn.~\eqref{eq:local-growing} comes from
choosing a partition for the whole path. It does not require
non-malleability or exponentially small extractor error.
For any fixed prediction advantage, the error of the sumset extractor
can be chosen as a sufficiently small constant.

\subsubsection{Directional Affine Extractors with Small Error}
\label{subsec:local-directional}

The directional derivative bias bound stated after
\Cref{def:directional-intro} allows one side of the separated partition
to be much smaller than the entropy threshold. We use the bound
\begin{equation}
 \left|\E(-1)^{f(X)+f(X+a)}\right|\le2\varepsilon
 \quad\text{for every }a\ne0
 \text{ and affine }X\text{ of dimension at least }k.
 \label{eq:local-directional-bias}
\end{equation}
Every nonzero output parity of a $(k,\varepsilon)$ directional affine
extractor satisfies Eqn.~\eqref{eq:local-directional-bias}.

\begin{lemma}[An asymmetric separated partition]
\label{lem:local-asymmetric-separator}
Fix $0<\rho<1/2$. For a collection of at most $d$ supports of
size at most $\ell$, set
\begin{itemize}
 \item $a_\rho=(1-\rho)/2$ and $\kappa_\rho=a_\rho(\ln4-3/4)$;
 \item $\lambda=1+d\ell/n$ and $p=a_\rho/(4\lambda)$;
 \item $t=\lfloor np^\lambda-ne^{-\kappa_\rho n/\ell}\rfloor$.
\end{itemize}
If $t>0$, there is a separated partition with
$|I|\ge t$ and $|J|\ge(1-a_\rho)n$.
\end{lemma}

\begin{proof}
Again add singleton supports for unused coordinates. Their sizes
$s_i$ satisfy $1\le s_i\le\ell$ and $\sum_i s_i\le\lambda n$.
Color each support red independently with probability $p$.
Define $I,J$ by all-red and all-blue incidence, respectively.
As before, $\E|I|\ge np^\lambda$.
Let $W$ be the sum of the sizes of the red supports. Then
$n-|J|\le W$, and independence gives
\[
 \E4^{W/\ell}
 =\prod_i(1+p(4^{s_i/\ell}-1))
 \le\exp\left(\frac{3p}{\ell}\sum_i s_i\right)
 \le e^{3a_\rho n/(4\ell)}.
\]
Here $4^x-1\le3x$ for $0\le x\le1$.
Markov's inequality yields
$\Pr[W>a_\rho n]\le e^{-\kappa_\rho n/\ell}$.
On the event $G=\{W\le a_\rho n\}$, the set $J$ has the
required size, and
\[
 \E[|I|\mathbf1_G]\ge np^\lambda-ne^{-\kappa_\rho n/\ell}.
\]
Some coloring in $G$ has $|I|$ at least this expectation,
which proves the assertion.
\end{proof}

\begin{lemma}[Directional derivative bounds on separated leaves]
\label{lem:local-directional-rectangle}
Suppose $f$ satisfies Eqn.~\eqref{eq:local-directional-bias}.
If each path of a depth-$d$ tree $T$ with $L$ leaves has a separated
partition with $|I|\ge t$ and $|J|\ge k$, then
\begin{equation}
 \operatorname{Corr}(f,T)^2
 \le L(2^{-t}+2\varepsilon)
 \le2^d(2^{-t}+2\varepsilon).
 \label{eq:local-directional-correlation}
\end{equation}
\end{lemma}

\begin{proof}
Fix a leaf and a slice $x_S=s$, with rectangle $A\times B$ of
densities $\alpha,\beta$. Write $\phi(u,v)=(-1)^{f(u,v,s)}$.
For distinct $u,u'$, the source $(u,U_J,s)$ has dimension
$|J|\ge k$, and translation by $(u+u',0_J,0_S)$ is nonzero.
Thus Eqn.~\eqref{eq:local-directional-bias} gives
\[
 \left|\E_v\phi(u,v)\phi(u',v)\right|\le2\varepsilon.
\]
Put $R(v)=\E_u\mathbf1_A(u)\phi(u,v)$.
Expanding the second moment, separating equal and unequal rows, gives
\[
 \E_vR(v)^2
 \le\alpha2^{-|I|}+2\varepsilon\alpha^2
 \le\alpha(2^{-t}+2\varepsilon).
\]
Cauchy--Schwarz therefore bounds the signed contribution of this
rectangle by
\[
 \left|\E_v\mathbf1_B(v)R(v)\right|
 \le\sqrt{\alpha\beta(2^{-t}+2\varepsilon)}.
\]
Average over $s$ and apply Cauchy--Schwarz again: a leaf of probability
$p_\tau$ contributes at most
$\sqrt{p_\tau(2^{-t}+2\varepsilon)}$.
Finally, $\sum_\tau\sqrt{p_\tau}\le\sqrt L$.
The derivative bound is used on the full $J$-cube before restriction
to $B$, so we do not need to assume that the conditional distribution
on the leaf is an affine source.
\end{proof}

Choose $t$ as in \Cref{lem:local-asymmetric-separator}. If $k\le(1-a_\rho)n$ and $t>0$, then \Cref{lem:local-directional-rectangle} applies. In particular, for any $b\ge0$ and error $\varepsilon=2^{-E}$, the correlation is at most $2^{-b}$ whenever $t,E\ge d+2b+2$.

\begin{theorem}[Local hardness from directional derivative bounds]
\label{thm:local-directional-logarithmic}
Fix $0<\rho<1/2$ and $a>0$.
Suppose $f_n$ satisfies Eqn.~\eqref{eq:local-directional-bias} with
$k=\lceil\rho n\rceil$ and $\varepsilon\le2^{-an}$.
There is $C_\rho>0$ such that, whenever
$\ell\to\infty$ and $\ell\le n/(C_\rho\log n)$, every
non-oblivious $\ell$-local tree of depth
\begin{equation}
 d\le\left\lfloor\frac{n}{4\ell}
                  \frac{\log\ell}{\log\log\ell}\right\rfloor
 \label{eq:local-directional-depth}
\end{equation}
has correlation at most $2^{-\Omega_{\rho,a}(n/\sqrt\ell)}$ with $f_n$.
Consequently,
\[
 \operatorname{LDT}_{\ell}(f_n)
 =\Omega_{\rho,a}\left(\frac n\ell
                         \frac{\log\ell}{\log\log\ell}\right).
\]
\end{theorem}

\begin{proof}
For the parameters of \Cref{lem:local-asymmetric-separator},
Eqn.~\eqref{eq:local-directional-depth} gives
\[
 \lambda\le1+\frac{\log\ell}{4\log\log\ell},
 \qquad
 \lambda\log(4\lambda/a_\rho)\le\tfrac12\log\ell
\]
for all sufficiently large $\ell$. Hence $p^\lambda\ge\ell^{-1/2}$.
Choose $C_\rho$ large enough that
$ne^{-\kappa_\rho n/\ell}\le n^{-4}$ in the stated range.
Every leaf path then has a separated partition with
\[
 |I|\ge t\ge(1-o(1))n/\sqrt\ell,
 \qquad |J|\ge(1+\rho)n/2\ge\lceil\rho n\rceil.
\]
Since $d=o(n/\sqrt\ell)$ and $\varepsilon\le2^{-an}$,
\Cref{lem:local-directional-rectangle} gives
$\operatorname{Corr}(f_n,T)^2
\le2^d(2^{-t}+2\varepsilon)=2^{-\Omega_{\rho,a}(n/\sqrt\ell)}$.
Taking square roots proves the correlation bound; exact computation
would have correlation one.
\end{proof}

\begin{corollary}[Non-malleable affine extractors]
\label{cor:local-anm-logarithmic}
Let $F_n:\F_2^n\to\F_2^m$ be a non-malleable affine extractor at
entropy $\lceil\rho n\rceil$ and error at most $2^{-an}$, for fixed
$0<\rho<1/2$ and $a>0$.
Every nonzero output parity $\langle u,F_n\rangle$ satisfies the
correlation and depth bounds of \Cref{thm:local-directional-logarithmic}.
Only single-tampering non-malleability is needed.
\end{corollary}

\begin{proof}
For every nonzero $c$, the affine map $x\mapsto x+c$ has no fixed
points. Non-malleability, followed by the character
$(z,z')\mapsto(-1)^{\langle u,z+z'\rangle}$, gives
Eqn.~\eqref{eq:local-directional-bias} with error $\varepsilon$.
Apply \Cref{thm:local-directional-logarithmic}.
\end{proof}

The same conclusion already applies to the published directional
affine extractors of Li and Zhong~\cite{LZdir}.
Our non-malleable affine extractor constructions supply further instances;
general affine tampering is not required by this reduction.
The correlation bounds in this subsection and in
\Cref{thm:local-sumset-logarithmic} also hold for randomized trees
whose depth is bounded for every choice of coins, by averaging over
the coins independently of the input.

\subsubsection{Parameters for Oblivious Local Trees}
\label{local:oblivious-parameters}

We record a parameter consequence of the sumset condenser and the
concentration theorem of Golovnev and Gurumukhani~\cite{GG26}.
This uses their construction and gives a more precise range than
their stated $n/\polylog n$ locality bound.

A $(k_1,k_2,k_{\rm out})$ sumset condenser with error $\eta$ maps
$X+Y$ to a distribution $\eta$-close to one of min-entropy
$k_{\rm out}$ whenever $X,Y$ are independent $n$-bit sources of
min-entropies at least $k_1,k_2$. By~\cite[Theorem~5.1]{GG26},
there are constants $C'>C>1$ and $\gamma>0$ such that, for every
constant $\delta>0$ and parameters satisfying
\[
 \begin{aligned}
 k_1&\ge k_2\ge\log^C n,\\
 m&\le(1-\delta)k_1,
 \end{aligned}
\]
such a condenser with output length $m$ is computable in polynomial
time. Its output entropy and error parameters are
\[
 \begin{aligned}
 k_{\rm out}&=m-\log^{C'}n,\\
 \eta&=n^{-\gamma}.
 \end{aligned}
\]
All logarithms in this subsection are binary.

\begin{corollary}[A parameter consequence of~\cite{GG26}]
\label{local:optimized-condenser}
For the constants $C'>C>1$ above, suppose
$\ell\log^{C-1}n=o(n)$. There is a polynomial-time computable
function $F:\{0,1\}^n\to\{0,1\}^m$ with
\begin{equation}
 m=O\!\left(
 \ell\log^{C-1}n\log\frac{en}{\ell\log^{C-1}n}
 +\log^{C'}n\right)=o(n)
 \label{local:optimized-output}
\end{equation}
such that every oblivious $\ell$-local decision tree computing
$F$ has depth at least $n-2m+1$.
\end{corollary}

\begin{proof}
Put $P=\lceil\log^{C-1}n\rceil$ and
$\Delta=\lceil\log^{C'}n\rceil$. Choose the remaining parameters in order:
\begin{itemize}
\item Set $\alpha=\lceil4\ell P\rceil$ and define
\[
 R=\alpha\left(\log\frac n\alpha+3\right).
\]
\item Set the output length to $m=\lceil4(R+\Delta)\rceil$.
\item Set $d=n-2m$.
\end{itemize}
These parameters satisfy $\alpha=o(n)$, $m=o(n)$ and
$d\ge n/2$ for sufficiently large $n$. In particular,
$2\ell d/n\le\alpha\le d$ and $m\le n$.
The concentration theorem~\cite[Theorem~5.2]{GG26} applies to
source entropies
\[
 \begin{aligned}
 k_1&=n-d=2m,\\
 k_2&=\frac{\alpha n}{2d\ell}
                  \log\frac{2d\ell}{\alpha},
 \end{aligned}
\]
and requires output entropy at least
$\alpha(\log(d/\alpha)+3)$. Since
$4\ell P\le\alpha\le5\ell P$, we have
\[
 k_2\ge2P\log\frac n{5P}\ge\lceil\log^C n\rceil
\]
for all sufficiently large $n$. We therefore use the condenser
above with second-source threshold $\lceil\log^C n\rceil$,
first-source threshold $2m$, and $\delta=1/2$. Its output entropy
is at least
\[
 m-\Delta\ge R\ge\alpha\left(\log\frac d\alpha+3\right),
\]
and its error is below $1/2$. Theorem~5.2 of~\cite{GG26} gives
depth at least $d+1$. The stated bound on $m$ follows from the
choice of $\alpha$.
\end{proof}

For example, locality
$\ell=\Theta(n/(\log^{C-1}n\log\log n))$ gives output length
\[
 m=O\!\left(\frac{n\log\log\log n}{\log\log n}
                   +\log^{C'}n\right)=o(n)
\]
and depth $n-o(n)$. The depth overhead $\ell d/n$ is then
$\Omega(n/(\log^{C-1}n\log\log n))$. These are consequences of
the technical theorems of~\cite{GG26}; they do not provide a new
condenser or resolve its linear-overhead question.

The output length is relevant to that question. The parameters
in~\cite[Theorem~5.2]{GG26} always satisfy
\begin{equation}
 m\ge k_{\rm out}
 =\alpha\left(\log\frac d\alpha+3\right)
 \ge3\alpha\ge\frac{6\ell d}{n}.
 \label{local:condenser-output-barrier}
\end{equation}
Thus linear locality and linear depth force linear output length
in this argument, even with an improved condenser. This limitation
concerns that reduction. The small-output formulation is essential:
the identity map already has depth $n$ by its number of outputs.

A selector does not generally preserve this vector lower bound.
For $F:\{0,1\}^n\to\{0,1\}^m$, put
$s=\lceil\log m\rceil$ and $f(x,i)=F(x)_i$, with a fixed output
on invalid indices. If $f$ has an oblivious $\ell$-local tree of
depth $t$, fix each valid value of $i$ and concatenate the resulting
query lists. This gives a tree for $F$ of depth at most $mt$.
Consequently a depth lower bound $D$ for $F$ yields only
$t\ge D/m$ by this conversion, on $n+s$ input bits. The loss can
matter: for $F(x)=x$ and $\ell>s$, partition the data coordinates
into blocks of size at most $\ell-s$. One local query per block
tests whether the selector points to a one in that block; their
OR computes $f$. Its oblivious depth is at most
$\lceil n/(\ell-s)\rceil$, despite the vector depth $n$.

The logarithmic-entropy sumset extractor of Li~\cite[Theorem~7.13]{Li23}
has one output bit and constant error. It does not directly replace
the strong sumset extractor of longer output used to supply the
seed in~\cite[Section~6]{GG26}. A longer strong output at this
entropy, or a different reduction avoiding
Eqn.~\eqref{local:condenser-output-barrier}, would be needed to
improve this route further.

\section{Lower Bounds for Resolution over Parities}
\label{sec:proof-complexity}

We first use non-malleable affine extractors to derandomize lossless
lifting and obtain near-maximal size lower bounds at depth $N$.
We then give overlapping extractor gadgets preserving a known
size--depth tradeoff, and a separate condensation construction that
improves the tradeoff for formulas with short ordinary-resolution proofs.

\subsection{Deterministic Lossless Lifting}
\label{sec:deterministic-lossless}

We now use the non-malleable affine extractor of
\Cref{thm:joint-affine-tuple} to derandomize the lossless lifting
construction of Efremenko and Itsykson~\cite{EI26}.
For every fixed loss in the exponent, the construction is uniformly
polynomial-time and counts all variables in the resulting CNF.

\begin{theorem}[Deterministic Lossless Lifting]
\label{thm:deterministic-lossless}
For every fixed $0<\xi<1$, there is a deterministic polynomial-time
construction of an infinite family of unsatisfiable CNFs on $N$
Boolean variables, with polynomially many clauses of width
$O_\xi(\log N)$, such that:
\begin{itemize}
\item every $\operatorname{Res}(\oplus)$ refutation has width at least
$(1-\xi)N$;
\item every $\operatorname{Res}(\oplus)$ refutation of resolution depth
at most $N$ has size at least $2^{(1-\xi)N}$.
\end{itemize}
The polynomial bounds may depend on the fixed $\xi$.
\end{theorem}

We use the width, size, and resolution-depth conventions from
\Cref{subsec:proof-complexity-prelim}.
The size bound also holds in the more general path-free fragment
specified in \Cref{formula:lifting}. It does not assert a size lower
bound for unrestricted-depth refutations.

The required gadget property is simultaneous surjectivity on affine
sources with sufficient rank in every set of blocks. For an affine
source $X=(X_1,\ldots,X_r)$ with $\ell$-bit blocks, assume that every
$I\subseteq[r]$ satisfies
\[
 H_\infty(X_I)\ge\eta\ell|I|.
\]
We call this the \emph{rank-spread condition}.
For any fixed output length $a$, we construct one family
$g_i:\F_2^\ell\to\F_2^a$, $i\in[r]$, such that, for every such
source and every nonzero coefficient tuple
$v=(v_1,\ldots,v_r)\in(\F_2^a)^r$,
\[
 \left|\E(-1)^{\sum_i v_i\cdot g_i(X_i)}\right|
 \le2^{-\gamma\ell|\{i:v_i\ne0\}|}.
\]
The decay with the number of nonzero blocks allows us to sum all
nontrivial Fourier coefficients. With $\ell=C\lceil\log r\rceil+C_0$,
the sum is below one, so every output pattern occurs.

To prove this decay, we must control dependencies among the input blocks.
\Cref{thm:universal-certificate} provides linearly many disjoint clusters,
each containing a bounded number of blocks and one distinguished block
with substantial entropy after the other selected clusters are fixed.
We call this distinguished block the \emph{honest block}, and its remaining
entropy its \emph{private rank}; neither its own cluster nor the blocks
outside the selected clusters are fixed here. This entropy allows us to
apply non-malleability within each cluster. The theorem also gives a joint
rank condition that bounds the probability of collisions in several
clusters at once. Together, these properties yield the required product
bound using only a fixed number of tamperings. We construct the gadgets and
choose their parameters below; the structural theorem and the rest of the
lifting argument are proved in \Cref{app:lossless-lifting}.

\subsubsection{A Deterministic Family of Lossless Gadgets}
\label{sec:deterministic-gadgets}

We now combine the universal rank certificate with the fixed-tampering
extractor. The gadgets are fixed independently of the source and Fourier
support; certificates may depend on both and are used only in the proof.

\begin{theorem}[Deterministic Simultaneous Surjectivity]
\label{thm:deterministic-gadgets}
Fix a rational $0<\eta\le1$ and an integer $a\ge1$. There are integers
$C,C_0\ge1$ and a uniform deterministic algorithm with the following
property. Given $r\ge1$, set $\ell=C\lceil\log r\rceil+C_0$.
The algorithm constructs functions $g_i:\F_2^\ell\to\F_2^a$ for
$i\in[r]$ in total time $r^{O_{\eta,a}(1)}$, with total truth-table
size $r^{O_{\eta,a}(1)}$. For every affine source
$X=(X_1,\ldots,X_r)$ satisfying
$H_\infty(X_I)\ge\eta\ell|I|$ for every $I\subseteq[r]$,
\[
 \{(g_1(x_1),\ldots,g_r(x_r)):x\in\Supp(X)\}=(\F_2^a)^r.
\]
One family satisfies this conclusion simultaneously for all such sources.
\end{theorem}

\begin{proof}
Fix the constants $\alpha,\beta,H,Q,\ell_0$ from
\Cref{thm:universal-certificate}, decreasing $\beta$ if necessary so
that $\beta\le1$. Apply \Cref{thm:joint-affine-tuple} with entropy rate
$\sigma=\beta/2$ and comparison count
$h=\max\{1,H2^Q-1\}$. Let $\anmExt_L$ denote this extractor on $L$-bit inputs, with error at most
$2^{-eL}$ for a fixed $e>0$ and all sufficiently large $L$.

Put $t=\lceil\log r\rceil$, assign distinct constants
$c_i\in\F_2^t$, and set $L=\ell+t$. Define
\begin{equation}
 g_i(x)=\pref_a\bigl(\anmExt_L(c_i\circ x)\bigr).
 \label{eq:deterministic-tagged-gadgets}
\end{equation}
Choose $C\ge1$ and $C_0\ge\ell_0$ sufficiently large that $L\le2\ell$ and
$\anmExt_L$ has at least $a$ output bits. The tags carry no entropy and
introduce no variables.

Let $v=(v_1,\ldots,v_r)\in(\F_2^a)^r$ be nonzero, and put
$S=\{i:v_i\ne0\}$. Rank spread is inherited when we keep only the
blocks indexed by $S$.
\Cref{thm:universal-certificate} therefore supplies a certificate
on these original blocks with at least $\alpha|S|$ honest blocks,
cluster size at most $H$, and at most $Q$ outside classes. Set
$k=\lceil\beta\ell\rceil$ and $k'=\beta\ell$. The private ranks,
being integers, are at least $k$; the joint-rank threshold $k'$ need
not be an integer. Every honest input to $\anmExt_L$ has entropy at least
$\beta\ell\ge\sigma L$ after the fixings in the cluster argument.
Set $\gamma=\alpha\min\{e,\beta\}/2^{Q+1}$.
\Cref{thm:lossless-cluster-phase}, applied to the nonzero characters
$v_i\cdot g_i$, gives
\begin{equation}
 \left|\E(-1)^{\sum_i v_i\cdot g_i(X_i)}\right|
 \le 2^{-\gamma\ell|S|},
 \label{eq:deterministic-phase-decay}
\end{equation}
provided $C_0$ is sufficiently large. Indeed, the quantity inside the
parentheses in Eqn.~\eqref{eq:lossless-product} is at most
$2^{-\min\{e,\beta\}\ell/2}$. Coincident comparison inputs are
allowed by \Cref{thm:joint-affine-tuple}; distinct tags separate
original blocks, but do not separate the corners belonging to the
same block.

There are $\binom r s(2^a-1)^s$ characters with support size $s$.
The sum of the absolute values of all nontrivial Fourier coefficients
is consequently at most
\begin{equation}
 \bigl(1+(2^a-1)2^{-\gamma\ell}\bigr)^r-1.
 \label{eq:deterministic-fourier-sum}
\end{equation}
Take $C\ge\max\{1,1/\gamma\}$ and enlarge $C_0$ so that
$\gamma C_0\ge\log(4(2^a-1))$. Then
Eqn.~\eqref{eq:deterministic-fourier-sum} is at most
$\exp(1/4)-1<1$. Fourier inversion gives positive probability to
every target in $(\F_2^a)^r$.

All parameters were fixed in the order
$\eta,a$; the certificate constants; the extractor constants;
and finally $C,C_0$. Evaluation of $\anmExt_L$ is uniformly polynomial in
$L$, with constants depending only on the fixed parameters. Evaluating
it on every tagged input takes $r2^\ell\poly(L)=r^{O_{\eta,a}(1)}$
time.
\end{proof}

By Eqn.~\eqref{eq:deterministic-safety}, the rank-spread condition is
equivalent to the safety condition in
\Cref{subsec:proof-complexity-prelim}. Thus
\Cref{thm:deterministic-gadgets} supplies simultaneous output
surjectivity on every $k$-safe state when $\ell-k\ge\eta\ell$.

\subsubsection{The Formula Parameters}

\begin{proof}[Proof of \Cref{thm:deterministic-lossless}]
Fix positive rationals $\varepsilon=\delta_0<\xi/16$. Use
\Cref{formula:sparse-base} at $\delta=\delta_0/2$, obtaining $Q=2^a$
and row width $s$. Apply \Cref{thm:deterministic-gadgets} with
$\eta=\varepsilon/2$ and output length $a$, and take
\begin{itemize}
\item block length $\ell=C\lceil\log_2r\rceil+C_0$;
\item legal-family capacity $w=\lfloor(1-\delta_0)r\rfloor$;
\item safeness parameter $k=\lfloor(1-\varepsilon/2)\ell\rfloor$;
\item initial variable count $N_0=r\ell$;
\item rank threshold $K_0=kw$;
\item path parameter $b_0=\lceil2(\varepsilon+\delta_0)N_0\rceil$.
\end{itemize}
For large $r$, $w<\gamma=\lceil(1-\delta_0/2)r\rceil$ and
$\varepsilon\ell/2\le\ell-k\le\varepsilon\ell$. Thus the gadgets
are surjective on safe systems and
\begin{equation}\label{formula:rank-threshold}
 K_0\ge(1-\varepsilon-\delta_0)N_0-O(\ell).
\end{equation}
By \Cref{formula:lifting}, the size exponent is at least
\begin{equation}\label{formula:final-exponent}
 K_0-b_0-(\ell-k)w
 \ge(1-4\varepsilon-3\delta_0)N_0-O(\ell)
 \ge(1-\xi)N_0.
\end{equation}
The last inequality holds eventually by the fixed slack in our choices.
Also $K_0+b_0>N_0$, so every depth-at-most-$N_0$ refutation is
$(K_0,b_0)$-path-free. The width bound follows from
Eqn.~\eqref{formula:rank-threshold}.

Enumerating violating raw tuples gives at most
$(r+1)2^{s\ell}=r^{O_\xi(1)}$ clauses, each of width
$s\ell=O_\xi(\log r)$, in deterministic polynomial time. Tags, field
symbols, and code coordinates introduce no Boolean variables.
If some designated raw variables never occur, omit them and write
$N\le N_0$ for the actual count. Both lower bounds remain valid with
$N$, and depth at most $N$ is covered. Since every CNF has a resolution
refutation of width at most its variable count, the width bound also
implies $N\ge(1-\xi)N_0$ eventually. Thus $\log r=\Theta(\log N)$,
completing the construction with its actual variable count.
\end{proof}

\subsection{Overlapping Gadgets and Resolution over Parities}
\label{sec:overlapping-lifting}

We now allow the gadgets in a substitution to share variables, with each
input block used by exactly two gadgets. Our first lifting argument uses
directional affine extractors to control a gadget while several of its
input blocks remain free. With distinct tags and suitable extractor
parameters, non-malleable affine extractors give the needed estimate even
when only one block remains free, under unrelated invertible affine
changes of coordinates at the gadgets.

Both versions incur an additive conditioning term in the size--depth
tradeoff. To remove it, we use non-malleability against a fixed number of
tamperings to find a restriction under which the gadgets become disjoint
three-bit majority functions. The lifting theorem of
\cite[Theorem~6.1]{IPS26} then applies. For a sharing graph of fixed degree
$D$, blocks of length at most $2^{CD}$ suffice, giving a coefficient
$2^{-O(D)}$ in the quadratic size--depth bound; see
\Cref{lem:overlap-block-length,cor:overlap-quantitative-product}.

\subsubsection{The Substitution}

We use the proof-system conventions and legal partial assignments from
\Cref{subsec:proof-complexity-prelim}. Let $\varphi$ be an unsatisfiable
CNF on $m$ variables and let $\mathcal P$ be a family of legal partial
assignments of capacity $w$.

Let $\Lambda$ be a simple $D$-regular graph on the variables of $\varphi$.
Introduce an independent block $z_e\in\F_2^t$ on every edge of
$\Lambda$, and replace variable $y_i$ by a Boolean function $g_i$ of the
blocks incident to $i$. Fix an order of the edges incident to each
vertex; $\operatorname{inc}(i)$ denotes this ordered list, and
$z_{\operatorname{inc}(i)}$ is the concatenation of its blocks.
The number of raw variables is
\begin{equation}
 N=|E(\Lambda)|t=\frac{mDt}{2}.
 \label{eq:overlap-variable-count}
\end{equation}
We define $\varphi\circ_\Lambda g$ as an ordinary CNF. For each clause
$C$ of $\varphi$, enumerate the assignments to the union of its incident
raw variables. For every assignment falsifying $C(g_1,\ldots,g_m)$,
include the clause excluding that assignment on the entire union. These
clauses represent $C(g_1,\ldots,g_m)$ exactly, so the substituted formula
is unsatisfiable. If $D$ and the base width are constant and
$t=O(\log m)$, the construction has polynomial size and is computable in
deterministic polynomial time whenever the gadgets are explicit. No
extension variables or preimage-sampling algorithm are required.

\subsubsection{Product Bounds from Directional Affine Extractors}

Let $f:\F_2^{Dt}\to\F_2$ be a $(k,\varepsilon)$
directional affine extractor, and let $A_i$ be an arbitrary invertible
affine map of $\F_2^{Dt}$ at each vertex. Set
\[
 g_i(z)=f(A_i(z_{\operatorname{inc}(i)})).
\]
Write $h_i(z)=(-1)^{g_i(z)}$.
We use the following consequence for every affine space $H$ of
dimension at least $k$ and every nonzero $a$:
\begin{equation}
 \left|\E_{X\sim U_H}(-1)^{f(X)+f(X+a)}\right|
 \le 2\varepsilon.
 \label{eq:overlap-directional-premise}
\end{equation}
Define the parameters as follows.
\begin{itemize}
\item $d_0:=\left\lceil k/t\right\rceil+1$;
\item $B:=\max_{v\in V(\Lambda)}
       |\{u:\operatorname{dist}_\Lambda(u,v)\le2\}|$;
\item $\delta:=(2\varepsilon+2^{-t})^{1/2}$;
\item $\gamma:=D/2-d_0+1$.
\end{itemize}
In particular, $B\le D^2+1$. We assume $\gamma>0$ and $\delta<1$.

\begin{lemma}[A restricted gadget]
\label{lem:overlap-restricted-gadget}
Fix any incident blocks of a gadget, leaving at least $d_0$ blocks free.
Write $u\in\F_2^t$ for one free block and $v$ for the others.
For all real functions $a,b$ bounded in absolute value by one,
\begin{equation}
 \left|\E_{u,v}h_i(u,v)a(u)b(v)\right|\le\delta.
 \label{eq:overlap-restricted-product}
\end{equation}
\end{lemma}

\begin{proof}
For distinct $u,u'$, the sections obtained by varying $v$ are parallel
affine spaces of dimension at least $(d_0-1)t\ge k$. Their displacement
is nonzero because $A_i$ is invertible. Hence
$|\E_v h_i(u,v)h_i(u',v)|\le2\varepsilon$; for $u=u'$ the
expectation is one. Cauchy--Schwarz gives
\[
 \left|\E_v b(v)\E_u h_i(u,v)a(u)\right|^2
 \le \E_{u,u'}
       \left|\E_v h_i(u,v)h_i(u',v)\right|
 \le 2\varepsilon+2^{-t}.
\]
\end{proof}

\begin{lemma}[Moments with a linear character]
\label{lem:overlap-twisted-moments}
Fix any raw blocks. Let $I$ be a set of vertices each retaining at least
$d_0$ free incident blocks. For $S\subseteq I$ and any linear character
$\chi_a$ on the remaining raw variables,
\begin{equation}
 \left|\E\chi_a(Z)\prod_{i\in S}h_i(Z)\right|
 \le\delta^{|S|/B}.
 \label{eq:overlap-twisted-moment-bound}
\end{equation}
The expectation is zero if $a$ is nonzero on a free block not incident
to any vertex of $S$.
\end{lemma}

\begin{proof}
Choose a maximal subset of $S$ whose distinct vertices have graph distance
at least three. Each chosen vertex excludes at most $B$ candidates, so
there are at least $|S|/B$ chosen vertices. Fix all free blocks not
incident to a chosen vertex. The remaining sets of blocks at the chosen
vertices are disjoint. Every unchosen gadget factor depends on at most
one remaining block: dependence on blocks at two chosen vertices would
put those vertices at distance two. The character also factors over the
remaining blocks. Thus the conditional expectation factors over the
chosen vertices, and \Cref{lem:overlap-restricted-gadget} bounds each
factor by $\delta$, absorbing the one-block factors into $a$ and $b$.
Average over the fixings. The last assertion follows by averaging a
nontrivial character on an unused block.
\end{proof}

\subsubsection{Affine Restrictions and a Monotone Closure}

After some raw blocks are fixed, call a gadget \emph{active} if it
still has a free incident block. A gadget whose incident blocks are
all fixed is \emph{closed}.
For a set $T$ of raw blocks, let $W_T$ be the space of linear forms
supported on those blocks. In the notation of
\Cref{subsec:proof-complexity-prelim}, a row space $U$ on the free raw
variables is $\kappa$-safe if every block set $T$ satisfies
\begin{equation}
 \dim(U\cap W_T)\le\kappa|T|.
 \label{eq:overlap-safeness}
\end{equation}

\begin{lemma}[Output probabilities on safe affine spaces]
\label{lem:overlap-safe-output-probabilities}
Suppose that, after some raw blocks are fixed, each active gadget has at
least $d_0$ free incident blocks. Let $A$ be a nonempty affine space on
the remaining variables with $\kappa$-safe equation row space. Put
\begin{equation}
 \begin{aligned}
 \alpha&=2^{\kappa D}\delta^{1/B},\\
 \theta&=(1+\alpha)^m-1.
 \end{aligned}
 \label{eq:overlap-safe-error}
\end{equation}
Every prescribed pattern on any $b$ active outputs has probability under
$U_A$ between $(1-\theta)2^{-b}$ and $(1+\theta)2^{-b}$. In particular,
if $\theta<1$, every pattern on all active outputs has a preimage in $A$.
\end{lemma}

\begin{proof}
Let $U$ be the equation row space, and let $\sigma(a)$ be the sign of
the consistent right-hand side for $a\in U$. Expanding the normalized
indicator of $A$ gives
\begin{equation}
 \E_{Z\sim U_A}\prod_{i\in S}h_i(Z)
 =\sum_{a\in U}\sigma(a)
       \E_Z\chi_a(Z)\prod_{i\in S}h_i(Z).
 \label{eq:overlap-affine-fourier}
\end{equation}
Only forms supported on the at most $D|S|$ free blocks incident to $S$
contribute. Safeness bounds their number by $2^{\kappa D|S|}$, so
\Cref{lem:overlap-twisted-moments} bounds
Eqn.~\eqref{eq:overlap-affine-fourier} in absolute value by $\alpha^{|S|}$.
Fourier inversion on $b$ outputs bounds each pattern's relative deviation
from $2^{-b}$ by
\[
 \sum_{\varnothing\ne S\subseteq[b]}\alpha^{|S|}
 =(1+\alpha)^b-1\le\theta.
\]
\end{proof}

The following closure depends only on the row space, independently of its
right-hand sides. This allows equations to be forgotten and proof paths
to be restarted.

\begin{lemma}[Monotone closure]
\label{lem:overlap-monotone-closure}
For every row space $U$ on all raw variables, there are a block set $C(U)$
and a vertex set $V(U)$ with the following properties:
\begin{enumerate}
\item $U\subseteq U'$ implies $C(U)\subseteq C(U')$ and
$V(U)\subseteq V(U')$.
\item Every vertex of $V(U)$ has all incident blocks in $C(U)$; every
other vertex has at least $d_0$ incident blocks outside $C(U)$.
\item The projection of $U$ outside $C(U)$ is $\kappa$-safe.
\item Writing $r=\dim U$, we have
\begin{equation}
 \begin{aligned}
 |V(U)|&\le\frac{2r}{\kappa\gamma},\\
 |C(U)|&\le\frac r\kappa\left(1+\frac{2D}{\gamma}\right).
 \end{aligned}
 \label{eq:overlap-closure-size}
\end{equation}
\end{enumerate}
\end{lemma}

\begin{proof}
Define the supermodular function
\[
 q_U(S)=\dim(U\cap W_S)-\kappa|S|.
\]
Supermodularity means that
$q_U(S)+q_U(T)\le q_U(S\cup T)+q_U(S\cap T)$.
Indeed, the intersection of $U\cap W_S$ and $U\cap W_T$ is
$U\cap W_{S\cap T}$, while their sum is contained in $U\cap W_{S\cup T}$.
Let $R_U(S)$ be the union of all sets $T\supseteq S$ maximizing
$q_U(T)$. The union of two maximizers is a maximizer, so $R_U(S)$
is the unique largest maximizer under set inclusion.
Supermodularity makes $R_U$ extensive and monotone in $S$.

It is also monotone in $U$. For $U\subseteq U'$, the difference
\[
 \dim(U'\cap W_T)-\dim(U\cap W_T)
\]
is nondecreasing in $T$, since it is the dimension of the image of
$U'\cap W_T$ in $U'/U$. If $T$ and $T'$ are the old and new largest
maximizers, these increasing differences and supermodularity imply
$q_{U'}(T\cup T')\ge q_{U'}(T')$. Since $T'$ contains every superset of $S$
maximizing $q_{U'}$, it contains $T\cup T'$, giving $T\subseteq T'$.
The same union argument applies when
the forced set increases.

Starting with no blocks, alternate $R_U$ with the following peeling
operation: whenever a vertex has fewer than $d_0$ incident blocks outside
the current set, adjoin all its incident blocks. Iterate to a common
fixed point $C(U)$, and let $V(U)$ consist of the vertices whose incident
blocks all lie in $C(U)$. Both operations are extensive and monotone,
which proves the first two assertions. For $C=C(U)$ and $T$ outside $C$,
\[
 \dim(\pi_{\overline C}U\cap W_T)
 =\dim(U\cap W_{C\cup T})-\dim(U\cap W_C)
 \le\kappa|T|.
\]
The equality follows by taking the preimage under projection, and the
inequality follows from $q_U(C\cup T)\le q_U(C)$.

Call a block first added by an $R_U$ operation a seed block. If such an
operation enlarges $S$ to $T$, then
\[
 \kappa|T\setminus S|
 \le\dim(U\cap W_T)-\dim(U\cap W_S).
\]
These dimension increases telescope to at most $r$, so there are at most
$r/\kappa$ seed blocks. Order the peeled vertices by when they close.
Each has at least $D-d_0+1$ previously fixed incident blocks. A seed edge
is charged at most twice. Every other edge charged in this way joins an
earlier peeled vertex to a later one and is charged at most once. If
there are $s$ seed blocks and $b=|V(U)|$ peeled vertices, there are at
most $Db/2$ such internal edges. Hence
\[
 (D-d_0+1)b\le2s+Db/2,\qquad \gamma b\le2s.
\]
Every final block is a seed or is incident to a peeled vertex. Thus
$|C(U)|\le s+Db$, proving Eqn.~\eqref{eq:overlap-closure-size}.
\end{proof}

\subsubsection{Extension, Forgetting, and the Rank Bound}

\begin{lemma}[The affine-state invariant]
\label{lem:overlap-affine-invariant}
Suppose $\varphi$ has a legal family of capacity $w$ and $\theta<1$.
Set
\begin{equation}
 L=\left\lfloor\frac{\kappa\gamma w}{2}\right\rfloor.
 \label{eq:overlap-rank-threshold}
\end{equation}
An adversary can consistently answer parity queries and forget equations
as long as the stored equation row space, including a newly queried pivot
before any forgetting, has rank at most $L$.
It maintains a nonempty affine set on which no initial clause of
$\varphi\circ_\Lambda g$ is identically false.
\end{lemma}

\begin{proof}
For a stored row space $U$, maintain its equation values, a fixing $\zeta$
of $C(U)$, and a legal coarse assignment $\rho$ on $V(U)$. Let $A$ be
the nonempty affine space satisfying the equations and $\zeta$. The
closed gadget outputs equal $\rho$. By
\Cref{lem:overlap-safe-output-probabilities} and \Cref{lem:overlap-monotone-closure}, every assignment to the active outputs
has a preimage in $A$.

When a parity $p$ is requested, first form $U'=U+\Span\{p\}$,
before selecting its answer. The closures grow monotonically and
$|V(U')|\le w$. Extend $\rho$ legally to $V(U')$. Surjectivity on the
old state supplies $z\in A$ with these newly prescribed outputs. Fix the
blocks in $C(U')\setminus C(U)$ to their values in $z$, and answer with
$p(z)$. The new affine space contains $z$, has the required closed
outputs, and again every assignment to the active outputs has a preimage
in it.

If equations are forgotten, so $U'\subseteq U$, discard the block
fixings outside $C(U')$ and restrict $\rho$ to $V(U')$. The old space
witnesses nonemptiness of the relaxed one. Monotonicity and
\Cref{lem:overlap-safe-output-probabilities} restore the invariant.
This also permits replacement by a basis for any subspace of the stored
row space.

Suppose an initial substituted clause were identically false on $A$.
It is implied by a substituted base clause, which would therefore also
be false throughout $A$. If one of that base clause's variables were
active, surjectivity could satisfy its literal. All its variables must
therefore be closed, contradicting legality of $\rho$.
\end{proof}

\begin{corollary}[Lower bound on clause rank]
\label{cor:overlap-rank-width}
Under \Cref{lem:overlap-affine-invariant}, every refutation of
$\varphi\circ_\Lambda g$ contains a clause of rank at least $L$.
\end{corollary}

\begin{proof}
If all clauses had rank at most $L-1$, follow falsified clauses backwards
from the empty clause while storing their falsifying equations. At a
resolution, query its pivot, select a falsified premise, and forget
unnecessary equations. The stored rank, including the temporary pivot,
is at most $L$. At backwards semantic weakening, the falsifying space
of the conclusion is contained in that of the premise, so only
forgetting is needed. Reaching an initial clause contradicts
\Cref{lem:overlap-affine-invariant}.
\end{proof}

\subsubsection{The Size--Depth Tradeoff}

\begin{theorem}[Directional overlapping-gadget lifting]
\label{thm:directional-overlapping-lifting}
Assume the parameters $d_0,B,\delta,\gamma$ defined above satisfy
$\gamma>0$. Choose an integer $\kappa\ge1$, put
$\theta=(1+2^{\kappa D}\delta^{1/B})^m-1$, and assume
\begin{equation}
 \theta\le\frac1{4m}.
 \label{eq:overlap-small-error}
\end{equation}
Let $\varphi$ have a legal family $\mathcal P$ of capacity $w$, and let
$L\ge1$ be as in Eqn.~\eqref{eq:overlap-rank-threshold}. Set $\nu=w$. If
$\mathcal P$ is affine, one may instead set
\begin{equation}
 \nu=\max_{|A|\le w}\bigl(|A|-\dim\mathcal P_A\bigr).
 \label{eq:overlap-legal-codimension}
\end{equation}
Every refutation of $\varphi\circ_\Lambda g$, of size $S$ and resolution
depth $d$, satisfies $d\ge L$ and
\begin{equation}
 (2d-L)\left(\log S+\nu\log\frac2{1-\theta}\right)\ge L^2.
 \label{eq:overlap-size-depth}
\end{equation}
More precisely, set
\[
 p=\left\lfloor\log S+\nu\log\frac2{1-\theta}\right\rfloor.
\]
With $J=\lceil L/p\rceil$, one has
\begin{equation}
 d\ge JL-\frac{pJ(J-1)}2
 \ge\frac{L^2}{2p}+\frac L2.
 \label{eq:overlap-staircase-bound}
\end{equation}
In particular, $p\le\nu+\lceil\log S\rceil$. The conclusions allow
arbitrary separate invertible affine maps $A_i$.
\end{theorem}

\begin{proof}
We first give a sampler for the invariant in
\Cref{lem:overlap-affine-invariant}. Start uniformly on a valid
affine state $A_0$, retain all additional parity equations, and stop
when their rank together with the starting query space reaches $L$.
When a query closes $b>0$ new gadgets, prescribe a legal extension by
a fixed deterministic policy. By
\Cref{lem:overlap-safe-output-probabilities}, the prescribed pattern
has probability at least $(1-\theta)2^{-b}$ in the current uniform
affine state. Condition on this pattern, and sample the newly fixed raw
blocks and the query answer from this conditional distribution. After
fixing these sampled values, the residual distribution is uniform on
the new affine state: the prescribed output pattern is determined by
the fixed blocks. If no gadget closes, sample the additional block values
and the query answer from the current distribution and condition on
these values. At most $w$ new
coarse variables are assigned, giving the following density bound for
every completed input $z$ in $A_0$:
\begin{equation}
 \frac{\Pr[Z=z]}{\Pr_{U_{A_0}}[z]}
 \le \left(\frac2{1-\theta}\right)^w
 \label{eq:overlap-density-general}
\end{equation}
after uniform completion in the stopped state.

For an affine legal family, write $h(A)=|A|-\dim\mathcal P_A$.
If $A\subseteq A'$ and $|A'|\le w$, the extension property makes the
restriction map $\mathcal P_{A'}\to\mathcal P_A$ surjective. Over a
fixed old assignment, the legal extensions therefore form an affine
space of size
\[
 2^{|A'\setminus A|-(h(A')-h(A))}.
\]
Condition on membership in this entire space, rather than on one
extension. Its probability is at least
$(1-\theta)2^{-(h(A')-h(A))}$. If $h(A')=h(A)$, every pattern is legal,
so no conditioning is required. The positive increments of $h$ telescope
to at most $\nu$, and their number is at most their sum. The bound in
Eqn.~\eqref{eq:overlap-density-general} consequently improves to the
following bound, where $D_\nu=(2/(1-\theta))^\nu$:
\begin{equation}
 \frac{\Pr[Z=z]}{\Pr_{U_{A_0}}[z]}\le D_\nu.
 \label{eq:overlap-density}
\end{equation}
Fixing the additional blocks to their sampled values determines the
chosen legal extension, so the residual distribution is again uniform
on the resulting affine state.

These density bounds hold for every completed input, even when the
queries depend on previous answers. The input determines every query answer, closure,
fixed block value, and coarse assignment on its stopped path. Thus it
is compatible with only one sequence of these choices. The product of
reciprocal conditioning probabilities bounds its likelihood ratio;
summing over the possible sequences introduces no additional factor.

At the beginning of phase $j$, maintain a proof node $v_j$, a query
space $U_j$ of rank $r_j$, its equation values, fixed blocks $C(U_j)$,
and a legal assignment on $V(U_j)$. Let $A_j$ be the resulting nonempty
affine state and put
\[
 W_j=U_j+W_{C(U_j)}.
\]
Require every point of $A_j$ to falsify the clause at $v_j$. Initially,
$v_0$ is the empty conclusion, $U_0=0$, and $A_0$ is the whole input
space. Its closure is empty, since $d_0\le D$ and $\kappa>0$.

Suppose $r_j<L$. Follow falsified premises backwards using the sampler
with initial distribution $U_{A_j}$. The temporary query space is
$U_j$ plus the span of pivots encountered in this phase; fixed raw
coordinates are not counted in this rank. Stop when it reaches $L$.
The invariant prevents reaching an initial clause first. Thus every
sampled path advances at least $L-r_j$ resolution steps. A pivot already
determined by fixed raw coordinates still increases query rank by at most
one and has no conditioning cost.

For a possible boundary node $v$, let $T_v$ be the row space defining
its falsifying assignments. This affine space either misses $A_j$ or
has relative uniform mass
\begin{equation}
 2^{-\dim((T_v+W_j)/W_j)}.
 \label{eq:overlap-boundary-mass}
\end{equation}
Stopping at $v$ implies membership in that space, so
Eqn.~\eqref{eq:overlap-density} bounds its sampled probability by $D_\nu$
times Eqn.~\eqref{eq:overlap-boundary-mass}. There are at most $S$ boundary
nodes. If every positively reached one had quotient rank at least
$p+1$, their probabilities would sum to at most
$SD_\nu2^{-(p+1)}<1$. Some positively reached $v$ therefore satisfies
\begin{equation}
 \dim((T_v+W_j)/W_j)\le p.
 \label{eq:overlap-boundary-rank}
\end{equation}
Here $S\ge2$, since the invariant excludes an empty initial clause;
hence $p\ge1$.

Fix a successful path to this node and let $Q$ be the span of its
pivots. The proof rules give
\begin{equation}
 T_v\subseteq W_j+Q,
 \label{eq:overlap-query-inclusion}
\end{equation}
with the corresponding right-hand sides. Indeed, $A_j$ falsifies the
starting clause. Querying a pivot and selecting a falsified premise
adds only that pivot equation, while backwards weakening drops implied
equations. This induction does not use the temporary block fixings.

Choose $t_1,\ldots,t_h\in T_v$ representing a basis modulo $W_j$, with
$h\le p$. By Eqn.~\eqref{eq:overlap-query-inclusion}, write
$t_i=w_i+q_i$, where $w_i\in W_j$ and $q_i\in Q$, and set
\[
 U_{j+1}=U_j+\Span\{q_1,\ldots,q_h\}.
\]
Then $r_{j+1}\le r_j+p$ and $U_{j+1}\subseteq U_j+Q$. Monotonicity
gives
\[
 C(U_j)\subseteq C(U_{j+1})\subseteq C(U_j+Q),
\]
and the analogous inclusions for $V$. Give the retained equations and
raw blocks their values in the successful final state, and restrict its
legal assignment to $V(U_{j+1})$. The new affine state $A_{j+1}$ contains
that final state, so it is nonempty. It retains every equation and raw
fixing of $A_j$, as well as the $q_i$ equations, and therefore implies
all falsifying equations in $T_v$. Restart at $v_{j+1}=v$ uniformly on
$A_{j+1}$.

The density losses are not multiplied across phases. Each phase proves
the existence of a path segment and a valid next state, after which the
distribution is reset. The segments concatenate to a syntactic path in
the proof DAG; a single input need not realize the entire concatenation.

Since $r_j\le jp$, we may run $J=\lceil L/p\rceil$ phases. Their
total length is at least
\[
 \sum_{j=0}^{J-1}(L-jp)=JL-\frac{pJ(J-1)}2
 \ge\frac{L^2}{2p}+\frac L2.
\]
The first phase also gives $d\ge L$. Since
$p\le\log S+\log D_\nu$, this proves
Eqn.~\eqref{eq:overlap-size-depth}. Finally,
\[
 \nu\log\frac1{1-\theta}
 \le m\log\frac1{1-1/(4m)}<1,
\]
so $p\le\nu+\lceil\log S\rceil$.
\end{proof}

\subsubsection{An Explicit Family of Formulas}

\begin{lemma}[Tseitin legal assignments]
\label{lem:overlap-tseitin-legal}
Let $G$ be a constant-degree graph on $v$ vertices with edge expansion
at least $h>0$: every vertex set of size at most $v/2$ has at least $h$
times its size edges leaving it. The CNF encoding an odd-charge Tseitin
system on $G$ has constant width and an affine legal family of capacity
$w=\lfloor hv/8\rfloor\ge1$. For a domain $A$ of assigned edges, $|A|\le w$,
the legal space has codimension
\begin{equation}
 c_G(A)=\#\operatorname{components}(G\setminus A)-1.
 \label{eq:overlap-tseitin-codimension}
\end{equation}
In particular, $w=\Omega(m)$ for $m=|E(G)|$, and the affine version of
\Cref{thm:directional-overlapping-lifting} applies with
\begin{equation}
 H_G(w)=\max_{|A|\le w}c_G(A).
 \label{eq:overlap-tseitin-cost}
\end{equation}
\end{lemma}

\begin{proof}
After at most $w$ edges are assigned, the unassigned graph has a unique
component with more than $v/2$ vertices. Otherwise a union of its
components has between $v/4$ and $v/2$ vertices, and its boundary has at
least $hv/4>w$ assigned edges. Uniqueness is immediate. Call this
component large. Move assigned edge values into the charges of their
endpoints. Declare an assignment legal if every component other than
the large one has even total residual charge. The total charge remains
odd, so the large component has odd residual charge.

The empty assignment is legal. Unassigning edges merges components and
preserves the residual charge of their union, proving restriction
closure. To assign an edge while fewer than $w$ are assigned, remove it
from its component. If the component stays connected, either value
preserves its total charge. If a small component splits, choose the edge
value making one new component even; the other is then even. If the
large component splits, one resulting component is still large, and
choose the edge value making the other even. A completely assigned
vertex is a singleton small component and has even residual charge, so
its constraint is satisfied. No legal assignment falsifies an initial
Tseitin clause.

For fixed $A$, the small-component conditions are affine equations in
the assigned edge values. Contract the components of $G\setminus A$.
The resulting multigraph of assigned edges is connected. Its incidence
rows indexed by all vertices except the large component are independent,
so the equations have rank $c_G(A)$. They are consistent because the
matrix has full row rank. This proves
Eqn.~\eqref{eq:overlap-tseitin-codimension}.
\end{proof}

If $G$ has edge-connectivity $\lambda$, summing the boundaries of the
small components gives
\[
 H_G(w)\le\min\{w,\lfloor2w/\lambda\rfloor\}.
\]
If a deletion disconnects the graph, the large component also has at
least $\lambda$ boundary edges, improving this to
$H_G(w)\le\min\{w,\max(0,\lfloor2w/\lambda\rfloor-1)\}$.
On the other hand, if the maximum degree is $\Delta$, isolate vertices
from an independent set by assigning all their incident edges. This
gives
\[
 H_G(w)\ge
 \min\left\{\left\lfloor\frac w\Delta\right\rfloor,
             \left\lfloor\frac v{\Delta+1}\right\rfloor\right\}.
\]
Thus $H_G(w)=\Theta(m)$ for a fixed-degree expander family with
$w=\Omega(m)$. The codimension refinement can improve the additive
constant without removing its linear scale in the coarse variables.

For example, replace each edge of an explicit expander by $b$ parallel
copies, for a fixed integer $b$. The degree and expansion scale by $b$,
the capacity becomes $\lfloor bhv/8\rfloor$, and the Tseitin argument is
unchanged. Connectivity can
change only when an entire bundle is assigned, so
$H_G(w)\le\lfloor w/b\rfloor$. The base width remains constant for
fixed $b$, and the same formula can use this cost instead of $w$.

\begin{corollary}[Explicit bounded-depth lower bounds]
\label{cor:overlap-explicit-lower-bounds}
There is a deterministic polynomial-time constructible family of
polynomial-size unsatisfiable CNFs $\Phi_N$ on $N$ variables and of width
$O(\log N)$ such that every $\operatorname{Res}(\oplus)$ refutation,
of size $S$ and resolution depth $d$, satisfies
\begin{equation}
 d\bigl(\log S+O(N/\log N)\bigr)=\Omega(N^2).
 \label{eq:overlap-explicit-tradeoff}
\end{equation}
For every fixed $a>0$, refutations of depth at most $aN$ therefore have
size $2^{\Omega_a(N)}$. This includes top-regular refutations, whose
pivot forms are linearly independent along each path and whose depth
is consequently at most $N$.
\end{corollary}

\begin{proof}
Take an explicit constant-degree expander family and its odd-charge
Tseitin formulas. \Cref{lem:overlap-tseitin-legal} gives legal
capacity $w=\Omega(m)$. Add at most four unused coarse variables so
that $m$ is divisible by five; extend the legal family by arbitrary
values on them, while retaining capacity $w$. Use a disjoint union of
$K_5$ graphs as the coupling graph. Then $D=4$ and $B=5$.

Set $t=C_0\lceil\log m\rceil$ and use the first output bit of the
directional affine extractor of Li and Zhong~\cite{LZdir} on $4t$ bits
at entropy threshold $k=t$. It has error $\varepsilon=2^{-\Omega(t)}$.
Thus $d_0=2$, $\gamma=1$, and $\delta\le2^{-\beta t}$ for a constant
$\beta>0$ and sufficiently large $t$. Take
$\kappa=\lfloor\beta t/40\rfloor$. Then
\[
 \alpha=2^{4\kappa}\delta^{1/5}\le2^{-\beta t/10}.
\]
Choosing $C_0$ sufficiently large makes $\alpha\le1/(8m^2)$ and hence
$\theta\le1/(4m)$. We have $\kappa=\Omega(t)$,
$N=2mt$, and $L=\lfloor\kappa w/2\rfloor=\Omega(N)$.
\Cref{thm:directional-overlapping-lifting}, with
$\nu=H_G(w)=O(m)$, proves Eqn.~\eqref{eq:overlap-explicit-tradeoff}.

If the base width is $b_0$, every substituted clause depends on at most
$4b_0t=O(\log m)$ raw bits. Its truth table and canonical CNF can thus
be generated in polynomial time. The closure and sampler are used only
in the lower-bound proof and need not be computed by the construction.
Arbitrary specified invertible affine maps at the gadget locations are
permitted by \Cref{thm:directional-overlapping-lifting}.
\end{proof}

The asymptotic statement concerns the sizes supplied by the expander
family. It is an overlapping-gadget transfer, rather than an improvement
over the strongest size--depth bounds for disjoint substitution. The
additive $O(N/\log N)$ term remains significant beyond linear depth.

\begin{remark}[A simpler gadget in fixed coordinates]
\label{rem:overlap-quadratic-gadget}
Without separate affine changes of coordinates,
\Cref{lem:overlap-restricted-gadget} also follows from
\[
 g(z_1,\ldots,z_D)
 =\Tr_{\F_{2^t}/\F_2}
       \left(\sum_{a<b}z_a z_b\right).
\]
After any fixing leaving two blocks free, distinct sections in one free
block have zero correlation: their product contains a nontrivial
character of a second free block. Cauchy--Schwarz gives
$\delta=2^{-t/2}$. Directional affine extractors supply robustness under
unrelated local affine maps, which need not arise from one global affine
change of the overlapping raw variables.
\end{remark}

\subsubsection{Tagged Affine Non-Malleability and One Free Block}

The directional argument leaves at least $d_0$ free incident blocks at
each active gadget. Affine non-malleability gives a stronger bound when
only one block remains. Distinct fixed tags exclude fixed points between
the two extractor inputs obtained from a shared block.

Fix distinct tags $c_i\in\F_2^a$, where $a=\lceil\log m\rceil$.
Let $F:\F_2^{a+Dt}\to\F_2$ be an affine extractor of error
$\varepsilon$ at entropy $K$, also non-malleable with error $\varepsilon$
against globally fixed-point-free invertible affine maps. Full affine
non-malleability implies this assumption. Suppose $0<2\varepsilon<1$
and
\begin{equation}
 K\le t-1.
 \label{eq:overlap-tagged-entropy}
\end{equation}
For arbitrary invertible affine maps
$A_i:\F_2^{Dt}\to\F_2^{Dt}$, set
\[
 g_i(z)=F(c_i,A_i(z_{\operatorname{inc}(i)})).
\]
Write $h_i(z)=(-1)^{g_i(z)}$.
The tags are constants and add no raw variables to the formula.

\begin{theorem}[Tagged matching bound]
\label{thm:overlap-tagged-matching}
Fix any raw blocks, and call a vertex active if it has at least one free
incident block. For a set $S$ of active vertices and any linear character
$\chi_b$ of the remaining raw variables,
\begin{equation}
 \left|\E\chi_b(Z)\prod_{i\in S}h_i(Z)\right|
 \le(2\varepsilon)^{|S|/(D+1)}.
 \label{eq:overlap-tagged-moment}
\end{equation}
More precisely, the exponent may be the size of a maximum matching in the
graph of free edges incident to $S$. The expectation vanishes if $b$ is
nonzero on a free block not incident to $S$.
\end{theorem}

\begin{proof}
Let $H$ be the graph consisting of the free edges incident to $S$ and all
their endpoints. It has maximum degree at most $D$, no isolated vertices,
and at least $|S|$ vertices. A maximum matching $M$ has
$|M|\ge |V(H)|/(D+1)$. For $D=1$ this is immediate. For $D\ge2$, the
unmatched vertices are independent. If both endpoints of a matched edge
had distinct unmatched neighbors, the resulting length-three alternating
path would augment $M$. Thus each matched edge has at most $D-1$
distinct unmatched neighbors: at most one if both endpoints have such a
neighbor, and at most $D-1$ if only one does. Assign each unmatched
vertex to a neighboring matched edge. Then
$|V(H)|-2|M|\le(D-1)|M|$, proving the bound.

Fix all free blocks outside $M$. Every selected gadget now depends on
at most one remaining block. Each matching block occurs in one or two
selected gadgets, and the character factors over matching blocks, so
the conditional expectation factors over these edges.

For a matching block $u\in\F_2^t$ appearing in one selected
gadget, the input to $F$, including its tag, is an injective affine
function of $u$. If its
character is nontrivial, condition on that character's value; each
fiber has dimension $t-1\ge K$. The affine extractor guarantee bounds its sign
expectation on each fiber by $2\varepsilon$, and hence gives the same
bound after inserting the character. For a trivial character, apply
the affine extractor guarantee without conditioning.

If both endpoints $i,j$ are selected, their inputs without the tags are
injective affine maps $p_i(u),p_j(u)$ into $\F_2^{Dt}$. The affine
isomorphism $p_i(u)\mapsto p_j(u)$ extends to an invertible affine map
$A$ of $\F_2^{Dt}$ by extending bases. Therefore
\[
 T(c,x)=(c+c_i+c_j,A(x))
\]
is an invertible affine map taking the first tagged input to the second.
It is globally fixed-point-free because $c_i\ne c_j$. After conditioning
on the character if necessary, the first tagged input is still an affine
source of dimension at least $t-1\ge K$. Non-malleability bounds the product of the two output
signs by $2\varepsilon$ on each fiber, and thus also after inserting
the character. Multiplying over matching edges and averaging the fixed
blocks proves Eqn.~\eqref{eq:overlap-tagged-moment}. The vanishing statement
follows by averaging a character on an unused block.
\end{proof}

\begin{corollary}[Tagged lifting]
\label{cor:overlap-tagged-lifting}
Let the base formula have a legal family of capacity $w$. For an
integer $\kappa\ge1$, define
\[
 \begin{aligned}
 \alpha&=2^{\kappa D}(2\varepsilon)^{1/(D+1)},\\
 \theta&=(1+\alpha)^m-1.
 \end{aligned}
\]
Suppose
\begin{equation}
 \theta\le\frac1{4m}.
 \label{eq:overlap-tagged-error}
\end{equation}
Put $L_{\mathrm{tag}}=\lfloor\kappa Dw/2\rfloor$ and assume
$L_{\mathrm{tag}}\ge1$. Every refutation
of the tagged substitution, of size $S$ and resolution depth $d$,
satisfies
\begin{equation}
 (2d-L_{\mathrm{tag}})
 \left(\log S+\nu\log\frac2{1-\theta}\right)
 \ge L_{\mathrm{tag}}^2,
 \label{eq:overlap-tagged-tradeoff}
\end{equation}
where $\nu=w$, or the value in
Eqn.~\eqref{eq:overlap-legal-codimension} for an affine legal family. The
exact bound Eqn.~\eqref{eq:overlap-staircase-bound} also holds with
$L_{\mathrm{tag}}$ in place of $L$. The coupling graph may have any
fixed degree $D\ge1$.
\end{corollary}

\begin{proof}
For safe equation spaces, insert
Eqn.~\eqref{eq:overlap-tagged-moment} into
Eqn.~\eqref{eq:overlap-affine-fourier}. This gives
Eqn.~\eqref{eq:overlap-tagged-error} and the same probability bound for
every prescribed output pattern. For arbitrary row spaces, let $C(U)$ be
the union of all sets $C$ maximizing
$\dim(U\cap W_C)-\kappa|C|$. The proof of
\Cref{lem:overlap-monotone-closure} gives monotonicity and safeness
outside $C(U)$. No peeling is needed: every vertex outside $V(U)$, the
set of fully fixed vertices, retains a free block. Since the maximum
is nonnegative,
\begin{equation}
 \begin{aligned}
 |C(U)|&\le\frac{\dim U}{\kappa},\\
 D|V(U)|&\le2|C(U)|,\\
 |V(U)|&\le\frac{2\dim U}{\kappa D}.
 \end{aligned}
 \label{eq:overlap-tagged-closure}
\end{equation}
Thus the extension and forgetting invariant holds through rank
$L_{\mathrm{tag}}$. The density and restart argument of
\Cref{thm:directional-overlapping-lifting} uses only this invariant,
monotonicity, and the bound for every output pattern, so it applies
unchanged.
\end{proof}

Suitable parameters require exponentially small error at a sufficiently
small fixed entropy rate. Writing $b=\log(1/(2\varepsilon))$, take
\[
 \kappa=\min\left\{
       \left\lfloor\frac t2\right\rfloor,
       \left\lfloor\frac b{2D(D+1)}\right\rfloor\right\}.
\]
Then $\alpha\le2^{-b/(2(D+1))}$. If $b=\Omega(t)$ and
$t=C\log m$ for sufficiently large fixed $C$, then
$\kappa=\Omega(t)$ and Eqn.~\eqref{eq:overlap-tagged-error} holds. An
extractor for every fixed positive entropy rate supplies
Eqn.~\eqref{eq:overlap-tagged-entropy} on input length $a+Dt$ by choosing
the rate sufficiently small. The canonical truth-table construction
is again deterministic polynomial time.

\begin{remark}[What affine non-malleability adds]
\label{rem:overlap-nonmalleability-scope}
The matching bound leaves only one free block per active gadget and
removes peeling from the closure. It holds under arbitrary unrelated
affine maps $A_i$. The denominator $D+1$ is sharp for the bound obtained by counting matching edges:
if only the edges of one star are free and $S$ contains its
center and all $D$ neighbors, then $|S|=D+1$ and the maximum matching
has size one.

Directional affine extractors also give this matching bound when each
shared block is inserted unchanged at the same block position in both
endpoint inputs. The resulting affine sections are parallel, and distinct
tags make their displacement nonzero. If the graph admits a proper
$D$-edge-coloring, use the edge color to choose the block position.
For a general simple $D$-regular graph, a $(D+1)$-edge-coloring gives
$D+1$ positions, with the unused position fixed at each vertex; adjust
the extractor input length accordingly. Full affine non-malleability
allows unrelated local affine changes of coordinates.

The refinement for affine families of legal assignments charges the
codimension of legal extensions
rather than the number of newly assigned coarse variables. Even perfectly
uniform outputs assign a codimension-$h$ affine fiber probability $2^{-h}$.
Thus reducing extractor error alone does not remove this conditioning
cost. The matching decomposition and safeness establish the displayed
global moments for the overlapping construction; they do not establish
non-malleability for arbitrary multipoint affine configurations or
simultaneous surjectivity of the entire gadget family on every safe
affine space, as required for lossless lifting.
We establish that property separately in
\Cref{thm:deterministic-gadgets} using the rank certificates of
\Cref{def:lossless-certificate}.
\end{remark}

\subsubsection{Transfer through Majority Restrictions without the Conditioning Term}
\label{sec:overlap-ips-transfer}

We remove the conditioning term in \Cref{eq:overlap-size-depth} by finding
a restriction under which the overlapping gadgets become disjoint
three-bit majority functions. Non-malleability against a fixed number of
affine tamperings guarantees that such a restriction exists, allowing us
to apply \cite[Theorem~6.1]{IPS26}.

\begin{lemma}[Prescribed joint output patterns]
\label{lem:overlap-joint-table}
Let $F$ be one output bit of the extractor in
\Cref{thm:joint-affine-tuple}, with $h=q-1$ and error $\varepsilon$.
Suppose that $(X^1,\ldots,X^q)$ is jointly affine, each source meets
the extractor's entropy threshold, and no two sources take the same
value in any outcome.
For every $b\in\F_2^q$,
\[
 \Pr[(F(X^1),\ldots,F(X^q))=b]
 \ge 2^{-q}\bigl(1-2\varepsilon(2^q-1)\bigr).
\]
\end{lemma}
\begin{proof}
For each nonempty $T\subseteq[q]$, choose one $j\in T$ and apply
the tuple theorem with $X^j$ as the original source. It bounds
$|\E(-1)^{\sum_{i\in T}F(X^i)}|$ by $2\varepsilon$, since replacing
the output bit $F(X^j)$ by an independent uniform bit makes the expectation
zero. Fourier inversion proves the claim.
\end{proof}

Fix a constant $D\ge1$. For this transfer, $\Lambda$ is a simple
$D$-regular graph with a specified perfect matching $M$. Give the two
endpoints of each matched edge distinct tags $c_i\in\F_2$. For a
sufficiently large constant block length $t$, take the first output
bit $F:\F_2^{1+Dt}\to\F_2$ of \Cref{thm:joint-affine-tuple} with
comparison count $h=127$ and entropy rate $\rho=1/(2D)$, requiring
\begin{equation}
 \varepsilon\le2^{-132}.
 \label{eq:overlap-majority-parameters}
\end{equation}
We choose $t\ge6$ large enough for this error bound. Each source of
dimension $t$ meets its entropy threshold. For arbitrary specified
invertible affine maps $A_i:\F_2^{Dt}\to\F_2^{Dt}$, put
\[
 g_i(z)=F(c_i,A_i(z_{\operatorname{inc}(i)})).
\]
The tags are constants, so there are still $N=mDt/2$ raw variables.

\begin{lemma}[Block length]\label{lem:overlap-block-length}
There is an absolute constant $C>0$ such that the preceding construction
can be instantiated with $6\le t=t(D)\le2^{CD}$.
\end{lemma}
\begin{proof}
Apply \Cref{cor:affine-error-tuning} with the preceding $h$ and $\rho$,
input length $n_0=1+Dt$, one output bit, and
\[
 E=\max\{132,\lceil\log_2 n_0\rceil\}.
\]
Its sufficient length condition is $G_{\rho,127}(1+E)\le n_0$,
where $G_{\rho,127}\le2^{C_0D}$ for an absolute $C_0$; this includes
the minimum lengths of the constituent extractors. Taking
$t=2^{\lceil C_1D\rceil}$ for a sufficiently large absolute $C_1$
satisfies this condition, since $E=O(C_1D+\log D+132)$.
The resulting error is at most $2^{-132}$, and the entropy threshold
$\rho n_0=t/2+1/(2D)$ is at most $t$. Enlarging one absolute constant
gives the stated bound.
\end{proof}

Write $\operatorname{MAJ}_3$ for the function that is one exactly when
at least two of its three input bits are one. In the canonical CNF
$\varphi\circ\operatorname{MAJ}_3$, every base variable receives its
own three input bits. A canonical axiom for a base clause $C$ forbids
one assignment to the triples for the variables of $C$ on which $C$
is false.

\begin{lemma}[Disjoint majority restrictions]
\label{lem:overlap-majority-restriction}
There is a restriction of $\varphi\circ_\Lambda g$, followed by
negations of free variables, whose remaining axioms are weakenings
of canonical axioms of $\varphi\circ\operatorname{MAJ}_3$.
The majority gadgets have pairwise disjoint inputs. The restriction
is computable in time polynomial in $m$.
\end{lemma}
\begin{proof}
Fix every nonmatching block to zero. On a matched edge $\{i,j\}$,
the endpoint inputs without their tags become injective affine maps
$p_i,p_j:\F_2^t\to\F_2^{Dt}$. Let $U$ be the span of the first six
coordinate vectors. For uniform $X\in\F_2^t$, consider the 128 affine
sources
\[
 (c_i,p_i(X+u)),\quad(c_j,p_j(X+u)),\qquad u\in U.
\]
They are jointly affine and have entropy $t$. Injectivity separates
different corners at one endpoint; the tags separate the endpoints.
By \Cref{lem:overlap-joint-table}, every assignment to their 128 output
bits has positive probability. Indeed, the parenthesized factor in that lemma exceeds
$7/8$ under Eqn.~\eqref{eq:overlap-majority-parameters}. Prescribe
\[
 \begin{aligned}
 F(c_i,p_i(X+u))&=\operatorname{MAJ}_3(u_1,u_2,u_3),\\
 F(c_j,p_j(X+u))&=\operatorname{MAJ}_3(u_4,u_5,u_6).
 \end{aligned}
\]
Choose an offset $x_e$ realizing these prescribed outputs. Fix coordinates outside
$U$ to those of $x_e$ and negate free coordinate $a$ when
$(x_e)_a=1$. The endpoints become majority functions on disjoint
triples. Apply this independently on each matched edge.

For a base clause $C$, a surviving truth-table axiom excludes a
falsifying assignment on all remaining raw coordinates incident to
$C$. Its literals on the majority inputs of variables in $C$ form
a canonical axiom of $C\circ\operatorname{MAJ}_3$. Any additional
literals lie on the triple of a matched partner outside $C$.
The surviving axiom is therefore a weakening of the canonical one.
Finally, exhaustive search over the $2^t$ offsets per matched edge
is polynomial time because $t$ is fixed.
\end{proof}

\begin{theorem}[Overlapping size--depth transfer]
\label{thm:overlap-ips-transfer}
Suppose every ordinary-resolution refutation of $\varphi$ has width
at least $w$ or depth at least $H$. Every $\operatorname{Res}(\oplus)$
refutation of the preceding $\varphi\circ_\Lambda g$, of size $S$
and resolution depth $d$, satisfies
\begin{equation}
 d=\Omega\!\left(\min\left\{H,
                      \frac{w^2}{\log(2S)}\right\}\right).
 \label{eq:overlap-ips-tradeoff}
\end{equation}
If $\varphi$ has polynomially many clauses of width $O(\log m)$,
the substituted CNF is constructible in polynomial time and has
polynomial size.
\end{theorem}
\begin{proof}
Use the parity-decision-DAG formulation of
\cite[Sections~2.5 and~6]{IPS26}. Restriction and variable negation
do not increase size or depth. At each surviving leaf, replace its
weakened axiom by the canonical majority axiom from
\Cref{lem:overlap-majority-restriction}; an assignment falsifying the
former also falsifies the latter. This gives a parity decision DAG
for the clause-search problem of
$\varphi\circ\operatorname{MAJ}_3$ with the same bounds.
Three-bit majority is 1-stifling: after any one input is left free,
fixing the other two inputs to a desired bit forces that output.
Thus
\cite[Theorem~6.1]{IPS26} gives Eqn.~\eqref{eq:overlap-ips-tradeoff}.
Its implied constant is absolute, since the restricted gadget is always
$\operatorname{MAJ}_3$.
Each base clause involves $O(Dt\log m)=O(\log m)$ raw variables,
which proves the construction and size assertions.
\end{proof}

\begin{corollary}\label{cor:overlap-quantitative-product}
For the condensed bases in \Cref{quad:main}, the preceding construction
gives, for every fixed $D\ge1$, polynomial-size CNFs with polynomial-size
ordinary-resolution proofs such that every
$\operatorname{Res}(\oplus)$ refutation satisfies
\begin{equation}
 d\log_2(2S)\ge2^{-CD}N^2
 \label{eq:overlap-ips-product}
\end{equation}
for an absolute constant $C>0$. Here $N=mDt/2$ counts all raw variables.
\end{corollary}
\begin{proof}
The bases have $m=\Theta(b)$ and width--depth thresholds
$(\Theta(b),\Theta(b^2))$, with absolute constants. Thus
\Cref{thm:overlap-ips-transfer} gives $d\log_2(2S)\ge c m^2$ for an
absolute $c>0$. Substituting $m=2N/(Dt)$ and using
\Cref{lem:overlap-block-length} proves Eqn.~\eqref{eq:overlap-ips-product}.
The bases have polynomial-size ordinary-resolution proofs of line width
$O(\log b)$. Such proofs remain polynomial
after our substitution: each inference involves $O(\log b)$ raw
variables and hence admits a polynomial-size exhaustive resolution
derivation of its encoded implication. A sharing graph can be formed
from $D$ cyclically shifted perfect matchings between two equal
parts; if necessary, first add a variable constrained by a unit
clause to make $m$ even.
\end{proof}

For each fixed $D$, this gives polynomial-time constructible overlapping
gadgets that preserve the size--depth tradeoff of
\cite[Theorem~6.1]{IPS26}, even under unrelated local affine coordinate
changes. By \Cref{lem:overlap-block-length}, the resulting bound is
$d\log(2S)\ge2^{-O(D)}N^2$.
The improvement for formulas with short ordinary-resolution proofs comes
from the parity substitutions in \Cref{quad:main}, which do not use
extractors.

This restriction retains only $3m$ of the $N=mDt/2$ input variables,
a fraction $6/(Dt)$. It therefore cannot transfer a near-maximal size
exponent measured in the full variable count $N$. For that purpose,
\Cref{thm:deterministic-lossless} uses simultaneous surjectivity on all
affine sources satisfying the rank-spread condition.

\textbf{}
\subsection{A Quadratic Size--Depth Tradeoff with Short Resolution Proofs}
\label{sec:quadratic-tradeoff}

We improve the size--depth tradeoff of Itsykson, Podolskii, and
Shekhovtsov~\cite{IPS26} by reducing the number of variables in the
base formula through parity substitutions (condensation). Each
original variable is replaced by a parity of new variables; these
parities may share variables. The construction in this subsection
does not use an extractor.

\begin{theorem}\label{quad:main}
There are absolute constants $c,C>0$ and a deterministic algorithm
which, for every sufficiently large integer $N$, constructs an
unsatisfiable CNF $\Psi_N$ on exactly $N$ variables in time $N^C$.
The formula has at most $N^C$ clauses and an ordinary-resolution
refutation of size at most $N^C$. Every $\operatorname{Res}(\oplus)$
refutation of $\Psi_N$ of size $S$ and depth $d$ satisfies
\begin{equation}\label{quad:product}
 d\log_2(2S)\ge cN^2.
\end{equation}
Every variable introduced by the construction or by padding is
included in $N$.
\end{theorem}

For formulas with polynomial-size ordinary-resolution refutations,
\cite[Theorem~6.4]{IPS26} gives the tradeoff
$d\log(2S)=\Omega(N^2/\log^2N)$. \Cref{quad:main} removes the
$\log^2N$ loss and gives \Cref{quad:product} by reducing the number of
variables before applying their lifting theorem. The condition that ordinary-resolution proofs remain
short is essential to this comparison:
\cite[Corollary~1.3]{IPS26} already gives quadratic size--depth bounds for
other formula families. The construction here is separate from the
deterministic lossless lifting in \Cref{thm:deterministic-lossless}.

\Cref{tab:short-proof-comparison} records the resulting size bounds
and depth ranges for the two families with short ordinary-resolution
proofs.
\begin{table}[!htbp]
\centering
\small
\setlength{\tabcolsep}{5pt}
\renewcommand{\arraystretch}{1.25}
\begin{tabularx}{\linewidth}{@{}P{.32\linewidth}>{\raggedright\arraybackslash}X>{\raggedright\arraybackslash}X@{}}
\toprule
Consequence & \cite[Theorem~6.4]{IPS26} & \Cref{quad:main} \\
\midrule
Lower bound on $\log(2S)$
& $\displaystyle\Omega\!\left(\frac{N^2}{d\log^2N}\right)$
& $\displaystyle\Omega\!\left(\frac{N^2}{d}\right)$ \\
\addlinespace
At depth $d=\Theta(N)$
& $S=2^{\Omega(N/\log^2N)}$
& $S=2^{\Omega(N)}$ \\
\addlinespace
Depth range forcing superpolynomial size
& $\displaystyle d=o\!\left(\frac{N^2}{\log^3N}\right)$
& $\displaystyle d=o\!\left(\frac{N^2}{\log N}\right)$ \\
\bottomrule
\end{tabularx}
\caption{Consequences for explicit polynomial-size CNFs with
polynomial-size ordinary-resolution proofs. Here $N$ is the variable
count, $S$ is refutation size, and $d$ is resolution depth. These are lower-bound
consequences of the two tradeoffs. Both constructions use \cite[Theorem~6.1]{IPS26}; the improvement here is in the condensed base and
does not require an extractor.}
\label{tab:short-proof-comparison}
\end{table}

\begin{corollary}\label{quad:simulation}
If $d(N)=o(N^2/\log N)$, depth-$d(N)$
$\operatorname{Res}(\oplus)$ does not polynomially simulate ordinary
resolution.
\end{corollary}
\begin{proof}
On the family in \Cref{quad:main}, this depth bound forces
$\log(2S)=\omega(\log N)$.
\end{proof}

Throughout this subsection, logarithms are binary. The depth of a
resolution proof counts resolution steps on a longest path;
weakening can be eliminated without increasing this measure.
For a substitution, we encode each current clause by its full
truth-table CNF on the union of its input supports. We perform this
operation at every stage, rather than replacing successive encodings
by a different CNF for the composite substitution.

\subsubsection{Depth-Preserving Partial Condensation}
\label{quad:condensation-section}

Let $A\in\F_2^{M\times n}$ and suppose that every $a$ with
$0<\wt(a)\le r$ satisfies
\begin{equation}\label{quad:expansion}
 \wt(A^{\mathsf T}a)\ge2\wt(a).
\end{equation}
The substitution $z=Ax$ replaces each coarse variable $z_i$ by the
parity $\sum_{j=1}^n A_{ij}x_j$ over $\F_2$. Different parities may
use the same variables $x_j$. A disjoint set $y$ of original variables
may be left unchanged; we call this a \emph{partial condensation}.
The substituted formula $F(y,Ax)$ is converted to a CNF by the
clausewise encoding described above.

\begin{lemma}[Partial condensation]\label{quad:partial}
If the clausewise encoding of $F(y,Ax)$ has a resolution refutation
of width $W$ and depth $d$, and $2(W+1)\le r$, then $F(y,z)$ has a
resolution refutation of width at most $W+1$ and depth at most $d$.
\end{lemma}

\begin{proof}
We use the usual query game for resolution: a prover remembers a
partial assignment, queries a variable, and may forget values. The
prover wins by falsifying a clause. A width-$W$, depth-$d$ refutation
gives a strategy using at most $W+1$ remembered variables and $d$
queries. Conversely, a winning strategy with memory $u$ and $d$
queries gives a width-at-most-$u$, depth-at-most-$d$ refutation, with
weakening. These implications follow by traversing a proof backwards
and, conversely, labelling a strategy by the negations of its
remembered assignments; see also~\cite[Lemma~5.1]{FPR22}.

For $H\subseteq[n]$, $|H|\le r$, define
\[
 \begin{aligned}
 \mathcal L_H&=\{a:\wt(a)\le r,
                   \supp(A^{\mathsf T}a)\subseteq H\},\\
 C(H)&=\bigcup_{a\in\mathcal L_H}\supp(a).
 \end{aligned}
\]
Then $\mathcal L_H$ is a linear space and
\begin{equation}\label{quad:closure}
 |C(H)|\le |H|/2.
\end{equation}
To prove the bound, every individual relation has support at most
$|H|/2$. If their union were larger, take the first finite union $U$
whose size exceeds $|H|/2$. Then $|U|\le |H|\le r$. In the span
of these relations a uniform vector has expected weight $|U|/2$,
whereas its image under $A^{\mathsf T}$ has expected weight at most
$|H|/2$. Averaging Eqn.~\eqref{quad:expansion} gives $|U|\le |H|/2$,
a contradiction. Sums of relations are now supported on $C(H)$
and therefore remain within the range $r$, proving linearity.
The closure is monotone in $H$. Also, for $v\notin H$,
\begin{equation}\label{quad:one-relation}
 \dim(\mathcal L_{H\cup\{v\}}/\mathcal L_H)\le1,
\end{equation}
since evaluation of $(A^{\mathsf T}a)_v$ has kernel $\mathcal L_H$.

An assignment $\rho$ on $H$ and an assignment $\sigma$ on $C(H)$
have a common extension satisfying $(Ax)_{C(H)}=\sigma$ exactly when
\begin{equation}\label{quad:compatibility}
 a\cdot\sigma=(A^{\mathsf T}a)\cdot\rho
 \qquad(a\in\mathcal L_H).
\end{equation}
Indeed, these are precisely the linear dependencies of the equations
$x_H=\rho$ and $(Ax)_{C(H)}=\sigma$. On adding one raw coordinate,
first query all new coarse variables in the closure. Their answers
preserve every old relation. By Eqn.~\eqref{quad:one-relation}, at most
one new relation remains; choose the raw answer to satisfy it.
Thus arbitrary answers of the coarse adversary can be accommodated.

Put $u=W+1$. Let $I$ be the currently remembered identity variables
and $R$ the currently remembered raw variables. Maintain a historical
raw set $H\supseteq R$, its assignment $\rho$, and a compatible
coarse assignment on $C(H)$, with
\[
 |H|\le2(u-|I|).
\]
The coarse memory is at most $|I|+|C(H)|\le u$.
Forgetting an identity variable forgets its value on both sides;
forgetting a raw variable need only shrink $R$.

Write $v=u-|I|$. A query already in $H$ uses its historical value.
Before a fresh raw query, if $|H|=2v$, reset $H$ to $R$ and forget
coarse variables outside $C(R)$. The query is legal only if
$|R|\le v-1$, so this reset discards at least $v+1$ historical
coordinates and at most $v$ coarse variables. Then add the queried
coordinate and use Eqn.~\eqref{quad:compatibility}.
Before a fresh identity query, the historical capacity decreases
from $2v$ to $2(v-1)$. If $H$ does not fit, reset it to $R$ first.
In this case $|H|\ge2v-1$ and $|R|\le v-1$, so at least $v$
historical coordinates and at most $v$ coarse variables are
discarded. The new identity query then fits the memory bound.
Restriction and monotonicity preserve compatibility at both resets.

If $Q_z$ counts all simulated coarse nonidentity queries, including
queries repeated after forgetting, telescoping the closure changes
gives
\[
 Q_z=|C(H_{\rm end})|+\sum_{\rm resets}\Delta_C
 \le |H_{\rm end}|+\sum_{\rm resets}\Delta_H
 \le\text{number of raw nonidentity queries}.
\]
Here $\Delta_C\le\Delta_H$ at either type of reset, as just shown.
Each identity query requires at most one coarse identity query.
The entire simulation therefore makes at most $d$ queries.

Finally, falsification of a full truth-table clause assigns all
raw support coordinates of the corresponding coarse clause. Each
coarse variable in that clause belongs to $C(R)\subseteq C(H)$,
and compatibility gives it the value of its substituted parity.
The coarse assignment falsifies the original clause. This proves
the game simulation and hence the lemma.
\end{proof}

The historical set is part of the simulation, not of the prover's
memory. Its sole purpose is to charge later closure losses to
discarded raw queries. In particular, \Cref{quad:partial} applies
separately at successive stages; it requires no expansion property
of their composite matrix.

\subsubsection{Two Constructive Matrix Lemmas}
\label{quad:matrices}

To keep the substituted clauses short, we need to control how many
new variables they contain. The first matrix construction bounds the
total number of output coordinates affected by an input group, and
the second has constant row weight. For a coefficient map $T$, the
variable substitution is $z=T^{\mathsf T}x$, so the required weight
inequality is Eqn.~\eqref{quad:expansion} with $A=T^{\mathsf T}$.

\begin{lemma}[Grouped matrix]\label{quad:grouped}
Fix $K=2^{16}$. For integers $B,q\ge2$, put
$J=\lceil\log q\rceil$ and $L_j=J-j+2$ for $0\le j\le J$, and set
\[
 n=KB\sum_{j=0}^{J}L_j.
\]
One can deterministically construct a map
$T:\F_2^{Bq}\to\F_2^n$ in time $\poly(B,2^q)$,
with a fixed exponent, such that $\wt(Ta)\ge2\wt(a)$ for
$0<\wt(a)\le B$.
The inputs are partitioned into $B$ groups of length $q$.
Outputs at scale $j$ are indexed by $(j,v,c)$ with $v\in[KB]$
and $c\in[L_j]$. A group reaches at most $Kd_j$ distinct labels
$v$ at this scale, where $d_j=\max\{1,\lfloor2^j/128\rfloor\}$.
In particular, one input group reaches $O(q)$ outputs in total.
\end{lemma}

\begin{proof}
At scale $t=2^j$, construct a biregular multigraph with $B$ left
vertices, $KB$ right vertices, and degrees $Kd$ and $d$, where
$d=\max\{1,\lfloor t/128\rfloor\}$. For $d=1$ use disjoint
stars. Otherwise use a Ramanujan graph with nontrivial singular
values at most
\[
 \lambda\le\sqrt{Kd-1}+\sqrt{d-1}.
\]
Existence and the interlacing construction are due to
Gribinski and Marcus~\cite[Theorems~1.2 and~1.4]{GM21}.
The uniformity in $d$ used here is justified below.
For a left set of size $s\le B/t$, the first and second moments
of its incidence vector give
\[
 |\Gamma(S)|\ge
 \frac{Kd\,s}{ds/B+\lambda^2/(Kd)}
 \ge(1-1/32)Kd\,s.
\]
Indeed, the denominator is at most
$1+1/128+2/\sqrt K+1/K$, and its excess over one is below $1/32$.
Consequently at most $Kd\,s/16$ selected edge slots reach
nonunique neighbors: a neighbor receiving $a\ge2$ slots contributes
$a\le2(a-1)$. This counts parallel edges with their multiplicities.
Repeat the graph on $L=L_j$ disjoint copies of its right side,
using the same labels and ports in every copy. Set $D=KdL$.

Set $\tau=\min\{4t,q\}$. There is a linear code
$C_t:\F_2^q\to\F_2^D$ such that every $u$ with
$1\le\wt(u)\le\tau$ satisfies
\begin{equation}\label{quad:local-code}
 \wt(C_tu)\ge D/4.
\end{equation}
Here $D\ge256tL$ and
$\sum_{a=1}^{\tau}\binom qa\le2^{4tL}$.
For a random binary matrix, the initial conditional estimator
\[
 \sum_{1\le\wt(u)\le\tau}
 2^{D/4}\E[2^{-\wt(C_tu)}\mid\text{fixed entries}]
\]
is at most $2^{4tL+D/4}(3/4)^D\le2^{4tL-D/8}<1$.
A row contributes $3/4$ when its parity is free, and $1$ or $1/2$
otherwise. Both choices of the next matrix entry can therefore be
evaluated exactly by enumerating at most $2^q$ vectors. Choose the
smaller estimator. Its final value excludes every violation of
Eqn.~\eqref{quad:local-code}. All denominators are powers of two with
polynomial bit length.

Apply $C_t$ within each input group, route its $D$ code bits through
the edge slots, and take their parity at each right output.
Concatenate all scales. For an input of weight $w\le B$ meeting
$s$ groups, choose
$j=\min\{J,\lfloor\log(B/s)\rfloor\}$.
If $j<J$, then $B<2ts$, and at least $s/2$ active groups have
weight at most $4t$. Their codewords contain at least $Ds/8$
ones. At most $Ds/16$ can lie on nonunique slots, including slots
from the other active groups. Thus at least $Ds/16\ge16ts>2w$
ones survive. If $j=J$, every active group is protected and at
least $3Ds/16\ge2qs\ge2w$ ones survive.
This proves the expansion claim. Finally
$\sum_jKd_jL_j=O(q)$, since
$\sum_j2^j(J-j+2)<12q$ and $\sum_jL_j=O(\log^2q)=O(q)$.
Reusing the routing labels gives the stronger per-scale statement.
\end{proof}

\paragraph{Uniform routing.}
Since $d$ grows with $q$, we need a running-time bound uniform in $d$.
We obtain it by evaluating the conditional polynomials in the construction
of~\cite{GM21} as follows.
Keep $K$ fixed. A \emph{$K$-claw} is a star with one left vertex
as its center and $K$ distinct right vertices as its leaves.
A \emph{complete $K$-claw matching} consists of $B$ vertex-disjoint
such stars covering both sides. Build $d$ such matchings. There
are $Bd$ whole-claw choices, each with at most $(KB)^{K+1}$
children. Advancing a whole claw is valid in the interlacing tree:
among its children is a path of $K$ successive edge choices whose
relevant root never increases.

At a whole-claw boundary let $A_0\in\mathbb Q^{KB\times B}$ be
the fixed adjacency matrix. If $a\ge1$ claws remain in the current
matching, let $I,J'$ be its unassigned rows and columns, of sizes
$Ka,a$. Define rational matrices
\[
 \begin{aligned}
 C&=A_0+\mathbf1_I\mathbf1_{J'}^{\mathsf T}/a,\\
 P&=\operatorname{diag}(\mathbf1_I)
      -\mathbf1_I\mathbf1_I^{\mathsf T}/(Ka),\\
 Q&=\operatorname{diag}(\mathbf1_{J'})
      -\mathbf1_{J'}\mathbf1_{J'}^{\mathsf T}/a.
 \end{aligned}
\]
Let $A^j_{p,v}$ be the coefficient of $x^{B-j}y^pz^v$ in
\begin{equation}\label{quad:gm-det}
 \det\bigl[xI_B+C^{\mathsf T}(I-P+yP)C(I-Q+zQ)\bigr].
\end{equation}
In bases adapted to the two projections, Cauchy--Binet identifies
this coefficient with the sum of squared $j$-minors using $p$
projected rows and $v$ projected columns. Such bases need not be
computed. With $s=Ka-1$ and $r'=a-1$, the partial-frame average
of~\cite[Proposition~4.2 and Lemma~4.3]{GM21} gives the expected
elementary Gram coefficient
\begin{equation}\label{quad:gm-coeff}
 e_i=\sum_{j,p,v} K^{i-j}
 \frac{\binom{s-p}{i-j}\binom{r'-v}{i-j}}
      {\binom{s}{i-j}}A^j_{p,v},
\end{equation}
where impossible indices are omitted. The remaining centered
matching has singular values $\sqrt K$, which accounts for the
power of $K$. For $a=1$ the completion is deterministic; for $a=0$
compute the Gram polynomial directly.

After completing the current, say $u$th, matching, remove its
fixed trivial factor $x-Ku^2$. The remaining whole matchings are
averaged by the rectangular convolution of~\cite[Section~2.1]{GM21}
on dimensions $KB-1,B-1$, with single-matching polynomial
$(x-K)^{B-1}$. Each convolution is linear in the current coefficient
vector; finally restore $x-Kd^2$. Thus at most $d$ coefficient
transforms account for all future matchings.

All three degrees in Eqn.~\eqref{quad:gm-det} are at most $B$, so
interpolation requires polynomially many rational determinants.
The rational coefficients have polynomial bit length. One direct
denominator bound comes from averaging integer graph polynomials
over at most $((KB)!B!)^d$ completions; its logarithm is
$O(dKB\log(KB))$. Binomial factors and determinant heights have
polynomial bit length as well. Exact real-root isolation and gcd
calculations compare candidate roots, including shared roots.
The resulting bit complexity is $(Bd)^{O_K(1)}$. This proves the
fixed-exponent uniformity required in \Cref{quad:grouped}.

\begin{lemma}[Constant row weight]\label{quad:constant-degree}
There are absolute constants $C,C_g,c_g>0$ and $\kappa<3$ with
the following property. For every sufficiently large integer $L$
and every $M\ge CL^6$, one can construct a binary matrix
$A\in\F_2^{M\times n}$ with $n\le C_gM/L$ such that
every $a$ with $0<\wt(a)\le c_gM/L^4$ satisfies
\[
 \wt(A^{\mathsf T}a)\ge2\wt(a).
\]
Every row has weight at most sixteen. The running time is
$\poly(M)+\exp(O(L^{2\kappa}\log L))$, with absolute constants.
Deleting any subset of rows preserves both the expansion range
on the retained rows and the bound on the number of outputs.
\end{lemma}

\begin{proof}
Set $\delta=2^{-21}L^{-2}$. For the degree $D$ chosen below, put
$m=\lceil D/(2L)\rceil$.
There is a bipartite graph on $D$ left vertices and $m$ right
vertices, of left degree eight, with at least $4|T|$ unique
neighbors whenever $|T|\le\delta D$.
To see this, choose eight distinct neighbors independently for
each left vertex. Failure on an $s$-set implies fewer than $6s$
neighbors, so a union bound is at most
\[
 \binom Ds\binom m{6s}(6s/m)^{8s}
 \le\left[4e^7 6^2L^2(s/D)\right]^s
 \le(1/8)^s.
\]
Here $6s\le m$, and $4e^7 6^2<2^{18}$.
The sum over $s\ge1$ is below one. Exhaustive enumeration of
neighbor lists and subsets constructs such a graph in
$\exp(O(D\log D))$ time.

Use the degree-$65$ nonbipartite Morgenstern graphs on
$v_k=64^{6k}-64^{2k}$ vertices, with nontrivial eigenvalues of
absolute value at most $16$~\cite[Theorems~5.11 and~5.13]{Mor91}.
Let $p$ be least with $(16/65)^p\le\delta/8$ and take the
bipartite double cover of a $p$th walk power. With
$\kappa=\log65/\log(65/16)<3$, its degree is
\[
 D=65^p=O(\delta^{-\kappa})=O(L^{2\kappa}).
\]
Choose the first outer size with $M'=v_kD\ge M$.
The sizes $v_k$ have a bounded multiplicative ratio; the condition
$M\ge CL^6$ ensures $M'=\Theta(M)$ and covers the minimum outer
size. Its normalized second singular value is at most $\delta/8$.

Use outer edges as the new left vertices. At each endpoint, route
the incident edge's port through the eight-neighbor inner graph;
right vertices are the pairs of an outer vertex and an inner
right vertex. The resulting left degree is sixteen and there
are $2v_km=O(M/L)$ outputs.
For $s\le\delta^2M'/8$ selected outer edges, call an endpoint
heavy if it sees more than $\delta D$ selected edges. If the two
heavy sets have total size $h$, then $h\le2s/(\delta D)$.
Expander mixing bounds the number of edges between them by
\[
 \frac{Dh^2}{4v_k}+\frac{\delta Dh}{16}
 \le \frac{s^2}{\delta^2Dv_k}+\frac{s}{8}\le s/4.
\]
At least $3s/4$ selected edges therefore have a light endpoint.
The inner graph supplies four unique neighbors per selected
incidence there, and different endpoints have disjoint output
sets. There are at least $3s\ge2s$ unique neighbors in total.
They are nonzero coordinates of the binary parity image.

Retain exactly $M$ left vertices. The same estimate applies to
every retained subset, and all outputs remain counted. This gives
range $\Omega(M/L^4)$. Finite-field enumeration, walk enumeration,
and port labelling take polynomial time in the output description;
the inner search has the displayed cost. More generally, any row
restriction preserves the parity inequality by extending a
retained coefficient vector with zeros.
\end{proof}

\subsubsection{Uniform Recursive Support Bounds}
\label{quad:recursion}

We next apply \Cref{quad:grouped} recursively. The issue is to
bound the support of an entire original group by $O(q)$, with a
constant independent of the recursion depth.
Put $Q_*=2^{40}$ and define
\[
 \begin{aligned}
 f(x)&=4096(\lceil\log x\rceil+3)^2,\\
 H(x)&=\begin{cases}0&x\le Q_*,\\1+H(f(x))&x>Q_*.
 \end{cases}
 \end{aligned}
\]
For integers $x>Q_*$, $f(x)\le x/2$. The function $H$ is
nondecreasing, $H(x)\le\log x$, and $H(x)=O(\log^*x)$.
For the last assertion, $f(f(x))=O((\log\log x)^2)\le\log x$
above an absolute threshold, and the remaining range is fixed.

At a node with $B$ groups of length $q>Q_*$, apply
\Cref{quad:grouped}. Partition the copy lengths $2,\ldots,J+2$
into bands, starting with $R=J+2$. While $R>64$, take the band
of integer lengths
\[
 \lceil4\log R\rceil\le L\le R
\]
and replace $R$ by $\lceil4\log R\rceil-1$. The final band consists
of the remaining lengths $2,\ldots,R$. For a band $\mathcal B$, put
$q_{\mathcal B}=\sum_{L\in\mathcal B}L$. Group its outputs by
routing label $v$, giving $KB$ groups of length $q_{\mathcal B}$,
and recurse. Nodes of length at most $Q_*$ are left unchanged.
All active nodes at the same depth are processed simultaneously.

\begin{lemma}\label{quad:uniform-recursion}
This construction has depth at most $H(q)$, and its final output
length is at most $BD(K+1)^{H(q)}$ for an absolute $D$.
At every prefix, an original group reaches at most $C_Pq$ current
coordinates, for an absolute $C_P$. Each active simultaneous map
satisfies Eqn.~\eqref{quad:expansion} through weight $B$, where $B$
is the initial group count. Its construction takes time
$\poly(B,2^q)$ with a fixed exponent.
\end{lemma}

\begin{proof}
For a nonterminal band with upper endpoint $R$, $q_{\mathcal B}\le R^2$.
Since $2^j=2^{J+2-L_j}<8q\,2^{-L_j}$,
\begin{equation}\label{quad:band-sum}
 \sum_{L_j\in\mathcal B}2^j q_{\mathcal B}
 \le16qR^2 2^{-\lceil4\log R\rceil}
 \le16q/R^2.
\end{equation}
Successive nonterminal upper endpoints at least halve. The terminal upper endpoint
is bounded by $64$. The terms coming from the $1$ in
$d_j\le1+2^j$ contribute at most $(J+1)(J+2)^2$.
It follows that, for an absolute $A_0$,
\[
 \sum_jKd_jq_{\mathcal B(L_j)}\le A_0q.
\]
For children that remain active, their band upper endpoints exceed $\sqrt{Q_*}$.
Eqn.~\eqref{quad:band-sum} and the unweighted estimate give
\begin{equation}\label{quad:active-mass}
 \frac1q\sum_{j:q_{\mathcal B(L_j)}>Q_*}
 Kd_jq_{\mathcal B(L_j)}
 \le\frac{64K}{3Q_*}+\frac{K43^3}{2^{40}}<\frac1{32}.
\end{equation}
For the second term, on $2^{40}<q\le2^{41}$ one has $J=41$;
the corresponding bound decreases on later dyadic intervals.

Let $P(q)$ be the largest path support of one original group.
A group reaches at most $Kd_j$ child groups at scale $j$, so
\[
 P(q)\le\sum_jKd_jP(q_{\mathcal B(L_j)}).
\]
Count the terminal children by their actual lengths and apply induction
to the active children. By Eqn.~\eqref{quad:active-mass},
$P(q)\le A_0q+C_Pq/32\le C_Pq$ for a sufficiently large fixed
$C_P$. The same induction applies to every unfinished prefix:
an unprocessed child contributes its identity support.

Every child length is at most $(J+2)^2\le f(q)<q$.
In descending band order, successive lengths $q_i,q_{i+1}$ obey
$q_{i+1}\le f(q_i)$. Indeed a nonterminal band with upper endpoint $R$ has
$q_i\ge R^2/4$, while the next has length at most
$16(\log R)^2$. Thus the positive child heights are distinct and
at most $H(q)-1$. There are at most $64$ terminal children: after
the first terminal length, the upper endpoint is at most $2\sqrt{Q_*}$,
and it continues to halve until the last band.
If $E(q)$ denotes final output length per input group, then
$E(q)=K\sum_iE(q_i)$ at an active node and $E(q)=q$ at a terminal
node. Set $D=64KQ_*$. Induction gives
\[
 E(q)\le D(K+1)^{H(q)}.
\]
The active child contributions sum over distinct smaller heights;
their geometric sum leaves an additive $D$ to cover all terminal
children. This also proves the depth bound.

Every descendant group count is at least $B$. A direct sum of
active maps therefore expands every input of total weight at most
$B$; terminal coordinates are handled by \Cref{quad:partial}.
Finally the sum of child input lengths at an active node is
\[
 KB\sum_{\mathcal B}q_{\mathcal B}
 =KB(J+1)(J+4)/2<Bq/2.
\]
Hence total input mass over the recursion tree is at most $2Bq$.
No group length increases, and the fixed-exponent algorithms of
\Cref{quad:grouped} give the claimed time bound.
\end{proof}

\subsubsection{Compressing and Merging the Terminal Parts}
\label{quad:accumulator-section}

We apply the recursion with $B=b$ and $q=5\log b$, for powers of
two $b$. The remaining output factor $(K+1)^{H(q)}$ is removed
by treating different root bands at different times.
Set $\theta_0=q$ and, as long as $H(\theta_j)>512$, set
\[
 \theta_{j+1}=Q_*(H(\theta_j)+1)^{16}.
\]
Let $J_*$ be the number of such steps and $h_j=H(\theta_j)$.
For $h_j\ge513$,
\begin{equation}\label{quad:height-halving}
 \begin{aligned}
 h_{j+1}&\le40+16\log(h_j+1)\le h_j/2,\\
 \theta_{j+1}&\le2^{h_j-1}\le\theta_j/2.
 \end{aligned}
\end{equation}
These inequalities hold at $513$ and their margins increase
thereafter. In particular the thresholds strictly decrease.

Assign a terminal component of the recursion to bucket $B_j$
if its first root band has length in
$(\theta_{j+1},\theta_j]$; put the remaining components in $Z$.
These are disjoint sets of allocated variables. The distinct
positive root-band heights and the output bound in
\Cref{quad:uniform-recursion} give
\begin{equation}\label{quad:bucket-initial}
 |B_j|\le C_0b(K+1)^{h_j},\qquad |Z|\le C_Zb.
\end{equation}
The second bound absorbs $(K+1)^{512}$ into its absolute constant.
If the root is terminal, or $J_*=0$, all its outputs belong to $Z$.

Before any further compression, an original group's support in
$B_j$ is at most
\begin{equation}\label{quad:bucket-pre-support}
 C_1\left(q/\theta_{j+1}+(\log q+3)^3\right).
\end{equation}
Indeed, a root band of length greater than $\theta_{j+1}$ has
upper endpoint greater than $\sqrt{\theta_{j+1}}$.
Summing Eqn.~\eqref{quad:band-sum} over these upper endpoints gives at most
$64q/(3\theta_{j+1})$. The unweighted contribution is bounded
by $(J+1)(J+2)^2$. The uniform support estimate in each descendant
group multiplies both terms by an absolute constant.

Compress each nonempty bucket separately using
\Cref{quad:constant-degree}. If its current size is $M=bR$ and
$R$ is larger than a fixed stopping ratio $R_*$, take
\[
 L=\left\lfloor(c_gR/2)^{1/4}\right\rfloor.
\]
Choose $R_*$ large enough to satisfy all fixed lower bounds on
$L$, and to make the next ratio at most $C_2R^{3/4}$ with
strict contraction. The map has expansion range at least $2b$.
By Eqn.~\eqref{quad:bucket-initial}, bucket $j$ reaches at most $R_*b$
variables in at most
\[
 k_j\le\log_{4/3}(h_j+1)+C_3
\]
stages. This follows by iterating
$\log R_{\rm next}\le(3/4)\log R+\log C_2$ about its fixed
point. With $\beta=\log16/\log(4/3)<10$, the support multiplication
is at most
\[
 16^{k_j}\le C_4(h_j+1)^\beta.
\]
Thus, after these compressions, each original group's support in
bucket $j$ is at most
\begin{equation}\label{quad:bucket-final-support}
 C_5\left(
 \frac{q}{(h_j+1)^{16-\beta}}
 + (\log q+3)^3(h_j+1)^\beta\right).
\end{equation}
The same upper bound holds during the bucket compression.
All bucket schedules run in parallel, with identity on $Z$ and
on completed buckets; they need $O(\log(H(q)+1))$ rounds.

The size requirement in \Cref{quad:constant-degree} is $M\ge CL^6$.
At every application, $L\le2^{O(H(q))}$ and $M\ge b$; since
$q=5\log b$ and $H(q)=O(\log^*q)$, the requirement is satisfied
for all sufficiently large $b$. Moreover
\[
 \exp(O(L^{2\kappa}\log L))
 =\exp(2^{O(H(q))})=b^{o(1)}.
\]
The number of calls and their output sizes are polynomial in $b$.

\begin{lemma}[Ordered accumulator]\label{quad:accumulator}
There is a uniformly polynomial-time continuation of the preceding
substitution sequence whose final variable count is at most $Ab$
for an absolute $A$. The total number of stages is
$T=O(1+H(q))$, each active map expands through weight $b$, and
at every prefix each original group reaches only $O(q)$ variables.
\end{lemma}

\begin{proof}
Fix an admissible integer $L_0\ge2C_g$, and then an integer
\[
 A\ge\max\{1,R_*,C_Z,L_0^4/c_g\}.
\]
On $2Ab$ input rows, \Cref{quad:constant-degree} supplies a
sixteen-neighbor map with at most $Ab$ outputs and expansion
range at least $2b$. Its size condition holds for large $b$
because $L_0$ is fixed.

Start with an empty accumulator. For $j=0,\ldots,J_*-1$, merge
it with $B_j$ by this map. Assign the at most $2Ab$ active
variables to distinct rows and delete the unassigned rows.
Keep every output coordinate; these form the next accumulator.
All other variables are left unchanged. Finally merge the
accumulator with $Z$ by the same procedure. If $J_*=0$, this
last map simply processes $Z$. Each intermediate accumulator
has at most $Ab$ variables, and after the last merge no untouched
part remains.

Row deletion preserves expansion by zero extension of coefficient
vectors. It adds neither input variables nor clauses. Outputs
adjacent only to deleted rows are still counted, and all retained
outputs are available as rows at later merges. A clause acquires
no path through an unused input row. This is also why unused
allocated coordinates do not enlarge its support.

The order of the merges controls support. Bucket $j$ traverses
$J_*-j+1$ new maps. By Eqn.~\eqref{quad:height-halving},
\[
 \begin{aligned}
 h_j&\ge513\cdot2^{J_*-1-j},\\
 16^{J_*-j+1}&\le256(h_j/513)^4=O((h_j+1)^4).
 \end{aligned}
\]
Multiply Eqn.~\eqref{quad:bucket-final-support} by this factor and sum.
The two contributions are
\begin{align*}
 q\sum_{j<J_*}(h_j+1)^{-(12-\beta)}&=O(q),\\
 (\log q+3)^3\sum_{j<J_*}(h_j+1)^{\beta+4}
 &\le O((\log q+3)^3(H(q)+1)^{\beta+4})=O(q).
\end{align*}
The first sum is geometric when read backwards and the second
when read forwards. Here $12-\beta>2$, $H(q)\le\log q$, and
$(\log q+3)^{17}=O(q)$. Their constants are absolute. The part
$Z$ has support $O(q)$ and traverses only the last map.
At an unfinished prefix, every processed bucket has traversed
no more maps than in this estimate; an unprocessed bucket has
traversed none. The same bound therefore holds at every prefix.
The estimate also bounds prefixes of the earlier parallel bucket
compression, before the additional factors are applied.

The recursion uses at most $H(q)$ synchronized stages, bucket
compression uses $O(\log(H(q)+1))$, and the accumulator uses
$J_*+1=O(1+H(q))$. The active ranges are respectively at least
$b$, $2b$, and $2b$. The new calls have fixed $L_0$ and size
$2Ab$, so their total construction time is polynomial.
\end{proof}

\subsubsection{The Base Formula, Clause Encodings, and Variable Count}
\label{quad:final-proof}

We finish the proof using the Buss--Thapen formula
$\Phi_{b,c,2}$~\cite{BT24}, where $c=2\log b$.
Its $b$ natural columns have $q=5\log b$ variables each.
Every refutation of width below $b/2$ has depth at least
$2^c=b^2$~\cite[Theorem~7]{BT24}.
We need the following additional locality of its short proof.

Write $\ell=\log b$ and $a=2^c=b^2$. Column $x$ contains
pointer bits $f(x)_t$, $t<\ell$, and variables $H^y_{e,x}$ for
$y<c$, $e\in\{0,1\}$. For the binary digits $i_y$ of $i$, put
\[
 P_{i,x}=\bigvee_{y<c}H^y_{i_y,x}.
\]
Write $B_{x,x'}=(f(x)\ne x')$.
The axioms are the units $\neg H^y_{0,0}$, the bottom clauses
$P_{a-1,x}$, and the transitions
\[
 B_{x,x'}\vee P_{i,x}\vee\neg H^y_{(i+1)_y,x'}
 \qquad(i<a-1, x,x'<b, y<c).
\]
Tautological transitions may be omitted. From $P_{i+1,x'}$,
resolve every target variable not already present in $P_{i,x}$
against its transition. This derives
$B_{x,x'}\vee P_{i,x}$. The target difference is nonempty,
including when $x=x'$, since consecutive row indices differ.
Resolving these clauses for all pointer targets $x'$ along the
binary pointer tree derives $P_{i,x}$. Iterate upwards from the
bottom row and resolve $P_{0,0}$ against the top units.
The proof uses $O(b^4\log b)$ lines, has width at most $q-1$,
and the union of the two parents and conclusion of each inference
lies in at most two columns. Every axiom has the same locality.

Apply the successive substitutions of \Cref{quad:accumulator}.
For each original clause $C$, let $\mathcal E_t(C)$ be its
descendant clause set after stage $t$, removing identical clauses
only within this original clause's descendants. The path-support
bound gives, for every $D\in\mathcal E_t(C)$,
\[
 \begin{aligned}
 \operatorname{vars}(D)&\subseteq R_t(C),\\
 |R_t(C)|&=O(q),
 \end{aligned}
\]
where $R_t(C)$ counts paths from the original variables before
parity cancellations. This follows by induction: the support of
the next truth-table encoding lies in the union of the next
neighbors of the current support. Identity coordinates contribute
one neighbor, and deleted input rows contribute no paths.
There are at most $3^{O(q)}$ distinct clauses on $R_t(C)$.
Each expansion costs $2^{O(q)}$, so all stages can be generated
and deduplicated in $b^{O(1)}$ time. The exponent does not grow
with $T$. This constructs the actual intermediate CNFs required
by \Cref{quad:partial}.

For the upper bound on proof size, define the same descendant
sets for all lines of the original short proof. Their conjunction
is equivalent to that line under the successive substitutions.
Thus encoded parents imply every encoded conclusion clause.
All variables of this implication lie in the path neighborhood
of two original columns, of size $O(q)$. A CNF on $r$ variables
which entails a clause $D$ has a resolution-with-weakening
derivation of $D$ of size $O(2^r)$ and width at most $r$: fix
$D$ false, branch on the remaining variables, take a falsified
input clause at every leaf, weaken it to the clause excluding
that assignment, and resolve up the tree. No assumption clauses
are used. Applying this to every encoded conclusion, and joining
the local derivations along the original proof, gives a
polynomial-size ordinary-resolution proof of width $O(q)$.
Weakening can then be eliminated.

Let $F_b$ be the final condensed CNF, with its complete allocated
universe of at most $Ab$ variables. If it had a proof of width
at most
\[
 w_b=\lfloor b/4\rfloor-T
\]
and depth below $b^2$, reverse all $T$ stages using
\Cref{quad:partial}. The resulting proof of $\Phi_{b,c,2}$ has
width at most $\lfloor b/4\rfloor<b/2$ and the same depth,
a contradiction. Every range condition holds, since throughout
the reversal $2(W+1)\le2\lfloor b/4\rfloor\le b$.
Here $T=O(1+\log^*q)=o(b)$, so $w_b=\Theta(b)$.

The lifting theorem of \cite[Theorem~6.1]{IPS26} states that a
width-or-depth premise $(w,h)$, lifted by a fixed 1-stifling
gadget, gives
\[
 d=\Omega\bigl(\min\{h,w^2/\log(2S)\}\bigr).
\]
Apply it to $F_b$ with disjoint three-bit majority gadgets,
which are 1-stifling. Since $w=\Theta(b)$ and $h=b^2$,
every refutation of the lifted formula satisfies
$d\log(2S)=\Omega(b^2)$.
The full truth tables of the two gadget polarities are admissible
CNF representations in \cite[Section~2.6]{IPS26}; distributing
their disjunction gives exactly our clausewise encoding.
The lifting multiplies each local support by at most three;
the same local simulation gives a polynomial ordinary-resolution
proof. Its formula has at most $3Ab$ allocated variables.

To obtain exactly $N$ variables, fix an integer $C_0\ge3A$ and
choose the largest power of two $b$ with $C_0b\le N$.
Then $N/(2C_0)<b\le N/C_0$. Give a unit clause to any allocated
final coordinate absent from the hard component, and add fresh
unit variables until the total is $N$. Restricting these variables
to their satisfying values recovers the hard component, without
increasing refutation size or depth. The ordinary proof ignores
the units. This proves \Cref{quad:main}.

Only outputs of the last accumulator and the final majority
gadgets survive in this count. Intermediate coordinates, local
code bits, graph ports, and deleted input rows are not retained
as propositional variables. The construction and the short proof
use no extension variables.

\section*{Acknowledgments}
The authors thank Ronen Shaltiel, Roei Tell, and Mohit Gurumukhani for
helpful discussions regarding~\cite{CHLLTV26}. This work was first
envisioned at the beginning of the second author's Ph.D. and has been
revisited on and off ever since, particularly during the authors' visit
to the Simons Institute for the Theory of Computing for the program
Pseudorandomness and High-Dimensional Expansion in Fall 2026. The authors thank
the institute for its hospitality and support.

\paragraph{AI Disclosure}

The authors used OpenAI's ChatGPT and Codex for mathematical
development, proof review, literature search, and drafting.
Initial exploration took place in ChatGPT; subsequent development
and checking used Codex, including parallel agents. The final contributions of this paper were
developed through iterative exchanges between ChatGPT and the authors, who have
reviewed the mathematical arguments and take responsibility for
the final manuscript.

\clearpage
\phantomsection
\addcontentsline{toc}{section}{References}
\begingroup
\small
\bibliographystyle{alpha}
\bibliography{nmExt}
\endgroup
\clearpage
\appendix
\section{Extensions and Instantiations for Linear Branching Programs}
\label{sec:lbp-appendix}

We prove the auxiliary-input, list-prediction, and higher-moment
extensions of \Cref{sec:lbp}, then give the rank estimates and parameter
calculations for its explicit Boolean functions.

\subsection{Independent Uniform Auxiliary Variables}
\label{subsec:lbp-auxiliary}

Both reductions remain valid when the program has additional independent
uniform input bits. Fixing these bits need not preserve the weakly
read-once condition. For inputs $(X,R)\in\F_2^n\times\F_2^t$, let
\[
 E_X=\{(\lambda,0):\lambda\in(\F_2^n)^*\}
\]
be the space of linear forms depending only on $X$. For a linear space
$W$ of query forms on $(X,R)$, define its \emph{pure-input query rank} by
\[
 \rho(W)=\dim(W\cap E_X).
\]
This counts independent linear combinations of queries in which all
coefficients on $R$ cancel. It is an intersection dimension, not the
dimension of the projection of $W$ onto the $X$-coefficients. For
example, $W=\langle x_1+r_1\rangle$ has $\rho(W)=0$, although its
projection onto the $X$-coefficients has dimension one. We use
$\rho$ to define the cut below.

A \emph{cut marker} is a designated point on an edge, not an additional
query node. Its prefix space is the span of the forms queried before
that point. In a full program, this span is the same on every path
reaching the marker. We place markers where the pure-input query rank
first reaches the prescribed threshold.

\begin{lemma}[A cut for the pure-input rank]
\label{lem:lbp-pure-input-cut}
Let $P$ be a size-$s$ weakly read-once linear branching program on $(X,R)\in\F_2^n\times\F_2^t$, and let $1\le q<n$. There is an equivalent full weakly read-once linear branching program
with a cut $V$ of at most $2s$ markers, each computation crossing exactly
one marker. At each marker $v$, let $W_v$ be its prefix space,
$d_v=\dim W_v$, and let $Q_v$ be the linear map whose coordinate forms are a basis for $W_v$. Then:
\begin{enumerate}
\item $\rho(W_v)=\dim(W_v\cap E_X)=q$.
\item Reachability is $Q_v(X,R)\in A_v$ for some $A_v\subseteq\F_2^{d_v}$, with probabilities $p_v=|A_v|/2^{d_v}$ summing to one.
\item There are affine sections $\sigma_b:\F_2^{d_v}\to\F_2^{n+t}$, indexed by $b\in\F_2^{n+t-d_v}$, as in \Cref{lem:lbp-future-sections}. Write $\pi_X$ for projection onto the $X$ coordinates. Their projected maps $B_b=\pi_X\sigma_b$ satisfy
\begin{equation}
Q_0B_b(a)=Ja
\label{eq:lbp-common-pure-coordinates}
\end{equation}
for a rank-$q$ map $Q_0:\F_2^n\to\F_2^q$ and a rank-$q$ map $J:\F_2^{d_v}\to\F_2^q$, independent of $b$.
\item For every fixed $a$ and independent uniform $b,b'$,
\begin{equation}
\Pr_{b,b'}[B_b(a)=B_{b'}(a)]=2^{-(n-q)}.
\label{eq:lbp-auxiliary-collision}
\end{equation}
\end{enumerate}
\end{lemma}

\begin{proof}
Put $N=n+t$. Adding one query increases $\rho$ by at most one. Normalize as in \Cref{lem:lbp-full-cut}, and extend each separate short terminal edge by ignored pure-$X$ queries until $\rho\ge q$. This is possible while $\rho<q<n$.

Mark the first transition from $q-1$ to $q$. If an original branching query crosses the threshold, give its two outgoing edges separate markers. If an inserted ignored query crosses, its answers have the same continuation and use one marker. This convention prevents a shared node from being marked on a path that has already crossed the cut. An original vertex contributes at most two markers, either after its own query or on its two outgoing replacement chains. Each chain crosses at most once, so $|V|\le2s$. At each marker, all reaching paths have the same prefix space; its
dimension $d_v$ may depend on the marker.

Reachability and the sections follow by the proofs of \Cref{lem:lbp-full-cut} and \Cref{lem:lbp-future-sections}. Choose the rows of $Q_0$ to be the $q$ independent pure-$X$ forms in $W_v$. They are independent linear combinations of the rows of $Q_v$, giving the rank-$q$ map $J$ and Eqn.~\eqref{eq:lbp-common-pure-coordinates}.

For fixed $a$, a uniform suffix transcript parametrizes the uniform distribution on $Q_v^{-1}(a)$. Its $X$-projection is uniform on an affine space of dimension $n-q$. Indeed, a form $\lambda$ on $\F_2^n$ annihilates $\pi_X\ker Q_v$ if and only if its pure-$X$ extension $(\lambda,0)$ lies in $W_v$. This annihilator has dimension $q$. Two independent uniform points in the projected fiber coincide with probability $2^{-(n-q)}$, proving Eqn.~\eqref{eq:lbp-auxiliary-collision}.
\end{proof}

\begin{theorem}[Directional hardness with independent uniform auxiliary variables]
\label{thm:lbp-directional-auxiliary}
Let $X\sim U_n$ and $R\sim U_t$ be independent. Suppose $f:\F_2^n\to\F_2$ has directional derivative bias at most $\beta$ at entropy $K$. If $q<n$ and $2q-n\ge K$, every size-$s$ weakly read-once linear branching program on $(X,R)$ satisfies
\begin{equation}
\left|\E_{X,R}(-1)^{f(X)+P(X,R)}\right|
\le\sqrt{2s\bigl(\beta+2^{-(n-q)}\bigr)}.
\label{eq:lbp-directional-auxiliary}
\end{equation}
The bound has no dependence on $t$.
\end{theorem}

\begin{proof}
Use the cut of \Cref{lem:lbp-pure-input-cut} and put $h=(-1)^f$. Eqn.~\eqref{eq:lbp-common-pure-coordinates} shows that every $B_b$ has rank at least $q$, while the displacement $D_{b,b'}=B_b+B_{b'}$ takes values in $\ker Q_0$, of dimension $n-q$. Conditioning on $D_{b,b'}(a)=c$ reduces the rank of $B_b(a)$ by at most $n-q$, leaving dimension at least $2q-n\ge K$. On a nonzero fiber the derivative bound applies; on the zero fiber the sign product is one. Therefore
\[
\left|\E_a h(B_b(a))h(B_{b'}(a))\right|
\le\beta+\Pr_a[B_b(a)=B_{b'}(a)].
\]
For the suffix output sign $g_v(b)$, put $H_v(a)=\E_b g_v(b)h(B_b(a))$. Expanding its second moment and averaging Eqn.~\eqref{eq:lbp-auxiliary-collision} gives
\[
\E_aH_v(a)^2\le\beta+2^{-(n-q)}.
\]
The contribution from marker $v$ is $\E_a\mathbf1_{A_v}(a)H_v(a)$. The two Cauchy--Schwarz applications in \Cref{thm:lbp-section-correlation}, using $|V|\le2s$ and $\sum_vp_v=1$, prove the result.
\end{proof}

Affine non-malleability needs only one bit of conditioning to handle intersecting projected sections.

\begin{lemma}[Intersecting projected sections]
\label{lem:lbp-intersecting-sections}
Let $F:\F_2^n\to\F_2^m$ be a $(k,\varepsilon)$ non-malleable affine extractor, let $u\ne0$, and put $h(x)=(-1)^{\langle u,F(x)\rangle}$. Suppose affine maps $B,B':\F_2^d\to\F_2^n$ satisfy $Q_0B=Q_0B'=J$, where $Q_0$ and $J$ are linear maps of rank $q\ge k+1$. For $c=\Pr_{a\sim U_d}[B(a)=B'(a)]$,
\begin{equation}
\left|\E_a h(B(a))h(B'(a))-c\right|
\le2\varepsilon(1-c).
\label{eq:lbp-intersecting-sections}
\end{equation}
Non-malleability against globally fixed-point-free affine involutions suffices.
\end{lemma}

\begin{proof}
Choose a linear right inverse $U$ of $J$. Write a uniform parameter as $a=Uz+v$, with $z$ uniform on $\F_2^q$ and $v$ independently uniform on $\ker J$. For each fixed $v$, the maps $\sigma(z)=B(Uz+v)$ and $\tau(z)=B'(Uz+v)$ are sections of the same $Q_0$.

Put $\Delta(z)=\sigma(z)+\tau(z)\in\ker Q_0$. For a binary linear functional $\lambda$, let $E_\lambda=\{z:\lambda(\Delta(z))=1\}$. A nonempty $E_\lambda$ is an affine space of dimension at least $q-1\ge k$. Choose $w\in\ker Q_0$ with $\lambda(w)=1$, which is possible because $E_\lambda$ is nonempty, and define
\[
\begin{aligned}
\Delta_\lambda(z)&=\Delta(z)+w\bigl(1+\lambda(\Delta(z))\bigr),\\
T_\lambda(x)&=x+\Delta_\lambda(Q_0x).
\end{aligned}
\]
Then $Q_0\Delta_\lambda=0$ and $\lambda(\Delta_\lambda)=1$ everywhere. Thus $T_\lambda$ is affine, globally fixed-point-free, and $T_\lambda^2=\operatorname{id}$. On $E_\lambda$ it sends $\sigma(z)$ to $\tau(z)$. If $Z$ is uniform on $E_\lambda$, the source $\sigma(Z)$ has dimension at least $k$, so non-malleability gives
\[
\left|\E_z\mathbf1_{E_\lambda}(z)h(\sigma(z))h(\tau(z))\right|
\le2\varepsilon\Pr[E_\lambda].
\]
This is also true for an empty event. Averaging over uniform $\lambda$, every nonzero displacement is selected with probability $1/2$ and a zero displacement is never selected. Consequently
\[
\left|\E_z\mathbf1_{\{\Delta(z)\ne0\}}h(\sigma(z))h(\tau(z))\right|
\le2\varepsilon\Pr[\Delta(z)\ne0].
\]
On a collision the sign product is one. Add this contribution and average over $v$ to obtain Eqn.~\eqref{eq:lbp-intersecting-sections}.
\end{proof}

\begin{theorem}[Affine hardness with independent uniform auxiliary variables]
\label{thm:lbp-affine-auxiliary}
Let $F:\F_2^n\to\F_2^m$ be a $(k,\varepsilon)$ non-malleable affine extractor, where $1\le k\le n-2$ and $0\le\varepsilon\le1/2$. Let $X\sim U_n$ and $R\sim U_t$ be independent. For every nonzero $u\in\F_2^m$ and every size-$s$ weakly read-once linear branching program $P$ on $(X,R)$,
\begin{equation}
\left|\E_{X,R}(-1)^{\langle u,F(X)\rangle+P(X,R)}\right|
\le\sqrt{2s\left(2\varepsilon+(1-2\varepsilon)2^{-(n-k-1)}\right)}.
\label{eq:lbp-affine-auxiliary}
\end{equation}
Non-malleability against globally fixed-point-free affine involutions suffices. The bound has no dependence on $t$.
\end{theorem}

\begin{proof}
Use \Cref{lem:lbp-pure-input-cut} with $q=k+1$. For each pair of projected sections, \Cref{lem:lbp-intersecting-sections} gives
\[
\left|\E_a h(B_b(a))h(B_{b'}(a))\right|
\le2\varepsilon+(1-2\varepsilon)\Pr_a[B_b(a)=B_{b'}(a)].
\]
Expanding $\E_aH_v(a)^2$ as in the preceding proof and using Eqn.~\eqref{eq:lbp-auxiliary-collision} bounds it by
\[
2\varepsilon+(1-2\varepsilon)2^{-(n-q)}.
\]
The same two Cauchy--Schwarz applications prove Eqn.~\eqref{eq:lbp-affine-auxiliary}. More generally, any $k+1\le q<n$ gives the collision term $2^{-(n-q)}$.
\end{proof}

\subsection{Full-Output and List Prediction}
\label{subsec:lbp-list-prediction}

The full vector guarantee also bounds prediction of an extractor output, or of a short list containing it. We first consider programs without auxiliary inputs.

\begin{theorem}[List prediction]
\label{thm:lbp-list-prediction}
Let $F:\F_2^n\to\F_2^m$ be a $(k,\varepsilon)$ non-malleable affine extractor, with $1\le k<n$. Let $P$ be a size-$s$ weakly read-once linear branching program whose sinks carry lists of at most $J$ distinct elements of $\F_2^m$, where $1\le J<2^m$. Write $p=J/2^m$ and $c=2^{-(n-k)}$, and define
\[
R_0=cp(1-p)+\varepsilon\bigl(c|1-2p|+(1-c)\max\{p,1-p\}\bigr).
\]
Then
\begin{equation}
\Pr[F(U_n)\in P(U_n)]\le p+\sqrt{2sR_0}.
\label{eq:lbp-list-prediction}
\end{equation}
For $J\le2^{m-1}$, this implies
\begin{equation}
\Pr[F(U_n)\in P(U_n)]
\le p+\sqrt{2s(1-p)\bigl(\varepsilon+p2^{-(n-k)}\bigr)}.
\label{eq:lbp-small-list-prediction}
\end{equation}
Non-malleability against globally fixed-point-free affine involutions suffices.
\end{theorem}

\begin{proof}
Pad each list to size exactly $J$, without changing the query structure or decreasing success. Use the cut and sections of \Cref{lem:lbp-full-cut} and \Cref{lem:lbp-future-sections}. The list on suffix $b$ is a fixed set $L_{v,b}$, independent of $a$. Put
\[
\begin{aligned}
Z_{v,b}(a)&=\mathbf1_{\{F(\sigma_b(a))\in L_{v,b}\}}-p,\\
H_v(a)&=\E_bZ_{v,b}(a).
\end{aligned}
\]
The ordinary affine extractor guarantee follows by marginalizing non-malleability against a nonzero translation. Thus the list probability on each section differs from $p$ by at most $\varepsilon$, giving
\[
\E_a Z_{v,b}(a)^2
\le p(1-p)+\varepsilon|1-2p|.
\]
For $b\ne b'$, the involution in \Cref{lem:lbp-nonmalleable-sections} gives the full vector guarantee
\[
(F(\sigma_b(U_k)),F(\sigma_{b'}(U_k)))
\approx_\varepsilon(U_m,F(\sigma_{b'}(U_k))).
\]
The product of the two centered list indicators has ideal expectation zero. Its possible values are $p^2$, $-p(1-p)$, and $(1-p)^2$, with range length $\max\{p,1-p\}$. Therefore
\[
\left|\E_a Z_{v,b}(a)Z_{v,b'}(a)\right|
\le\varepsilon\max\{p,1-p\}.
\]
Since $\Pr[b=b']=c$, expansion gives $\E_aH_v(a)^2\le R_0$. The excess success probability is $\sum_v\E_a\mathbf1_{A_v}(a)H_v(a)$, so two applications of Cauchy--Schwarz give Eqn.~\eqref{eq:lbp-list-prediction}. For $p\le1/2$, the coefficient of $\varepsilon$ in $R_0$ is $1-p-cp\le1-p$, proving Eqn.~\eqref{eq:lbp-small-list-prediction}.
\end{proof}

For a single full-output prediction, $p=2^{-m}$, so the collision contribution includes the factor $2^{-m}$. For $m=1$ and $J=1$, the exact bound is
\[
\Pr[P(U_n)=F(U_n)]
\le\frac12+\frac12\sqrt{2s\bigl(c+2\varepsilon(1-c)\bigr)},
\]
which retains the off-diagonal factor from Eqn.~\eqref{eq:lbp-section-second-moment}.

The list-prediction bound also holds with independent uniform auxiliary
inputs, with a one-bit loss in the collision exponent.

\begin{theorem}[List Prediction with Independent Uniform Auxiliary Inputs]
\label{thm:lbp-list-auxiliary}
Let $F:\F_2^n\to\F_2^m$ be a $(k,\varepsilon)$ non-malleable affine
extractor, where $1\le k\le n-2$ and $0\le\varepsilon\le1/2$.
Let $X\sim U_n$ and $R\sim U_t$ be independent. Suppose $P$ is a
size-$s$ weakly read-once linear branching program on $(X,R)$ whose
sinks carry lists of at most $J$ distinct elements of $\F_2^m$,
where $1\le J<2^m$.
Write $p=J/2^m$ and $c=2^{-(n-k-1)}$, and define
\[
 R_0=cp(1-p)+\varepsilon
 \bigl(c|1-2p|+(1-c)\max\{p,1-p\}\bigr).
\]
Then
\begin{equation}
 \Pr[F(X)\in P(X,R)]\le p+\sqrt{2sR_0}.
 \label{eq:lbp-list-auxiliary}
\end{equation}
For $J\le2^{m-1}$, this implies
\begin{equation}
 \Pr[F(X)\in P(X,R)]
 \le p+\sqrt{2s(1-p)
                   \bigl(\varepsilon+p2^{-(n-k-1)}\bigr)}.
 \label{eq:lbp-small-list-auxiliary}
\end{equation}
The bounds do not depend on $t$. Non-malleability against globally
fixed-point-free affine involutions suffices.
\end{theorem}

\begin{proof}
Pad each list to size exactly $J$. Apply
\Cref{lem:lbp-pure-input-cut} with $q=k+1$.
At a cut marker $v$, let $a$ be the uniform prefix coordinates and
$b$ the uniform suffix answers. Suppress $v$ from the notation.
The projected affine maps satisfy $Q_0B_b(a)=J_0a$, with
$\operatorname{rank}Q_0=\operatorname{rank}J_0=q$.
The output list $L_b$ is independent of $a$. Put
\[
 \begin{aligned}
 I_b(a)&=\mathbf1_{\{F(B_b(a))\in L_b\}},\\
 Z_b(a)&=I_b(a)-p,\\
 H(a)&=\E_b Z_b(a).
 \end{aligned}
\]
We prove $\E_aH(a)^2\le R_0$.

For fixed $a$, a uniform suffix parametrizes the full input fiber
uniformly. Its $X$-projection is uniform on an affine space of
dimension $n-q$. Hence, for every fixed $a,b$,
\begin{equation}
 \Pr_{b'}[B_{b'}(a)=B_b(a)]=c.
 \label{eq:lbp-list-conditional-collision}
\end{equation}
Let $C$ denote this collision event. By symmetry and
$zz'\le(z^2+z'^2)/2$,
\[
 \E_{a,b,b'}\mathbf1_C Z_bZ_{b'}
 \le\E_{a,b,b'}\mathbf1_C Z_b^2
 =c\,\E_{a,b}Z_b^2.
\]
Each $B_b$ has rank at least $q\ge k$. The ordinary affine extractor guarantee
therefore gives $|\E_a I_b-p|\le\varepsilon$.
Since $Z_b^2=p^2+(1-2p)I_b$, the collision contribution is at most
\begin{equation}
 c\bigl(p(1-p)+\varepsilon|1-2p|\bigr).
 \label{eq:lbp-list-collision-contribution}
\end{equation}
When $p>1/2$, this uses the lower bound $\E_a I_b\ge p-\varepsilon$.

To bound the remaining contribution, fix $b,b'$ and choose a linear
right inverse $U$ of $J_0$. Write $a=Uz+w$, with independent uniform
$z\in\F_2^q$ and $w\in\ker J_0$. For fixed $w$, the maps
\[
 \sigma(z)=B_b(Uz+w),\qquad
 \tau(z)=B_{b'}(Uz+w)
\]
are affine sections of $Q_0$.
For a linear functional $\lambda$, let
$E_\lambda=\{z:\lambda(\sigma(z)+\tau(z))=1\}$.
A nonempty $E_\lambda$ has dimension at least $q-1=k$.
The involution constructed in the proof of
\Cref{lem:lbp-intersecting-sections} maps $\sigma(z)$ to $\tau(z)$
on $E_\lambda$ and is globally fixed-point-free.
Apply non-malleability to the uniform source on $\sigma(E_\lambda)$.
The test
\[
 \phi(y,y')=(\mathbf1_{y\in L_b}-p)
             (\mathbf1_{y'\in L_{b'}}-p)
\]
has ideal expectation zero and range length $M_p=\max\{p,1-p\}$.
Thus
\[
 \left|\E_z\mathbf1_{E_\lambda}(z)
       \phi(F(\sigma(z)),F(\tau(z)))\right|
 \le\varepsilon M_p\Pr[E_\lambda].
\]
Averaging over uniform $\lambda$ selects each nonzero displacement
with probability $1/2$ and never selects a zero displacement.
The triangle inequality, followed by averaging over $w,b,b'$, gives
\[
 \left|\E_{a,b,b'}\mathbf1_{\neg C}Z_bZ_{b'}\right|
 \le\varepsilon M_p(1-c).
\]
Together with Eqn.~\eqref{eq:lbp-list-collision-contribution}, this
proves $\E_aH(a)^2\le R_0$.

The excess success probability is
$\sum_v\E_a\mathbf1_{A_v}(a)H_v(a)$.
Two applications of Cauchy--Schwarz, using at most $2s$ cut markers
and $\sum_v\Pr[A_v]=1$, bound it by $\sqrt{2sR_0}$.
For $p\le1/2$, the coefficient of $\varepsilon$ in $R_0$ is
$1-p-cp\le1-p$, giving
Eqn.~\eqref{eq:lbp-small-list-auxiliary}.
\end{proof}

For full-output prediction, take $J=1$. For lists with $p\le1/2$,
success probability at least $p+\eta$ implies
\[
 s\ge
 \frac{\eta^2}
 {2(1-p)\bigl(\varepsilon+p2^{-(n-k-1)}\bigr)}.
\]
Consequently, for constant $\eta>0$ and $\varepsilon\le2^{-E}$,
\begin{equation}
 s\ge 2^{\min\{E,\,n-k-1+m-\log J\}-O_\eta(1)}.
 \label{eq:lbp-list-auxiliary-size}
\end{equation}
This concerns full-output or list prediction. The additional term
$m-\log J$ does not improve the Boolean prediction bound.

\subsection{Higher Moments for Several Affine Tamperings}
\label{subsec:lbp-higher-moments}

The $h$-non-malleability guarantee improves the collision term in the
preceding reductions. It makes the original output nearly uniform
and independent of all $h$ tampered outputs together; the tampered maps
may coincide with one another. We use the affine-tampering formulation
in \Cref{thm:fixed-multi-affine}, equivalent to the jointly affine
formulation by \Cref{pre:affine-nm-equivalence}.

\subsubsection{Boolean Prediction}

\begin{theorem}[Higher moments]
\label{thm:lbp-higher-moment}
Let $F:\F_2^n\to\F_2^m$ be a $(k,\varepsilon)$ $h$-non-malleable affine extractor, where $1\le k<n$.
Let $r\ge1$ be an integer with $2r-1\le h$. For every nonzero output
parity $f_u(x)=\langle u,F(x)\rangle$ and every size-$s$ weakly
read-once linear branching program $P$,
\begin{equation}
 \operatorname{Corr}(f_u,P)
 \le\left[2s\left(2\varepsilon+(2r-1)!!\,2^{-r(n-k)}\right)\right]^{1/(2r)}.
 \label{eq:lbp-higher-moment}
\end{equation}
It suffices to assume $h$-non-malleability against globally
fixed-point-free affine involutions.
\end{theorem}

\begin{proof}
Use the cut and sections of
\Cref{lem:lbp-full-cut} and \Cref{lem:lbp-future-sections}. Put $B=2^{n-k}$
and $\chi=(-1)^{f_u}$. Write $g_v(b)=(-1)^{P_v(b)}$ for the suffix output sign and define
\[
 H_v(a)=\E_b g_v(b)\chi(\sigma_b(a)).
\]
First consider $t+1\le2r$ distinct sections
$\sigma_0,\ldots,\sigma_t$ of the common map $Q_v$. The maps
\[
 T_j(x)=x+\sigma_0(Q_vx)+\sigma_j(Q_vx),\qquad 1\le j\le t,
\]
are globally fixed-point-free affine involutions and satisfy
$T_j\sigma_0=\sigma_j$, as in
\Cref{lem:lbp-nonmalleable-sections}. $h$-non-malleability therefore gives
\begin{equation}
 \left|\E_a\prod_{j=0}^t\chi(\sigma_j(a))\right|\le2\varepsilon.
 \label{eq:lbp-joint-section-products}
\end{equation}
The first parity is uniform independently of the others in the ideal
distribution, so its product with the other signs has expectation zero.

Expand $\E_aH_v(a)^{2r}$ over independent suffix indices
$b_1,\ldots,b_{2r}$. Every index occurring with even multiplicity
cancels. If any indices remain, their number is even and at most $2r$,
so Eqn.~\eqref{eq:lbp-joint-section-products} bounds the absolute
expectation by $2\varepsilon$. Otherwise the product of all the
$g_v$ and $\chi$ signs is one. Let $\alpha_r(B)$ be the probability
that all multiplicities among $2r$ uniform indices from $[B]$ are even.
Then
\begin{equation}
 \E_a|H_v(a)|^{2r}
 \le\alpha_r(B)+2\varepsilon(1-\alpha_r(B))
 \le2\varepsilon+(2r-1)!!B^{-r}.
 \label{eq:lbp-higher-even-moment}
\end{equation}
For the last inequality, an all-even tuple admits a perfect matching
of its positions in which matched entries agree. There are
$(2r-1)!!$ matchings, and a fixed matching is respected with probability
$B^{-r}$.

More generally, if $\E_a|H_v(a)|^{2r}\le M_r$ at every cut node,
H\"older's inequality gives
\begin{equation}
 \begin{split}
 \operatorname{Corr}(f_u,P)
 &\le\sum_{v\in V}
       \left|\E_a\mathbf1_{A_v}(a)H_v(a)\right|\\
 &\le M_r^{1/(2r)}\sum_{v\in V}p_v^{1-1/(2r)}
 \le(2sM_r)^{1/(2r)}.
 \end{split}
 \label{eq:lbp-moment-cut}
\end{equation}
Here $\sum_vp_v=1$ and $|V|\le2s$. Apply
Eqn.~\eqref{eq:lbp-higher-even-moment}.
\end{proof}

For $r=1$, the sharper upper bound in
Eqn.~\eqref{eq:lbp-higher-even-moment} is
$2^{-(n-k)}+2\varepsilon(1-2^{-(n-k)})$.
For $r=2$, its collision probability is
$\alpha_2(B)=3B^{-2}-2B^{-3}$.
If agreement is at least $1/2+\eta$, then
\begin{equation}
 s\ge\frac{(2\eta)^{2r}}
 {2\left(2\varepsilon+(2r-1)!!\,2^{-r(n-k)}\right)}.
 \label{eq:lbp-higher-moment-size}
\end{equation}
For fixed $r$, constant $\eta>0$, and $\varepsilon=2^{-E}$, this is
$s\ge2^{\min\{E,r(n-k)\}-O_{r,\eta}(1)}$.
Thus a moment of order $2r$ can improve the collision-limited exponent
from $n-k$ to $r(n-k)$. It does not multiply the extractor error exponent.
To reach the error-limited exponent up to an additive constant, it
suffices that $r\ge E/(n-k)$ and $h\ge2r-1$, with $r$ fixed.
For small advantage, one should also compare the available lower
moments, since Eqn.~\eqref{eq:lbp-higher-moment-size} contains the
factor $(2\eta)^{2r}$.

\subsubsection{List Prediction}

For Boolean prediction, even repetitions of a sign cancel in the moment expansion. Centered list indicators do not have this property. In the ideal distribution with independent uniform outputs, the moment of their average is a scaled central moment of a binomial random variable.

\begin{theorem}[Higher moments for list prediction]
\label{thm:lbp-list-higher-moment}
Under the extractor hypotheses of \Cref{thm:lbp-higher-moment}, let
$P$ be a size-$s$ weakly read-once linear branching program whose sinks
carry lists of at most $J$ distinct elements of $\F_2^m$, where
$1\le J<2^m$. Set
\begin{itemize}
\item $p=J/2^m$, the success probability of a fixed list on a uniform output;
\item $B=2^{n-k}$, the number of suffix sections;
\item $a_p=\max\{p,1-p\}$, the largest absolute centered indicator value.
\end{itemize}
For independent Bernoulli-$p$ variables
$Y_1,\ldots,Y_B$, define
\[
 \mathcal M_{2r}(B,p)
 =\E\left(\frac1B\sum_{i=1}^B(Y_i-p)\right)^{2r}.
\]
Then
\begin{equation}
 \Pr[F(U_n)\in P(U_n)]
 \le p+\left[2s\left(\mathcal M_{2r}(B,p)
                     +2r\varepsilon a_p^{2r-1}\right)\right]^{1/(2r)}.
 \label{eq:lbp-list-higher-moment}
\end{equation}
There is a constant $C_r$, depending only on $r$, such that
\begin{equation}
 \mathcal M_{2r}(B,p)
 \le C_r\left[
       \left(\frac{p(1-p)}B\right)^r
       +\frac{p(1-p)}{B^{2r-1}}\right].
 \label{eq:lbp-list-central-moment}
\end{equation}
\end{theorem}

\begin{proof}
Pad each list to size exactly $J$ and use the same cut and sections.
At a fixed cut node put
\[
 \begin{aligned}
 Z_b(a)&=\mathbf1_{\{F(\sigma_b(a))\in L_b\}}-p,\\
 H_v(a)&=\E_bZ_b(a).
 \end{aligned}
\]
Consider a tuple in the expansion of $\E_aH_v(a)^{2r}$. Suppose its
distinct suffix indices have multiplicities $e_1,\ldots,e_t$.
The $h$-non-malleability guarantee allows us to replace the output on any one
section by a uniform output independent of all the others. This
requires at most $t-1\le2r-1$ tamperings. Write
\[
 \mu_j=p(1-p)^j+(1-p)(-p)^j=\E(Y_1-p)^j.
\]
Successively replacing the section outputs compares the product moment
with $\prod_{i=1}^t\mu_{e_i}$.

A product of powers of centered indicators of total degree $q$ has
range length at most $a_p^{q-1}$. For $p\le1/2$, its positive values
are at most $(1-p)^q$, and any negative value has absolute value at
most $p(1-p)^{q-1}$. The case $p\ge1/2$ follows by complementing the
indicators. Since $|\mu_j|\le a_p^j$, each replacement contributes
error at most $\varepsilon a_p^{2r-1}$. The last replacement uses
the ordinary affine extractor guarantee, which follows by marginalizing non-malleability.
Thus
\[
 \E_aH_v(a)^{2r}
 \le\mathcal M_{2r}(B,p)+2r\varepsilon a_p^{2r-1}.
\]
Apply the H\"older argument of Eqn.~\eqref{eq:lbp-moment-cut} to
the excess success probability
$\sum_v\E_a\mathbf1_{A_v}(a)H_v(a)$.

For Eqn.~\eqref{eq:lbp-list-central-moment}, put $v=p(1-p)$.
The independent moment expansion has no contribution from a suffix
index occurring exactly once, since $\mu_1=0$. A tuple with
$\ell\le r$ distinct indices, each occurring at least twice,
contributes at most $v^\ell$ in absolute value, since
$|\mu_j|\le v$ for $j\ge2$. There are at most
$C_rB^\ell$ such tuples. Hence
\[
 \mathcal M_{2r}(B,p)
 \le C_r\sum_{\ell=1}^r v^\ell B^{\ell-2r}.
\]
For fixed $r$, this sum is bounded by a constant depending only on
$r$ times its two endpoint terms, proving the claim.
\end{proof}

For example, with $v=p(1-p)$,
\[
 \mathcal M_2(B,p)=\frac vB,
 \qquad
 \mathcal M_4(B,p)=\frac{3v^2}{B^2}+\frac{v(1-6v)}{B^3}.
\]
The second term in Eqn.~\eqref{eq:lbp-list-central-moment} is needed
when $Bp$ is small. If $p\le1/2$, put $\ell=\log_2(1/p)$.
For fixed $r$, constant excess success $\gamma>0$, and
$\varepsilon=2^{-E}$, \Cref{thm:lbp-list-higher-moment} gives
\begin{equation}
 s\ge2^{\min\{E,\,r(n-k+\ell),\,(2r-1)(n-k)+\ell\}
                 -O_{r,\gamma}(1)}.
 \label{eq:lbp-list-higher-moment-size}
\end{equation}
For $r=1$, \Cref{thm:lbp-list-prediction} has a sharper error
coefficient. The direct sign argument in
\Cref{thm:lbp-higher-moment} is also sharper when $p=1/2$.

\subsubsection{Higher Moments with Auxiliary Variables}

Projected sections may intersect when auxiliary variables are present.
We handle this by controlling the part of a moment tuple in which
some projected input occurs exactly once.

\begin{theorem}[Higher moments with independent uniform auxiliary variables]
\label{thm:lbp-auxiliary-higher-moment}
Under the extractor hypotheses of \Cref{thm:lbp-higher-moment}, let
$X\sim U_n$ and $R\sim U_t$ be independent. Let $P$ be a size-$s$
weakly read-once linear branching program on $(X,R)$. Put $L=2r$,
and let $D_L$ be the number of partitions of $[L]$ whose blocks all
have size at least two. For every integer $q$ with
$k+\binom L2\le q<n$ and every nonzero $u\in\F_2^m$,
\begin{equation}
 \left|\E_{X,R}(-1)^{f_u(X)+P(X,R)}\right|
 \le\left[2s\left(D_L2^{-r(n-q)}
                    +2(2^L-1)\varepsilon\right)\right]^{1/(2r)}.
 \label{eq:lbp-auxiliary-higher-moment}
\end{equation}
The bound has no dependence on $t$, and queries may mix the original
and auxiliary variables arbitrarily.
\end{theorem}

\begin{proof}
Use \Cref{lem:lbp-pure-input-cut} at pure-input rank $q$.
At each marker, the projected sections satisfy $Q_0B_b(a)=Ja$,
where $Q_0,J$ have rank $q$. Put $\chi=(-1)^{f_u}$ and
$H_v(a)=\E_b g_v(b)\chi(B_b(a))$ as before.

Fix a tuple $b_1,\ldots,b_L$, abbreviate $B_i=B_{b_i}$, and let
\[
 U_i=\{a:B_i(a)\ne B_j(a)\text{ for every }j\ne i\}.
\]
We first prove that for every nonempty $I\subseteq[L]$,
\begin{equation}
 \left|\E_a\mathbf1_{\cap_{i\in I}U_i}(a)
                        \prod_{j=1}^L\chi(B_j(a))\right|
 \le2\varepsilon\Pr_a\left[\bigcap_{i\in I}U_i\right].
 \label{eq:lbp-unique-section-intersection}
\end{equation}
Choose a linear right inverse $U$ of $J$ and write $a=Uz+w$,
with independent uniform $z\in\F_2^q$ and $w\in\ker J$.
For fixed $w$, the maps $\sigma_i(z)=B_i(Uz+w)$ are affine
sections of $Q_0$.

Let $\mathcal E_I$ be the undirected pairs of indices having at
least one endpoint in $I$. For each $e=\{i,j\}\in\mathcal E_I$,
put $\Delta_e(z)=\sigma_i(z)+\sigma_j(z)$ and choose an independent
uniform linear functional $\lambda_e$ on $\F_2^n$. Consider
\[
 E_{\boldsymbol\lambda}
 =\{z:\lambda_e(\Delta_e(z))=1\text{ for every }e\in\mathcal E_I\}.
\]
If nonempty, this is an affine space of dimension at least
$q-|\mathcal E_I|\ge q-\binom L2\ge k$.
Moreover, for every $z$,
\begin{equation}
 \E_{\boldsymbol\lambda}\mathbf1_{E_{\boldsymbol\lambda}}(z)
 =2^{-|\mathcal E_I|}
   \mathbf1_{\{\Delta_e(z)\ne0\ \forall e\in\mathcal E_I\}}.
 \label{eq:lbp-random-separation-tests}
\end{equation}

Fix a nonempty $E_{\boldsymbol\lambda}$ and choose $i_0\in I$.
For every $j\ne i_0$, its pair with $i_0$ belongs to $\mathcal E_I$.
Write $\Delta_j=\sigma_{i_0}+\sigma_j$ and
$\lambda_j=\lambda_{\{i_0,j\}}$.
Choose $w_j\in\ker Q_0$ with $\lambda_j(w_j)=1$; such a vector
exists because the corresponding constraint is nonempty. Define
\[
 \begin{aligned}
 D_j(z)&=\Delta_j(z)+w_j(1+\lambda_j(\Delta_j(z))),\\
 T_j(x)&=x+D_j(Q_0x).
 \end{aligned}
\]
Then $Q_0D_j=0$ and $\lambda_j(D_j)=1$ everywhere. Each $T_j$ is
a globally fixed-point-free affine involution, and
$T_j\sigma_{i_0}(z)=\sigma_j(z)$ on $E_{\boldsymbol\lambda}$.
Apply $h$-non-malleability to the source
$\sigma_{i_0}(U_{E_{\boldsymbol\lambda}})$ and these $L-1\le h$
maps. It bounds the absolute expectation of the sign product on this
source by $2\varepsilon$. Averaging over the functionals in
Eqn.~\eqref{eq:lbp-random-separation-tests} cancels the common factor
$2^{-|\mathcal E_I|}$ on both sides. Averaging over $w$ proves
Eqn.~\eqref{eq:lbp-unique-section-intersection}.

Let $N(a)$ count the indices whose projected value occurs exactly once
in the fixed tuple. Inclusion--exclusion, followed by
Eqn.~\eqref{eq:lbp-unique-section-intersection}, gives
\[
 \left|\E_a\mathbf1_{\{N(a)>0\}}
                      \prod_{j=1}^L\chi(B_j(a))\right|
 \le2\varepsilon\E_a(2^{N(a)}-1).
\]
The product of the suffix signs $g_v(b_j)$ is fixed here and does
not change this bound. On the remaining event the absolute product
is at most one.

For every fixed $a$, independent uniform suffixes give independent
uniform projected inputs in the same affine space of dimension
$n-q$, by the proof of \Cref{lem:lbp-pure-input-cut}. Put
$B=2^{n-q}$. Let $N$ count singleton indices among $L$ independent
uniform draws from $[B]$, and write
$\beta_L(B)=\Pr[N=0]$ and $\gamma_L(B)=\E(2^N-1)$.
Expanding the moment and averaging the preceding bounds yields
\begin{equation}
 \E_a|H_v(a)|^{2r}\le\beta_L(B)+2\varepsilon\gamma_L(B).
 \label{eq:lbp-auxiliary-occupancy-moment}
\end{equation}
Every tuple with no singleton has an equality partition with at most
$r$ blocks. A fixed such partition occurs with probability at most
$B^{-r}$, so $\beta_L(B)\le D_LB^{-r}$.
Also $\gamma_L(B)\le2^L-1$.
Apply Eqn.~\eqref{eq:lbp-moment-cut}, whose H\"older argument also
applies when the prefix dimensions vary between markers.
\end{proof}

For fixed $r$ and $k+\binom{2r}{2}<n$, take
$q=k+\binom{2r}{2}$ in
\Cref{thm:lbp-auxiliary-higher-moment}. Prediction with constant advantage $\eta>0$
then requires
\[
 s\ge2^{\min\{E,r(n-k)\}-O_{r,\eta}(1)}
 \qquad(\varepsilon=2^{-E}).
\]
The additive term includes the cost $r\binom{2r}{2}$ in the collision
exponent. For $r=1$, the sharper constants are those in \Cref{thm:lbp-affine-auxiliary}.

For constant prediction advantage, higher moments improve the collision term in the size bound but leave the extractor error exponent unchanged. Applying a higher moment also requires non-malleability against more tamperings, which may worsen the output and error constants in \Cref{thm:fixed-multi-affine}. The moment order should therefore be chosen together with the extractor parameters.

\subsection{Explicit Instantiations}
\label{subsec:lbp-instantiations}

The cubic trace function of~\cite[Theorem~15]{GPT22} admits a stronger
bound than the one obtained from its worst-case derivative bias.
We average the ranks of its derivatives over the possible displacements
between two affine sections. All dimensions below are over $\F_2$
unless a different field is specified.

\begin{lemma}[Counting displacements]
\label{lem:lbp-cubic-quotient-count}
Let $F=\F_{2^m}$, and let $U,K\le F^3$ satisfy $U\cap K=\{0\}$ and
$\dim K=d$. Suppose $S\subseteq K$ is an affine space of dimension
$r\le d$ and $\dim(F^3/U)=d+r$. For the quotient map
$\pi:F^3\to F^3/U$, put $a(c)=\dim\pi(Fc)$. Then
\begin{equation}
\E_{c\sim S}2^{-a(c)}\le\frac32\,2^{-r/2}.
\label{eq:lbp-cubic-quotient-average}
\end{equation}
\end{lemma}

\begin{proof}
For an integer $j\ge0$, let $T_j=\{c\in S:a(c)\le j\}$.
Choose a maximal $F$-linearly independent subset
$c_1,\ldots,c_h$ of $T_j\setminus\{0\}$, and put
$L=\operatorname{span}_F\{c_1,\ldots,c_h\}$. Thus $T_j\subseteq L$,
$h\le3$, and
\[
\dim\pi(L)=\dim\bigl(\pi(Fc_1)+\cdots+\pi(Fc_h)\bigr)\le hj.
\]
If $3j<d+r$, then $h=3$ is impossible: it would give $L=F^3$
and contradict $\dim\pi(F^3)=d+r$. Hence $h\le2$.
Since $\pi$ is injective on $K$, it is injective on $T_j$, and
\begin{equation}
\Pr_{c\sim S}[a(c)\le j]\le2^{2j-r}
\qquad\text{when }3j<d+r.
\label{eq:lbp-cubic-low-rank-count}
\end{equation}
The argument also covers $T_j=\varnothing$ or $T_j=\{0\}$.

For every nonnegative integer-valued random variable $a$,
\[
\E2^{-a}=\frac12\sum_{j\ge0}2^{-j}\Pr[a\le j].
\]
If $r=2h$, apply Eqn.~\eqref{eq:lbp-cubic-low-rank-count}
for $j<h$ and the bound $\Pr[a\le j]\le1$ thereafter. Since $r\le d$,
all these values of $j$ satisfy $3j<d+r$. This gives
\[
\E_{c\sim S}2^{-a(c)}
\le\frac12\sum_{j=0}^{h-1}2^{j-2h}
   +\frac12\sum_{j=h}^{\infty}2^{-j}
=\frac32\,2^{-h}-\frac12\,2^{-2h}.
\]
If $r=2h+1$, use Eqn.~\eqref{eq:lbp-cubic-low-rank-count}
for $j\le h$, obtaining
\[
\E_{c\sim S}2^{-a(c)}
\le\frac12\sum_{j=0}^{h}2^{j-(2h+1)}
   +\frac12\sum_{j=h+1}^{\infty}2^{-j}
=2^{-h}-2^{-(2h+2)}
\le\sqrt2\,2^{-r/2}.
\]
Both estimates imply Eqn.~\eqref{eq:lbp-cubic-quotient-average}.
\end{proof}

\begin{lemma}[Cubic trace on affine sections]
\label{lem:lbp-cubic-sections}
Let $n=3m$ and $f(x,y,z)=\Tr_{\F_{2^m}/\F_2}(xyz)$.
Let $Q:\F_2^n\to\F_2^q$ have rank $q$, put $d=n-q$, and let
$\sigma,\tau:\F_2^q\to\F_2^n$ be affine sections of $Q$.
For $p_0=\Pr_a[\sigma(a)=\tau(a)]$,
\begin{equation}
\left|\E_a(-1)^{f(\sigma(a))+f(\tau(a))}-p_0\right|
\le\frac32\,2^{-(q-2m-d/2)}.
\label{eq:lbp-cubic-section-bound}
\end{equation}
The sections need not be disjoint.
\end{lemma}

\begin{proof}
Write $F=\F_{2^m}$ and identify $\F_2^n$ with $F^3$.
For $c=(c_1,c_2,c_3)\ne0$, the derivative
$D_cf(w)=f(w+c)+f(w)$ is quadratic. Its polar form is
\[
B_c(u,v)=\Tr\bigl(c_1(u_2v_3+u_3v_2)
                   +c_2(u_1v_3+u_3v_1)
                   +c_3(u_1v_2+u_2v_1)\bigr).
\]
The corresponding alternating matrix over $F$ has rank two and
kernel $Fc$. Nondegeneracy of the trace pairing therefore gives
$\operatorname{rad}(B_c)=Fc$.

Let $U\le F^3$ have dimension $t$, and put
$a(c)=\dim\pi(Fc)=m-\dim(U\cap Fc)$, where $\pi$ is the quotient
map modulo $U$. The image of $U$ in the nondegenerate alternating
space $F^3/Fc$ has dimension $t-m+a(c)$. In an alternating space
of dimension $2m$, the restriction to a subspace of dimension $b$
has rank at least $2b-2m$. Consequently
\[
\rank(B_c|_U)\ge2(t-2m+a(c)).
\]
The absolute bias of a quadratic polynomial is at most two to
minus half the rank of its polar form, as follows by squaring the
bias and averaging over differences. Thus, for every translate $w+U$,
\begin{equation}
\left|\E_{z\sim w+U}(-1)^{D_cf(z)}\right|
\le2^{-(t-2m+a(c))}.
\label{eq:lbp-cubic-restricted-bias}
\end{equation}

Now put $K=\ker Q$, $\Delta=\sigma+\tau$, and
$S=\operatorname{Im}\Delta$. Let $\sigma_0$ and $\Delta_0$ denote
the linear parts of $\sigma$ and $\Delta$. Then $S$ is an affine
subspace of $K$ of dimension $r=\rank\Delta_0\le d$.
Every nonempty displacement fiber under $\sigma$ is a translate of
\[
U=\sigma_0(\ker\Delta_0).
\]
Since $Q\sigma_0=\operatorname{id}$, we have $\dim U=q-r$ and
$U\cap K=\{0\}$. In particular, $\dim(F^3/U)=d+r$, so
\Cref{lem:lbp-cubic-quotient-count} applies.
Conditional on $\Delta(a)=c$, the point $\sigma(a)$ is uniform
on a translate of $U$. The displacement is uniform on $S$.
Its zero value contributes exactly $p_0$. Applying
Eqn.~\eqref{eq:lbp-cubic-restricted-bias} only to nonzero displacements,
we obtain
\begin{align*}
\left|\E_a(-1)^{f(\sigma(a))+f(\tau(a))}-p_0\right|
&\le2^{-(q-r-2m)}\E_{c\sim S}
       \mathbf1_{\{c\ne0\}}2^{-a(c)}\\
&\le\frac32\,2^{-(q-2m-r/2)}
\le\frac32\,2^{-(q-2m-d/2)}.
\end{align*}
\end{proof}

\begin{proof}[Proof of \Cref{cor:lbp-cubic-trace}]
We prove a bound for every $n=3m$ and any number $t$ of independent
uniform auxiliary bits. Let $1\le q<n$, put $d=n-q$, and apply
\Cref{lem:lbp-pure-input-cut} to $P(X,R)$, where $X\sim U_n$ and
$R\sim U_t$ are independent. At a cut marker $v$, the projected maps
$B_b:\F_2^{d_v}\to\F_2^n$ satisfy $Q_0B_b=J$, where both
$Q_0$ and $J$ have rank $q$.
Choose a linear right inverse $L$ of $J$ and write a uniform parameter
as $a=Lz+u$, where $z\sim U_q$ and $u$ is independently uniform
on $\ker J$. For each fixed $u$, the maps
\[
z\longmapsto B_b(Lz+u),\qquad
z\longmapsto B_{b'}(Lz+u)
\]
are sections of $Q_0$. Applying \Cref{lem:lbp-cubic-sections} and
averaging over $u$ gives, for $h=(-1)^f$,
\[
\left|\E_a h(B_b(a))h(B_{b'}(a))\right|
\le\Pr_a[B_b(a)=B_{b'}(a)]+\frac32\,2^{-(q-2m-d/2)}.
\]
Let $g_v(b)$ be the suffix output sign and put
$H_v(a)=\E_b g_v(b)h(B_b(a))$.
Expanding its second moment and using
Eqn.~\eqref{eq:lbp-auxiliary-collision}, we obtain
\[
\E_a H_v(a)^2\le2^{-d}+\frac32\,2^{-(q-2m-d/2)}.
\]
As in \Cref{thm:lbp-section-correlation}, the contribution from marker
$v$ is $\E_a\mathbf1_{A_v}(a)H_v(a)$. Cauchy--Schwarz bounds its
absolute value by $\sqrt{p_v\E_a H_v(a)^2}$.
Summing over the at most $2s$ markers and using $\sum_vp_v=1$ gives
\begin{equation}
\left|\E_{X,R}(-1)^{f(X)+P(X,R)}\right|^2
\le2s\left(2^{-d}+\frac32\,2^{-(m-3d/2)}\right).
\label{eq:lbp-cubic-general}
\end{equation}
There is no dependence on $t$.

For $m\ge3$, take $d=\lfloor2m/5\rfloor$. Then
$m-3d/2\ge d$, so Eqn.~\eqref{eq:lbp-cubic-general} is at most
$5s\,2^{-d}$. When $15\mid n$, we have $d=2n/15$, proving
Eqn.~\eqref{eq:lbp-cubic-trace}. Exact computation has correlation
one, and the stated average-case bound follows by substituting the
size of the program.

For an arbitrary input length $N\ge9$, put $m=\lfloor N/3\rfloor$
and apply the cubic trace function to three $m$-bit blocks, ignoring
the remaining at most two bits. Treating these bits as auxiliary inputs
gives
\begin{equation}
\operatorname{Corr}(f_N,P)
\le\sqrt{5s}\,2^{-\frac12\lfloor2\lfloor N/3\rfloor/5\rfloor}.
\label{eq:lbp-cubic-all-lengths}
\end{equation}
In particular, exact computation requires size at least
$2^{\lfloor2\lfloor N/3\rfloor/5\rfloor}/5
=2^{2N/15-O(1)}$.
\end{proof}

\begin{proof}[Proof of \Cref{cor:exponential-average-hardness}]
Use the cubic trace function extended to all input lengths as above.
Eqn.~\eqref{eq:lbp-cubic-all-lengths} gives
\[
\operatorname{Corr}(f_n,P)\le\sqrt{5s}\,2^{-n/15+O(1)}.
\]
For $s\le2^{n/36}$ and sufficiently large $n$, this is at most
$2^{-n/36}$. Agreement advantage is half the correlation, proving
the assertion with $c_1=c_2=1/36$.

Alternatively, fix $0<\rho<1$ and use a nonzero output parity of the directional affine extractor of Li and Zhong~\cite[Theorem 8]{LZdir}. Its entropy threshold is $K=\lceil\rho n\rceil$ and its error is at most $2^{-a_\rho n}$ for some $a_\rho>0$. Put $\lambda=\min\{a_\rho,(1-\rho)/2\}>0$. For sufficiently large $n$,
\[
2\varepsilon+2^{-\lfloor(n-K)/2\rfloor}\le6\cdot2^{-\lambda n}.
\]
For $s\le2^{\lambda n/4}$, \Cref{thm:lbp-directional-hardness} bounds the correlation by $\sqrt{12}\,2^{-3\lambda n/8}$, and hence by $2^{-\lambda n/4}$ for sufficiently large $n$.

One may also use the non-malleable affine extractor of Li~\cite[Theorem 1.12]{Li23}, with constants $\gamma,c>0$, entropy threshold $K=\lceil(1-\gamma)n\rceil$, linear output length, and error at most $2^{-cn}$. For $\lambda'=\min\{\gamma,c\}>0$,
\[
2\varepsilon+2^{-(n-K)}\le4\cdot2^{-\lambda'n},
\]
and Eqn.~\eqref{eq:lbp-affine-hardness} gives the same qualitative conclusion. Finally, \Cref{thm:affine-linear} supplies a non-malleable affine extractor at every fixed positive entropy rate, to which the same reduction applies. \Cref{thm:lbp-affine-auxiliary} permits independent uniform auxiliary inputs in these affine instantiations as well.
\end{proof}

\section{Proof of Deterministic Lossless Lifting}
\label{app:lossless-lifting}

We first prove the rank-certificate theorem underlying the gadget
construction in \Cref{sec:deterministic-lossless}. We then construct the
sparse inconsistent system and give the closure and counting arguments
needed for the lifting theorem. The only extractor guarantee used is
\Cref{thm:joint-affine-tuple}; we prove the remaining ingredients here.

\subsection{Rank Certificates and the Fourier Bound}
\label{sec:lossless-structure}

Let $Z$ be uniform on a binary vector space $V$, and write
$X_i=a_i+A_iZ\in\F_2^\ell$, for $i\in[N]$. Put
$U_i=\operatorname{row}(A_i)\le V^*$ and
$U_T=\sum_{i\in T}U_i$, where $V^*$ is the space of linear forms on
$V$. For a row subspace $Y\le V^*$, its annihilator is
$\operatorname{Ann}(Y)=\{z\in V:f(z)=0\text{ for every }f\in Y\}$.
For row spaces $A,C$, write
\[
 r(A\mid C)=\dim(A+C)-\dim C.
\]
For sets of block indices $T,B$, abbreviate $r(U_T\mid U_B)$ as $r(T\mid B)$.
A single block index in a set argument denotes its singleton set.
All partitions below are partitions of complete original blocks. A
subspace of $U_i$ used in a proof is not an additional original block.

\begin{definition}[Rank certificate]
\label{def:lossless-certificate}
A certificate with parameters $(M,H,q,k,k')$ consists of a partition
\[
 [N]=\Gamma_1\mathbin{\dot\cup}\cdots\mathbin{\dot\cup}\Gamma_M
       \mathbin{\dot\cup}C_1\mathbin{\dot\cup}\cdots
       \mathbin{\dot\cup}C_q
\]
and indices $i_j\in\Gamma_j$, where $1\le|\Gamma_j|\le H$, such that
\begin{align}
 r\left(U_{i_j}\mid U_{\bigcup_{v\ne j}\Gamma_v}\right)&\ge k
       &&(j\in[M]),\label{eq:lossless-private}\\
 r\left(\sum_{j\in T}U_{i_j}\mid U_{C_c}\right)&\ge k'|T|
       &&(T\subseteq[M],\ c\in[q]).\label{eq:lossless-joint}
\end{align}
The sets $C_1,\ldots,C_q$ are the \emph{outside classes}; they partition
the blocks outside the selected clusters $\Gamma_1,\ldots,\Gamma_M$.
We call $X_{i_j}$ the \emph{honest block} of $\Gamma_j$: its tagged
input serves as the original source in the non-malleability guarantee.
The left-hand side of Eqn.~\eqref{eq:lossless-private} is its
\emph{private rank}. It equals the entropy remaining in $X_{i_j}$
after any feasible fixing of all blocks in the other clusters
$\Gamma_v$, $v\ne j$. Neither its own cluster nor the outside classes
$C_c$ are included in this fixing.
Empty outside classes are allowed. The private threshold $k$ is a
positive integer and the joint threshold $k'>0$ may be real; the second
condition is vacuous when $q=0$.
\end{definition}

Fix distinct tags $c_i\in\F_2^t$ and put
$g_i(x)=\anmExt(c_i\circ x)$, where $\anmExt$ is a non-malleable
affine extractor. We state the next result for a one-bit output.
The same proof allows a nonzero output character at each index when
$\anmExt$ has several output bits.

\begin{theorem}[Cluster product bound]
\label{thm:lossless-cluster-phase}
Suppose the jointly affine input blocks $X_1,\ldots,X_N$ have a
certificate with parameters $(M,H,q,k,k')$. If $\anmExt$ satisfies
\Cref{thm:joint-affine-tuple} with entropy threshold $k$ for the original
source, against at most $H2^q-1$ comparisons, and with error
$\eps\le1/2$, then
\begin{equation}
 \left|\E\prod_{i=1}^N(-1)^{g_i(X_i)}\right|
 \le\left(2\eps+(2^q-1)2^{-k'}\right)^{M/2^q}.
 \label{eq:lossless-product}
\end{equation}
The comparison maps may be singular and may coincide with one another.
\end{theorem}

\begin{proof}
Write $\Lambda=\E\prod_{i=1}^N(-1)^{g_i(X_i)}$ for the expectation
to be bounded.
For each outside class $C_c$, set
$K_c=\bigcap_{i\in C_c}\ker A_i$. The product of the factors indexed
by $C_c$ is invariant under translation by $K_c$. Successive applications of Cauchy--Schwarz over
cosets of $K_1,\ldots,K_q$ give
\begin{equation}
 |\Lambda|^{2^q}\le
 \E_{h_c\sim U_{K_c}}\E_Z
 \prod_{j=1}^M\prod_{i\in\Gamma_j}
 \prod_{\omega\in\F_2^q}
 (-1)^{g_i(X_i+A_i\sum_c\omega_ch_c)}.
 \label{eq:lossless-cube}
\end{equation}
Here the directions are independent. At each step, retain the earlier
directions, apply Jensen before Cauchy--Schwarz on the next quotient,
and use translation invariance to remove the factors in that outside class. Each step replaces
each remaining gadget factor by two shifted copies, proving the
displayed inequality by induction.

By Eqn.~\eqref{eq:lossless-private}, there is a $k$-dimensional space
$V_j\subseteq\bigcap_{i\in\bigcup_{v\ne j}\Gamma_v}\ker A_i$
on which $A_{i_j}$ is injective. The $V_j$ have a direct sum: applying
$A_{i_j}$ to a relation among them isolates its $j$th summand. Choose
a complement $V_0$. After fixing the coordinate in $V_0$ and all directions in
Eqn.~\eqref{eq:lossless-cube}, the cluster products depend on independent
uniform coordinates in $V_1,\ldots,V_M$.

In cluster $j$, the distinguished input $X_{i_j}$ at $\omega=0$ is an
injective affine function of the uniform vector in $V_j$, and hence has
entropy $k$. Its tagged version serves as the original source in the
non-malleability guarantee. Each of the other at most $H2^q-1$ tagged
inputs is an affine function of this source on its support. Different
block indices have different tags. For $i=i_j$, an input with
$\omega\ne0$ can equal the original input only if
\begin{equation}
 A_{i_j}\sum_c\omega_ch_c=0
 \quad\text{for some }0\ne\omega\in\F_2^q.
 \label{eq:lossless-collision}
\end{equation}
Call this event bad. On a good cluster, non-malleability bounds the
absolute phase mean by $2\eps$; on a bad cluster use one. The tagged inputs form
a jointly affine tuple, so \Cref{thm:joint-affine-tuple} applies directly.

For every $T\subseteq[M]$,
\begin{equation}
 \Pr[\text{all }j\in T\text{ are bad}]
 \le\bigl((2^q-1)2^{-k'}\bigr)^{|T|}.
 \label{eq:lossless-joint-collisions}
\end{equation}
Indeed, choose a nonzero witness $\omega_j$ for each bad event and
group the indices by the last nonzero coordinate $c$ of $\omega_j$.
After fixing the earlier directions, the equations in group $T_c$
prescribe the joint image of $h_c$. Their conditional probability is
at most $2^{-k'|T_c|}$ by Eqn.~\eqref{eq:lossless-joint}. Multiply over
$c$ and sum over the $(2^q-1)^{|T|}$ witness choices.

Set $a=2\eps$ and $b=(2^q-1)2^{-k'}$. Expanding a nonnegative product
and applying Eqn.~\eqref{eq:lossless-joint-collisions}, we obtain
\[
 \E a^{\#\mathrm{good}}
 =\E\prod_j\left(a+(1-a)\mathbf1_{\{j\text{ bad}\}}\right)
 \le(a+(1-a)b)^M\le(a+b)^M.
\]
Together with Eqn.~\eqref{eq:lossless-cube}, this proves the theorem.
\end{proof}

\subsection{Three Linear-Algebra Lemmas}
\label{sec:matrix-tools}

For direct-sum decompositions $A=\bigoplus_{i=1}^m A_i$ and
$W=\bigoplus_{j=1}^n B_j$, a Boolean mask $a\in\F_2^m$ defines
the projection $Q(a)$ that is the identity on $A_i$ when $a_i=1$
and zero on $A_i$ otherwise. Define $P(b)$ on $W$ in the same way
for $b\in\F_2^n$. Write $B_S=\bigoplus_{j\in S}B_j$ for
$S\subseteq[n]$, abbreviate $B_{\supp(b)}$ by $B_b$, and write
$b^c=\mathbf1-b$. Independent subspaces below mean subspaces whose
sum is direct.

\begin{lemma}[Rank capacities for two conditioning spaces]
\label{lem:lossless-two-capacity}
Let $A_1,\ldots,A_m$ have dimension at most $\ell$, let $C_0,C_1$
be two subspaces, and let $w\ge1$ be an integer.
For $T\subseteq[m]$, put $A_T=\sum_{i\in T}A_i$, and define
\[
 R=\min_{T_*\mathbin{\dot\cup}T_0\mathbin{\dot\cup}T_1=[m]}
 \left(w|T_*|+r(A_{T_0}\mid C_0)+r(A_{T_1}\mid C_1)\right).
\]
There are integers $0\le x_i\le w$ with $\sum_i x_i=R$ and
$\sum_{i\in T}x_i\le r(A_T\mid C_c)$ for every $T$ and $c$.
If the $A_i$ are independent, $W=\bigoplus_jB_j$, and
$C_0=B_b,C_1=B_{b^c}$, then, for $w=\ell$,
\begin{equation}
 R=\min_a\rank\bigl(P(b)J-JQ(a)\bigr).
 \label{eq:lossless-projector-capacity}
\end{equation}
Here $J:\bigoplus_iA_i\hookrightarrow W$, and $P,Q$ are the
coordinate projections in the two decompositions.
\end{lemma}

\begin{proof}
Give each index $i$ a set of $w$ interchangeable slots. For
$c\in\{0,1\}$, a set of slots is independent if one can assign to
each slot of index $i$ a vector from $A_i$ so that their images
modulo $C_c$ are linearly independent. These sets form the linear
transversal matroid of the spaces $A_i$ modulo $C_c$.
On a slot set $E$, its rank is
\[
 r_c(E)=\min_{T\subseteq[m]}
 \left(r(A_T\mid C_c)+|E\setminus E_T|\right),
\]
where $E_T$ contains all slots of $T$~\cite{Rado42}. Matroid
intersection~\cite{Edmonds70} gives
the largest common independent set as
$\min_E(r_0(E)+r_1(E_{[m]}\setminus E))$.
If $z_i=|E\cap E_{\{i\}}|$, each rank is a minimum of affine
functions of $z$, hence is concave on $[0,w]^m$. Rounding each
$z_i$ to $0$ or $w$ does not increase the sum. Expanding the two
ranks at such a corner gives exactly the partition minimum $R$.
The slot counts of a maximum common set give the $x_i$.

For the last assertion, each index in $T_*$ can be moved to $T_0$ or
$T_1$ without increasing the expression, since $\dim A_i\le\ell=w$.
The kernel of $P(b)J-JQ(a)$ is the direct sum of
$A_{\supp(a)}\cap B_{\supp(b)}$ and
$A_{\supp(a)^c}\cap B_{\supp(b)^c}$. Subtract its dimension from
$\dim\bigoplus_iA_i$ and minimize over $a$.
\end{proof}

\begin{lemma}[Boolean projection rounding]
\label{lem:lossless-boolean-rounding}
Suppose $A_1,\ldots,A_m$ are independent nonzero spaces, $W=\bigoplus_{j=1}^nB_j$,
and $J:\bigoplus_iA_i\hookrightarrow W$. If
\[
 \max_b\min_a\rank(P(b)J-JQ(a))\le d,
\]
then there is a map $\pi:[m]\to[n]\cup\{0\}$ such that
\begin{equation}
 \rank\bigl(P(b)J-JQ(b\circ\pi)\bigr)\le72d
 \quad\text{for every }b.
 \label{eq:lossless-label-rounding}
\end{equation}
Here $(b\circ\pi)_i=b_{\pi(i)}$ when $\pi(i)\ne0$, and it is zero
when $\pi(i)=0$. Thus the label zero denotes the zero coordinate function.
\end{lemma}

\begin{proof}
Choose witnesses $a(b)$ and put $a_i^*=\dim A_i$.
Products $xy$ of Boolean vectors in this proof are coordinatewise.
Injectivity of $J$ gives
$\rank JQ(v)=\sum_i a_i^*v_i$. Additivity of $P,Q$ bounds the
expected weighted number of coordinates where $a(x+y)\ne a(x)+a(y)$
by $3d$ for independent uniform $x,y$, with coordinate $i$ weighted
by $a_i^*$. For its $i$th coordinate,
let $\epsilon_i=\Pr[a_i(x+y)\ne a_i(x)+a_i(y)]$ and
$f_i(b)=(-1)^{a_i(b)}$. Fourier expansion gives
\[
 1-2\epsilon_i=\sum_\xi\widehat f_i(\xi)^3
 \le\max_\xi\widehat f_i(\xi).
\]
Thus there is a homogeneous linear form $L_i$ with
$\Pr[a_i(b)\ne L_i(b)]\le\epsilon_i$. It follows that the
additive matrix map $E(b)=P(b)J-JQ(L(b))$ has mean rank at most
$4d$. Since $E(z)=E(x)+E(x+z)$, its rank is at most $8d$ everywhere.

The projection identities imply
\[
 J\bigl(Q(L(xy))-Q(L(x))Q(L(y))\bigr)
 =-E(xy)+P(x)E(y)+E(x)Q(L(y)).
\]
Hence the expected weighted number of coordinates where
$L(xy)\ne L(x)L(y)$ is at most $24d$. A parity
on at least two coordinates fails multiplication on at least $3/8$
of uniform pairs $(x,y)$: its defect has bilinear matrix $I+\mathbf1
\mathbf1^{\mathsf T}$, of rank at least two. The total weight of
these coordinates is therefore at most $64d$. Replace them by zero.
Each remaining coordinate is zero or one input bit, and the resulting
label map has error at most $8d+64d$.
\end{proof}

\begin{lemma}[Rectangular block completion]
\label{lem:lossless-block-completion}
Partition the rows and columns of a matrix $M$ into the same set of
block labels, allowing unequal or empty groups. Put
$t=\max_S\rank M[S,S^c]$. There is a block-diagonal matrix $D$
with $\rank(M-D)\le5t$.
\end{lemma}

\begin{proof}
Choose a cut $(I,J)$ locally maximizing $r=\rank M[I,J]$ under
one-block moves and reversal. Choose $r$ pivot rows $R$ in $I$ and
pivot columns $C$ in $J$, with invertible pivot matrix $P$. Set
$L=M[:,C]P^{-1}M[R,:]$ and $E=M-L$, taking $L=0$ if $r=0$.
Then $\rank L=r$, $E[I,J]=0$, and $E$ vanishes on all pivot
rows and columns. Let $D$ be the diagonal blocks of $E$.

For $i\in I$, let $r_i$ be the number of pivot rows in that block.
The columns of $E[I\setminus\{i\},i]$ lie in the column space of
$M[I\setminus\{i\},C\cup i]$, of dimension at most $r$ by local
optimality. Evaluation on the remaining pivot rows has rank at least
$r-r_i$. Since the residual columns vanish there, their rank is at
most $r_i$. Summing over $i$ gives $\rank(E-D)[I,I]\le r$.
The transposed argument gives $\rank(E-D)[J,J]\le r$.
Reversal and $\rank L=r$ give
$\rank(E-D)[J,I]\le2r$. The fourth rectangle is zero.
Consequently $\rank(M-D)\le r+r+r+2r\le5t$.
\end{proof}

\subsection{Certificates for Every Rank-Spread Source}
\label{univ:sec}

\begin{theorem}[Universal rank certificate]
\label{thm:universal-certificate}
For every $0<\eta\le1$, there are constants $\alpha,\beta>0$ and
integers $H,Q,\ell_0\ge1$ with the following property.
Let $U_1,\ldots,U_N$ be binary row spaces of dimension at most
$\ell\ge\ell_0$, and put $U_I=\sum_{i\in I}U_i$ for $I\subseteq[N]$.
If
\begin{equation}
 \dim U_I\ge\eta\ell|I|\quad\text{for every }I\subseteq[N],
 \label{univ:eq:spread}
\end{equation}
then the arrangement has a rank certificate with at least $\alpha N$ honest blocks,
cluster size at most $H$, at most $Q$ outside classes, and both
thresholds at least $\beta\ell$. The constants also apply to every
nonempty subset of the blocks.
\end{theorem}

All clusters and classes below consist of entire original blocks.
A \emph{core} of a block with row space $U_i$ is a chosen linear
subspace $A_i\le U_i$. We call cores \emph{independent} when their
sum is direct: $\sum_i A_i=\bigoplus_i A_i$. The cores serve only as
proof witnesses; they are not additional input blocks. After a
restriction of the source, we consider the corresponding restricted
images of these subspaces.

\subsubsection{Normalization and Repair}

Say that an arrangement has intersection density at most $\chi$ if
\begin{equation}
 \sum_i\dim(U_i\cap Y)\le\chi\dim Y\qquad\text{for every row space }Y.
 \label{univ:eq:density}
\end{equation}

\begin{lemma}[Normalization and coordinate enlargement]
\label{univ:lem:normalization}
Suppose Eqn.~\eqref{univ:eq:spread} holds, and let $\chi,q\ge1$ be
integers. There are an actual source restriction $X_0$ and an
extension $X'\supseteq X_0$ such that:
\begin{itemize}
 \item Both arrangements have intersection density at most $\chi$.
 For $0<\sigma<\eta$, at least $N-N/[\chi(\eta-\sigma)]$ blocks
 have rank at least $\sigma\ell$ in $X_0$.
 \item $X'$ has the same individual block images as $X_0$, and its
 blocks partition into $q$ internally independent families: the row
 spaces of the blocks within each family have a direct sum.
 \item For $f(S)=\dim\operatorname{proj}_S X_0$ and
 $p(S)=\dim\operatorname{proj}_S X'$, the difference $d=p-f$ is
 nonnegative and monotone, and there are integers $w_i\ge0$ such that the weights
 $w(S)=\sum_{i\in S}w_i$ and total weight $W=\sum_iw_i$ satisfy
 \begin{equation}
 \begin{aligned}
 d(S)&\le w(S),\\
 W&=\dim X'-\dim X_0\le\frac{(\chi-1)N\ell}{2q}.
 \end{aligned}
 \label{univ:eq:coordinate-defect}
 \end{equation}
\end{itemize}
\end{lemma}

\begin{proof}
Choose $Z$ maximizing $\Phi(Z)=\sum_i\dim(U_i\cap Z)-\chi\dim Z$.
Then $\dim Z\le N\ell/\chi$. The quotient spaces
$\overline U_i=(U_i+Z)/Z$ have density at most $\chi$: if
$\widetilde Y$ is the inverse image of a quotient space $Y$, maximality gives
\[
 \sum_i\dim(\overline U_i\cap Y)
 =\sum_i\bigl(\dim(U_i\cap\widetilde Y)-\dim(U_i\cap Z)\bigr)
 \le\chi\dim Y.
\]
For $B=\{i:\dim\overline U_i<\sigma\ell\}$, rank spread gives
$\eta\ell|B|\le\dim U_B\le N\ell/\chi+\sigma\ell|B|$.
The quotient is realized by restricting latent directions to
$\operatorname{Ann}(Z)$; call this source $X_0$.

Put $L=\sum_i\dim\overline U_i$ and
$F(t)=\E\dim\overline U_{S_t}$, where each index belongs to $S_t$
independently with probability $t$. Define
\[
 D(t)=\sum_i\E\dim(\overline U_i\cap\overline U_{S_t\setminus\{i\}}).
\]
Differentiating the displayed expectation $F(t)$ and applying density
to $\overline U_{S_t}$ give
\[
 \begin{aligned}
 F'(t)&=L-D(t),\\
 tL+(1-t)D(t)&\le\chi F(t).
 \end{aligned}
\]
Solving $F'(t)\ge(L-\chi F(t))/(1-t)$, with $F(0)=0$, yields
\begin{equation}
 F(t)\ge\frac L\chi\bigl(1-(1-t)^\chi\bigr).
 \label{univ:eq:color-rank}
\end{equation}
Assign each block independently and uniformly to one of $q$ classes.
There is therefore a partition $C_1,\ldots,C_q$ with
\[
 D:=L-\sum_c\dim\overline U_{C_c}
 \le L\left[1-\frac q\chi\bigl(1-(1-1/q)^\chi\bigr)\right]
 \le\frac{\chi-1}{2q}L.
\]
The last inequality is the two-term inclusion--exclusion bound.

Represent $X_0$ in $E=\bigoplus_iE_i$, with $E_i$ its individual block
image. For $F_c=\operatorname{proj}_{C_c}X_0$, choose a complement
$N_c$ spanned by coordinate axes of $E_{C_c}$, and set
$N_* =\bigoplus_cN_c$, $X'=X_0+N_*$. If $n\in N_*\cap X_0$,
its projection onto each class $C_c$ belongs to $N_c\cap F_c=0$.
Thus $\dim X'-\dim X_0=D$, the projection onto each class is all of $E_{C_c}$,
and the individual images stay unchanged.

Let $w_i$ count the chosen axes in block $i$. Then $d(S)\le w(S)$.
If $\pi:(X')^*\to X_0^*$ is restriction, $d(S)$ is the dimension
of the block row span on $S$ intersected with $\ker\pi$, so it
is monotone. The map $\pi$ is injective on each individual row space
$U_i'$. Consequently
\[
 \sum_i\dim(U_i'\cap Y)
 \le\sum_i\dim(\overline U_i\cap\pi Y)
 \le\chi\dim\pi Y\le\chi\dim Y.
\]
This proves all assertions.
\end{proof}

\begin{lemma}[Repair with an additive defect bound]
\label{univ:lem:repair}
Let $p,f$ give the dimensions of the spans of block row spaces in
two arrangements on the same original block indices. Write
$f(T\mid C)=f(T\cup C)-f(C)$, and similarly for $p$.
Suppose $d=p-f$ is
nonnegative and monotone, and $d(S)\le w(S)$ for nonnegative block
weights of total mass $W$. If $p$ has a certificate with $m$ honest
blocks, cluster bound $H$, $Q$ outside classes, and integer threshold
$k$, then for every integer $1\le\lambda<k$, $f$ has a certificate
using surviving original clusters, threshold $\lambda$, at most
$Q+2$ outside classes, and at least
\begin{equation}
 m-\frac{W}{k-\lambda+1}-\frac{W}{k-\lambda}
 \label{univ:eq:repair-count}
\end{equation}
honest blocks.
\end{lemma}

\begin{proof}
Delete a cluster $\Gamma$ if its honest block $i$ has $f$-rank below
$\lambda$ conditioned on the other surviving clusters $E\setminus\Gamma$.
The private-rank inequality for the certificate in $p$ gives
\[
 d(i\cup(E\setminus\Gamma))-d(E\setminus\Gamma)\ge k-\lambda+1.
\]
By monotonicity, deleting $\Gamma$ decreases $d(E)$ by at least this
amount. There are at most $W/(k-\lambda+1)$ deletions; let $B$ be their
union. The remaining clusters have private threshold $\lambda$ in $f$
and joint threshold $k$ against $B$ in $p$, by expanding rank in a chain.

Process the original outside classes and $B$ one set at a time.
For each such set $C$, repeatedly delete a nonempty set $T$ of current
honest blocks with $f(T\mid C)<\lambda|T|$. Let $R_C$ be all blocks
deleted in this step. Conditional subadditivity and the joint-rank
inequality for $p$ give, when $R_C\ne\varnothing$,
\[
 (k-\lambda)|R_C|<d(R_C\cup C)-d(C)\le w(R_C)+w(C).
\]
All processed sets $C$ and all sets $R_C$ are pairwise disjoint.
Thus these deletions cost fewer than $W/(k-\lambda)$ targets.
Put their whole clusters in one further outside class. The private
inequalities established after the first step give joint threshold
$\lambda$ against this class by a rank chain. Deleting targets only
improves the other inequalities.
\end{proof}

In particular, $k\ge2$ and $\lambda=\lfloor k/2\rfloor$ give loss
at most $4W/k$ and threshold at least $k/4$.

\subsubsection{Alignment and Selection}

\begin{lemma}[Alignment to original family blocks]
\label{univ:lem:alignment}
Let $A=\bigoplus_{i=1}^m A_i$ be independent nonzero cores of dimension
at most $\ell$, with inclusion $J:A\hookrightarrow W$, and let
$B_1,\ldots,B_n$ be an internally independent family of nonzero
original block spaces, $n\ge1$. Put
\begin{equation}
 g=\max_{S\subseteq[n]}\min_{I_0\mathbin{\dot\cup}I_1=[m]}
 \bigl(r(A_{I_0}\mid B_S)+r(A_{I_1}\mid B_{S^c})\bigr).
 \label{univ:eq:real-capacity}
\end{equation}
There are labels $\pi(i)\in[n]$ and a map $F:A\to B_{[n]}$ satisfying
\begin{equation}
 \begin{aligned}
 F(A_i)&\subseteq B_{\pi(i)},\\
 \rank(J-F)&\le361g.
 \end{aligned}
 \label{univ:eq:alignment}
\end{equation}
\end{lemma}

\begin{proof}
Write $W=B\oplus C$, where $B=\bigoplus_jB_j$, and let
$h=\rank(J\bmod B)\le g$, using the empty cut. Here $J\bmod B$
is $J$ followed by the quotient map from $W$ to $W/B$.
Complete the family by a decomposition of $C$. Its cut capacity is
at most $g$, since additional conditioning only decreases ranks.
The projector identity and Boolean rounding
(\Cref{lem:lossless-two-capacity} and \Cref{lem:lossless-boolean-rounding})
therefore give original, auxiliary, or zero labels satisfying
\begin{equation}
 \rank\bigl(P(x)J-JQ(x\circ\pi)\bigr)\le72g\qquad\text{for every }x.
 \label{univ:eq:rounded-family}
\end{equation}
Let $P_B$ project onto $B$. Group the columns of $P_BJ$ by original
labels, collecting auxiliary- and zero-labeled columns into a group
$0$ with no rows. Every off-diagonal cut of this rectangular matrix
has rank at most $72g$: in Eqn.~\eqref{univ:eq:rounded-family}, select
the original labels on the row side and restrict to the complementary
column groups. If the row side contains $0$, omit that empty row
group and further restrict the columns.
By \Cref{lem:lossless-block-completion}, a block-diagonal map $F$
satisfies $\rank(P_BJ-F)\le360g$. Its group-$0$ columns vanish,
so assign them any original label. Finally
$\rank(J-F)\le h+360g\le361g$.
\end{proof}

\begin{lemma}[Simultaneous selection]
\label{univ:lem:selection}
Let $A_1,\ldots,A_m$ be independent cores of dimension at least $b$.
Suppose proposed clusters $\Delta_i$ contain their respective core
blocks, have at most $B$ blocks of rank at most $\ell$, and each
intersects at most $\Delta$ other proposed clusters. For an integer
$1\le k\le b$, put $a=b-k+1$. There are at least
\begin{equation}
 \frac{m}{4(\Delta+B\ell/a)}
 \label{univ:eq:selection}
\end{equation}
pairwise disjoint proposed clusters with honest private rank at least
$k$ against the other selected clusters.
\end{lemma}

\begin{proof}
Independence and conditional subadditivity give
\begin{equation}
 a\,|\{i:r(A_i\mid Y)<k\}|\le\dim Y.
 \label{univ:eq:bad-cores}
\end{equation}
Indeed, for these indices $I$, $b|I|\le\dim A_I\le\dim Y+(k-1)|I|$.
Sample clusters independently with probability $u$. Remove each
sampled cluster meeting another sampled cluster or having core rank
below $k$ against the others. Intersections cost at most
$u^2\Delta m$ in expectation. Each private-rank test depends only on the
other sampling coins; its failure probability is at most the failure probability
against the full sampled span $Y_S$. By Eqn.~\eqref{univ:eq:bad-cores},
the expected number of sampled clusters failing the private-rank bound
is at most $u\E\dim Y_S/a\le u^2B\ell m/a$.
Taking $u=1/[2(\Delta+B\ell/a)]$ proves the bound.
\end{proof}

\subsubsection{Assembly of the Independent Families}

\begin{lemma}[Assembly]
\label{univ:lem:assembly}
Suppose an old arrangement has a certificate with $m$ honest blocks,
cluster bound $H$, $Q_0$ outside classes, and common integer threshold
$\kappa$. Suppose its honest spaces contain independent cores of
ranks between $\lceil\rho\ell\rceil$ and $\ell$, where $0<\rho\le1$.
Add $q\ge1$ fresh, mutually disjoint families of original blocks, each internally
independent. Suppose their ranks are at most $\ell$ and their combined
intersection density is at most $\chi\ge1$. Set
\begin{itemize}
 \item $d_* =\lceil12\chi/\rho\rceil$, $D_*=H+2d_*$;
 \item $\delta=\rho^3/(2^{20}\chi D_*)$;
 \item $R=\lceil(2/\delta)\ln(4q)\rceil$.
\end{itemize}
For $\ell\ge8(R+1)/\delta$, the enlarged arrangement has a certificate
with honest blocks among the original ones and
\begin{align}
 M&\ge\frac{\rho m}{64D_*},
 &H_{\rm new}&\le H+d_*,\label{univ:eq:assembly-count}\\
 Q_{\rm new}&\le Q_0+q2^R+1,
 &k_{\rm new},k'_{\rm new}&\ge
 \min\{\kappa,\delta\ell/[8(R+1)]\}.
 \label{univ:eq:assembly-rank}
\end{align}
\end{lemma}

\begin{proof}
Put $b=\lceil\rho\ell\rceil$.
Write $B_{t,1},\ldots,B_{t,n_t}$ for the block spaces in fresh family
$t$, and let $J$ be the inclusion of the direct sum of the cores
into the ambient row space. For a set $S$ of indices in family $t$,
write $B_{t,S}=\sum_{j\in S}B_{t,j}$.

\paragraph{Peeling.}
For each fresh family $t$, start with all core indices $I=[m]$.
Compute its capacity $G_t(I)$ from Eqn.~\eqref{univ:eq:real-capacity},
always cutting the entire original family. If $G_t(I)\le\delta\ell|I|$,
stop and call $L_t=I$ the \emph{low-capacity remainder}. The alignment
lemma gives a map $F_t:A_{L_t}\to B_{t,[n_t]}$.
Writing $p_t=\rank(J-F_t)$, its guarantees are
\begin{equation}
 \begin{aligned}
 F_t(A_i)&\subseteq B_{t,\pi_t(i)},\\
 p_t&\le361\delta\ell|L_t|.
 \end{aligned}
 \label{univ:eq:low-maps}
\end{equation}
In the high-capacity case $G_t(I)>\delta\ell|I|$, a maximizing cut
and \Cref{lem:lossless-two-capacity}, with
cap $\ell$, give weights $0\le x_i\le\ell$ of total $G_t(I)$ whose
sum on every subset is bounded by its joint rank against either side.
Allowing the part $T_*$ in that lemma does not lower the two-part
minimum for the chosen cut: each index in $T_*$ can be moved to
either of the other parts without increasing the expression,
since $\dim A_i\le\ell$. Remove those indices with
$x_i\ge\lceil G_t(I)/(2|I|)\rceil$ and call the removed set a
\emph{high group}, meaning a group removed by the high-capacity case.
The other weights sum to less
than $G_t(I)/2$, so at least $\delta|I|/2$ indices are removed.
Every subset of this group has rank greater than $\delta\ell/2$
per index against both sides of the recorded cut.

If the process has not stopped in the low-capacity case after $R$
removals, discard its remaining at most $m/(4q)$ indices and set
$L_t=\varnothing$. Zero-rank family blocks are irrelevant; a wholly
zero family always falls in the high-capacity case on nonempty $I$.
Use the zero map when $L_t=\varnothing$.
Delete all discarded remainders globally, leaving $I_*$ of size
$m_*\ge3m/4$. Restrict the maps and high groups to $I_*$; all rank
inequalities persist. For each family, $I_*$ partitions into the
surviving part of $L_t$ and at most $R$ high groups.

\paragraph{Total rank discrepancy.}
Initially use the low-capacity remainders before global deletion.
Let $S=\sum_t|L_t|$ count their indices with multiplicity, and let
\[
 P=\sum_t p_t=\sum_t\rank(J|_{A_{L_t}}-F_t)
\]
be the total rank discrepancy between the inclusion maps and the maps $F_t$. Also put $A=\bigoplus_{i=1}^m A_i$ and
$Z_*=\sum_t\operatorname{im}(J|_{A_{L_t}}-F_t)$.
The grouped images of each $F_t$ lie in individual family blocks and
in $A+Z_*$. Internal independence and intersection density imply
\begin{align}
 bS-P&\le\sum_t\rank F_t(A_{L_t})
 \le\chi(\dim A+\dim Z_*)\le\chi(m\ell+P),
 \label{univ:eq:self-bound}\\
 P&\le361\delta\ell S.
 \label{univ:eq:map-error-sum}
\end{align}
Since $361\delta(\chi+1)\le\rho/2$,
\begin{equation}
 \begin{aligned}
 S&\le2\chi m/\rho,\\
 P&\le(722\chi\delta/\rho)m\ell
 <\frac{\rho^2m\ell}{1024D_*}.
 \end{aligned}
 \label{univ:eq:absorbed-error}
\end{equation}
Thus the bound on the total rank discrepancy $P$ is independent of $q$. Restricting the maps to
$I_*$ only decreases $P$.

\paragraph{Disjoint private clusters.}
Let $d_i$ count the low-capacity remainders containing $i\in I_*$. The same density
argument gives
$b\sum_i d_i\le\chi m_*\ell+(\chi+1)P$.
Since $P\le m_*\ell$, deleting indices with $d_i>d_*$ costs at most
$m_*/4$ indices. Within each family $t$, group the cores indexed by
$L_t$ according to their labels $\pi_t(i)$.
The sum of these groups' kernel dimensions is at most $p_t$.
A group of $s>2\ell/b$ cores has kernel dimension at least
$bs-\ell>bs/2$, so deleting such groups costs fewer than $2p_t/b$
indices. Deleting individual cores indexed by $L_t$ whose image rank
is below $b/2$
costs fewer than another $2p_t/b$ indices. Here the individual kernel
spaces are independent and lie in $\ker F_t$.

Eqn.~\eqref{univ:eq:absorbed-error}, together with $m_*\ge3m/4$, gives
\begin{equation}
 P\le\frac{\rho^2m_*\ell}{128D_*}.
 \label{univ:eq:selection-budget}
\end{equation}
Hence at least $m_*/2$ candidates remain. Propose
$\Delta_i=\Gamma_i\cup\{B_{t,\pi_t(i)}:i\in L_t\}$.
Each has at most $H+d_*$ blocks and conflicts with at most
$2d_*\ell/b$ others. For $k=\lfloor b/2\rfloor$, the coefficients
in the two deletion bounds from the proof of \Cref{univ:lem:selection}
sum to at most
\[
 \frac{2d_*\ell}{b}+\frac{(H+d_*)\ell}{b-k+1}
 \le\frac{2D_*\ell}{b}.
\]
Sampling with $u=b/(4D_*\ell)$ therefore retains disjoint clusters
with private rank $k$ and at least
\begin{equation}
 M_s\ge\frac{\rho m_*}{16D_*}
 \label{univ:eq:selected-count}
\end{equation}
honest blocks. For each family $t$, the labels $\pi_t(i)$ assigned to
the selected indices $i\in L_t$ are distinct.

\paragraph{Joint rank against unassigned blocks.}
Fix a family $t$ and restrict latent directions to
$\operatorname{Ann}(\operatorname{im}(J-F_t))$, of codimension $p_t$.
On this restriction, each core indexed by $L_t$ agrees with its image
under $F_t$.
Taking a complement to its individual kernel gives a matched space
$E_{t,i}\subseteq B_{t,\pi_t(i)}$ of dimension at least $b/2$.
Write $f_t$ for the restricted family rank function and
$e_j=\dim B_{t,j}$ before restriction. Set
\[
 \varphi_t(S)=\sum_{j\in S}e_j-f_t([n_t])+f_t([n_t]\setminus S).
\]
This is nonnegative and monotone, with $\varphi_t([n_t])\le p_t$,
because the original family is independent.

Let $I_{\rm sel}$ be the set of currently selected indices, and put
\[
 S_t=\{\pi_t(i):i\in I_{\rm sel}\cap L_t\}.
\]
Thus $S_t$ consists of the labels assigned to selected indices in the
low-capacity remainder $L_t$. Call blocks $B_{t,j}$ with $j\notin S_t$
\emph{unassigned} in family $t$. Set $\lambda=\lfloor b/4\rfloor$.
Whenever a nonempty $T\subseteq I_{\rm sel}\cap L_t$ has restricted
joint core rank below $\lambda|T|$ against the unassigned blocks,
delete $T$ from $I_{\rm sel}$ and update the assigned-label sets in
every family. Let $S_t$ denote the set before this deletion and put
$J=\pi_t(T)$. Then
\begin{align*}
 \varphi_t(S_t)-\varphi_t(S_t\setminus J)
 &=\sum_{j\in J}e_j-r_t(B_{t,J}\mid B_{t,[n_t]\setminus S_t})\\
 &\ge\sum_{i\in T}\dim E_{t,i}
       -r_t(E_{t,T}\mid B_{t,[n_t]\setminus S_t})
 >\frac b4|T|.
\end{align*}
For the middle inequality, restrict the kernel of the conditional
observation map to the smaller sum of matched spaces. Their restricted
rank is at most that of the failing full cores. A global deletion only
removes labels assigned to the deleted indices in other families, so
none of their potentials
increases. The sum of potentials, initially at most $P$, thus pays for
all deletions. Fewer than $4P/b$ selected indices are lost, leaving at least
\[
 \frac{\rho m_*}{16D_*}-\frac{4P}{b}
 \ge\frac{\rho m_*}{32D_*}\ge\frac{\rho m}{64D_*}.
\]
For each family $t$, every $T\subseteq I_{\rm sel}\cap L_t$ now has
joint core rank at least $\lambda|T|$ against that family's unassigned
blocks, on its fixed restriction. These inequalities hold on the
original source as well: for latent spaces $L\subseteq V$,
\begin{equation}
 r_L(A\mid C)=\dim A(L\cap\ker C)\le\dim A(V\cap\ker C)=r_V(A\mid C).
 \label{univ:eq:restriction-monotonicity}
\end{equation}
The family restrictions need not coincide.

\paragraph{Outside classes.}
Partition the blocks of each family by membership in its at most
$R$ recorded cuts, and restrict this partition to the unassigned blocks.
This gives at most $2^R$ classes, each lying on one side of every cut.
Because the cuts were taken in the complete original family, this
classification also applies to blocks that became unassigned when
selected clusters were discarded.
For one such class $C$, a set $T$ of surviving selected indices splits
into its part in $L_t$ and its parts in at most $R$ high groups. With $h=\lfloor\delta\ell/2\rfloor$,
\[
 r(U_T\mid U_C)\ge\max_j r(U_{T_j}\mid U_C)
 \ge\frac{\min\{\lambda,h\}}{R+1}|T|.
\]
Keep old outside classes, and put all discarded old clusters in one
additional class. The original certificate's private-rank inequalities
give joint threshold $\kappa$ against that class by a rank chain.
Unassigned fresh blocks stay in their family classes, so this is a partition of all original blocks.
The private-rank bounds remain valid after the deletions. Since $\ell\ge8(R+1)/\delta$,
$\lambda\ge h\ge2(R+1)$; the common integer threshold
$\min\{\kappa,\lfloor\min\{\lambda,h\}/(R+1)\rfloor\}$ is at least
$\min\{\kappa,\delta\ell/[8(R+1)]\}$.
\end{proof}

\subsubsection{Proof of the Universal Certificate Theorem}

\begin{proof}[Proof of \Cref{thm:universal-certificate}]
Choose the constants in the following order; logarithms here are natural.
\begin{itemize}
 \item For normalization and the core ranks, set
 $\chi=\lceil4/\eta\rceil$ and $\rho=\eta/(4\chi)$.
 \item For cluster selection, set $d_*=\lceil12\chi/\rho\rceil$
 and $D_*=1+2d_*$. For assembly, then set
 \[
 \delta=\frac{\rho^3}{2^{20}\chi D_*},\qquad
 \alpha_p=\frac{\rho^2}{1024D_*}.
 \]
 \item Choose the number of independent families using
 $M_0=128(\chi-1)/(\alpha_p\delta^2)$ and $q=\lceil M_0^2\rceil$.
 The number of peeling rounds and the rank threshold before repair are
 \[
 R=\left\lceil\frac{2}{\delta}\ln(4q)\right\rceil,\qquad
 \beta_p=\frac{\delta^2}{32\ln(4q)}.
 \]
 \item For the final certificate, set $\alpha=\alpha_p/2$ and
 $\beta=\beta_p/4$. Its bounds on cluster size and the number of outside classes are
 $H=1+d_*$ and $Q=q2^R+3$. Choose the minimum block length so that
 \[
 \ell_0\ge\max\{4/\rho,8(R+1)/\delta,2/\beta_p\}.
 \]
\end{itemize}
Apply normalization with $\sigma=\eta/2$. At least $N/2$ blocks of
$X_0$ and $X'$ have rank at least $\eta\ell/2$. For every set $I$ of
these blocks, density implies
\[
 \dim U_I'\ge\chi^{-1}\sum_{i\in I}\dim U_i'
 \ge2\rho\ell|I|.
\]
Put $b=\lceil\rho\ell\rceil\le2\rho\ell$.
By Rado's theorem on linearly independent representatives~\cite{Rado42},
applied to $b$ copies of each $U_i'$, we may choose a $b$-dimensional
subspace of each $U_i'$ such that their sum is direct.
Apply selection to singleton clusters with $B=1$, $\Delta=0$, and
$k_0=\lfloor b/2\rfloor$. It retains at least $\rho N/16$ honest
singletons, with private threshold $k_0\ge\rho\ell/4$.
Treat these singletons as the old arrangement, with no outside classes,
and the remaining blocks as the fresh families. Removing the old
singletons from the $q$ families preserves their independence.

Assembly gives at least $\alpha_pN$ honest blocks, cluster bound
$1+d_*$, and at most $q2^R+1$ outside classes. Since
$R+1\le(4/\delta)\ln(4q)$, both thresholds are at least
$\beta_p\ell$, hence at least the integer $k=\lceil\beta_p\ell\rceil$.
Our choice $M_0\ge16$ ensures
\[
 \begin{aligned}
 \frac q{\ln(4q)}&\ge M_0=\frac{128(\chi-1)}{\alpha_p\delta^2},\\
 W&\le\frac{(\chi-1)N\ell}{2q}
 \le\frac{\alpha_p\beta_pN\ell}{8}.
 \end{aligned}
\]
Indeed $M_0\ge\ln(4M_0^2)$ and $x/\ln(4x)$ is increasing in this
range. Repair with $\lambda=\lfloor k/2\rfloor$ loses at most
$4W/k\le\alpha_pN/2$ targets, preserves the cluster bound, adds at
most two classes, and leaves threshold at least $\beta_p\ell/4$.
This gives the certificate in $X_0$. Transfer it to the original source
using Eqn.~\eqref{univ:eq:restriction-monotonicity}.
Finally, Eqn.~\eqref{univ:eq:spread} is hereditary under deleting
blocks, so the same proof and constants apply to every nonempty support.
\end{proof}

\subsection{Deterministic Formulas and Lossless Lifting}
\label{formula:section}

We prove \Cref{thm:deterministic-lossless} by combining
\Cref{thm:deterministic-gadgets} with a sparse inconsistent system over
a constant-size field. The closure and counting argument follows
Efremenko and Itsykson~\cite{EI26}; we include its proof.

\subsubsection{A Sparse Base}

\begin{lemma}\label{formula:sparse-base}
For every fixed $0<\delta<1/4$ and sufficiently large $r$, one can
deterministically construct, in time $r^{O_\delta(1)}$, a system
$Ez=b$ over a field $\F_Q$ of characteristic two such that:
\begin{itemize}
\item $E$ has $r+1$ rows, $r$ columns, and row weight at most $s$, where
$Q$ and $s$ depend only on $\delta$;
\item for integers $\alpha,\beta=\Theta_\delta(r)$ with
$2\le\alpha$ and $2\alpha-2\le\beta$, every set of at most $\beta$
rows is independent;
\item $\wt(E^{\mathsf T}\lambda)\ge(1-\delta)r$ whenever
$\alpha\le\wt(\lambda)\le\beta$;
\item $Ez=b$ is inconsistent.
\end{itemize}
Here $\wt$ counts nonzero coordinates.
\end{lemma}

\begin{proof}
The lossless-expander theorem of Capalbo, Reingold, Vadhan, and
Wigderson~\cite[Theorem~7.1]{CRVW02} supplies the following graphs.
For any fixed power of two $R$ and sufficiently large power of two $n'$,
there is a deterministically constructible left-$D_0$-regular graph
with $Rn'$ left vertices and $n'$ right vertices satisfying
\begin{equation}\label{formula:lossless-expansion}
 |\Gamma(S)|\ge\tfrac78D_0|S|\quad\text{for }|S|\le c n'/D_0.
\end{equation}
The constant $c>0$ is independent of $R$, and $D_0$ depends only on $R$.
Fix $0<\rho\le c/16$ and choose constants in this order:
\begin{itemize}
\item a regular graph of degree $d$ whose normalized adjacency operator
has norm at most $\theta$ on vectors orthogonal to the constants,
with $\theta^2\le\rho\delta/8$;
\item a power of two $B\ge d$ with $(d-1)/B\le\delta/8$, and $Q=B$;
\item a power of two $R\ge4B$, and the corresponding degree $D_0$.
\end{itemize}
For the spectral graph, a fixed power of the lazy
Margulis--Gabber--Galil graph gives exactly $u^2$ vertices for every
integer $u$, with a uniform spectral gap~\cite[Construction~8.1 and
Theorem~8.2]{HLW06}. Neighbor lists may repeat vertices and include
the vertex itself, so loops and parallel edges are allowed.

Choose the graph sizes in order as follows:
\begin{equation}\label{formula:base-sizes}
 \begin{aligned}
 u&=\lfloor\sqrt{r/B}\rfloor,\\
 n&=u^2,\\
 r_0&=Bn.
 \end{aligned}
\end{equation}
Let $n'$ be the largest power of two at most $n$, and set
$K=\lfloor c n'/D_0\rfloor$. Thus $n/2<n'\le n$
and $r-r_0=O_\delta(\sqrt r)$. Let
$H:\F_Q^{Rn'}\to\F_Q^n$ be the incidence map of the
lossless graph, padded with zero rows. Each column corresponds to a
left vertex, and its entry at a right vertex counts their incident
edges in $\F_Q$. If $0<|S|\le K$, at least
$2|\Gamma(S)|-D_0|S|\ge3D_0|S|/4$ right vertices receive exactly one
edge from $S$. Each gives a nonzero image coordinate, even in
characteristic two. Therefore every $\lambda$ with
$0<\wt(\lambda)\le K$ satisfies
\begin{equation}\label{formula:incidence-weight}
 \wt(H\lambda)\ge\tfrac34D_0\wt(\lambda).
\end{equation}
Take $\beta=K$ and $\alpha=\lfloor K/2\rfloor+1$. Since
$D_0K\ge c n'/2$ eventually, this gives
\begin{equation}\label{formula:constant-density}
 \alpha\le\wt(\lambda)\le\beta
 \quad\Longrightarrow\quad \wt(H\lambda)\ge3cn/32\ge\rho n.
\end{equation}

At each vertex of the spectral graph, regard its ordered $d$-neighbor
tuple as coefficients of a polynomial of degree at most $d-1$ and
evaluate at all $B$ field elements. This defines
$T:\F_Q^n\to\F_Q^{Bn}$. A nonzero tuple gives at least
$B-d+1$ nonzero evaluations. Since $B\ge d$ and every input coordinate
occurs in a neighbor tuple, $T$ is injective; its columns have weight
at most $dB$.
For $S=\supp(y)$ of density $p\ge\rho$, the normalized adjacency
operator $P$ satisfies
\begin{equation}\label{formula:spectral-variance}
 \E_v(P\mathbf1_S(v)-p)^2\le\theta^2p(1-p).
\end{equation}
Thus at most a $\theta^2(1-p)/p\le\delta/8$ fraction of vertices have
no neighbor in $S$, and
\begin{equation}\label{formula:amplified-weight}
 \wt(Ty)\ge(1-\delta/8)(B-d+1)n\ge(1-\delta/4)r_0.
\end{equation}

Keep any $r+1$ columns of $TH$ and append $r-r_0$ zero rows, obtaining
$A\in\F_Q^{r\times(r+1)}$. There are enough columns because
$Rn'>2r_0$ and $r_0/r\to1$. Every column has weight at most
$s=D_0dB$, every set of at most $K$ columns is independent, and
Eqn.~\eqref{formula:constant-density} and
Eqn.~\eqref{formula:amplified-weight} give the required medium-weight
bound, since $(1-\delta/4)r_0\ge(1-\delta)r$ eventually.
Set $E=A^{\mathsf T}$. Find a nonzero $\lambda_*\in\ker A$, choose
$j$ with $\lambda_{*,j}\ne0$, and set
$b=\lambda_{*,j}^{-1}e_j$, where $e_j$ is the $j$th standard basis
vector. Then $\lambda_*^{\mathsf T}b=1$, proving
inconsistency. All graph parameters and auxiliary searches are constant
for fixed $\delta$; generation and Gaussian elimination take polynomial
time in $r$.
\end{proof}

\subsubsection{Legal Assignments and Closure}

\begin{lemma}\label{formula:legal-sets}
Let $E\in\F_Q^{m\times r}$, and fix parameters
$\alpha,\beta,\gamma$ satisfying the following conditions:
\begin{itemize}
\item $2\le\alpha$ and $2\alpha-2\le\beta$;
\item every subsystem of at most $\beta$ rows of $Ez=b$ is consistent;
\item $\wt(E^{\mathsf T}\lambda)\ge\gamma$ whenever
$\alpha\le\wt(\lambda)\le\beta$.
\end{itemize}
For $|D|<\gamma$, define
\begin{align}
 K_D&=\{\lambda\in\F_Q^m:\wt(\lambda)<\alpha,\quad
       \supp(E^{\mathsf T}\lambda)\subseteq D\},
       \label{formula:multiplier-space}\\
 H_D&=\{z_D:(E^{\mathsf T}\lambda)_D\cdot z_D=\lambda\cdot b
                 \text{ for all }\lambda\in K_D\}.
       \label{formula:legal-assignment-set}
\end{align}
Then $K_D$ is linear, $H_D$ is a nonempty affine subspace, and
\begin{equation}\label{formula:projection-identity}
 \begin{aligned}
 \pi_C(H_D)&=H_C&&\text{for every }C\subseteq D,\\
 \operatorname{codim}H_D&=\dim K_D-\dim K_\varnothing.
 \end{aligned}
\end{equation}
Every member of $H_D$ satisfies every original equation supported on $D$.
\end{lemma}

\begin{proof}
The sum of two members of $K_D$ has weight at most $2\alpha-2\le\beta$
and image supported on $D$. Since $|D|<\gamma$, its weight is below
$\alpha$. Scalar closure is immediate. An inconsistency in the equations
for $H_D$ would give $\lambda\in K_D$ with
$E^{\mathsf T}\lambda=0$ and $\lambda\cdot b\ne0$, contradicting
consistency of the fewer than $\alpha$ rows in its support. The kernel
of the coefficient map on $K_D$ is $K_\varnothing$, proving the
codimension formula. Eliminating $D\setminus C$ retains precisely
those combinations in $K_D$ whose coefficient images are supported on
$C$, namely $K_C$, with their original right sides. This proves the
projection identity. Finally, the unit multiplier of each row supported
on $D$ belongs to $K_D$.
\end{proof}

Partition the raw Boolean variables into $r$ blocks of length $\ell$.
Let $V_S$ be the space of forms supported on blocks in $S$ and let
$F[\bar S]$ denote projection of a row space $F$ outside $S$.
For an integer $0<k<\ell$, the safety condition from
\Cref{subsec:proof-complexity-prelim} is
\begin{equation}\label{formula:safety}
 \dim(F\cap V_S)\le k|S|\quad\text{for every }S.
\end{equation}
By annihilator duality,
\begin{equation}\label{formula:safety-rank-spread}
 \dim\pi_S(F^\perp)=\ell|S|-\dim(F\cap V_S).
\end{equation}
Thus \Cref{thm:deterministic-gadgets}, with rate $(\ell-k)/\ell$ and
$a=\log_2Q$ output bits, makes the gadget image of every safe affine
system equal to $\F_Q^r$. This also holds on a subset of blocks:
extend by independent uniform missing blocks and project the outputs.

Define $h_F(S)=\dim(F\cap V_S)-k|S|$, and let $C(F)$ be its
inclusion-minimal maximizer. Equivalently, $C(F)$ is the intersection
of all maximizing sets; the following proof shows that this intersection
also maximizes $h_F$.
\begin{lemma}\label{formula:closure}
This maximizer is unique; $F[\overline{C(F)}]$ is safe; and
$k|C(F)|<\dim F$ if $C(F)\ne\varnothing$.
If $F\subseteq F'$, then $C(F)\subseteq C(F')$.
For $F'=F+\langle f\rangle$ with $f\notin F$, put $C=C(F)$ and
$C'=C(F')$. If $C'\ne C$, then
\begin{equation}\label{formula:closure-increment}
 \dim F'[\bar C']=\dim F[\bar C]-k|C'\setminus C|.
\end{equation}
Otherwise, the projected rank increases by zero or one.
\end{lemma}

\begin{proof}
The function $h_F$ is supermodular: $(F\cap V_S)+(F\cap V_T)$ lies in
$F\cap V_{S\cup T}$ and the summands intersect in $F\cap V_{S\cap T}$.
Its maximizers are consequently closed under intersection and union,
giving the unique minimal one $C$. For $T\cap C=\varnothing$,
\[
 \dim(F[\bar C]\cap V_T)
 =\dim(F\cap V_{C\cup T})-\dim(F\cap V_C)\le k|T|.
\]
If $C\ne\varnothing$, then $h_F(C)>0$; otherwise the empty set would
also maximize $h_F$. This proves the first assertions.

For monotonicity it suffices to adjoin one independent form. Then
$a(S)=\dim(F'\cap V_S)-\dim(F\cap V_S)$ is increasing and takes
values in $\{0,1\}$: it is one exactly when the nonzero coset in
$F'/F$ has a representative supported on $S$. For any new maximizer $D$,
\[
 h_F(C\cap D)+h_{F'}(C\cup D)\ge h_F(C)+h_{F'}(D).
\]
Both left terms are bounded by their corresponding maxima, so $C\cap D$
is an old maximizer. Minimality gives $C\subseteq D$.
If $C'\ne C$, necessarily $a(C)=0$ and the maximum increases by one;
otherwise $C$ remains a new maximizer. Therefore
$a(C')=1$ and $h_F(C')=h_F(C)$, whence
$\dim(F'\cap V_{C'})=\dim(F\cap V_C)+k|C'\setminus C|+1$.
Subtract from $\dim F'=\dim F+1$ to obtain
Eqn.~\eqref{formula:closure-increment}.
\end{proof}

Write $G(x)=(g_i(x_i))_{i\in[r]}$ for the joint gadget map, and fix
an integer $w<\gamma$. A consistent affine system $\Phi$, with
row space $F$, is \emph{viable} if $|C(F)|\le w$ and some solution $x$
satisfies
\begin{equation}\label{formula:viability}
 (g_i(x_i))_{i\in C(F)}\in H_{C(F)}.
\end{equation}
Monotonicity of closure and Eqn.~\eqref{formula:projection-identity}
preserve viability under semantic weakening.

\begin{lemma}\label{formula:activation}
If the gadgets are surjective on every safe affine system, every viable
$\Phi$ of rank below $kw$ admits a viable answer to every parity query.
\end{lemma}
\begin{proof}
For a queried form $f$, put $C=C(F)$ and $C'=C(F+\langle f\rangle)$.
Then $C\subseteq C'$ and $|C'|\le w$. Let $x^0$ witness viability.
Extend $G(x^0)_C$ to $z_{C'}\in H_{C'}$ by
Eqn.~\eqref{formula:projection-identity}. Fix the raw coordinates in
$C$ to those of $x^0$, retaining all equations of $\Phi$.
The remaining consistent system has safe row space $F[\bar C]$,
so surjectivity realizes $z_{C'\setminus C}$. The resulting full
solution $x$ realizes $z_{C'}$ and answering $f$ by $f(x)$ is viable.
\end{proof}

\subsubsection{The Boolean Encoding and Counting Argument}

For each base row and each raw tuple on its support whose gadget
outputs violate it, include the full clause forbidding that tuple.
The resulting CNF $\Psi$ is unsatisfiable, uses only raw variables,
and has at most $(r+1)2^{s\ell}$ clauses of width at most $s\ell$.

Use the conversion from a refutation to its reversed query DAG and the
$(a,b)$-path-free convention from \Cref{subsec:proof-complexity-prelim}.
The conversion preserves size and resolution depth, and each node's
affine label has rank at most the width of the corresponding clause.

\begin{theorem}\label{formula:lifting}
Under \Cref{formula:legal-sets}, suppose $w<\gamma$ and the gadgets
are surjective on every safe affine system. Put $K_0=kw$.
Then $\Psi$ has $\operatorname{Res}(\oplus)$ width at least $K_0$,
and every $(K_0,b_0)$-path-free refutation has size at least
\begin{equation}\label{formula:path-free-bound}
 2^{K_0-b_0-(\ell-k)w}.
\end{equation}
\end{theorem}

\begin{proof}
No viable system of rank at most $K_0$ falsifies an initial clause.
Indeed, falsifying a full clause fixes all raw coordinates of its
blocks. Each such block must belong to the closure: adjoining it
otherwise increases $h_F$ by $\ell-k>0$. A viability witness would
then violate a base row supported on its legal domain, contradicting
\Cref{formula:legal-sets}.
Starting at the empty system, \Cref{formula:activation} and weakening
therefore permit a walk through viable labels in any proof of width
below $K_0$. Such a walk must reach a sink, a contradiction. This
proves the width bound.

For size, retain the complete affine query history with row space $F$
until rank $K_0$ is reached, and put $C=C(F)$. Dependent queries are
forced. An independent query is \emph{free} if closure is unchanged
and its projection lies outside $F[\bar C]$. Either answer then
preserves viability: fix the old witness's raw closure coordinates
and solve the remaining affine system with that answer.
At every other independent query, choose a viable answer using
\Cref{formula:activation}, and call the query \emph{paid}.
Fix deterministic choices whenever several answers are possible.

The potential
\begin{equation}\label{formula:query-potential}
 P(F)=k|C(F)|+\dim F[\overline{C(F)}]
\end{equation}
starts at zero and increases by one exactly at free queries. At a paid
query with unchanged closure, the outside rank is unchanged; when
closure grows, Eqn.~\eqref{formula:closure-increment} cancels its
contribution to $P$. Hence every prefix pays at most
\begin{equation}\label{formula:paid-bound}
 \dim F-P(F)=\dim(F\cap V_C)-k|C|
 \le(\ell-k)|C|\le(\ell-k)w
\end{equation}
queries. No sink is reached before the threshold. A uniform raw
assignment gives an unbiased answer to each independent query,
conditional on its complete affine history. It agrees with all paid
answers, and therefore reaches the threshold viably, with probability
at least $2^{-(\ell-k)w}$.

At a successful stopping node $v$, its label is a weakening of the
history and hence viable. Write $d$ for its rank. If $d<K_0-b_0$,
restart the strategy from this label, keeping the new suffix history.
It cannot reach a sink before rank $K_0$, and so yields at least
$K_0-d>b_0$ independent forms modulo the starting label. The suffix
has at least that rank by itself, contradicting path-freeness.
Thus $d\ge K_0-b_0$. A consistent rank-$d$ label is satisfied by
exactly a $2^{-d}$ fraction of assignments. Taking a union bound over
stopping nodes gives
\[
 2^{-(\ell-k)w}\le |\text{refutation}|\,2^{-(K_0-b_0)}.
\]
The full history is used only to count paid queries; the restart begins
from the actual node label. Thus forgetting is permitted throughout,
and every rank and query cost is binary, without a factor of $\log Q$.
\end{proof}

\end{document}